\documentclass[12pt]{article}
\usepackage{amsfonts}
\usepackage{amssymb}
\usepackage{amstext}
\usepackage{amsmath,bm}
\usepackage{caption}
\usepackage{graphicx}
\usepackage[all]{xy}
\usepackage{color}
\usepackage{float}

\begin{document}

%%%%%%%%%%%%%%%%%%%%%%%%%%%%%%%%%%%%%%%%%%%%%%%%%%%%%%%%%%%%%%%%%%%

\textwidth=2cm
\textheight=26cm
\hoffset=-1cm
\voffset=-4cm

%%%%%%%%%%%%%%%%%%%%%%%% (AMS)LATEX MACROS %%%%%%%%%%%%%%%%%%%%%%%%%

\def\theequation{\thesection.\arabic{equation}}
\def\be{\begin{equation}}
\def\ee{\end{equation}}
\def\ba{\begin{eqnarray}}
\def\ea{\end{eqnarray}}
\def\lb{\label}
\def\nn{\nonumber}

\def\a{\alpha}
\def\b{\beta}
\def\g{\gamma}
\def\d{\delta}

\def\e{\varepsilon}
\def\l{\lambda}
\def\r{\rho}
\def\s{\sigma}
\def\t{\tau}
\def\o{\omega}
\def\x{\xi}

\def\D{\Delta}
\def\G{\Gamma}
\def\O{\Omega}
\def\L{\Lambda}

\def\fU{\mathfrak A}
\def\bo{\mathfrak b}
\def\fp{\mathfrak p}
\def\hp{\hat p}

\def\E{{\cal E}}
\def\Vp{{\cal V}_p}
\def\Hp{{\cal H}_p}

\def\bq{\overline{q}}
\def\bM{\bar M}
\def\bz{\bar z}

\def\bU{\overline{U}_q}
\def\bbU{{\overline {\overline U}}_q}
\def\tU{{\widetilde{U}}_q}

\def\bD{\overline {\cal D}}

\def\fl{\mathfrak l}
\def\fr{\mathfrak r}

\def\bd{\bf d}
\def\sJ{^*\!\!{\cal J}}

\def\ux{\underline x}
\def\up{\underline p}
\def\uq{\underline q}
\def\uz{\underline z}
\def\pz{\Pi_{23}\;\uz}

\def\id{\mbox{\em 1\hspace{-3.4pt}I}}
\def\idd{\scriptsize{\mathit 1 \hspace{-4.2pt}I}}
\newcommand{\ID}[2]{\id^{| #1 {\cal i}}_{\;\;\; {\cal h} #2 |}}

\def\p{\hat p}

\def\Z{\mathbb Z}
\def\R{\mathbb R}
\def\C{\mathbb C}
\def\F{\mathbb F}
\def\H{\mathbb H}

\def\1{1\!\!{\rm I}}

\def\eod{\phantom{a}\hfill \rule{2.5mm}{2.5mm}}

\def\hR{\hat{R}}
\def\Rp{\hat{R}(p)}
\def\Rcp{\hat{R}^c(p)}

\def\hp{\hat p}
\def\fp{\mathfrak p}
\def\sp{\slash\!\!\!{p}}
\def\sfp{\slash\!\!\!{\mathfrak p}}

\def\subbbc{{\rm C}\kern-3.3pt\hbox{\vrule height4.8pt width0.4pt}\,}
\def\qd{\stackrel{.}{q}}
\def\pl{\partial}
\def\vac{\mid \! 0 \rangle}
\def\lvac{\langle 0 \!\mid}

\def\Ba{\bar a}
\def\Bp{\bar p}
\def\Fd{{\cal F}^{diag}}
\def\Fp{{\cal F}'}
\def\bm{\mathbf m}

\begin{center}

%%%%%%%%%%%%%%%%%%%%%%%%  T I T L E  %%%%%%%%%%%%%%%%%%%%%%%%

{\Huge\bf Canonical approach}\\[3 mm]
{\Huge\bf to the WZNW model}\\[18 mm]

%%%%%%%%%%%%%%%%%%%%%%%%%% AUTHORS %%%%%%%%%%%%%%%%%%%%%%%%%%

{\large{\bf
P. Furlan$^{a,b}$
\footnote[1]{e-mail address: furlan@trieste.infn.it},
L. Hadjiivanov$^{c,b}$
\footnote[2]{e-mail address: lhadji@inrne.bas.bg},
I. Todorov$^{c}$
\footnote[3]{e-mail address: todorov@inrne.bas.bg}\\}}
\vskip 0.5 cm
$^a$Dipartimento di Fisica dell' Universit\`a degli Studi di Trieste,\\
Strada Costiera 11, I-34014 Trieste, Italy\\

$^b$Istituto Nazionale di Fisica Nucleare (INFN),\\
Sezione di Trieste, Trieste, Italy\\

$^c$Institute for Nuclear Research and Nuclear Energy,\\
Bulgarian Academy of Sciences,\\
Tsarigradsko Chaussee 72, BG-1784 Sofia, Bulgaria
%% $^d$The Abdus Salam International Centre
%% for Theoretical Physics,\\
%% Strada Costiera 11, I-34014 Trieste, Italy\\

\end{center}

%\date{today}

\vspace{2cm}

%%%%%%%%%%%%%%%%%%%%%% ABSTRACT %%%%%%%%%%%%%%%%%%%%%%%%%%%%

\begin{abstract}

{\normalsize
\noindent
The chiral Wess-Zumino-Novikov-Witten (WZNW) model
provides the simplest class of rational conformal field theories which
exhibit a non-abelian braid-group statistics and an associated "quantum
symmetry".  The canonical derivation of the Poisson-Lie symmetry of the
classical chiral WZNW theory (originally studied by Faddeev,
Alekseev, Shatashvili and Gaw\c{e}dzki, among others) is reviewed along
with subsequent work on a covariant quantization of the theory which
displays its quantum group symmetry.}

\end{abstract}

\newpage

{\scriptsize\tableofcontents}

\newpage

%%%%%%%%%%%%%%% I N T R O D U C T I O N %%%%%%%%%%%%%%%%%

\section{Introduction}

\setcounter{equation}{0}
\renewcommand\theequation{\thesection.\arabic{equation}}

The WZNW model is a conformally invariant theory of a Lie group valued field
on the 2-dimensional ($2D$) space-time, $g(x^0,x^1)\in G\,.$
We shall concentrate exclusively in this paper on the case when the group $G\,$ is a connected
and simply connected compact Lie group and ${\cal M}\,,$ the integration domain of the classical
action of the model, is the compactified Minkowski space (see Eqs. (\ref{MinkD2}) and (\ref{Swznw0}) below).
In modern parlance, one can say that the model describes then a closed string moving on a compact group manifold \cite{GW}.
The state space of the quantized model provides a representation of the {\em current} ({\em affine Kac-Moody}) algebra,
associated with the group $G\,.$ The resulting quantum field theory (QFT) has a unique vacuum if one fixes ($G\,$ and) a
positive integer $k\,,$ the eigenvalue of the central charge of the current algebra, called the {\em level}.

Although the WZNW model was originally formulated in terms of a multivalued classical action
\cite{W} (exploiting ideas of \cite{WZ} and \cite{N}), it was first solved in a quantum axiomatic framework
\cite{KZ, T} using the theory of highest weight representations of affine Lie algebras \cite{Kac, KR}
and ended up as a textbook example of a rational {\em conformal field theory}
(CFT) \cite{DFMS}. Following the original ideas of \cite{BPZ, DF},
the correlation functions of the theory have been written as sums of products of chiral
conformal blocks which carry a monodromy representation of the braid group
\cite{TK, Kohno}. The {\em braid group statistics}
is associated with a {\em quantum group symmetry} \cite{A-GGS, GP, PS}
or some of its  generalizations \cite{MS, BNS, PZ}.
We point out that the appearance of such non-trivial features is not just an
artifact of the ambiguity in the splitting of a local $2D$ field into
chiral components. In fact, the above peculiarities of {\em chiral vertex
operators} (CVO) show up in the non-group-theoretic {\em fusion rules} of
$2D$ fields and the associated non-integer {\em statistical dimension} (for
background and further references -- see \cite{FRS, Lo, FrG} as well as
more recent overviews in \cite{TH, Sch05}).

The canonical approach to the WZNW model, triggered by work of Babelon
\cite{B} and Blok \cite{Bl} which related it to the {\em Yang-Baxter
equation} (YBE), shed new light on the problem. After the initial push in
\cite{Bl} the classical theory was developed by Faddeev et al.
(\cite{F1, AS1, AS, AF, F2, AFS, AF2} as well as in \cite{BDF} and, in a sense,
completed by Gaw\c{e}dzki et al. \cite{Gaw, FG, FG1} although further work
in both the classical and the quantum problem is still going on
(\cite{ChuG, AT, BFP1, BFP, BFP2, BFPquasi, FHT1, FHT2, FHT3, CL, DT,
HIOPT, Goslar, FHIOPT, FHT6}). More recently it has also included the
boundary WZNW model (\cite{ASch, FFFS, GTT, G1, GR, G2}).

\smallskip

The idea of how one exhibits the hidden quantum symmetry is quite simple. The general solution of the classical equations
of motion for the periodic group-valued field $g(x^0, x^1+2{\pi})=g(x^0,x^1)\,$
(the field configurations for fixed time being elements of the {\em loop group} \cite{PrS} ${\tilde G}$ of $G$)
is given by a product of chiral multivalued fields,
\be
\lb{LR}
g(x^0, x^1) \, \equiv \, g (x^+ ,x^- ) = g_L (x^+ )\, g_R^{-1} (x^- )\ ,\qquad x^\pm = x^1 \pm x^0\ ,
\ee
which satisfy a twisted periodicity condition,
\be
\lb{cM}
g_C (x+2\pi ) = g_C (x)\, M\ ,\qquad C = L , R\,,\quad M\in G\ ,
\ee
implying that the $2D$ field is periodic:
\be
g(x^+ +2\pi , x^- +2\pi ) = g(x^+ , x^- )\ .
\lb{gper}
\ee
The chiral components $g_C\,$ are not uniquely determined: Eq.(\ref{LR})
is respected by any transformation $g_C (x)\rightarrow g_C (x)\, S\,$
where $S\,$ is an $x$-independent invertible matrix. In
particular, we do not have to assume that $g_C\,$ are unitary, albeit
$g(x^+ , x^-)\,$ is. Moreover, as we shall see, the elements of the
{\em monodromy matrix} $M\,$ carry dynamical degrees of freedom (they have
non-vanishing Poisson brackets among themselves and with $g_C (x)$) and
it is natural to allow for "dynamical matrices" $S\,$ describing the ambiguity in the definition of $g_C\,.$
We use the resulting freedom to impose a {\em Poisson-Lie} symmetry
on the chiral theory, the classical counterpart of a {\em quantum group} symmetry.
Requiring that the left and right components $g_L\,$ and $g_R\,$ Poisson commute yields a further extension
of the phase space of the theory consisting in introducing independent left and right monodromy matrices $M_C\,.$
This allows the introduction of quantum group covariant {\em chiral zero
modes} (in whose treatment, both classical and quantum, in particular for
$G=SU(n)\,,$ the authors have taken part \cite{AT, FHT1, FHT2, FHT3, DT, HIOPT, Goslar,
FHIOPT, FHT6, AFH, FHT7, TH10}). In the present paper we combine the phase spaces
of zero modes and {\em "Bloch waves"} (chiral fields with {\em diagonal} monodromy $M_p\,$)
to derive the {\em Poisson brackets} of the covariant chiral fields $g_C\,,$ thus preparing
the ground for the subsequent discussion of a quantum group invariant quantization.

There is a price to pay for achieving manifest quantum group covariance of the chiral theory.
While the unitary $2D$ WZNW model only involves a finite number of weights
(e.g., for $G=SU(2)\,,$ those not exceeding the level), we are led to allow all weights,
thus ending with an infinite (non-unitary) extension of the chiral state space.
The resulting theory is related to a logarithmic CFT of the type studied systematically
by B.L. Feigin, A.M. Gainutdinov, A.M. Semikhatov, I.Yu. Tipunin, and others
\cite{FGST1, FGST2, FGST3, FGST4, FT, FHST, GT, S1, S2, S3, S4}.
(We review relevant part of this work in Section 5.)
An alternative possibility, weakening the requirement of quantum group invariance but only allowing for a
finite dimensional unitary extension of the chiral state space has been developed in the framework of
boundary CFT (for a review and references – see \cite{PZ}). It would be interesting to work out a canonical
formulation also of this approach starting with the classical theory.

\smallskip

A few words about the organization of the material, summarized in the table of content.

We begin in Section 2.1 by showing that the invariance of a $2D$ sigma model type action with respect to infinite dimensional
chiral loop group "gauge transformations" requires a {\em Wess-Zumino (WZ) term} \cite{WZ, N, W}.
In Section 2.2 we introduce the relevant first order canonical formalism \cite{G, JS}.
For a field theory in a $D$-dimensional space-time, it is based on a $(D+1)$-dimensional closed differential form $\o\,.$
This approach has at least two advantages, compared to the standard one that starts with a
Lagrangean $D$-form ${\mathbf L}$ whose integral gives the classical action:

\smallskip

\noindent
(i) $\o = {\mathbf d}\,{\mathbf L}$ does not change if we add a full derivative term to ${\mathbf L}$
(that would not affect the equations of motion);

\noindent
(ii) $\o$ may exist in theories with no single-valued classical action, in particular, in the WZNW model of interest.

\smallskip

\noindent
The integral of $\o$ over an equal time surface (a circle, in our case) gives rise to a {\em symplectic form}.
We study in Section 2.3 its splitting into monodromy dependent {\em chiral symplectic forms}
$\O (g_C , M_C)\,,\ C = L,R $ for $g$ given by (\ref{LR}). The expression for $\O$ involves a $2$-form $\rho (M)\,,$ like
(\ref{ro}), defined on an open dense neighbourhood of the identity of the complexification $G_{\mathbb C}$ of our
compact Lie group $G$ (using, for $G_{\mathbb C} = SL(n,{\mathbb C})\,,$ a Gauss type factorization of $M$).
Section 2.4 is devoted to a study of the symmetries of the chiral theory. We demonstrate, in particular,
that the symmetry of $\O$ with respect to (constant) right shifts of the chiral field $g$ is of Poisson-Lie type
\cite{D1, S-T-S}.

Section 3 deals with the classical theory of chiral zero modes which diagonalize the monodromy matrix.
They display the Poisson-Lie symmetry in a finite dimensional context (Section 3.1; cf. \cite{AF}).
In Section 3.2 we recall some facts from the theory of the semisimple Lie algebras and prepare the ground
for obtaining the chiral Poisson brackets. Section 3.3.1 reviews the result of Gaw\c{e}dzki and Falceto \cite{G, FG1}
that establishes a one-to-one correspondence between $2$-forms $\rho (M)$ such that
\be
\d \rho (M) = \frac{1}{3}\,{\rm tr}\, (M^{-1} \d M )^{\wedge 3} =: \theta (M)
\lb{rh-th}
\ee
and non-degenerate solutions of the (modified) {\em classical Yang-Baxter equation}, see Proposition 3.2.

The Schwinger-Bargmann theory of angular momentum \cite{Sch, B62} gives rise to a model of the finite dimensional
irreducible representations of $SU(2)\,$ by quantizing the $2$-dimensional complex space ${\mathbb C}^2\,$
equipped with the K\"ahler symplectic form $i\, (d z^1 \wedge d \bar z^1 + d z^2 \wedge d \bar z^2)\,.$
It yields the Fock space of a pair of creation and annihilation operators.
In Section 3.3.2 we first present the classical $4$-dimensional phase space involved in this construction as a
submanifold of codimension two in a $6$-dimensional space consisting of a $2\times 2\,$ matrix $a = (a^i_\a)\,$
and a $2$-dimensional weight vector $p_i\,,\ i=1,2\,.$ Then we generalize this construction to the case of $SU(n)\,$
in which the classical phase space is a submanifold of codimension two in a $n(n+1)$-dimensional space.
Finally, we construct a $q$-deformation of the resulting algebra, corresponding to the classical counterpart
of a model space construction for the finite dimensional irreducible representations of the quantum universal
enveloping algebra $U_q(s\ell (n))\,$ for generic $q\,.$
The computation of the Poisson (and Dirac) brackets of the
Poisson-Lie covariant zero modes involves the full complication of a theory
with a non-local Wess-Zumino term. It is dealt with in Section 3.3.3.

The Poisson brackets (PB) for the infinite dimensional Bloch waves $u(x)\,$ (Section 3.4) are simpler to compute.
A peculiarity of our treatment is the fact that the determinant  of $u(x)\,$ depends on the weights $p\,$ (and is so
chosen that only the product of $\det u(x)\,$ and $\det a\,$ is equal to $1$). The resulting PB for the
Poisson-Lie covariant chiral field $g(x) = u(x)\, a\ (\, = ( u^A_i(x) a^i_\a ))\,$ are spelled out in Section 3.5
where the reconstruction of the $2D\,$ model is also explained.

Chapter 4 is devoted to the study of the quantum chiral WZNW model. The quantization of the
current algebra ${\widehat{\cal G}}_k\,$ (Section 4.1) involves the renormalization of the level $k\ \to\ h=k+g^\vee\,$
(where $g^\vee\,$ is the dual Coxeter number of the Lie algebra ${\cal G}\,$ of $G$)
in the Sugawara formula \cite{Sugawara, Sommerfield}.
The state space construction reproduces the representation theory
of affine Kac-Moody algebras supplemented with a derivation of the Knizhnik-Zamolodchikov equation. The exchange algebra
of the chiral field $g(x)\,$ is expressed (Section 4.2) in terms of the constant $SL(n,\mathbb C)\,$ quantum $R$-matrix.
The exchange relations for the monodromy matrix $M\,$ (Section 4.3) acquire a more transparent form
in terms of its Gauss components $M_\pm\,$ which give rise to
the quantum universal enveloping algebra $U_q(s\ell(n))\,.$ A hallmark of our approach is the treatment \cite{HIOPT, FHT6, FHIOPT}
of the zero modes' algebra (Section 4.4) which also involves the {\em quantum dynamical $R$-matrix} $R(p)\,.$
Section 4.5 is devoted to the study of the chiral state space.
For generic $q\,$ (i.e. $q\ne 0\,,$ not a root of unity) the Fock space ${\cal F}\,$ of the zero modes' algebra
provides a model for the finite dimensional representations of $U_q(s\ell(n))\,$ (Section 4.5.1).
The problems arising for $q\,$ a root of unity (fully resolved for $n = 2\,$ only) are discussed in Section 4.5.2.
We show here that the exchange relations involving bilinear combinations of
zero modes $a^i_\a a^j_\b = S^{ij}_{\a\b} + A^{ij}_{\a\b}\,$ can be written in a very simple,
yet equivalent, form in terms of their $q$-symmetric and $q$-antisymmetric parts
($S^{ij}_{\a\b} = q^{\epsilon_{\a\b}} S^{ij}_{\b\a}\,$ and $A^{ij}_{\a\b} = - q^{-\epsilon_{\a\b}} A^{ij}_{\b\a}\,,$
respectively). The braiding properties of chiral quantum fields are displayed in Section 4.5.3.

The study of the $\widehat{su}(2)_k\,$ quantum WZNW model and of its (non-unitary) chiral extension is pursued further in Chapter 5.
It is facilitated by the fact that (only) for $n=2\,$ the determinant condition is quadratic in the zero modes.
A canonical basis of the Fock space ${\cal F}\,$ is introduced in Section 5.1. In the following Section 5.2
(based on \cite{FHT7}) we exploit the fact that the quantum group acting on ${\cal F}\,$ is the finite ($2 h^3$-)dimensional
quotient Hopf algebra, the {\em restricted quantum group} $\bU\,.$ In Section 5.2.1 we display the structure of the
zero modes' Fock space as an indecomposable $\bU\,$ module. The {\em universal monodromy matrix}
${\cal M} \in \bU \otimes \bU\,$ and the corresponding {\em Drinfeld map}, introduced in Section 5.2.2, allow to describe
the {\em Grothendieck ring} of $\bU\,$ (Section 5.2.3). The $\widehat{su}(2)_k\,$ invariant extended chiral state space
appears as an example of a logarithmic CFT. The Lusztig's extension $\tU\,$ of $\bU\,$ (Section 5.3.1) and the monodromy
representation of the braid group in the space of $4$-point conformal blocks (Section 5.3.2) play a dual role in this setting.

In Section 6 we complete the description of the $2D\,$ WZNW model in terms of its chiral components.
We first display the exchange relations of the right chiral field (in Sections 6.1.1 and in Section 6.1.2,
for those involving the constant and the dynamical $R$-matrix, respectively).
To avoid subtleties with matrix inversion in the quantum case, we work with "bar" right sector variables
in terms of which $g (x, \bar x ) = g (x )\, \bar g (\bar x)\,, \ (x, \bar x) = (x^+,x^-)\,,$ cf. (\ref{LR}).
(Their description does not follow automatically from the knowledge of the left sector since the quantum chiral field
$g_C (x)\,,$ unlike the classical one, is not invertible.)
% The right sector zero modes and their Fock space are considered in full detail for $n=2\,$ in Section 6.1.3.
In Section 6.2 we return to the reconstruction of the two dimensional WZNW model.
It is demonstrated in Section 6.2.1 that the $2D\,$ field, expressed in terms of
products of left and right components, is local and quantum group invariant.
The zero modes' part of the extended model (studied for $n=2\,$ in Section 6.2.2) gives rise to a finite dimensional
quotient space harbouring the unitary model. The construction (see \cite{FHT2, DT, HF3}) is reminiscent to the
cohomological (BRS \cite{BRS1, BRS2}) treatment of a gauge theory \cite{Str78, Araki, Str, BLOT},
the quantum group playing the role of a generalized gauge symmetry
(only acting on the internal degrees of freedom). Some first steps towards the extension of this picture to $n \ge 3\,,$
a problem of considerable combinatorial difficulty, are outlined in Section 6.2.3.

Section 7 summarizes the main features of the present approach and indicates some possible steps for future work.
We also mention some (of the numerous) WZNW-related topics left out of the scope of our investigation.

To make the paper self-contained, the main text is supplemented by three Appendices.
In Appendix A we recall the basic facts about complex semisimple Lie algebras and fix conventions
used in the present paper. Appendix B deals with the quantum group side. Appendix B.1 introduces
the Hopf algebra $\, U_q(s\ell(n))\,,$ while Appendices B.2 and B.3 contain supplementary
information needed in Section 5.2. Appendix C is concerned with the quantum determinant of the monodromy matrix
(with {\em non-commuting} entries) $M\,$ \cite{HF2, FH2}.

The paper contains a rather extensive (albeit far from complete) list of references.

\newpage

\section{{\boldmath$2D$} and chiral WZNW model. Symplectic densities}

\setcounter{equation}{0}
\renewcommand{\theequation}{\thesection.\arabic{equation}}

\subsection{Chiral symmetry requires a Wess-Zumino term}

\noindent
The dynamics of the group valued WZNW field $g$ is, in effect, determined by the symmetry of the WZNW model.
Combining the conformal invariance with the internal symmetry generated by the currents one ends up,
as we shall see, with an infinite dimensional left and right {\em chiral symmetry}.

\smallskip

We proceed in two steps, beginning with the natural (non-linear) sigma model action on a compact Lie group $G$
\be
S_0 [g] = \lambda\, \int_{\cal M} {\rm tr}\, (g^{-1}\partial_\mu g) (g^{-1}\partial^\mu g)\, d x^0 d x^1
\equiv - \lambda\, \int_{\cal M} {\rm tr}\, (\partial_\mu g) (\partial^\mu g^{-1})\, d x^0 d x^1
\lb{S0}
\ee
where the world sheet is oriented, $d x^0 d x^1 \equiv d x^0 \wedge d x^1 = - d x^1 \wedge d x^0\,$
(we omit the wedge sign for exterior products of differentials) and $\l > 0\,.$
We are denoting by ${\rm tr}\, (X Y)$ the Killing form $(X , Y )$ on the Lie algebra,
proportional to the matrix trace (see Appendix A). In a second step, we shall complement $S_0 [g]$
with a non-local term that will ensure the infinite chiral symmetry.

\smallskip

It is appropriate to carry the integration in (\ref{S0}) over the compactified two dimensional Minkowski space
${\cal M}\ ( \equiv {\bar M}_2)$ which we proceed to describe in some detail.
${\cal M}$ is a somewhat degenerate special case of the $D$-dimensional compactified Minkowski space
\ba
{\bar M}_D &:=& \{\, z = (z^\a)\,,\ \a=1,2,\dots , D \, \mid\, z^\a = e^{it} u^\a\,,\ t , u^\a \in {\mathbb R}\,;\ u^2 = 1 \} =\nn\\
&\,=& {\mathbb S}^1 \times {\mathbb S}^{D-1} / \{ 1, -1 \}\qquad\qquad (\, u^2 := \sum_{\a =1}^D (u^\a)^2 \,)
\lb{MinkD2}
\ea
equipped with a real $O(2)\times O(D)$-invariant metric of Lorentzian signature
\be
d s^2 = \frac{d z^2}{z^2} = d u^2 - d t^2\ ,\qquad {\rm where}\quad u . d u := \sum_{\a =1}^D u^\a d u^\a = 0 \ .
\lb{ds2}
\ee
The universal cover of ${\bar M}_D$ for $D>2$ is the cylinder $\widetilde{\cal M}_D = {\R} \times {\mathbb S}^{D-1}\,.$
For $D=2\,,\ {\bar M}_2 = {\cal M}$ is diffeomorphic to the {\em flat Lorentzian torus} (with identified opposite points)
\be
{\cal M} = \{ \, z^1 = e^{ix^0} \sin x^1\,,\ z^2 = e^{ix^0} \cos x^1\,;\ ds^2 = (d x^1)^2 - (d x^0)^2 \,\}
\lb{clM}
\ee
which can be obtained from its universal cover ${\R}^2$ factoring by the relations
\be
(x^0 , x^1 )\ \sim \ (x^0 , x^1 + 2 \pi)\ ,\qquad (x^0 , x^1 )\ \sim \ (x^0 + \pi , x^1 + \pi)\ .
\lb{clM1}
\ee
Eqs. (\ref{clM1}) are equivalent to $2\pi$-periodic boundary conditions
\be
(x^+,\,x^-) \sim (x^+ + 2\pi n^+ ,\, x^- + 2\pi n^-)\ ,\quad n^\pm\in\Z
\lb{clM2}
\ee
in each of the cone variables $x^\pm$ defined in (\ref{LR}),
\be
x^\pm = x^1 \pm x^0\ ,\qquad \partial_\pm= \frac{1}{2} (\partial_1 \pm \partial_0 )\ ,\qquad
d x^+ d x^- = 2 \, d x^0 d x^1\ .
\lb{conev}
\ee

\smallskip

We are looking for an action invariant with respect to the infinite dimensional group of chiral "gauge transformations" of the type
\be
g(x^+, x^-)\ \to\ {\fl} (x^+) .\, g(x^+, x^-) . {\fr} (x^-)
\lb{infgr}
\ee
where both ${\mathfrak l}$ and ${\mathfrak r}$ are loop group ($G$-valued, periodic) functions
of the corresponding light cone variables.
Computing the variation of the sigma model action (\ref{S0})
\ba
&&\d S_0 [g] = 2\l\,\int_{\cal M} {\rm tr}\, \d (g^{-1}\partial_\mu g) (g^{-1}\partial^\mu g) d x^0 d x^1 =\nn\\
&&= - 2\l\,\int_{\cal M} {\rm tr}\, \left( g^{-1}\d g \,\partial_\mu(g^{-1}\partial^\mu g)
- \partial_\mu ( g^{-1}\d g \,g^{-1}\partial^\mu g) \right) d x^0 d x^1 = \nn\\
&&= - 2\l\,\int_{\cal M} {\rm tr}\,g^{-1}\d g \left( \partial_+ (g^{-1} \partial_- g) + \partial_- (g^{-1} \partial_+ g) \right) dx^+ d x^- \qquad
\lb{varS0}
\ea
(the boundary term can be neglected due to (\ref{gper}) and (\ref{clM2})), we see that $\d \, S_0 [g]$ does {\em not} vanish, in general, for
\be
g^{-1} \d g = g^{-1} \d {\fl} (x^+) g + \d {\fr}(x^-)
\lb{inf-conf}
\ee
(here $\d {\fl} (x^+)$ and $\d {\fr}(x^-)$ are assumed to be ${\cal G}$-valued periodic
functions of the respective chiral variables).

The possibility of obtaining an invariant theory found by Witten \cite{W} amounts to adding to $S_0 [g]$ (\ref{S0})
a WZ term\footnote{The possible continuations of the form $\theta (g)$ from the $2D$ compactified Minkowski
space ${\cal M}$ (\ref{clM}) to the $3$-dimensional real compact manifold with boundary, the bulk
$${\cal B} := \{\, ( z^\a , \rho ) \,,\ \a = 1,2\,\mid\, ( z^\a ) = z  \in {\cal M}\,,\ 0\le\rho \le 1\, \}\ ,\quad\partial {\cal B} = {\cal M}\ ,$$
split into equivalence classes labeled by the elements of the third homotopy group
$\pi_3 (G) \simeq {\Z}\,$ (see \cite{N, M, Schw, TH10}).} proportional to
\be
\Gamma [g] := \frac{1}{12\pi}\, \int_{\cal M} \,d^{-1} {\rm tr}\, (g^{-1}\ d g)^3 =
\frac{1}{12\pi}\, \int_{\cal B} \,{\rm tr}\, (g^{-1}\ d g)^3 \ \in 2 \pi {\mathbb Z}
\lb{GWZ}
\ee
which has a single valued variation due to the relation
\be
\d\, d^{-1} \frac{1}{3}\,{\rm tr}\, (g^{-1}\ d g)^3 = {\rm tr}\, ( g^{-1} \d g\, ( g^{-1} d g )^2 )\ .
\lb{totdiff0}
\ee
Using (\ref{totdiff0}) and
\be
\lb{dxdx}
d x^\mu d x^\nu = -\, \varepsilon^{\mu\nu} d x^0 d x^1
\quad (\varepsilon^{\mu \nu} = - \varepsilon^{\nu \mu }\,,\ \mu,\nu = 0,1\ ,\ \ \varepsilon^{01}=-1\ ,\ \
\varepsilon^{\mu\s} \varepsilon_{\s\nu} = \d^\mu_\nu )\ ,
\ee
we obtain
\ba
&&\d \, \Gamma [g] = \frac{1}{4\pi}\,\int_{\cal M} {\rm tr}\,g^{-1}\d g \,(g^{-1}\partial_\mu g)\, (g^{-1}\partial_\nu g)\, dx^\mu\, d x^\nu = \nn\\
&&= - \frac{1}{4\pi}\,\int_{\cal M} {\rm tr}\,g^{-1}\d g \,\varepsilon^{\mu \nu} (g^{-1}\partial_\mu g)\, (g^{-1}\partial_\nu g)\, d x^0 d x^1 =\nn\\
&&= \frac{1}{4\pi}\,\int_{\cal M} {\rm tr}\,g^{-1}\d g \,\varepsilon^{\mu \nu} \partial_\mu (g^{-1}\partial_\nu g)\, d x^0 d x^1 = \nn\\
&&= \,\frac{1}{4\pi}\,\int_{\cal M} {\rm tr}\,g^{-1}\d g \left( \partial_- (g^{-1} \partial_+ g)
- \partial_+ (g^{-1} \partial_- g) \right) dx^+ d x^- \ .\quad
\lb{varGWZ}
\ea
The partition function, the exponent $e^{i S [g]}$ of the action functional, which determines the correlation functions in the Feynman path
integral formulation, is single valued if we set the coefficient of the WZ term equal to an {\em integer},
\be
S [g] = S_0 [g]+ k\, \Gamma [g]\ ,\qquad k\in{\mathbb Z}
\lb{SWZ}
\ee
so that
\ba
&&\d\,S [g] = \, - (2\,\l + \frac{k}{4\pi})\,\int_{\cal M} {\rm tr}\,g^{-1}\d g \, \partial_+ (g^{-1} \partial_- g)\, dx^+ d x^- -\nn\\
&&\qquad\quad\ \ \, -\, (2\,\l - \frac{k}{4\pi})\,\int_{\cal M} {\rm tr}\,g^{-1}\d g \, \partial_- (g^{-1} \partial_+ g)\, dx^+ d x^-\ .
\lb{SWZ1}
\ea
Now, for $g^{-1} \d g$ given by (\ref{inf-conf}), the first term vanishes, due to
$\partial_+ (g^{-1} \partial_- g) = g^{-1} \partial_- ((\partial_+ g) g^{-1}  )\, g$ and
\ba
&&{\rm tr}\,g^{-1}\d g \, \partial_+ (g^{-1} \partial_- g) =\nn\\
&&= {\rm tr}\, \left( g^{-1} \d\,{\fl} (x^+) g \ g^{-1} \partial_- ((\partial_+ g) g^{-1} )\, g +
\d\,{\fr} (x^-) \partial_+ (g^{-1} \partial_- g) \right) = \nn\\
&&= \partial_-\, {\rm tr}\, (\d\,{\fl} (x^+) (\partial_+ g) g^{-1} ) +
\partial_+\, {\rm tr}\, (\d\,{\fr} (x^-) ( g^{-1} \partial_- g)\ ,
\lb{varS1}
\ea
while vanishing of the second term implies $\l = \frac{k}{8\pi}\,.$ Thus we end up with the WZNW action functional
which is invariant with respect to (\ref{infgr}),
\be
S [g] = \frac{k}{4\pi} \int_{\cal M}
{\rm tr}\, \left( \frac{1}{2}\,(g^{-1}\partial_\mu g) (g^{-1}\partial^\mu g)\, dx^0 d x^1
+ \frac{1}{3}\, d^{-1} {\rm tr}\, (g^{-1}\ d g)^3 \right)
\lb{Swznw0}
\ee
(with $k$ a {\em positive} integer).

\smallskip

In order to get around the absence of a single valued WZ term we proceed to formulating
the dynamics of the WZNW model in terms of a canonical $3$-form.

\subsection{First order canonical formalism with a basic $(D+1)$-form}

\noindent
The first order Lagrangean and covariant Hamiltonian formalism has been applied to the WZNW model by Gaw\c{e}dzki (see \cite{G}
where the reader can also find early references; for more recent developments and further applications,
cf. \cite{JS}). Here we shall give a brief introduction to the subject and shall then apply this truly canonical
approach to the $2D$ WZNW theory of interest.

In general, a field theory lives on a fibre bundle ${\cal E}\,$ described locally by a collection of charts
${\cal U}^i\times{\cal F}\,,$ where $\cup_i\ {\cal U}^i$ forms an atlas of the $D$-dimensional (base) space-time manifold
${\cal M}\,$ and the values of the fields belong to the fiber ${\cal F}\,.$
We shall use, correspondingly, two exterior differentials, a {\em horizontal} one, $d\,,$ acting on ${\cal M}\,,$ and
a {\em vertical} one (the variation) $\d\,,$ acting on ${\cal F}\,$ so that the exterior
differential on the total space ${\cal E}\,$ will appear as their sum:
\be
{\bd}\, = \, d\, +\, {\d}\,,\quad {d}^2\, =\, 0\, =\, \d^2\,,\quad
{\bd}^2\, =\, 0\,  = [\, d\, ,\, \d\, ]_+ \
\lb{bd}
\ee
(note that, in contrast with the convention adopted in \cite{JS}, $d\,$ and $\d\,$
necessarily anticommute in order to have their sum satisfying the condition ${\bd}^2 = 0\,$ for an exterior differential).
Each differential form can be decomposed into homogeneous $(a , b)\,$ forms of degrees $a\,$ in $d$ and $b$ in $\d\,.$

\smallskip

If an action density ${\bf L}\,$ (a $D$-form) exists, in the first order formalism it is assumed to be
a sum of $(D , 0)\,$ and $(D-1 , 1)\,$ forms. The exterior differential
\be
\o\, :=\, {\bd}\,{\bf L}\ \,
\lb{om}
\ee
(which does not change if we substitute ${\bf L}\,$ by ${\bf L} + {\bd} \, {\bf K}\,$ for any
$(D-1)$-form ${\bf K}\,$) provides an invariant characterization of
the system: equating to zero the pull-back of its contraction with vertical vector fields (like $\frac{\d}{\d \phi_i}\,,$ in a discrete basis) such that
\be
\frac{\hat \d}{\d \phi_i}\, \d \phi_j + \d \phi_j \frac{\hat \d}{\d \phi_i} = {\d}^i_j\ ,
\lb{vfd}
\ee
one reproduces the equations of motion, while the integral of $\o\,$ over a
$(D-1)$ dimensional space-like (or, for non-relativistic systems, just equal time)
surface in ${\cal M}\,$ defines the symplectic form of the system. A closed $(D+1)$-form $\o\,$ may exist, however, even when there
is no single-valued action density.
The resulting more general framework is the only one appropriate for classical formulation of the WZNW model.

\smallskip

Before going to the model of interest we shall display the role of the form $\o\,$ in the simple example of
a classical mechanical system for which ${\cal M} = \R\,$ is the time axis (i.e., $D=1\,$), and ${\cal F}\,$
is a $2 f$-dimensional phase space parametrized by coordinates $q = (q^1 , \dots , q^f )\,$ and momenta
$p = (p_1 , \dots , p_f )\,.$ We shall write the action density $1$-form as a Legendre transform,
\ba
&&{\bf L} = p \,\, {\bd}\, q - H(p,q)\, dt\ ,\qquad p \,\, {\bd}\, q := \sum_{i=1}^f p_i \,{\bd}\, q^i\ ,\nn\\
&&\o = {\bd}\, {\bf L} = {\bd}\, p\, {\bd}\, q - \d H(p,q)\, dt = {\bd}\, p\, {\bd}\, q -
( \frac{\partial H}{\partial q}\, \d q + \frac{\partial H}{\partial p}\, \d p )\, dt \equiv\lb{D1}\\
&&\equiv\d p \, \d q + ( \stackrel{.}{q} - \frac{\partial H}{\partial p} )\, \d p\ dt
- ( \stackrel{.}{p} + \frac{\partial H}{\partial q} )\, \d q\, d t\qquad
( d p \,\equiv\, \stackrel{.} p  d t\ ,\ \ d q\, \equiv\, \stackrel{.} q d t )\quad\nn
\ea
(we omit throughout the wedge sign $\wedge\,$ for exterior products of differentials).
It is clear that for $d t = 0\,,\ \, \o\,$ reduces to the standard canonical symplectic form $\O = \d p\, \d q\,.$
Contracting, on the other hand, $\o\,$ with $\frac{\d}{\d q^i}$ and $\frac{\d}{\d p_i}$ (using (\ref{vfd})) and
equating to zero the pull-back of the result (which amounts to setting $\d p = 0 = \d q$), we obtain the Hamiltonian equations of motion
\be
{\stackrel{.}{p}}_i + \frac{\partial H}{\partial q^i}  = 0\ ,\qquad
{\stackrel{.}{q}}^i - \frac{\partial H}{\partial p_i}  = 0\ ,\qquad i=1,\dots , f\ .
\lb{Ham0}
\ee
In general, to any function $h\,$ on the phase space one associates a vertical {\em Hamiltonian vector field} $X_h\,$ such that
its contraction with the symplectic form ${\hat X}_h\, \O\ (\equiv i_{X_h}\, \O ) := \O (X_h , . )\,$ equals $\d h$:
\be
{\hat X}_h\, \O = \d h\quad\Leftrightarrow\quad
X_h = \frac{\partial h}{\partial q}\,\frac{\d }{\d p} - \frac{\partial h}{\partial p}\,\frac{\d }{\d q}\qquad
(\,X_{q^i} = \frac{\d}{\d p_i}\,,\ X_{p_j} = - \frac{\d}{\d q^j}\,)\ .
\lb{defOX}
\ee

A Poisson structure on (a smooth manifold) ${\cal N}\,$ is a skew symmetric bilinear map $\{ ~,~ \}\,:
C^\infty ({\cal N}) \times C^\infty ({\cal N}) \rightarrow C^\infty ({\cal N})$ satisfying
the Jacobi identity and the Leibniz rule. This is equivalent to defining a bivector
(a skew symmetric contravariant $2$-tensor) $\,{\cal P}\in T {\cal N}\wedge T{\cal N}\,$ such that
$\{ g , h \}\,= {\cal P} (g,h) \equiv \hat{{\cal P}}\, ( \d g\otimes \d h)\,.$
A covariant tensor defining a symplectic form gives always rise to a Poisson tensor defined by its inverse;
in general, the Poisson tensor may not be invertible.

In the above case of a finite dimensional mechanical system
${\cal P} = \frac{\d }{\d q}\wedge\frac{\d }{\d p} = - \frac{\d }{\d p}\wedge\frac{\d }{\d q}\,$
and, for any pair of functions $g = g(p,q)\,,\ h = h (p,q)\,,$ the
PB $\{ g , h \}\,$ is given in terms of the symplectic dual vector fields (\ref{defOX}) by
\be
\{ g , h \}\, = X_g\, h \equiv {\hat X}_g\, \d h \  (\, = - {\hat X}_h \, \d g\, ) =
\frac{\partial g}{\partial q}\,\frac{\partial h}{\partial p} - \frac{\partial g}{\partial p}\,\frac{\partial h}{\partial q}\
\quad\Rightarrow\quad \{ q^i , p_j \}\, = \d^i_j
\lb{PBdef}
\ee
(here $\d h = \frac{\partial h}{\partial p} \, \d p + \frac{\partial h}{\partial q}\, \d q$ is the total variation of $h$).
It follows from (\ref{Ham0}), (\ref{defOX}) and (\ref{PBdef})
that the time evolution of any phase space variable $g(p,q)$ is governed by its PB with the Hamiltonian:
\be
\stackrel{.}{g}\, = \frac{\partial g}{\partial p} \stackrel{.}{p} + \frac{\partial g}{\partial q} \stackrel{.}{q} =
\left( \frac{\partial H}{\partial p} \frac{\partial}{\partial q} - \frac{\partial H}{\partial q} \frac{\partial}{\partial p} \right)\! g =
-\, X_H\, g = \{ g , H \}\ .
\lb{time-evol}
\ee

\smallskip

\noindent
{\bf Remark 2.1~}
The definition of a Hamiltonian vector field in the first equation (\ref{defOX}) is not universal. Many authors set instead
${\hat L}_h\, \O = - \d h\,$ (see e.g. \cite{Blau}) so that $L_h = - X_h\,,$ leading to the opposite sign of the PB and, correspondingly,
to equations of motion $\stackrel{.}{g} = L_H\, g\,.$ Both choices, however, provide a representation of the Lie algebra of
Poisson brackets that is an ingredient in the {\em prequantization} (see e.g. \cite{TAE, WZ05, ZZ}). We have, in particular,
\be
[ X_g , X_h ] = X_{\{ g , h \}}\ .
\lb{repPBalg}
\ee

\smallskip

We proceed now to defining the classical WZNW model. We shall only consider the case when the Lie group $G\,$ is {\em compact}
and the corresponding quantized theory is {\em rational} \cite{CIZ1, FrSh, AM, MS2}.
(These two requirements single out combinations of WZNW models on compact semi-simple groups and "lattice vertex algebras" \cite{KT}.)
Albeit we only provide details for our main example $G = SU(n)\,,$ most results remain valid in the general case.

\smallskip

In the first order formalism the fiber ${\cal F}\,$ consists of a pair of periodic in $x^1\,$ maps $( g , {\cal J} )\,$
such that, for $x = (x^0 , x^1 ) \in {\widetilde{\cal M}}_2$
\ba
&&g (x) \in  G\,,\qquad g(x^0, x^1+2\pi) = g (x^0, x^1)\equiv g(x)\,,\lb{gJ}\\
&&{\cal J} (x) = j_\mu (x)\, dx^\mu \,,\quad
j_\mu (x) \in i\, {\cal G}\,,\quad j_\mu (x^0, x^1+2\pi) = j_\mu (x^0,x^1) \equiv j_\mu (x)\,,\quad
\nn
\ea
where ${\cal G}\,$ is the Lie algebra of $G\,$ (our conventions are such that, for $G$ compact, the current is Hermitean).
Note that the $i\, {\cal G}$-valued $1$-form ${\cal J} (x)\,$ is {\em horizontal}.

\smallskip

We define the basic $3$-form $\o\,$ by
\be
{4\pi}\, \o = {\bd}\,{\rm tr}\, ((i g^{-1} {\bd} g + \frac{1}{2 k}\, {\cal J} )\,
\sJ ) + \, k \,\theta (g)\ ,\qquad
\theta (g) := \frac{1}{3}\,{\rm tr}\, (g^{-1} {\bd} g )^3\ .
\lb{omWZW}
\ee
Here ${\rm tr}$ is the Killing form (\ref{Kill})
on ${\cal G}\,,\ k\,$ is the real "coupling constant" that will be ultimately restricted
to (positive) integer values to ensure the single valuedness of the
exponential of the action, and $\sJ\,$ is the Hodge dual to ${\cal J}$,
\be
\sJ (x) = \varepsilon_{\mu\nu} j^\mu (x) d x^\nu\qquad (\, \varepsilon_{01} = 1\,)\ .
\lb{sJ}
\ee
To identify (\ref{omWZW}) with the more customary (component) expressions, one uses (\ref{dxdx}) and
\be
{\cal J}\, \sJ = j_\mu j^\mu d x^0 d x^1 = - \sJ \,{\cal J}\,.
\lb{JsJ}
\ee

For compact $G$ we shall use the physicist's convention
introducing a {\em Hermitean basis} $T_a \in i \, {\cal G}\,$ for which
\be
\frac{1}{i}\,[T_a , T_b ] = f_{ab}^{~~c}\, T_c\,,\qquad {\rm tr}\,( T_a T_b ) = \eta_{ab}
\lb{etaab}
\ee
with {\em real} structure constants $f_{ab}^{~~c}$ and a {\em positive} metric $( \eta_{ab} )$
(see Appendix A). The tensor $f_{abc}\,,$ defined by
\be
\frac{1}{i}\,{\rm tr}\,( T_a\, [ T_b , T_c ] ) = \eta_{ad} f_{bc}^{~~d} =: f_{bca} = f_{abc}
\lb{fabc}
\ee
is totally antisymmetric (due to the cyclicity of the trace).
For $x$-independent $\gamma \in G\,$ so that $d \g =0\,$ and $\g^{-1} {\d} \g = i\, \Gamma^a\, T_a\,$ where $\Gamma^a$ are basic
left-invariant ${\cal G}$-valued $1$-forms, the WZ term $\theta (\g)$ (\ref{omWZW})
is just the invariant $3$-form on $G$ corresponding to the tensor $f_{abc}$ (see e.g. \cite{Schw}):
\be
\theta(\g) = \frac{1}{3}\,{\rm tr}\, (\g^{-1} \d \g )^3 =
\frac{1}{3!}\,\Gamma^a \, \Gamma^b \, \Gamma^c\, \frac{1}{i}\, {\rm tr}\, (T_a \, [ T_b , T_c ] )
= \frac{1}{3!}\, f_{abc}\, \Gamma^a\, \Gamma^b\, \Gamma^c\ .
\lb{can3}
\ee

The $3$-form $\o\,$ (\ref{omWZW}) is well defined and single valued while the
corresponding WZNW action density $2$-form
\be
4\pi\, {\bf L} =  {\rm tr}\, ((i g^{-1} {\bd} g + \frac{1}{2 k}\, {\cal J} )\,
\sJ ) + k\, {\bd}^{-1} \theta (g)
\lb{acdenWZW}
\ee
cannot be globally defined on $G\,$ since the $3$-form $\theta (g)\,,$ albeit closed, ${\bd}\, \theta (g) = 0\,,$ is not exact.
(Accordingly, the corresponding WZ term in the WZNW action in the second order formalism (\ref{Swznw0})
is multivalued.)

\smallskip

If we identify $ig^{-1} \pl_\mu g\,$ with the velocity on the group manifold,
then $j_\mu$ plays the role of covariant canonical momentum (cf. (\ref{gJ}) -- (\ref{sJ})), and the coefficient to the
space-time volume form $dx^0 \frac{d x^1}{2\pi}\,$ (with a minus sign) in (\ref{acdenWZW}) is the {\em covariant Hamiltonian} $H = H (j )\,,$
just as $- H\,$ was the coefficient to $d t\,$ in the classical mechanical action density ${\bf L}\,$
(\ref{D1}). Note that the only such term in the right-hand side of (\ref{acdenWZW}) comes from
\be
\frac{1}{8 \pi k}\, {\rm tr}\, ( {\cal J}\, \sJ ) = \frac{1}{8 \pi k}\,
{\rm tr}\, j_\mu j^\mu d x^0 d x^1 =: - H (j)\, dx^0 \frac{d x^1}{2\pi}\,.
\lb{covH}
\ee

It is remarkable that the $3$-form (\ref{omWZW}) contains the full information about the model:
it allows to derive both the equations of motion and the symplectic structure. To begin with, we note that
\be
{{\bd}}\, {\rm tr}\,( {\cal J}\, \sJ ) = \d \, {\rm tr}\, ( {\cal J}\, \sJ ) = 2\, {\rm tr}\, ( j_\mu \d j^\mu )\, d x^0 d x^1\ .
\lb{dJsJ}
\ee
We shall denote the pull-back of a form by $g^*\,;$ by definition,
\be
g^*  \left(\,  f (d g, d {\cal J}, d \,{\sJ}\, ;\, \d g , \d {\cal J} , \d {\sJ} )\,\right)  = f ( d g, d {\cal J}, d\, \sJ\, ;\, 0,0,0)\ .
\lb{pullb}
\ee
Introduce, for arbitrary $Y \in i\,{\cal G}\,$ (in particular, for any $n\times n$ Hermitean traceless matrix,
for ${\cal G} = su(n)$), the vertical vector field
${Y}_{j^\mu} := \,{\rm tr}\left( Y \frac{{\d}}{\d j^\mu} \right)$ so that
\be
{\hat Y}_{j^\mu} ( \d j^\nu ) = Y \d^\nu_\mu\qquad (\,{\hat Y}_{j^\mu} ( \d {\cal J} ) = Y d x_\mu\ ,\quad
{\hat Y}_{j^\mu} ( \d {\sJ} ) = Y \varepsilon_{\mu\nu}\, d x^\nu\, )\ .
\lb{rels0}
\ee
Using (\ref{dxdx}), we derive the first equation of motion:
\ba
&& g^* \left( {\hat Y}_{j^\mu}\, \o \right) = \frac{1}{4\pi}\, {\rm tr}\, Y
( i g^{-1} \partial_\mu g + \frac{1}{k}\, j_\mu  )\, d x^0 d x^1 = 0\ ,\qquad {\rm or}\nonumber\\
&&j_\mu = - i k \, g^{-1} \partial_\mu g\qquad \Leftrightarrow \qquad {\cal J} = - i k \, g^{-1} d g\ .
\lb{plbk1}
\ea
To obtain the remaining equations, we introduce the vector field
$Y_g := i\, {\rm tr}\left( g\, Y \frac{{\d}}{\d g} \right)\,$ satisfying
\be
{\hat Y}_g\, (g^{-1} {{\bd}}\, g) = i\, Y\qquad\Rightarrow\qquad
{\hat Y}_g \, \theta (g) = i\, {\rm tr}\left( Y (g^{-1} {{\bd}} g )^2 \right)\ .
\lb{rels1}
\ee
Equating to zero the pull-back of ${\hat Y}_g \,\o\,,$
\be
g^*\, ( {\hat Y}_g \,\o ) = \frac{1}{4\pi}\,{\rm tr}\, Y \left( d\, \sJ + i k\,(g^{-1} d g)^2 + [ g^{-1} d g , \sJ ]_+ \right) = 0
\lb{seeqmot}
\ee
together with the first equation of motion (\ref{plbk1}) and the anticommutativity relation (\ref{JsJ})
\be
[ g^{-1} d g,\, \sJ \, ]_+ = \frac{i}{k}\, [ {\cal J},\, \sJ\, ]_+\, = 0
\lb{[]+}
\ee
implies the second equation of motion which can be written entirely in terms of currents:
\ba
&&d\, \sJ = \frac{i} {k} \,{\cal J}^2
\qquad \Leftrightarrow\qquad \partial_\mu j^\mu = -\, \frac{i}{2k}\,\e^{\mu\nu} [j_\mu , j_\nu ]\nn\\
&&{\rm i.e.,}\quad  \partial_1 j^1 + \partial_0 j^ 0 = - \frac{i}{k}\, [j^0 , j^1 ]\ .
\lb{eqnmot2}
\ea
Next, we compare the result with the horizontal ($d$-) differential (the curl) of (\ref{plbk1}),
\ba
&&d {\cal J} = i k\, (g^{-1} d g )^2 = - \frac{i}{k}\, {\cal J}^2
\qquad\Leftrightarrow\qquad \e^{\mu\nu} \partial_\mu j_\nu = -\,\frac{i}{2k}\,\e^{\mu\nu} [j_\mu , j_\nu ]\nn\\
&&\hspace{16mm}{\rm i.e.,}\quad \partial_1 j^0 + \partial_0 j^1 = \frac{i}{k}\, [ j^0 , j^1 ]\ .
\lb{takecurl}
\ea
This yields the easily solvable equation
\be
d\, ( {\cal J} + {\sJ} ) = 0\qquad \Leftrightarrow\qquad
(\partial_0 + \partial_1 ) (j^0 + j^1 ) = 0\ .
\lb{clos}
\ee
In order to write down its general solution we introduce the light cone variables (and the corresponding derivatives) (\ref{conev}).
We can summarize the result as
\be
\partial_+ j_R = 0\qquad{\rm for}\qquad
j_R := \frac{1}{2}\, (j^0 + j^1 )\ = - i k\,  g^{-1} \partial_- g \ .
\lb{eqsmR}
\ee
This (second order in $g=g(x^+,x^-)$) equation is equivalent to
\be
\partial_- j_L = 0\qquad{\rm for}\qquad
j_L := \frac{1}{2}\, g (j^0 - j^1 ) g^{-1}\ = i k\, (\partial_+ g )\, g^{-1}\ ,
\lb{eqsmL}
\ee
since $\partial_+ j_R = - g^{-1} (\partial_- j_L)  g\,,$ or alternatively, to the closedness of the corresponding
current $1$-forms
\ba
&&{\cal J}_L :=\, i k\, ( \partial_+ g) g^{-1} d x^+\ ,
\qquad {\cal J}_R := - i k\, ( g^{-1} \partial_- g )\, d x^-\nn\\
&&(\, {\cal J} = {\cal J}_R - g^{-1} {\cal J}_L g \ ,\quad
{\sJ} = {\cal J}_R + g^{-1} {\cal J}_L g  \,)\ ,\nn\\
&&d\, {\cal J}_L  = 0 = d\, {\cal J}_R\ .
\lb{ediff1}
\ea

\smallskip

\noindent
{\bf Remark 2.2~} In the pioneer paper \cite{W} on non-abelian bosonization Witten starts with the observation that a set of vector currents
\be
j_a^\mu (x) = i\, {\tilde{\psi}}(x) \gamma^\mu\, T_a \psi (x)\ ,\quad \g_1^2 = 1 = - \g_0^2\ ,\quad [\g_0 , \g_1 ]_+ = 0
\lb{Wjj}
\ee
where $\psi$ is a ($2$-component) free massless fermion field with values in the fundamental representation of ${\cal G}\,,$
splits into conserved left and right components obtained by substituting $\g^\mu$ with $\frac{1}{2} \g^\mu (1\mp\g_5 )\,,\
\g_5 := \g^0 \g^1\,$ and depending on $x^\pm\,,$ respectively.
Demanding such a splitting into chiral components for the Lie algebra valued current $j_\mu$
(\ref{plbk1}), one comes to the necessity of adding to the "standard" action,
given by the first term in the right-hand side of (\ref{Swznw0}), the second, Wess-Zumino term.

\smallskip

The definition of the (conserved and traceless) stress energy tensor $T^\mu_{~\nu}\,$ is encoded in
the first order action density (\ref{acdenWZW}). Its form illustrates the observation that
the WZ term only influences the symplectic structure, respectively the PB relations,
while the stress energy tensor is determined by just the coefficient $H$ to the space-time volume.
Expressing $T^\mu_{~\nu}\,$ in terms of the covariant Hamiltonian (\ref{covH}) and its functional derivatives,
\be
T^\mu_{~\nu} (x) = {\rm tr}\, \left( \frac{\d H}{\d j_\mu (x)} j_\nu (x) \right)
- H \d^\mu_\nu = \frac{1}{2 k}\, {\rm tr}\, \left( \frac{1}{2}\, j^2 (x)
\d^\mu_\nu - j^\mu (x) j_\nu (x) \right)\,,
\lb{stren}
\ee
we recover the classical Sugawara formula\footnote{The "Sugawara formula" has in fact
many authors -- see, e.g. the bibliographical notes to Section 4 of \cite{FSoT}, p.75 and references cited there.}.

The same expression can be obtained by Hilbert's variational
principle varying the action density \be - H(j, h) {\sqrt{- h}} =
\frac{1}{4k}\, h^{\a\b} \, {\rm tr}\, j_\a j_\b \,\sqrt{-h}\qquad
(\, h = \det (h_{\a\b})\,,\ \, h^{\a\s} h_{\s\b} = \d^\a_\b\,)
\lb{Hjh} \ee with respect to $h^{\mu\nu}$ in the neighbourhood of
the flat Minkowski space metric $h_{\mu\nu} = \eta_{\mu\nu}\,.$
Using the Jacobi formula \be \d h = h\, h^{\mu\nu}\d h_{\mu\nu} =
- h\, h_{\mu\nu}\d h^{\mu\nu}\ , \lb{Jach} \ee we find \be
\frac{1}{\sqrt{- h}}\, \d \,( H(j, h) {\sqrt{- h}} ) = \frac{1}{2}
\, T_{\mu\nu} \, \d h^{\mu\nu}\qquad (\,T^{\mu}_{~\mu} =
h^{\mu\nu} T_{\mu\nu} = 0\,) \lb{fder-Hjh} \ee which reproduces
(\ref{stren}) for $h_{\mu\nu} = \eta_{\mu\nu}$.

The two independent chiral components of $T^\mu_{~\nu}\,$ are quadratic in the
corresponding chiral components of the current:
\ba
&&T_L := \frac{1}{2}\, (T^0_{~0} - T^1_{~0} ) =
\frac{1}{8 k}\, {\rm tr}\, (j^0 - j^1 )^2 = \frac{1}{2 k}\, {\rm tr}\, j_L^2\ ,\nonumber\\
&&T_R := \frac{1}{2}\, (T^0_{~0} + T^1_{~0} ) =
\frac{1}{8 k}\, {\rm tr}\, (j^0 + j^1 )^2 = \frac{1}{2 k}\, {\rm tr}\, j_R^2\ .
\lb{Tchir}
\ea
The conservation of $T^\mu_{~\nu}$ follows trivially from the chirality of $j_L = j_L (x^+)$ and $j_R = j_R(x^-)$
(cf. (\ref{conev}), (\ref{eqsmR}), (\ref{eqsmL})):
\be
\partial_- T_L \pm \partial_+ T_R = 0\qquad\Leftrightarrow\qquad \partial_\mu T^\mu_{~\nu} = 0\ .
\lb{dT0}
\ee

The traditional derivation of the equations of motion from the multivalued action density (\ref{acdenWZW})
is based on the easily verifiable relation
\be
\d\, \frac{1}{3}\, {\rm tr}\, (g^{-1} d g )^3 = - \,d \, {\rm tr}\, ( g^{-1} \d g \, ( g^{-1} d g )^2 )
\lb{totdiff}
\ee
implying that the vertical ("variational") differential of the multivalued WZ term
$d^{-1} g^* (\theta (g) )\,$ is single valued,
\be
\d\, d^{-1} g^* (\theta (g) ) = {\rm tr}\, ( g^{-1} \d g\, ( g^{-1} d g )^2 )
\lb{totdiff1}
\ee
(cf. (\ref{totdiff0})). Taking $\d\,$ of
the pull-back of the action density (\ref{acdenWZW}) and using (\ref{dJsJ}), we thus obtain
\ba
&&\d \, g^* ({\bf L}) = - \, d\,\a- \frac{1}{4\pi}\, {\rm tr}\, \{\d\, \sJ \, (ig^{-1} d g + \frac{1}{k} {\cal J} ) \} -
\nn\\
&&-\frac{ i}{4\pi}\,{\rm tr}\,\{ g^{-1} \d g \,( d\, \sJ + i k \, (g^{-1} d g )^2 + [ g^{-1} d g ,\,  \sJ ]_+) \}
\lb{varL}
\ea
where $\a\,$ is the {\em Noether form} \cite{JS} (of degree $(a,b) = (D-1,1) = (1,1)$)
\be
\a = \frac{i}{4\pi}\,{\rm tr}\, ( g^{-1} \d g \,\sJ ) \ .
\lb{Noetherf}
\ee
The vanishing of $\d \, g^* ({\bf L})\,,$ up to the boundary term $d\, \a\,,$ reproduces
(after using(\ref{[]+})) the equations of motion (\ref{plbk1}) and (\ref{eqnmot2}).

\smallskip

In the second order formalism the equations of motion are expressed directly in terms of $g\,$ and its derivatives.
From (\ref{SWZ1}) we get
\ba
\d S [g] &=& - \frac{k}{2\pi} \, \int_{\cal M} {\rm tr}\, \{\,\d g\, g^{-1} \partial_- ((\partial_+ g) g^{-1} ) \}\, d x^+ d x^- \nn\\
&\equiv& - \frac{k}{2\pi}\, \int_{\cal M} {\rm tr}\, \{\, g^{-1} \d g \, \partial_+ (g^{-1} \partial_- g ) \}\, d x^+ d x^-\ ,
\lb{varL2}
\ea
and equating (\ref{varL2}) to zero for arbitrary variations $\d g$ reproduces (\ref{eqsmL}) and (\ref{eqsmR}).

In accord with the general rules formulated in the beginning of this section, the true {\em symplectic density} $\o_0\,$
for the WZNW model is obtained \cite{G} by restricting the form $\o\,$ (\ref{omWZW}) to an equal time surface,
i.e. taking the coefficient of $d x^1\,.$
Noting that ${\sJ}\mid_{dx^0=0} = j^0 d x^1\,,$ we see that the resulting $(1, 2)\,$ form differs from
${\d\,\a}\!\mid_{dx^0 =0}= \frac{i}{4\pi}\,\d\, {\rm tr}\, ( j^0\, g^{-1} \d g )\, d x^1\,,$ cf. (\ref{Noetherf})
(which is a special case of the $(D-1 , 2)\,$ symplectic density
considered in \cite{JS}) by a contribution from the WZ term:
\be
\o_0 = {\d\, \a}\mid_{dx^0 =0} + \frac{k}{4\pi}\, {\rm tr}\,
\left( g^{-1} g'(g^{-1} \d g )^2 \right) d x^1\,,\qquad
g' := \partial_1 g\ .
\lb{omega0}
\ee
The symplectic form $\Omega^{(2)}\,$ of the theory is obtained by integrating
$\o_0\,$ (\ref{omega0}) over a constant time circle i.e., over a period in $x^1:$
\ba
{\Omega}^{(2)} &=& \int\limits_{-\pi}^{\pi} \o_0\, dx^1 =\nn\\
&=&\frac{1}{4\pi}\int\limits_{-\pi}^{\pi}\, dx^1\,{\rm
tr}\, \left(  i \,\d \left( j^0 g^{-1} \d g \right) + k\, g^{-1} g' \left( g^{-1} \d g \right)^2 \right) =\quad\lb{OmegaWZ0}\\
&=&\frac{1}{2\pi} \int\limits_{-\pi}^{\pi}\, dx^1\, {\rm tr}\,
\left( i\, \d \left( j_R\, g^{-1} \d g \right) +
\frac{k}{2}\, g^{-1} \d g {\left( g^{-1} \d g \right)}' \right) = \quad\lb{OmegaWZR}\\
&=& \frac{1}{2\pi} \int\limits_{-\pi}^{\pi}\, dx^1\, {\rm tr}\,
\left( i\, \d \left( j_L\, \d g g^{-1} \right) -
\frac{k}{2}\, \d g g^{-1} {\left( \d g g^{-1} \right)}' \right)\ .\lb{OmegaWZL}
\ea
In verifying the equivalence between these three forms of ${\Omega}^{(2)}$ we use the relations
\be
j^0 = 2 j_R + i k \, g^{-1}\, g' = 2\, g^{-1} j_L\, g - i k \, g^{-1}\, g'   \ ,
\lb{abc1}
\ee
cf. (\ref{eqsmL}), (\ref{eqsmR}).

\subsection{Splitting $g(x^+ , x^- )\,$ into chiral components}

\noindent
Given the equations of motion, the classical phase space ${\cal S}\,$ of the $2D$ WZNW model can be identified with the
manifold of their initial data,
\be
{\cal S}\ = \ T^* {\tilde G}\ \simeq {\tilde G} \times {\tilde{\cal G}}\ ,
\lb{T*}
\ee
where ${\tilde G}\,$ is the loop group corresponding to $G\,,$ and ${\tilde{\cal G}}$ -- its Lie algebra.
We can choose, for example, the parametrization in terms of $g\,$ and $j_L\,,$ see (\ref{OmegaWZL}),
so that
\be
\lb{Ph}
{\cal S} = \{\, g (x)\mid_{x^0=0}\, \in {\tilde G}  \,,\ j_L(x)\mid_{x^0=0}\, \in {\tilde{\cal G}} \,\}\,.
\ee
${\cal S}$ can be viewed, alternatively, as the
space of solutions of the equation of motion (\ref{eqsmR}) (or, equivalently, of (\ref{eqsmL}))
\be
\pl_+ (g^{-1}\pl_- g ) \, =\, 0\qquad\left(\ \Leftrightarrow\
\pl_- ((\pl_+ g ) g^{-1} ) = 0\,\right)\ .
\lb{eqmotion}
\ee
The general solution of (\ref{eqmotion}) is given by the factorized expression
$g(x^+,x^-) = g_L (x^+)\, g^{-1}_R(x^-)\,$ (\ref{LR}), where the chiral components
$g_C\,,\ C=L,R\,$ satisfy the twisted periodicity condition $g_C (x+2\pi ) = g_C (x)\, M\,,\ M\in G$ (\ref{cM})\footnote{To
simplify notation, we shall often denote, in what follows, by $x\,$ the
single argument of any of the chiral fields. It should not be confused
with the vector $x=(x^0 , x^1)$ which only appears in the $2D$ field $g$ (\ref{LR}).}.
Note that the currents $j_C\,$ can be expressed in
terms of the corresponding chiral components of $g\,,$
\be
j_L (x^+ ) = i k \, g'_L (x^+) g_L^{-1} (x^+)\ ,\qquad
j_R (x^- ) = i k \, g'_R (x^-) g_R^{-1} (x^- ) \ .
\lb{jLR}
\ee

The space of pairs of twisted-periodic maps with equal monodromies from the light rays to the group,
\be
{\tilde{\cal S}}=\{ \left( g_L (x^+)\,, g_R (x^-) \right),\, x^\pm\in{\mathbb R}\ \mid
\ g_C^{-1} (x)\, g_C (x+2\pi ) = M\in G\}
\lb{extPh}
\ee
is an extension of ${\cal S}\,.$ More precisely, ${\tilde{\cal S}}\,$ can be
viewed as a principal fibre bundle over ${\cal S}\,$ \cite{BFP} with respect
to the free\footnote{I.e., without fixed points, for $h\ne e\in G$.} right action of $G\,$ on ${\tilde{\cal S}}\,$
\be
(g_L ,g_R )\ \rightarrow\ (g_L h\, , g_R h )\ ,\quad M \ \rightarrow\  h^{-1} M h \qquad ( h \in G )\ ,\nonumber\\
\lb{free-act}
\ee
the projection $pr :\, {\tilde{\cal S}} \ \longrightarrow\ {\cal S}\,$ being defined as
\be
{\tilde{\cal S}}\ni ( g_L (x^+ ) \,,\, g_R (x^- ) )\quad \stackrel{pr}{\longrightarrow}\quad
( g_L (x) g_R^{-1} (x)\,,\, i k\, g'_L (x) g_L^{-1} (x) ) \in {\cal S}\,.
\lb{princB}
\ee

By rewriting the symplectic form $\Omega^{(2)}$ (\ref{OmegaWZL}) on ${\cal S}$ in terms of the chiral fields $g_L\,,\, g_R\,$
it is extended to a closed (but degenerate) form $\Omega^{(2)} (g_L , g_R)$ on ${\tilde{\cal S}}\,.$

\vspace{3mm}

\noindent
{\bf Proposition 2.1~} (Gaw\c{e}dzki \cite{G}; Falceto \& Gaw\c{e}dzki \cite{FG1})~ {\em
One can present $\Omega^{(2)} (g_L , g_R)\,$ as the difference of two chiral $2$-forms:
\be
\Omega^{(2)} (g_L , g_R) \, =\, \Omega_c (g_L,M ) \, - \, \Omega_c (g_R,M ) \ ,
\lb{O-O}
\ee
\ba
&&\Omega_c (g_C,M ) =  \frac{k}{4\pi} {\rm tr}\,\{
\int\limits_{-\pi}^{\pi} g_C^{-1} \d g_C (x)\, (g_C^{-1} \d g_C (x))'\,dx   +
\, \d g_C g_C^{-1} (-\pi ) \,\d g_C g_C^{-1} (\pi )  \,\} \equiv \nn\\
&&\equiv \frac{k}{4\pi} {\rm tr}\,\{
\int\limits_{-\pi}^{\pi} g_C^{-1} \d g_C (x)\, (g_C^{-1} \d g_C (x) )'\,dx  +
\, b_C^{-1} \d b_C \,\d M M^{-1} \,\}\ ,
\lb{Oc}
\ea
$C = L , R\,,$ where $b_C := g_C (-\pi )\,$ and $g_C (x+2\pi ) = g_C (x)\, M$ so that the monodromy
\be
\lb{bM}
M = b_C^{-1} g_C (\pi )
\ee
is independent of the chirality $C\,$.}

\medskip

\noindent
{\bf Proof~} From the expressions for $g\,$ (\ref{LR}) and $j_L\,$ (\ref{jLR}) we get
\ba
&&\d g g^{-1} = g_L \,(g_L^{-1} \d g_L - g_R^{-1} \d g_R )\, g_L^{-1}\ ,\nn\\
&&{\rm tr} ( j_L \d g g^{-1} ) = i k\, {\rm tr}\,\left( g_L^{-1} g_L'  (g_L^{-1} \d g_L - g_R^{-1} \d g_R ) \right)\ ,
\ea
so that
\ba
&&i\, \d\, {\rm tr} ( j_L \d g g^{-1} ) = k\, {\rm tr}\,
\left( (g_L^{-1} \d g_L - g_R^{-1} \d g_R ) (g_L^{-1} \d g_L' - g_L^{-1} g_L'\, g_R^{-1} \d g_R )\right)\ ,\nn\\
&& \\
&&{\rm tr}\, \left( \d g g^{-1} (\d g g^{-1} )'\right) = 2\, {\rm tr}\,
\left( (g^{-1}_L \d g_L - g^{-1}_R \d g_R ) ( g_L^{-1} \d g_L' - g_L^{-1} g_L'\, g_R^{-1} \d g_R )\right) -\nn\\
&&-\,{\rm tr}\,\left( (g^{-1}_L \d g_L - g^{-1}_R \d g_R )(( g_L^{-1} \d g_L )' + ( g_R^{-1} \d g_R )' ) \right)\ .
\nn
\ea
Hence, $\Omega^{(2)} (g_L , g_R ) \,$ (\ref{O-O}) is expressed as
\be
\Omega^{(2)} (g_L , g_R ) = \frac{k}{4\pi}\,\int\limits_{-\pi}^{\pi} {\rm tr}\,\left\{ (g^{-1}_L \d g_L (x) - g^{-1}_R \d g_R (x) )
(( g_L^{-1} \d g_L (x) )' + ( g_R^{-1} \d g_R (x) )' ) \right\}\,dx\ .
\lb{Olr}
\ee
To complete the proof, it remains to note that the two mixed terms in (\ref{Olr}) combine to
\ba
&&\int\limits_{-\pi}^{\pi} dx\,{\rm tr}\, (g_L^{-1} \d g_L (x) g_R^{-1} \d g_R (x) )'
\equiv {\rm tr}\, ( g_L^{-1} \d g_L (\pi) g_R^{-1} \d g_R (\pi) - b_L^{-1} \d b_L b_R^{-1} \d b_R ) =\nn\\
&&=\, {\rm tr}\, \left( (b_L^{-1} \d b_L - b_R^{-1} \d b_R )\, \d M M^{-1} \right) \equiv\nn\\
&&\equiv\, {\rm tr}\, \left( \d g_L g^{-1}_L (-\pi) \,\d g_L g^{-1}_L (\pi) -
\d g_R g^{-1}_R (-\pi) \,\d g_R g^{-1}_R (\pi)\right) \ ,
\ea
since $\, g^{-1}_C \d g_C (-\pi) \equiv b^{-1}_C \d b_C\ ,\ \ g_C(\pi) = b_C M\ ,\ \ {\rm tr}\left( \d M M^{-1}\right)^2 =0\ ,$ and
\be
g^{-1}_C \d g_C (\pi) = M^{-1} b_C^{-1} \d ( b_C M ) = M^{-1} ( b_C^{-1} \d b_C + \d M M^{-1} ) M
\lb{gpi}
\ee
or, conversely,
\be
\d M M^{-1} = \d (b_C^{-1} g_C(\pi))\,  g_C(\pi)^{-1} b_C = - b_C^{-1} \d b_C + b_C^{-1}\, \d g_C g_C^{-1} (\pi)\, b_C\ .
\lb{gpi-conv}
\ee
\eod

\smallskip

As already mentioned, as a $2$-form on ${\tilde{\cal S}}$ (\ref{extPh}), $\Omega^{(2)} (g_L , g_R )$ (\ref{O-O}) is still closed but is degenerate. The closedness
follows from the fact that, for $g_L$ and $g_R$ having the same monodromy $M\,,$ one has
$\d\, \Omega_c (g_L,M) = \d\, \Omega_c (g_R,M) :$
\ba
&&\d\, \Omega_c (g_C,M ) = - \frac{k}{4\pi} {\rm tr}\,\{\,
\int\limits_{-\pi}^{\pi} dx\, ( g_C^{-1} \d g_C (x))^2\, (g_C^{-1} \d g_C (x))'  +\nn\\
&&+ ( b_C^{-1} \d b_C + \d M M^{-1} )\, b_C^{-1} \d b_C\, \d M M^{-1} \,\} =\nn\\
&&=\frac{k}{4\pi}\, \{\, \int\limits_{-\pi}^{\pi} d\, \theta (g_C(x)) -
{\rm tr}\,( b_C^{-1} \d b_C + \d M M^{-1} )\, b_C^{-1} \d b_C\, \d M M^{-1} \,\} =\nn\\
&&=\frac{k}{4\pi}\, \{\, \theta (b_C M) - \theta (b_C) -
{\rm tr}\,( b_C^{-1} \d b_C + \d M M^{-1} )\, b_C^{-1} \d b_C\, \d M M^{-1} \,\} =\nn\\
&&= \frac{k}{12\pi} \,{\rm tr}\, (M^{-1} \d M)^3 = \frac{k}{4\pi} \, \theta (M)
\lb{deltaO}
\ea
(we have used again (\ref{gpi}); note that the $3$-form $\theta (M)$ is purely vertical
since $M$ is $x$-independent). The degeneracy of $\Omega^{(2)} (g_L , g_R)$ on ${\tilde{\cal S}}\,$
is due to its invariance with respect to simultaneous equal right shifts of $g_L$ and $g_R\,,$
see (\ref{free-act}); accordingly,
if $Y_r$ is the vertical vector field generating the $1$-parameter group
\ba
&&g_L\ {\rightarrow}\ g_L\, e^{i t Y}\ ,\quad g_R\ {\rightarrow}\ g_R\, e^{i t Y}\qquad (\,i\, Y\in {\cal G}\,)\ ,\lb{g-l-r}\\
&&{\hat Y}_r\, \d g_C \equiv Y_r\, g_C := \frac{d}{dt} (g_C\, e^{i t Y} )|_{t=0}= i\, g_C Y\ ,\quad {\hat Y}_r (g^{-1}_C \d g_C) = i\,Y
\nn
\ea
for $C=L,R\,,$ it follows immediately from (\ref{Olr}) that ${\hat Y}_r\, \O^{(2)} (g_L,g_R) = 0\,.$

\vspace{1mm}

In order to define symplectic forms on each of the chiral phase spaces, one has to extend further
${\tilde{\cal S}}\,$ introducing {\em independent} chiral monodromies $M_C\,,\ C=L,R\,$ \cite{G}.
In such a way the left and the right sectors ${\cal S}_L ,\, {\cal S}_R\,,$ where
\be
{\cal S}_C =\{  g_C (x)\ ,\, x\in{\mathbb R}\ \mid \ g_C^{-1}(x)\, g_C (x+2\pi ) = M_C\in G \}\ ,\quad C = L \,,\, R\ ,
\lb{PC}
\ee
fully decouple. To avoid overcounting variables, we shall consider
each of the chiral phase spaces ${\cal S}_C\,$ as being parametrized by
the smooth functions $g_C (x)\,,\, -\pi < x < \pi\,$ and their boundary data,
$b_C = g_C(-\pi)\,$ and $M_C = b_C^{-1} g_C(\pi)\,.$
Due to (\ref{deltaO}), it appears  natural to set
\be
\Omega (g_C , M_C ) = \Omega_c (g_C,M_C ) - \frac{k}{4\pi}\, \rho (M_C)\ ,
\lb{O}
\ee
demanding that the $2$-form $\rho (M)$ (defined in some neighbourhood of the unit element) satisfies
\be
\d\,\rho (M)\, =\,\theta (M)\ .
\lb{drho}
\ee
The resulting $\Omega (g_C , M_C )$ is closed and non-degenerate (we shall see in what follows that it is invertible),
thus equipping each ${\cal S}_C$ with a true symplectic structure.

\smallskip

Unless not being explicitly specified otherwise, by "the chiral WZNW model" we shall understand below the theory with
\begin{itemize}
\item phase space ${\cal S}_C\,$ (\ref{PC}),
\item symplectic form $\Omega (g_C , M_C )$ (\ref{O}) (for certain $\rho (M_C)$ satisfying (\ref{drho})),
\item and (conformal) Hamiltonian $T_C$ (\ref{Tchir}), (\ref{jLR})
\end{itemize}
coinciding with the {\em left}\, WZNW sector described above, and shall omit in most cases the chirality index.
(The only difference between the two sectors is in the opposite signs of the corresponding symplectic forms;
recall that the one of the right sector is $\ - \,\Omega (g_R , M_R)$.)
We shall return to the problem of reconstructing the $2D\,$ theory from the chiral ones at the end of the next Section.

\smallskip

The $2$-form (\ref{O-O}) on ${\tilde{\cal S}}\,$ is thus recovered by
imposing the constraint of equal chiral monodromies
\be
\Omega^{(2)} (g_L , g_R) = (\Omega (g_L , M_L) - \Omega (g_R , M_R)) \mid_{M_L \approx M_R}\ .
\lb{O2alt}
\ee
The sign difference between the left and right symplectic forms forces us
to distinguish between left and right monodromy since
the resulting Poisson brackets for $M_L\,$ and $M_R\,$ will also
differ in sign. The monodromy invariance of the $2D$ theory will have to be
restored at a later stage as a constraint on the observable quantities.
Hence, recovering the $2D$ WZNW model from the extended phase space
(the product of two independent chiral spaces with different
monodromies) requires a {\em gauge theory framework}, cf. Section 3.5.4 below\footnote{In
the quantum theory, imposing the constraint of equal left and right monodromy corresponds
to singling a physical quotient of the extended state space; see Section 6.2.}.
The $2D$ {\em observables} are functions of the periodic (i.e.,
monodromy free) $2D$ field $g\,$ (\ref{LR}). The projection of the observable
algebra on a chiral (say, left mover's) phase space is generated by the chiral
currents $j_C\,,\ C=L,R$ which can be expressed, according to
(\ref{jLR}), in terms of the corresponding chiral variable $g_C\,$ and allow
to write down the chiral components (\ref{Tchir}) of the stress energy tensor.

\smallskip

As already noted, the WZNW form $\theta$ is not exact,
hence there is no globally defined smooth $2$-form on $G\,$ satisfying (\ref{drho}).
However, a form $\rho\,$ with this property can be
constructed locally, on an open dense neighbourhood of the identity
$\stackrel{\circ}{G}\,$ of $G\,.$ For example, if the monodromy matrix can be factorized \cite{S-T-S, RS}  as
\be
M\, =\, M_+\, M_-^{-1}\ ,\quad M_\pm \, \in \, G_{\subbbc}\,
\lb{M+-}
\ee
where $G_{\subbbc}\,$ is the complexification of $G\,,$ one can prove directly that the $2$-form
\be
\rho (M)\, =\, {\rm tr}\, (M_+^{-1} \d M_+ M_-^{-1} \d M_- )
\lb{ro}
\ee
satisfies (\ref{drho}) provided that
\be
\theta (M_\pm) \equiv \frac{1}{3}\,{\rm tr}\, (M_\pm^{-1} \d M_\pm )^3\, =\, 0\ .
\lb{prov}
\ee
Indeed, a simple computation using (\ref{prov}) gives
\ba
&&\theta (M) = \frac{1}{3}\, {\rm tr}\,(M^{-1} \d M )^3 =
\frac{1}{3} {\rm tr}\, (M_+^{-1} \d M_+ - M^{-1}_- \d M_- )^3 =\nonumber\\
&&= {\rm tr}\, \left(
M_+^{-1} \d M_+ (M_-^{-1} \d M_- -
M_+^{-1} \d M_+ ) M_-^{-1} \d M_-\right) = \d\, \rho (M)\ .\qquad
\lb{comp}
\ea
According to the {\em Cartan criterium for solvability} (see e.g. \cite{FS}), a Lie algebra $\,{\cal K}\,$ is solvable iff its Killing form satisfies
\be
X\in {\cal K}\ ,\quad Y\in [{\cal K} , {\cal K} ] \quad\Rightarrow\quad {\rm tr}\, (X Y) \equiv (X , Y ) = 0\ .
\lb{Ksolv}
\ee
By (\ref{can3}), Eqs. (\ref{prov}) follow automatically if $M^{-1}_\pm \d M_\pm$ take their values in a solvable Lie subalgebra of $G_{\subbbc}\,.$
We shall assume that these are the Borel subalgebras ${\mathfrak b}_\pm\,,$ in which case we shall call $M_\pm$ (\ref{M+-}) the
{\em Gauss components} of $M\,$ (other possibilities are considered in \cite{CP}).

\smallskip

For $G=SU(n),\,$ our main example in this paper, $G_{\subbbc} = SL(n)\,$ and we choose
$\stackrel{\circ}{G}\,$ to be the set of the matrices $M = (M^\a_\b )\in G\,$ such that
$M^n_n\ne 0\ne {\rm det}\,\begin{pmatrix}M^{n-1}_{n-1}&M^{n-1}_n\cr M^n_{n-1}&M_n^n\end{pmatrix}\,$ etc., while
$M_\pm\,$ belong to the Borel subgroups $B_\pm\,$ of $SL(n)\,$ of upper and lower triangular unimodular matrices, respectively.
The uniqueness of the decomposition (\ref{M+-}) is ensured by the relation
\be
{\rm diag}\, M_+  \, = \, {\rm diag}\, M_-^{-1}\, = D\, = ( d_\a \d^\a_\b )
\lb{diagMM}
\ee
where the diagonal matrix $D\,$ has unit determinant, $\prod_{\a =1}^n d_\a  = 1\,.$

Being a function of the monodromy matrix $M\in \stackrel{\circ}{G}$ only, the $2$-form $\rho (M)$ can
be presented in terms of an ($M$-dependent) operator $K_M\in End\,\,{\cal G}$ as
\be
\rho (M) = \frac{1}{2}\,{\rm tr}\, (\d M M^{-1} K_M (\d M M^{-1}))
\lb{defrhoK}
\ee
(without loss of generality, $K_M$ can be assumed to be skew symmetric with respect to the Killing form defined by the trace).
For $\rho(M)$ given  by (\ref{ro}) in terms of the Gauss components (\ref{M+-}) of $M\,,$ so that
\be
\d M M^{-1} = \d M_+ M_+^{-1} -  Ad_M\, ( \d M_- M_-^{-1} )\qquad (\, Ad_M (X) := M X M^{-1}\,)\ ,
\lb{dMM+-}
\ee
the corresponding $K_M$ acts simply as
\be
K_M (\d M M^{-1} ) = \d M_+ M_+^{-1} +  Ad_M\, ( \d M_- M_-^{-1} )\ .
\lb{KMMM}
\ee
Indeed, inserting (\ref{dMM+-}) and (\ref{KMMM}) into (\ref{defrhoK}), we recover (\ref{ro}):
\be
\lb{rhoKGauss}
\rho (M) = \,{\rm tr}\, \left(\d M_+ M_+^{-1} Ad_M\, ( \d M_- M_-^{-1} )\right)
= {\rm tr}\, \left( M_+^{-1} \d M_+ M_-^{-1} \d M_- \right)\ .
\ee

\subsection{$2D\,$ and chiral gauge symmetries}

\noindent
It is readily seen that the basic $3$-form $\o\,$ (\ref{omWZW})
of the $2D$ WZNW model is invariant with respect to both left and right {\em constant} group translations,
\ba
&&L:\ g\to h\, g\quad (\, g^{-1} {\bd} g \to g^{-1} {\bd} g \,,\ {\cal J} \to {\cal J}\,,\ {\sJ} \to {\sJ}\, )\,,\lb{LR-om}\\
&&R:\ g\to gh\quad (\, g^{-1} {\bd} g \to h^{-1} (g^{-1} {\bd} g ) h \,,\ {\cal J} \to h^{-1} {\cal J} h\,,
\ {\sJ} \to h^{-1} \, {\sJ} h\, )\,.\nn
\ea
It follows trivially from the transformation properties of the currents (\ref{eqsmR}), (\ref{eqsmL}),
\be
j_L \stackrel{L}{\rightarrow} h j_L h^{-1} \ , \quad j_R \stackrel{L}{\rightarrow} j_R\ ,\qquad
j_L \stackrel{R}{\rightarrow} j_L\ ,\quad j_R \stackrel{R}{\rightarrow} h^{-1} j_R\, h
\lb{LR-SO}
\ee
that the same applies to the stress energy tensor $T^\mu_{~\nu}\,$ and its chiral counterparts $T_C\,,\ C=L,R\,$ (\ref{Tchir}).

\smallskip

A canonical way of displaying the symmetries consists in letting the corresponding vector fields act on the symplectic form.
In particular, the vector fields implementing the left and right group translations,
\ba
&&g \stackrel{L}{\rightarrow} e^{i t Y} g \, ,\quad j_L \stackrel{L}{\rightarrow} e^{i t Y} j_L e^{-i t Y}\qquad (\,i\, Y\in {\cal G}\,)\,,\lb{YL}\\
&&{\hat Y}_L \d g \equiv Y_L\, g = i\, Y g\, ,\quad {\hat Y}_L (\d g g^{-1}) = i\,Y\,,\quad
{\hat Y}_L\, \d j_L \equiv Y_L\, j_L = i\, [Y,j_L]
\nn
\ea
and
\ba
&&g \stackrel{R}{\rightarrow}g\,  e^{i t Y} \,,\quad j_R \stackrel{R}{\rightarrow} e^{-i t Y}  j_R\,  e^{i t Y} \,,\lb{YR}\\
&&{\hat Y}_R\, \d g \equiv Y_R\, g = i\, g\, Y \,,\quad {\hat Y}_R (g^{-1} \d g) = i\,Y\,,\quad
{\hat Y}_R\, \d j_R \equiv Y_R\, j_R = i\, [j_R , Y]
\nn
\ea
acting on $\O^{(2)}$ give rise to the left and right (zero mode) charges.
Indeed, from (\ref{YL}) and (\ref{OmegaWZL}) we obtain
\ba
&&{\hat Y}_L \,\O^{(2)} = - \frac{1}{2\pi} \,{\rm tr}\,\int_{-\pi}^\pi \{\, [ Y , j_L ]\, \d g g^{-1} - \d j_L Y + j_L\, [ Y , \d g g^{-1} ]\, \}\, d x^1 =\nn\\
&&= \frac{1}{2\pi}\,{\rm tr}\, (Y\, \d\int_{-\pi}^\pi j_L\, d x^1 )= \, {\rm tr}\, (Y \d j^L_0)\quad {\rm for} \quad
j_L = \sum_{r\in{\mathbb Z}} j^L_r e^{-irx^1}\qquad
\lb{YOL}
\ea
(the contribution from the second term under the integral in (\ref{OmegaWZL}) vanishes,
as the $2D$ field $g$ is periodic in $x^1\,$ and $Y$ is constant).
Similarly, using now (\ref{OmegaWZR}), (\ref{YR}), we get
\ba
&&{\hat Y}_R \,\O^{(2)} = - \frac{1}{2\pi} \,{\rm tr}\,\int_{-\pi}^\pi \{\, [ j_R , Y ]\, g^{-1} \d g  - \d j_R Y - j_R\, [ Y , g^{-1} \d g ]\, \}\, d x^1 =
\nn\\
&&= \frac{1}{2\pi}\,{\rm tr}\,(Y \d \int_{-\pi}^\pi j_R\, d x^1 ) = \, {\rm tr}\, (Y \d j^R_0)\ ,\quad j_R = \sum_{r\in{\mathbb Z}} j^R_r e^{-irx^1} .
\qquad\quad
\lb{YOR}
\ea
In the case of the more general infinite dimensional symmetry (\ref{infgr}) which corresponds to periodic
(rather than constant) $Y=Y(x^1) = \sum_{r\in{\mathbb Z}} Y_r e^{-irx^1}$ in (\ref{YL}) and (\ref{YR}),
the vector fields $Y_L$ and $Y_R$ now act on the basic $1$-forms as
\ba
&&{\hat Y}_L (\d g g^{-1}) = i\,Y\ ,\quad {\hat Y}_L\, \d j_L = i\, [Y , j_L] - k\, Y'\ ,\nn\\
&&{\hat Y}_R (g^{-1} \d g) = i\,Y\ ,\quad {\hat Y}_R\, \d j_R = i\, [j_R , Y] + k\, Y'\ ,
\lb{Yx}
\ea
and their contractions with $\O^{(2)}$ involve {\em all} current modes:
\ba
&&{\hat Y}_L \,\O^{(2)} = \frac{1}{2\pi}\,{\rm tr}\, \int_{-\pi}^\pi Y\, \d j_L\, d x^1 = \sum_{r\in{\mathbb Z}} {\rm tr}\, (Y_r \d j^L_{-r})\ ,
\lb{YOL1}\\
&&{\hat Y}_R \,\O^{(2)} = \frac{1}{2\pi}\,{\rm tr}\, \int_{-\pi}^\pi Y\, \d j_R\, d x^1 = \sum_{r\in{\mathbb Z}} {\rm tr}\, (Y_r \d j^R_{-r})\ .
\lb{YOR1}
\ea
Of course, Eqs. (\ref{YOL}) and (\ref{YOR}) are special cases of (\ref{YOL1}) and (\ref{YOR1}), respectively (for $Y=Y(x^1) = Y_0$).

\smallskip

Eqs. (\ref{YOL}) and (\ref{YOR}), as well as (\ref{YOL1}) and (\ref{YOR1}),
have the standard Hamiltonian form (\ref{defOX}). The same is true for the periodic (or constant) {\em left} shifts of
the {\em chiral} field (we shall take $g \equiv g_L$ for concreteness).
Let $g_1 := g(-\pi)\,,\ g_2 := g(\pi)\,;$ then, from $M = g_1^{-1}g_2$ and
${\hat Y}_L \d g = i\, Y g$ we find
\ba
&&\d M M^{-1} = g_1^{-1} \d g_2\, g_2^{-1} g_1 - g_1^{-1} \d g_1\ ,\qquad {\rm hence}\lb{Mg12}\\
&&{\hat Y}_L (\d M M^{-1} ) = i\, g_1^{-1} Y (\pi)\, g_1 - i\, g_1^{-1} Y(-\pi)\, g_1 = 0 \qquad \Rightarrow\quad {\hat Y}_L\, \rho(M) = 0\nn
\ea
(cf. (\ref{defrhoK})). A simple computation using (\ref{jLR}) allows to reproduce the chiral counterpart of (\ref{YOL1})
(or of (\ref{YOL}), for constant $Y$):
\ba
&&{\hat Y}_L\, \O(g,M) = {\hat Y}_L\,\O_c(g,M) = \nn\\
&&= \frac{ik}{4\pi}\,\,{\rm tr}\, \{\int_{-\pi}^\pi \left( g^{-1} Y g \,(g^{-1} \d g)' - g^{-1} \d g \, (g^{-1} Y g)'\right) dx +
g_1^{-1} Y g_1 \d M M^{-1} \} =\nn\\
&&= \frac{ik}{2\pi}\,\,{\rm tr}\,\d\int_{-\pi}^\pi Y\, g' g^{-1} \, d x = \frac{1}{2 \pi}\, {\rm tr}\,\int_{-\pi}^\pi Y \d j (x)\, dx \ .
\lb{YLOc}
\ea

\smallskip

By contrast, the symmetry with respect to constant {\em right} shifts of the chiral field is of a rather
different nature. To begin with, we note that ${\hat Y}_R\, \d g = i\, g\, Y$ implies
\be
{\hat Y}_R (\d M M^{-1} ) = i\,g_1^{-1} g_2 Y g_2^{-1} g_1 - i\,Y = i\, ( M Y M^{-1} - Y) \equiv i\,(Ad_M - 1) Y\ .
\lb{YRM}
\ee
As a result, the contraction ${\hat Y}_R\, \O(g,M)$ of $Y_R$ with the chiral symplectic form
$\O(g,M) = \O_c(g,M) - \frac{k}{4\pi}\rho(M)\,$ (\ref{O}) depends crucially on $\rho(M)\,.$
Eqs. (\ref{Oc}) and (\ref{YRM}) give
\ba
&&{\hat Y}_R\, \O_c(g,M) = \frac{ik}{4\pi}\,\,{\rm tr}\, \{\int_{-\pi}^\pi Y (g^{-1} \d g )' dx + Y \d M M^{-1} - g_1^{-1} \d g_1 (Ad_M - 1) Y \}= \nn\\
&&= \frac{ik}{4\pi}\,\,{\rm tr}\,Y \{g_2^{-1} \d g_2 + \d M M^{-1} - Ad_M^{-1} (g_1^{-1} \d g_1) \} =\nn\\
&&= \frac{ik}{4\pi}\,\,{\rm tr}\,Y \{\d M M^{-1}+M^{-1}\d M \} \ ;
\lb{YROc}
\ea
for the last equality we have used (\ref{Mg12}) implying
\be
g_2^{-1} \d g_2 = M^{-1} g_1^{-1} (\d g_1 M + g_1 \d M) \equiv Ad_M^{-1}(g_1^{-1} \d g_1) + M^{-1} \d M\ .
\lb{MgM}
\ee
Evaluating ${\hat Y}_R\,$ on $\rho(M)\,$ (\ref{defrhoK}), we obtain
\ba
&&{\hat Y}_R\,\rho(M) = \nn\\
&&= \frac{i}{2}\,{\rm tr}\,\{ (\, (Ad_M-1)Y\, )\, (\, K_M(\d M M^{-1})\,) - \d M M^{-1} \, K_M (\, (Ad_M -1)Y \,)\} =\nn\\
&&= i\,{\rm tr}\,Y\, (Ad_M^{-1} - 1) K_M (\d M M^{-1} ) \ .
\lb{YRrho}
\ea
Note that both expressions (\ref{YROc}) and (\ref{YRrho}) only depend on the monodromy matrix.
Combining them, we get
\ba
&&{\hat Y}_R\, \O (g, M) = {\hat Y}_R\, \O_c(g,M) - \frac{k}{4\pi}\,{\hat Y}_R\,\rho(M) = \nn\\
&&=\frac{ik}{4\pi}\,{\rm tr}\,Y \{\left( Ad_M^{-1} + 1 - (Ad_M^{-1} - 1) K_M \right) ( \d M M^{-1} )\}\ .\lb{YRO1}
\ea

For $\rho(M)$ given  by (\ref{ro}) in terms of the Gauss components (\ref{M+-}) of $M\,,$
the general expression (\ref{YRO1}) leads, taking into account (\ref{dMM+-}) and (\ref{KMMM}), to
\ba
&&{\hat Y}_R\, \O (g, M) =\frac{ik}{4\pi}\,{\rm tr}\,Y \{\left( Ad_M^{-1} + 1 - (Ad_M^{-1} - 1) \right) ( \d M_+ M_+^{-1} ) -\nn\\
&&- \left( Ad_M  + 1 - (Ad_M - 1) \right) ( \d M_- M_-^{-1} ) \} =\nn\\
&&= \frac{ik}{2\pi}\,{\rm tr}\,Y (\d M_+ M_+^{-1} - \d M_- M_-^{-1} )\, .\lb{YRO}
\ea

We thus see that in the case of (e.g., constant) left translations the $1$-form $Z = \d\int_{-\pi}^\pi j (x)\, dx = 2\pi\d j_0$ (cf. (\ref{YLOc}))
is exact (and hence, closed) so that the corresponding symmetry is of Hamiltonian type.
By contrast, the forms $Z_\pm = \d M_\pm M_\pm^{-1}$ in (\ref{YRO}) satisfy the {\em Maurer-Cartan} (non-abelian flat connection)
equation $\d Z_\pm = Z_\pm^2\,,$ a fact which signals a {\em Poisson-Lie (PL) symmetry} (\cite{D1, S-T-S, D}) with respect to constant right
translations. (An infinite dimensional generalized PL symmetry with respect to non-constant translations satisfying
special boundary conditions has been found in \cite{AT}.)

\smallskip

We recall the definition of a PL group and of its Poisson action \cite{D1, S-T-S}. One introduces first
(cf. Chapter 1 of \cite{CP}) the notion of a {\em Poisson map} between two Poisson manifolds,
$\phi \,: {\cal L} \rightarrow {\cal N}\,,$ as a smooth map that preserves the Poisson bracket:
$\{ f , g \}_{\cal N} \circ \phi = \{ f \circ \,\phi , g \circ\, \phi \}_{\cal L}\quad\forall
f,g\in C^\infty ({\cal N})\,.$ Now a {\em PL group} is a Lie group $G\,$ with a Poisson structure
$\{ f , g \}_G (x)$ on it ($x\in G ,\ f,g \in C^\infty (G)$)
such that the group multiplication $m\,: G \times G \rightarrow G\,$ is a Poisson map.
(In the terminology of Lu and Weinstein \cite{LW}, a PL group is a Lie group
equipped with a {\em multiplicative} Poisson structure.) Further, a (left) {\em Poisson action} of a PL group $G\,$
on a Poisson manifold ${\cal N}\,$ is a Poisson map $\phi\,: G \times {\cal N} \rightarrow {\cal N}\,.$
In both cases the product Poisson structure on $G\times {\cal N}\ni (x,y)\,,$ is defined by
\be
\{ f , g \}_{G\times {\cal N}}\, (x,y)= \{ f (.\,,y), g (.\,,y) \}_G\, (x) + \{ f (x , . ), g (x , . ) \}_{\cal N}\,(y)
\lb{prodPB}
\ee
(in the case of a PL group ${\cal N} = G$).

\vspace{1mm}

So a PL group action preserves the Poisson bracket (PB) provided one takes into account
the {\em non-trivial PB on the group} as well. Indeed, we shall see below that the Poisson bracket
$\{ g_1 (x_1) , g_2 (x_2) \}\,,$ obtained by inverting the chiral symplectic form (\ref{O})
with $\rho (M)$ defined by (\ref{ro}),
is invariant with respect to the right shift $g(x) \to g(x)\, T\ \ (\,T\in G\, )\,$
provided that the matrix elements of $T\,$ (Poisson commuting with $g(x)$) are viewed as dynamical variables
with a non-trivial PB given by the {\em Sklyanin bracket} \cite{Sk}
\be
\{ T_1 , T_2 \}\, =\, \frac{\pi}{k}\, [ r_{12}\,, T_1 \, T_2 \,]
\lb{PBSkl}
\ee
where $r_{12}\,$ is a {\em classical $r$-matrix}.

\medskip

\noindent
{\bf Remark 2.3~} In (\ref{PBSkl}) we introduce the familiar Faddeev's shorthand notation \cite{FRT} for operations on
multiple tensor products of a (finite dimensional) vector space $V$.
(A similar notation is used sometimes for tensors in $V\otimes V\otimes \dots \otimes V\,.$)
The subscript $i = 1, 2, \dots\,$ refers to the $i$-th tensor factor: if, e.g.
$A_{12} = \sum_i X_i \otimes Y_i\otimes \id\,$ where $X_i\,,\, Y_i \,\in End\, V\,,$ then
$A_{13} = \sum_i X_i\otimes \id\otimes Y_i\,$ while $A_{21} = \sum_i Y_i \otimes X_i\otimes \id\,,$ etc.
If $P_{12} = P_{21}\  (\,P_{12}^2 = \id\,)\,$ is the permutation operator acting on $V\otimes V\,$ as
$P_{12} \, x\otimes y = y \otimes x\,,$ then $A_{21} = P_{12} A_{12} P_{12}\,.$
The Kronecker product of the operator matrices in a given basis of $V\,$ relates the compact notation with the multi-index one,
e.g. the matrix of $A_1 B_2 = A\otimes B\,$ for $A = (A^i_j)\,,\, B = (B^\ell_m)\,$ is $(A\otimes B)^{i\ell}_{jm} = A^i_j B^\ell_m\,$
(we shall always assume the lexicographic order of indices).\footnote{
Note that the relation $A_1 B_2 = B_2 A_1\,$ means that the entries of $A\,$ and $B\,$ commute,
$A^i_j B^\ell_m = B^\ell_m A^i_j \,.$ In particular, $A_1 A_2\,$ is not equal to $A_2 A_1\,$ for a matrix $A\,$ with non-commuting matrix elements.
This remark will be especially important for the quantum case, see below.}

\medskip

Respecting the unitarity of the monodromy matrix $M$ (for the general case of non-diagonal monodromy) forces one to consider
quadratic PB $ \{ g(x_1), g(x_2) \}\,$ involving a monodromy dependent $r$-matrix $r(M)\,$ \cite{BFP1, BFP}.
Thus the non-uniqueness of the splitting of the group valued field $2D$ field $g(x^0 , x^1)$ (\ref{LR}) into chiral components
and the associated freedom in the choice of the monodromy manifolds and of the $2$-form $\rho(M)\,$ satisfying (\ref{drho})
leave room for different types of symmetry of the chiral field under right shifts.
Allowing for more general non-unitary $M$, we shall be able to end up with PB involving constant $r$-matrices (for $- 2 \pi < x_1 - x_2 < 2 \pi$).
Their PL symmetry with respect to transformations satisfying (\ref{PBSkl}) is the classical counterpart
of the {\em Hopf algebraic} (quantum group) symmetry of the corresponding quantum exchange relations considered in Section 4.

\medskip

\noindent
{\bf Remark 2.4~} The above considerations apply to the case of {\em general} monodromy matrix $M$.
One can restrict, alternatively, the chiral phase space ${\cal S}_C\,$ to a subspace ${\cal S}^d_C$ of chiral fields $u(x)\,$ with
diagonal monodromy $M_p\,$ (such fields are called {\em Bloch waves} \cite{Ba, BFP}). Since the $3$-form $\theta (M_p)\,$ vanishes on the
Cartan subgroup\footnote{This follows from (\ref{can3}) applied to the (commutative) Cartan subalgebra. In general,
$\theta (M) = 0\,$ iff $\, M^{-1} \d M$ takes value in a solvable Lie subalgebra of $G_{\subbbc}\,$, cf. (\ref{prov}).},
the chiral form $\Omega_c (u,M_p)\,$ itself is already closed, in view of (\ref{deltaO}).
Hence, the freedom introduced by the chiral splitting is reduced in this case to an arbitrary
{\em closed} $2$-form $\rho (M_p)$ in (\ref{O}), $\O = \O_c - \frac{k}{4\pi}\, \rho (M_p)\,.$
Further, since $ \d M_p M_p^{-1} = M_p^{-1} \d M_p = \d \log M_p\,,$ it follows from (\ref{YROc}) that
the symmetry of such fields with respect to constant right shifts is still {\em Hamiltonian}.

So it is meaningful to denote a chiral field with a diagonal monodromy matrix $M_p$ by a different letter, $u(x)\,.$
As we shall see in the next section, the PB of the Bloch waves contain singularities depending on the eigenvalues of the monodromy matrix $M_p\,.$
Thus, at the classical level, the intertwining map $a\,$ between $u(x)$ and the chiral field $g(x)$ defined by $g(x) = u(x)\, a$
can only be regular in a restricted domain of diagonal monodromies. We shall face a similar problem when considering the quantization
in Section 4 where the above mentioned feature manifests itself in the vanishing of the quantum determinant $\det (a)\,.$

\section{Chiral phase spaces and Poisson brackets}

\setcounter{equation}{0}
\renewcommand{\theequation}{\thesection.\arabic{equation}}

\subsection{Diagonalizing the monodromy matrix}

\noindent
As anticipated in the preceding section, we shall write down the chiral group valued, twisted periodic field (\ref{PC})
\be
g(x) = (g^A_\a (x) )\ , \quad g(x+2\pi) = g(x) M
\lb{ggM}
\ee
as a product
\be
g^A_\a (x) = u^A_j (x)\, a^j_\a
\lb{gua}
\ee
of an ($x$-dependent) Bloch wave $u(x) = (u^A_j (x))\,$ and a (constant) {\em zero mode} matrix $a =  ( a^j_\a )\,.$
(We identify in this paper the Lie groups and the Lie algebras with their {\em defining} representations. Thus,
for $G= SU(n)\,$ all the indices $A, j, \a $ take values from $1$ to $n\,.$)

The Bloch waves are defined to be twisted-periodic fields with {\em diagonal} (i.e., belonging to the
subgroup corresponding to the chosen Cartan subalgebra ${\mathfrak h}$) monodromy $M_p$:
\be
u (x+2\pi ) = u(x) M_p\ ,\qquad M_p = e^{\frac{2\pi i}{k}\, \sp}\ ,\qquad \sp \in {\mathfrak h}\ .
\lb{uuMp}
\ee
(More generally, we may assume that $M_p$ has a normal Jordan form.)
Comparing (\ref{ggM}) and (\ref{uuMp}), we see that $M_p\,$ and $M\,$ are related by
\be
M_p\, a \, =\, a \, M\ .
\lb{aintertw}
\ee
Hence, if the zero modes' matrix $a$ is invertible, then $M$ is diagonalizable and its diagonal form is $M_p\,.$
To guarantee this, we have to restrict $\sp\,$ to belong to the interior $A_W$ of the positive Weyl alcove defined in Eq.(3.13) below
(for a discussion on this point, see e.g. \cite{FG1} and Section 3 of \cite{GR}).

\smallskip

The separation of variables (\ref{gua}) is analogous to the so called "vertex-IRF
(interaction-round-a-face) transformation" originally used in lattice models, see \cite{Ba}.
As the current $j(x)$ which generates the left group translations
is the same for $g(x)\,$ and $u(x)\,,$ it follows from (\ref{jLR}) that each of them satisfies the {\em classical Knizhnik-Zamolodchikov (KZ) equation}
\be
i k \frac{d g}{d x} (x) = j (x)\, g (x) \ ,\qquad i k \frac{d u}{d x} (x) = j (x)\, u (x)\ .
\lb{clKZ}
\ee
The corresponding solutions (given by ordered exponentials) can only differ by their initial values,
say at $x = -\pi\,.$ Hence, the zero modes' matrix in (\ref{gua}) is just $a = u(-\pi)\, g^{-1}(-\pi)\,.$

\medskip

We now proceed to introducing individual symplectic forms on the infinite dimensional manifold of Bloch waves
and on the zero modes' phase space.

There is an ambiguity in splitting the chiral symplectic form
$\Omega (g,M)$ (\ref{O}) into a Bloch wave and a finite dimensional (zero modes') part.
The following statement is verified by a straightforward computation.

\medskip

\noindent
{\bf Proposition 3.1~} {\it For $g(x)$ given by (\ref{gua}) and for every choice of the closed $2$-form $\omega_q (p)\,,$
the chiral symplectic form $\Omega (g,M)$} (\ref{O}) {\it splits into a sum of two closed forms, a Bloch wave form}
\ba
&&\Omega_B (u,M_p) = \Omega (u,M_p) + \omega_q (p) \ ,
\lb{OB}\\
&&\Omega (u,M_p)=
\frac{k}{4\pi} \, {\rm tr} \left\{ \int_{-\pi}^{\pi} dx \, u^{-1} (x) \, \delta u(x) (u^{-1} (x) \, \delta u(x))' + b^{-1} \delta b \,
\delta M_p \, M_p^{-1} \right\}\quad \nonumber
\ea
({\it with} $b := u(-\pi)$) {\it and a finite dimensional one},
\ba
&&\Omega (a,M_p) = \Omega_q (a,M_p) - \frac{k}{4\pi} \,\rho \,(a^{-1} \, M_p \, a) - \omega_q (p) \ ,
\label{Oq}\\
&&\Omega_q (a,M_p) = \frac{k}{4\pi} \, {\rm tr}\,
\{ \delta a \, a^{-1}  (M_p \, \delta a \, a^{-1} \, M_p^{-1} + 2\, \delta M_p \, M_p^{-1} ) \}  \ .\quad \nonumber
\ea

\smallskip

The proof of Proposition 3.1 is based on the following observations.
The $2$-form $\O (u, M_p)\,$ (\ref{OB}) is just (\ref{Oc}), with $g_C\,$ replaced by $u\,$ and $M\,$ by $M_p\,.$
In view of (\ref{deltaO}), to conclude that it is closed it is sufficient to note that $\theta (M_p )$ vanishes.
On the other hand, computing $\theta (M)\,$ for $M = a^{-1} M_p\, a\,,$ we obtain
\ba
&&\frac{k}{4\pi}\, \d \rho\,  (a^{-1} M_p\, a ) = \frac{k}{4\pi}\, \theta \,(a^{-1} M_p\, a )  = \lb{omegaaMp}\\
&&= \frac{k}{4\pi}\,{\rm tr}\, \{
(\d a a^{-1})^2 (2\, \d M_p\, M_p^{-1} + M_p\, \d a  a^{-1} M_p^{-1} - M_p^{-1} \d a  a^{-1} M_p ) -\nn\quad\\
&&- \,\d a\,  a^{-1} \d M_p M_p^{-1} (M_p \,\d a a^{-1} M_p^{-1} + M_p^{-1} \d a  a^{-1} M_p ) \}\ ,
\nn
\ea
which is equal to $\d\, \O_q (a, M_p )\,,$ so that $\O (a, M_p )$ (\ref{Oq}) is closed as well.
\eod

\bigskip

It is not difficult to verify that for infinitesimal right shifts of $a$ (leaving $M_p$ invariant) the finite dimensional form
$\O (a, M_p)$ (\ref{Oq}) transforms in the same way as the infinite dimensional one $\O_c (g, M)$ (\ref{Oc}).
Indeed, if ${\hat Y}_R\,\d\, a = i\, a Y\,,\ {\hat Y}_R\,\d M_p = 0\,,$ we find
\be
{\hat Y}_R\, \O_q (a, M_p) = \frac{ik}{4\pi}\,\,{\rm tr}\,Y \{\d M M^{-1}+M^{-1}\d M \}
\qquad {\rm for} \quad M \equiv a^{-1} M_p\, a\ ,
\lb{YROf}
\ee
thus reproducing the right-hand side of (\ref{YROc}). Taking further into
account (\ref{YRrho}), (\ref{dMM+-}) and (\ref{KMMM}), we verify the PL symmetry
of the zero mode symplectic form $\O (a, M_p)\,$ (\ref{Oq}) with respect to right shifts:
\be
{\hat Y}_R\, \O (a, M_p) = \frac{ik}{2\pi}\,{\rm tr}\,Y (\d M_+ M_+^{-1} - \d M_- M_-^{-1} )\ ,\qquad
M_+ M_-^{-1} = a^{-1} M_p\, a\ .
\lb{YROa}
\ee
There is also a Hamiltonian symmetry with respect to transformations $\ a \to e^{i t \a(p)} a\ $
with {\em diagonal} $\a(p)\,$ ($\in {\mathfrak h}$), that do not change the monodromy:
\ba
&&{\hat D}_L ( \d  a\, a^{-1} ) = i\, \a(p) \ ,\quad {\hat D}_L ( \d  M_p\, M_p^{-1} ) = 0\quad\Rightarrow\quad
{\hat D}_L\, \rho\, (a^{-1} M_p\, a) = 0\ ,\nn\\
&&{\hat D}_L \, \O (a, M_p) = -\, {\rm tr}\, ( \a(p)\, \d \sp )\ .
\lb{YLDp}
\ea

\medskip

\noindent
{\bf Remark 3.1~} In order to have the infinite and the finite dimensional parts fully decoupled, we should further extend the
chiral phase space, distinguishing the diagonal monodromy of the zero modes and that of the Bloch waves.
After doing this, the symplectic forms (\ref{OB}) and (\ref{Oq}) become completely independent.
As a corollary, on the extended phase space $M_{\mathfrak p} := u^{-1} (x) u (x + 2\pi )\,$ automatically Poisson commutes
with $a^i_\a\,$ (while $M_p\,$ and $M\,,$ related by (\ref{aintertw}), do not);
on the other hand, both $M\,$ and $M_p\,$ Poisson commute with $u (x)\,.$
To recover the original $g(x)\,,$ one has to make a reduction of the extended phase space, imposing the relations
$M_{\mathfrak p} \approx M_p\,$ as (first class) constraints and accordingly, after quantization,
$( M_{\mathfrak p} - M_p )\,{\cal H} = 0$ as a gauge condition characterizing the chiral state space ${\cal H}\,.$

\medskip

It is easy to see in the $SU(n)$ case that both $\O_B (u ,M_p)\,$ (\ref{OB}) and $\O (a, M_p)\,$ (\ref{Oq})
remain invariant with respect to multiplication of $u(x)\,,$ resp. $a\,,$ with scalar functions of $p\,;$
of course, such a transformation breaks the unimodularity property so one should further extend the corresponding phase spaces.
We shall make use of the resulting freedom as well of the one in choosing the form $\o_q\,$
to fit the quasi-classical limit of the (dynamical) $R$-matrix exchange relations
conjectured earlier in \cite{F1, F2, HIOPT, FHIOPT} and derived (by exploring the braiding
properties of the chiral correlation functions in the quantum model) in \cite{HST}.
To this end, we need the PB of the chiral zero modes and of the Bloch waves which are obtained by
inverting the corresponding symplectic forms.

\subsection{Basic right invariant $1$-forms for $G$ semisimple}

\noindent
Both the $2$-form $\O_q(a,M_p)\,$ (\ref{Oq}) and the $3$-form $\theta \,(a^{-1} M_p\, a )$ (\ref{omegaaMp})
are expressed in terms of Lie algebra valued right invariant $1$-forms. In this section we shall present
$\O_q(a,M_p)\,$ in terms of "ordinary" (${\mathbb C}$-valued) basic right invariant $1$-forms.
(The relevant notions and conventions about semisimple Lie algebras are collected for convenience in Appendix A.)

\smallskip

We shall identify, by duality, the fundamental Weyl chamber $C_W$ and the (interior $A_W$ of the) level $k$ positive Weyl alcove
with the following subsets of the Cartan subalgebra ${\mathfrak h} \ni {\sp} = \sum_{i=1}^r p_{\a_i}\, h^i\,:$
\be
C_W = \{ \sp \in {\mathfrak h}\,,\ p_{\a_i} > 0\}\ ,\quad A_W = \{ \sp \in C_W \,,\ \sum_{i=1}^r a_i^\vee p_{\a_i}  < k\ \}
\lb{CAG}
\ee
($\{ a^\vee_i \}_{i=1}^r$ are the dual Kac labels, cf. (\ref{dCL})). One can show that $\sp$ in (\ref{uuMp})
is fixed unambiguously, for a given $M \in G\,,$ by (\ref{aintertw}) iff it belongs to $A_W$ (\ref{CAG})
(see Section 3 of \cite{GR} for a detailed explanation).
In the case of $s\ell (n)\,,\ a_i^\vee\equiv 1$ and $A_W$ is just the set
\be
A^{s\ell(n)}_W = \{ \sp = \sum_{i=1}^{n-1} p_{\a_i} h^i\ ,\  p_{\a_i} > 0 \ ,\  \sum_{i=1}^{n-1} p_{\a_i} < k\, \}\ .
\lb{AWn}
\ee

The finite dimensional manifold  ${\cal M}_q\,$ with coordinates $\{ a^i_\a \,, p_{\a_j} \}\,$ and symplectic form
$\O_q (a, M_p )\,$ (\ref{Oq}) can be viewed as a {\em deformation} \cite{AF, AT}
of the symplectic manifold ${\cal M}_1\,$ obtained in the limit $k\to\infty\,.$ The role of the deformation parameter is played by
${\frac{\pi}{k}}\,$ or, better, by its exponential
\be
q = q_k := e^{-i \frac{\pi}{k}}\qquad( \, q\, \bq = 1 \ , \quad \lim_{k\to\infty} q = 1\,)\ .
\lb{qcl}
\ee
To show this, let the diagonal monodromy matrix be expressed as in (\ref{uuMp}) with ${\sp} = \sum_{j=1}^r p_{\a_j} h^j\in A_W\,,$ and
$\Theta^i\,,\, \Theta^{\pm\a}\,$  be the right invariant $1$-forms in $T^* G_{\C}\,$ corresponding to the Cartan-Weyl basis (\ref{CWbasis}), so that
\be
-i\, \d a\, a^{-1} = \sum_{j=1}^r \Theta^j h_j + \sum_{\a >0} (\Theta^\a e_\a + \Theta^{-\a} e_{-\a})
\lb{Thetas}
\ee
and, conversely,
\be
\Theta^j \, =\, -i\, {\rm tr}\, ( \d a\, a^{-1} h^j)\ ,\qquad \Theta^{\pm\a} \, =\,
-i\,\frac{(\a | \a )}{2}\, {\rm tr}\, (\d a\, a^{-1} e_{\mp\a})\ .
\lb{converse}
\ee
For a compact group $G\,$ and $a\,$ given by an unitary matrix, $a^{-1} = a^*\,$ the forms $\Theta^j\,$
are real, while $\Theta^{-\a}\,$ is complex conjugate to $\Theta^\a\,.$
We note that the matrix valued form (\ref{Thetas}) is not closed but satisfies the Maurer-Cartan relations
(defining thus a flat connection) which lead to corresponding equations for the basic $1$-forms (\ref{converse}).
We shall use, in particular,
\be
\d \,\Theta^j \, =\, i\, \sum_{\a >0} {\rm tr}\, (h^j\, [ e_\a , e_{-\a} ])\, \Theta^\a\, \Theta^{-\a}  =
\, i\, \sum_{\a >0}\, (\Lambda^j | \a^\vee )\, \Theta^\a\, \Theta^{-\a}\ ,
\lb{CM}
\ee
cf. (\ref{hee}), (\ref{CCWC}), (\ref{h-a}).

\smallskip

Inserting the expression (\ref{uuMp}) for $M_p\,$ into the second term
of $\O_q (a, M_p)$ (\ref{Oq}), we get
\be
\frac{k}{2\pi}\, {\rm tr}\, \d a a^{-1} \d M_p M_p^{-1} \,
=\, i\, {\rm tr}\, (\d a a^{-1} \d {\sp} ) =
\sum_{j=1}^r {\rm tr}\,( h_j\, \d {\sp} )\, \Theta^j =
\sum_{j=1}^r \d p_{\a_j} \Theta^j\,.
\lb{Oq1}
\ee
The first term of $\O_q (a, M_p)$ is expressed as a sum of products of conjugate off-diagonal forms
$\Theta^{\pm\a}\,,$
\be
\frac{k}{4\pi}\,{\rm tr} ( \d a a^{-1} M_p\, \d a a^{-1} M_p^{-1} ) =
\frac{k}{4\pi}\,(\bq -q) \sum_{\a >0} \frac{2}{(\a | \a )}\, [2 p_\a ]\, \Theta^\a \Theta^{-\a}
\lb{2term}
\ee
$(  [x] :=\frac{q^x - \bq^x}{q - \bq}  ) \,.$
Here we are using $[ h^j , e_{\pm\a} ] = \pm (\L^j | \a )\, e_{\pm\a}$ to derive
\ba
&&M_p \,e_{\pm \a}\, M_p^{-1}  \equiv Ad_{M_p}\, e_{\pm\a} = q^{\mp 2 p_\a} e_{\pm\a}\ ,
\lb{Adone}\\
&&p_\a := \sum_{j=1}^r (\L^j | \a )\, p_{\a_j} \equiv (\L \,|\, \a )\ ,\quad\sp \in A_W \quad
\Rightarrow\quad 0 < p_\a < k\quad \forall\ \a > 0\ ,
\nn
\ea
as well as (\ref{converse}). Combining (\ref{Oq1}) and (\ref{2term}), we arrive at
\be
\O_q (a , M_p) = \sum_{j=1}^r \d p_{\a_j} \Theta^j -
\frac{k}{4\pi} (q-q^{-1} ) \sum_{\a >0} \frac{2}{(\a | \a )}\, [2 p_\a ]\, \Theta^\a \Theta^{-\a}\ .
\lb{Ofvar}
\ee

As the weight manifold is simply connected, the closed $2$-form $\o_q (p)$ is actually exact:
\ba
&&\o_q (p) = \d \Upsilon^j (p)\, \d p_{\a_j}\ (\,\equiv \d \sum_{j=1}^r \Upsilon^j (p)\, \d p_{\a_j} \, )
= \frac{1}{2} \sum_{i,j=1}^r \o^{ij} (p)\, \d p_{\a_i} \d p_{\a_j}\ ,\nn\\
&&\o^{ij} = \frac{\partial \Upsilon^j}{\partial p_{\a_i}} - \frac{\partial \Upsilon^i}{\partial p_{\a_j}} = -\, \o^{ji} \ .
\lb{oijY}
\ea
One can therefore express the difference $\O_q - \o_q$ in (\ref{Oq}) as a kind of a gauge transformation of $\O_q$
(cf. \cite{BFP}):
\be
\O_q (a , M_p) - \o_q (p) = \O_q ( e^{i \Upsilon (p)} a , M_p )\ ,\qquad
\Upsilon (p) = \Upsilon^i (p)\, h_i \in {\mathfrak h}\ .
\lb{OeYa}
\ee
Taking further into account that the monodromy $M = a^{-1} M_p\, a\,$ (and hence the $2$-form $\rho\,$) is
invariant under the substitution $a = e^{-i \Upsilon (p)} a'\,,$ one can compute the PB
of $a$ from those of $a'$ obtained for $\o_q =0\,.$

\smallskip

The WZNW term vanishes in the undeformed limit $q\to 1$ ($k \to \infty$).
Indeed, taking into account the definition of $p_\a$ in (\ref{Adone}) and Eq.(\ref{CM}), we derive that
\ba
&&\O_1 (a , \sp) = \lim_{q\to 1} \O_q (a , M_p) =\nn\\
&&= \sum_{j=1}^r \d p_{\a_j} \Theta^j + \lim_{k\to\infty} \frac{i k }{2\pi}
\sum_{\a > 0} \frac{2}{(\a | \a )}\,\sin{\frac{2\pi p_\a}{k}}\, \Theta^\a \, \Theta^{-\a} = \lb{O1}\\
&&= \sum_{j=1}^r \d p_{\a_j} \Theta^j + i \sum_{\a > 0} \frac{2}{(\a | \a )}\,p_\a \Theta^\a \, \Theta^{-\a} =
\d \sum_{j=1}^r p_{\a_j} \Theta^j\ \equiv
- i\,\d\, {\rm tr}\,( \sp \,\d a\, a^{-1} )\qquad\quad
\nn
\ea
is not only closed but even exact by itself. As $A_W$ (\ref{CAG}) "expands" to $C_W$ for $k\to\infty\,,$
(\ref{O1}) is defined on the phase space $G\times C_W\,$ of dimension $(\dim G + {\rm rank}\, {\cal G} )$
which, after complexification, coincides with that of the (symplectic)
cotangent bundle $T^* (B)\,$ of a Borel subgroup $B\subset G_{\mathbb C}\,,$ considered in \cite{BF}.

\smallskip

The symplectic form $\O_1 (a , \sp)$ (\ref{O1}) can be readily inverted to obtain the corresponding Poisson bivector
field
\be
{\cal P}_1 = \sum_{j=1}^r V_j \wedge \frac{\d}{\d p_{\a_j}} + i \sum_{\a > 0} \frac{1}{p_\a}\, V_\a \wedge V_{-\a}\ ,
\lb{P1}
\ee
where the vector fields are dual to the corresponding basic $1$-forms (e.g. ${\hat V}_j \,\Theta^i = \d^i_j\,,\
{\hat V}_j\, \d p_{\a_i} = 0 = {\hat V}_j\, \Theta^\a\,,$ etc.; note that $p_\a$ (\ref{Adone}) is positive
for $\sp \in C_W$ and $\a >0$). The corresponding PB of the zero modes follow simply from here,
as (\ref{Thetas}) implies
\be
{\hat V}_j\,\d a = i\, h_j\, a\ ,\qquad {\hat V_\a}\, \d a = i\, e_\a\, a\ .
\lb{hatVa}
\ee

\smallskip

The expression (\ref{Ofvar}) looks very similar to (\ref{O1}), but one should remember
that $\O_q (a , M_p)$ is not closed (and is degenerate for $\sp \in A_W$ as
$[2 p_\a] = \frac{\sin \frac{2\pi}{k} p_\a}{\sin \frac{\pi}{k}}$ may vanish).
To find the PB of the zero modes, we have to invert the true symplectic form $\O (a , M_p)$ (\ref{Oq}),
taking into account the presence of the additional $2$-form $\rho\, (a^{-1}\, M_p\, a)\,.$

\subsection{PB for the zero modes}

\subsubsection{WZ $2$-forms and the classical Yang-Baxter equation}

\noindent
The correspondence between the WZ $2$-forms $\rho (M)$ satisfying $\d\rho (M) = \theta(M)$ (\ref{drho}) and
the non-degenerate constant solutions of the {\em classical Yang-Baxter equation} ("$r$-matrices")
has been first described by Gaw\c{e}dzki \cite{G} (see also \cite{FG1}) and considered in detail in \cite{FHT1}.
We proceed to review this relation, taking subsequent work, especially \cite{BFP, FehG}, into account.

We saw in Section 2.3 that the possibility of presenting $\rho (M)\,$ in the form (\ref{ro}) for a given
factorization of the monodromy matrix $M = M_+ M_-^{-1}$ implies PL symmetry with respect to right shifts
of the chiral field, see Eq.(\ref{YRO}) (or of the zero modes, Eq.(\ref{YROa})).
The so called {\em classical $r$-matrix}
gives rise to a solution of an infinitesimal version of the factorization problem \cite{D1, S-T-S}.

We shall briefly recall the basic facts about the PL symmetry \cite{CP}. The Lie algebra of a PL group $G$ possesses
a natural {\em Lie coalgebra} structure (and is, thus, a {\em Lie bialgebra} $({\cal G}, \d_{\cal G})$),
the {\em cocommutator} $\d_{\cal G} : {\cal G} \to {\cal G}\wedge{\cal G}\,$ being a (skew symmetric) linear map
satisfying the $1$-cocycle condition
\be
\d_{\cal G} ([X , Y]) = [\,\d_{\cal G}(X)\,,\, Y_1 + Y_2\, ] + [\,X_1 + X_2 \,,\, \d_{\cal G}(Y)\,] \quad\,\forall X,Y \in {\cal G}\ .
\lb{coc}
\ee
(The crucial fact is that the PB on $G\,$ induces a Lie bracket on the dual of
${\cal G}\,,\ \d_{\cal G}^* : {\cal G}^*\!\otimes\,{\cal G}^* \to {\cal G}^*\,;$
one defines, for any $\xi, \eta \in {\cal G}^*$ obtained
as differentials of appropriate functions $f , h \in C^\infty (G)$ at the identity element
$e\in G\,,\ (d\, f)_e = \xi\,,\, (d\, h)_e = \eta\,,$
\be
[\xi , \eta ]_{{\cal G}^*} \equiv \d_{\cal G}^* \, (\xi \otimes \eta ) = ( d \,\{f , h\} )_e\ .
\ee
Then the cocommutator is just $\d_{\cal G} = (\d_{\cal G}^*)^{\!*}\,,$ Eq.(\ref{coc})
being implied by the invariance of the PB with respect to the multiplication map in $G\,.$)
{\em Coboundaries} are those $1$-cocycles for which there exists a (not necessarily skew symmetric) element
$r_{12}\in {\cal G}\otimes\,{\cal G}$ such that
\be
\d_{\cal G} (X) = [X_1 + X_2 \,,\, r_{12} ]\ ;
\lb{cob}
\ee
skew symmetry of $\d_{\cal G}$ implies that $r_{12}+ r_{21}$ has to be $ad\, ({\cal G})$ invariant,
while (\ref{coc}) requires $ad$-invariance of
\be
[[ r ]]_{123} := [r_{12}, r_{13} ] + [r_{12}, r_{23} ] +
[r_{13}, r_{23} ] \in {\cal G}\otimes{\cal G}\otimes{\cal G}\ .
\lb{mCYBE-0}
\ee

If the Lie algebra ${\cal G}$ is semisimple (complex or compact), every $1$-cocycle $\d_{\cal G}$ on it is a coboundary.
Besides, then there is a one-to-one correspondence between elements
$A_{12}\,$ of $\,{\cal G}\otimes {\cal G} \,$ and linear operators $A \in End \,\,{\cal G}\,,$
\be
A_{12} \ \leftrightarrow\ A \ ,\qquad A\, X =  {\rm tr}_2\, ( A_{12} X_2 )\qquad \forall X \in {\cal G}\ ,
\lb{A-A}
\ee
the element corresponding to $^t\! A\,$ (where
${\rm tr}\, ( X A Y ) = {\rm tr}\, ( Y \, ^t\! A X )\ \ \forall X,Y \in {\cal G}$)
being just $A_{21}\,.$ The {\em polarized} Casimir operator $C_{12} \in {\rm Sym}\, ({\cal G}\otimes {\cal G} )\,$
corresponding to the quadratic invariant (\ref{CasCW}) is
\be
C_{12}\  ( = C_{21} ) = \eta^{ab}\, T_{a1}\, T_{b2}
=  h^\ell_1 h_{\ell 2} + e^\a_1\, e_{\a_2} \ .
\lb{Cas-Fadd}
\ee
The invariance of $C_{12}\,$ with respect to the $ad$-action of ${\cal G}\,$ on ${\cal G}\otimes{\cal G}\,,$
\be
[X_1 + X_2 \,,\, C_{12} ] = 0\qquad \forall X \in {\cal G}
\lb{ad-inv12}
\ee
follows from the antisymmetry of the structure constants $f_{abc}\,$ (\ref{fabc}), since
$[T_{a1} + T_{a2}\, ,\, C_{12} ] = i\, (f_{abc} + f_{acb})\, t_1^b\, t_2^c = 0\ .$
One also finds the following identities in the triple tensor product of ${\cal G}\,,$
\be
[C_{12} , C_{13} ] = [C_{13} , C_{23} ] = - [C_{12} , C_{23} ] = i f_{abc}\, t_1^a t_2^b t_3^c \ ,
\lb{CCrel}
\ee
the right hand side of (\ref{CCrel}) being the (unique, up to normalization)
${\cal G}$-invariant tensor in ${\cal G}\wedge {\cal G}\wedge {\cal G}\,.$
As the operator $C : {\cal G}\to{\cal G}$ corresponding, by (\ref{A-A}),
to $C_{12} \in  {\cal G}\otimes {\cal G}$ is just the identity operator on ${\cal G}$ since
\be
C \,T_a = {\rm tr}_2\, ( C_{12} T_{a2} ) =
\eta^{bc}\, T_b\, {\rm tr}\, ( T_c T_a ) = \eta^{bc} \eta_{ca} T_b = T_a\ ,
\lb{C-id}
\ee
the relation (\ref{A-A}) assumes the following convenient form:
\be
A_{12}\, =\, A_1\,C_{12}\qquad (\, \Leftrightarrow\  A_{21}\, =\, A_2\, C_{12}\, )\ .
\lb{A-A12}
\ee
Following \cite{S-T-S}, we shall use an operator formalism to introduce
the classical $r$-matrix. For any Lie algebra ${\cal G}\,$ and a {\em skew symmetric}
$\fr \in End \,\,{\cal G}\,, \ ^t \fr = - \fr$ (so that $r_{21} = - r_{12}\in {\cal G} \wedge{\cal G}$)
one defines the following two linear maps ${\cal G} \wedge{\cal G}\ \to \ {\cal G}\,,$
\be
[ X , Y ]_\fr\, := [\fr X , Y ] + [ X , \fr Y ] = - [ Y , X ]_\fr
\lb{XYr}
\ee
and
\be
B_\fr (X, Y) := [ \fr X , \r Y ] - \fr [X , Y ]_\r = - B_\fr (Y, X)\ .
\lb{Br}
\ee
It is easy to prove that the Jacobi identity for $[ X , Y ]_\fr\,$ is equivalent to the $2$-cocycle condition
\be
[ B_\fr (X, Y) , Z ] + [ B_\fr (Y, Z) , X ] + [ B_\fr (Z, X) , Y ] = 0\ ,
\lb{B-Jac}
\ee
hence Eq.(\ref{XYr}) defines a {\em second Lie bracket} on ${\cal G}\,$ (one denotes ${\cal G}\,$ equipped with
it by ${\cal G}_\fr$) whenever (\ref{B-Jac}) holds. An obvious (bilinear) sufficient condition this to happen is
the validity of (the operator version of) the {\em modified} classical Yang-Baxter equation (CYBE)
\be
B_\fr (X, Y) = \a^2\, [ X , Y ]
\lb{MCYBEa}
\ee
for some constant $\a\,.$ If $\a\ne 0\,,$ in the {\em complex} case
one can always reduce (\ref{MCYBEa}), by rescaling $\fr\,,$ to
\be
B_\fr (X, Y) = -\, [ X , Y ] \quad\Leftrightarrow\quad
\fr^\pm [ X , Y ]_\fr  = [ \fr^\pm X , \fr^\pm Y ] \ , \quad \fr^\pm := \fr\pm \id\quad
\lb{req}
\ee
(the minus sign in the right-hand side of the first equation is crucial for what follows).
Hence, the maps $\fr^\pm : {\cal G}_\fr \to {\cal G}\,$ are Lie algebraic homomorphisms,
their images ${\cal G}_\pm := \fr^\pm\, {\cal G}_\fr\,$ are Lie subalgebras of ${\cal G}\,$
and, since $\frac{1}{2}\, (\fr^+ - \fr^-) = \id\,,$ any $X\in {\cal G}$ can be decomposed in a unique way as
\be
X = X_+ - X_-\ ,\quad X_\pm := \frac{1}{2}\,\fr^\pm X \in {\cal G}_\pm\quad{\rm so \ that}\quad \fr X = X_+ + X_-
\lb{rX}
\ee
(this is the {\em infinitesimal form of the factorization theorem} of \cite{S-T-S})).
One can prove, using (\ref{A-A12}) and (\ref{CCrel}), that the modified CYBE (\ref{req})
is equivalent to the following equation (in $\,{\cal G}\otimes {\cal G}\otimes {\cal G}\,$)
for the classical $r$-matrix $r_{12}= - r_{21} \in {\cal G}\wedge {\cal G}\,:$
\be
[[ r ]]_{123} = [r_{12}, r_{13} ] + [r_{12}, r_{23} ] + [r_{13}, r_{23} ] = [C_{12}, C_{23} ]\ .
\lb{mCYBE}
\ee
The matrices corresponding to the operators $\fr^\pm\,$ are,  accordingly,
\be
r_{12}^\pm = r_{12}  \pm C_{12}\ .
\lb{rpmcl}
\ee
Applying (\ref{ad-inv12}), it is straightforward to show that they both satisfy the ordinary CYBE:
\be
[[ r^\pm ]]_{123} = 0\ .
\lb{CYBE}
\ee

\medskip

\noindent
{\bf Remark 3.2~}
In general, (non-skew-symmetric) solutions $r_{12} \in {\cal G}\otimes{\cal G}$
of the CYBE $[[ r ]]_{123} = 0$ (\ref{CYBE}) are called {\em non-degenerate}\, if their symmetric part,
$\frac{1}{2}\,(r_{12} + r_{21})\,$ is such. In this case the corresponding Lie bialgebra $({\cal G}, \d_{\cal G})$
(cf. (\ref{cob})) is called {\em factorizable}. The other extreme case $r_{12} + r_{21} = 0\,$ is
usually referred of as "the classical unitarity condition" \cite{RS}.

\medskip

As we shall see below, Eqs. (\ref{mCYBE}) (or (\ref{CYBE})) imply the Jacobi identity of the chiral PB.

\smallskip

The operator formalism described above implies the following

\smallskip

\noindent
{\bf Proposition 3.2~}
{\em Let $\rho (M) = \frac{1}{2}\,{\rm tr}\, (\d M M^{-1} K_M (\d M M^{-1}))\,$ (\ref{defrhoK}),
where $K_M\in End\,\,{\cal G}\,$ is defined in terms of the skew symmetric operator $\fr\,$
(for $M\,$ such that $(\fr^+ - Ad_M\, \fr^-)$ is invertible) by
\be
K_M = ( \fr^+ + Ad_M\,  \fr^-)\, (\fr^+ - Ad_M\, \fr^- )^{-1}\ .
\lb{KofM}
\ee
Then $\rho (M)\,$ satisfies $\d\, \rho (M)=\theta (M)$ (\ref{drho}) whenever $r\,$ solves the modified CYBE (\ref{req}).}

\smallskip

Note that
$K_{\1} = (\fr^+ + \fr^-)(\fr^+ - \fr^-)^{-1} = \fr\,;$ the skew symmetry of $K_M ,\ ^t\!K_M = - K_M\,$
follows from that of $\fr\,,$ taking into account the orthogonality of $Ad_M\,,$ $^t(Ad_M) = Ad^{-1}_M\,$ and the equality
\be
(\fr^- + \fr^+ Ad_M^{-1}) (\fr^+ - Ad_M \fr^-) = -\, (\fr^- - \fr^+ Ad_M^{-1}) (\fr^+ + Ad_M \fr^-)\ .
\ee

\smallskip

The proof of Proposition 3.2 can be obtained by adapting a more general statement in \cite{FehG}
to the case of monodromy independent $\fr\,.$

\smallskip
The importance of (\ref{KofM}) stems from the fact that the $r$-matrix $r_{12} \in {\cal G}\wedge\,{\cal G}\,$
corresponding to the same operator
$\fr$ appears in the PB of the the zero modes as well in those of the chiral field $g(x)$ \cite{BFP};
we shall provide a proof in Section 3.5 below.
For ${\cal G}\,$ {\em compact}, the modified CYBE (\ref{MCYBEa}) only has solutions for {\em real}
$\a\,,$ see \cite{CGR}. Thus Eq.(\ref{req}) cannot hold in this case.
The problem can be overcome by a more general Ansatz for $\rho (M)\,,$ still of the type (\ref{KofM}),
but allowing the operator $\fr\,$ to depend on $M\,$ \cite{BFP1, BFP}. Then the Jacobi identity for the emerging
PB is equivalent to a generalized version of the modified {\em dynamical}
CYBE (see below), including differentiation in the group parameters, for $\fr(M)\,.$

Alternatively, if we insist on working with monodromy independent $r$-matrices, we have to
extend the chiral phase space and its symplectic form (\ref{O}) to monodromy
(and hence, due to (\ref{aintertw}), zero mode) matrices
belonging to the {\em complexified} group, $M\in G_{\mathbb C}\,.$

\smallskip

The fact that $\rho (M)\,,$ given by (\ref{defrhoK}) and (\ref{KofM}), is a solution of (\ref{drho})
follows also from the factorization (\ref{M+-}) of the monodromy matrix $M\,$
into Gauss components, see \cite{G, FG, FHT1}. Indeed, if $M=M_+ M_-^{-1}$ (so that (\ref{diagMM}) holds), the $1$-forms
$X_\pm := \d M_\pm M_{\pm}^{-1}\,$ and
$Y_\pm = Ad_{M_\pm}^{-1} (\d M_\pm M_{\pm}^{-1}) = M_{\pm}^{-1} \d M_\pm\,$
take values in the respective Borel subalgebras ${\cal G}_\pm\,.$
Then (\ref{dMM+-}), (\ref{rX}) and (\ref{KofM}), which implies
\ba
&&K_M \, (\fr^+ - Ad_M\, \fr^- ) = \fr^+ + Ad_M\, \fr^-\quad \Leftrightarrow\lb{KMMrM}\\
&&K_M \, Ad_{M_+}\, (Ad^{-1}_{M_+} \fr^+ - Ad^{-1}_{M_-}\, \fr^- ) =
Ad_{M_+}\, (Ad^{-1}_{M_+} \fr^+ + Ad^{-1}_{M_-}\, \fr^- )\ ,\nn
\ea
lead to (\ref{KMMM}), proving thus (\ref{ro}) and hence, (\ref{drho}).
Comparing the second relation in (\ref{KMMrM}) and (\ref{rX}), we see that $K_M\,$
can be presented in the following simple form \cite{FHT1}:
\be
K_M = Ad_{M_+} \fr\, Ad_{M_+}^{-1}\ .
\lb{altKM}
\ee
The factorization of $M$ into Gauss components is related to a special solution of (\ref{req}) given by
\be
\fr\, h_i = 0\ ,\quad \fr\, e_{\pm\a} = \pm\, e_{\pm \a}\ ,\quad \a > 0 \ .
\lb{re1}
\ee
Using (\ref{A-A12}), (\ref{Cas-Fadd}) and (\ref{CasCW}), we obtain the corresponding solution of (\ref{mCYBE}),
the {\em standard} classical $r$-matrix:
\be
r_{12}\,\equiv\, \fr_1\, C_{12}\, = \sum_{\a >0} ( e_{\a 1} e_{-\a 2} - e_{-\a 1} e_{\a 2} )\quad (\, = - r_{21}\,)\ .
\lb{rstandard}
\ee
We shall restrict ourselves in what follows to $G=SU(n)$ (so that ${\cal G}_{\subbbc} = s\ell (n)$) and to the
$2$-form $\rho\,$ (\ref{ro}) corresponding to the factorization of $M$ into Gauss components
(thus related to $r_{12}\,$  (\ref{rstandard})).
In this case ${\cal G}_\pm\,$ are just the upper and lower triangular traceless matrices, respectively,
the uniqueness of the decomposition being guaranteed by the additional condition that the diagonal elements of
$X_+\,$ and $- X_-\,$ are equal (cf. (\ref{diagMM})). This choice is dictated by the quasi-classical correspondence,
if we postulate exchange relations for the quantized chiral field $g(x)\,$ in terms of the
standard \cite{D, J, FRT} constant $U_q s\ell (n)\,$ {\em quantum $R$-matrix}.
It is appropriate, assuming that the complexification only concerns the zero modes $a^j_\a\,$
and does not affect the properties of the $2D$ "gauge invariant" field $g(x^+ , x^-) \in G\,$
(which should still transform covariantly, in the usual sense, under both left and right shifts of the compact group $G$).

\subsubsection{Extending the zero modes' phase space}

\noindent
For the sake of simplicity we begin by exploring the PB for the undeformed ($q=1$) case corresponding to the symplectic form
\be
\O (a, \sp) = \lim_{q\to 1}\, ( \O_q (a , M_p) - \o_q (p))  = \O_1 (a , \sp) - \o_1 (p)
\lb{OG1}
\ee
where $\O_1 (a , \sp)$ is given by (\ref{O1}), and $\o_1 (p)$
is the limit of $\o_q(p)$ (\ref{oijY}). This is readily done using the Poisson bivector field (\ref{P1})
and the prescription after (\ref{OeYa}):
\ba
&&\{ p_{\a_j} , p_{\a_\ell} \} = 0\ ,\qquad \{ a^j_\a , p_{\a_\ell} \} = i\,(h_\ell)^j_s\, a^s_\a\ ,
\lb{PBapG}\\
&&\{ a_1 , a_2 \} = \left(\, \sum_{j\ne \ell} \o^{j \ell}(p)\, h_{j1} h_{\ell 2}  -
i \sum_\a \frac{e_{\a 1} e_{-\a 2}}{p_\a}\, \right)\, a_1 a_2\qquad\quad
\lb{PBa1a2G}
\ea
(note that the last summation goes over {\em all, positive and negative}, roots $\a$).

\smallskip

Going to the special case $G=SU(n)$ we first observe that the assumption $\det a = 1$ (as part of the requirement $a = (a^j_\a ) \in G$)
is more restrictive than what is needed to ensure that the classical chiral field $g\,$ (\ref{gua}) belongs to $G\,,$ i.e. that
$\det u\, . \det a = 1\,.$ We shall use the ensuing freedom to impose a Weyl invariant relation between $a$ and the weight variables $p\,.$
This can be done most conveniently in the barycentric parametrization of the $s\ell (n)$ roots and weights
presenting the simple roots as $\a_\ell = \e_\ell - \e_{\ell +1}$ for $(\e_i |\, \e_j ) = \d_{ij}$ so that the
root space is the hyperplane in the auxiliary $n$-dimensional Euclidean space spanned by $\{ \e_i \}_{i=1}^n$
orthogonal to $\e := \sum_{i=1}^n \e_i$ (see Appendix A). A linear combination of the weights can be expressed,
accordingly, in terms of barycentric coordinates $p_i \,,\ i = 1,..., n\,$ as
\be
p = \sum_{i=1}^n p_i\, \e_i\ ,\quad ( p \,| \e ) = 0\qquad\Rightarrow\qquad \sum_{i=1}^n p_i =: P = 0\ .
\lb{bary}
\ee
Using (\ref{al}), we find, for $p = \sum_{\ell =1}^{n-1} p_{\a_\ell} \L^\ell$
\be
p_i = \sum_{\ell =i}^{n-1} p_{\a_\ell} - \frac{1}{n}\,\sum_{\ell =1}^{n-1} \ell\, p_{\a_\ell}
\quad\Rightarrow\quad  p_{\a_i} \, ( \, \equiv p_{\a_{i\, i +1}} \, ) = p_i - p_{i+1}\ .
\lb{slnweights1}
\ee
Further, from (\ref{slnroots}) and (\ref{Adone}) it follows that in general
\be
p_{\a_{ij}} := \sum_{\ell = 1}^{n-1} ( \L^\ell |\, \a_{ij} )\, p_{\a_\ell} = p_i - p_j \equiv p_{ij} \ .
\lb{slnweights2}
\ee
The action of the $s\ell(n)$ Weyl group ${\cal S}_n$ in the orthonormal basis is easy to describe:
the reflection $s_i$ with respect to the root $\a_i$ ($i=1,\dots ,n-1$) is equivalent to
the transpositions $\e_i \leftrightarrow \e_{i+1}\,,\ p_i \leftrightarrow p_{i+1}$.
It is natural to assume that ${\cal S}_n$ also permutes the rows $a^j = (a^j_\a )$ of the matrix $a$, as
the upper index $( j )$ refers to the weights, cf. (\ref{PBapG}). We shall equate the determinant of $a$
which changes sign under odd permutations of rows to a natural pseudoinvariant of the weights $p_i\,:$
\be
D(a) := \det a = \prod_{1\le i<j\le n} p_{ij} =: {\cal D}(p)\ .
\lb{DaDp1}
\ee
We shall exhibit the effect of this constraint in the simplest (rank $r=1$) case corresponding to $G=SU(2)$
in which $\o_q(p) = 0$ so that the form (\ref{OG1}) involves no ambiguity. To see what is going on, we
parametrize the matrix $a$ by a $2$-component spinor $z = (z_1 , z_2 )$ and its complex conjugate ${\bar z}\,:$
\ba
\lb{ap1}
a = \begin{pmatrix}z_1&z_2\cr - {\bar z}_2& {\bar z}_1\end{pmatrix}\,,\
a^{-1} = \frac{1}{D(a)}\,\begin{pmatrix}{\bar z}_1&-z_2\cr {\bar z}_2& z_1\end{pmatrix}\,,\quad
D(a) = {\bar z} z  := {\bar z}_1 z_1 + {\bar z}_2 z_2\,.\quad
\lb{apD2}
\ea
For $D(a) = p_{12} \equiv p\,$ (according to (\ref{DaDp1})) the (exact) $2$-form $\O_1\,$ (\ref{O1})
can be written as
\be
\O_1 = \d \phi\ ,\quad \phi = \frac{1}{2i}\,{\rm tr}\,\left\{\begin{pmatrix}p&0\cr 0& -p\end{pmatrix} \d a a^{-1} \right\}
= p\,\frac{{\bar z} \d z - z \d {\bar z}}{2 i\,{\bar z} z} = \frac{1}{2 i}\, ({\bar z} \d z - z \d {\bar z} )\ .
\lb{Ophi}
\ee
Thus, for $D(a)\, (= {\bar z} z ) = p\,,\ \O_1\,$ coincides with the standard K\"ahler form on ${\C}^2$:
\be
\O_1 (a , \sp) = i\, \d z \d {\bar z} \qquad (\,{\rm for}\ {\bar z} z = p )\ .
\lb{On2}
\ee
The non-trivial PB,
\be
\{ z_\a , {\bar z}_\b \} = i\, \d_{\a\b}\quad\Rightarrow\quad \{ z_\a , p \} = i\, z_\a\ ,\quad
\{ {\bar z}_\a , p \} = - i\, {\bar z}_\a\ ,
\lb{classCCR2}
\ee
reproduce the classical limit of the canonical commutation relations for a pair of
$SU(2)$ spinors of creation ($z_\a$) and annihilation (${\bar z}_\a$) operators \cite{Sch, B62}
($p=z{\bar z}$ playing the role of the classical weight equal to twice the isospin).

\bigskip

\noindent
{\bf Remark 3.3~} Note that, had we set $D(a) = 1$ (instead of (\ref{DaDp1})), we would have obtained the awkward PB
$\,\{ z_1, {\bar z}_1 \} = \frac{i}{p}\, | z_2 |^2\ ,\ \ \{ z_2, {\bar z}_2 \} = \frac{i}{p}\, | z_1 |^2$ $( | z_\a |^2 = z_\a {\bar z}_\a )$
instead of (\ref{classCCR2}).

\bigskip

We shall use in what follows the $n \times n$ Weyl matrices
$\{ e^{~j}_i \}\,,\, i,j=1,\dots ,n\,,$ $(e^{~j}_i )^\ell_{~k} = \d^\ell_i \d^j_k\,$ satisfying
\be
e^{~j}_i\, e^{~\ell}_k \, = \, \d^j_k \, e^{~\ell}_i\ ,\qquad
{\rm tr}\,( e^{~j}_i e^{~\ell}_k )\, = \, \d^\ell_i \d^j_k\ ,\qquad
\sum_{i=1}^n e^{~i}_i = \id_n\ .
\lb{eij}
\ee
In the $n$-dimensional fundamental representation, the Cartan algebra duals of the $s\ell (n)$ roots and weights,
cf. (\ref{cdual}), are expressed in terms of the diagonal Weyl matrices $e^{~i}_i$
by replacing in (\ref{al}) $\e_i \to\, e^{~i}_i$ and $\,\a_\ell \to h_\ell\,,\ \L^j \to h^j$:
\ba
&&h_\ell = e^{~\ell}_\ell - e^{~\ell +1}_{\ell +1}\ ,\quad h^j =
(1-\frac{j}{n}) \sum_{r=1}^j e^{~r}_r - \frac{j}{n} \sum_{r=j+1}^n e^{~r}_r\ ,\nn\\
&&{\rm tr}\,(h_\ell h^j) = \d^j_\ell\ ,\quad 1\le j,\ell \le n-1\ .
\lb{al-d}
\ea
The condition that $\sp\,$ belongs to the interior of the level $k$ positive Weyl alcove (\ref{AWn}) becomes
\be
A^{s\ell(n)}_W = \{ \sp\
(= \sum_{\ell =1}^{n-1} p_{\ell \ell+1} h^\ell)
= \sum_{i=1}^n p_i\, e^{~i}_i \ \mid\, P=0\ ;\ 0 < p_{ij} < k\ ,\ \forall\, i<j\, \}\ ,
\lb{AWn1}
\ee
and the raising (lowering) operators are $e_{\a_{ij}} = e^{~j}_i\,$ for $i<j\,$ ($j<i\,$).
From (\ref{al}) and (\ref{Cas-Fadd}) we get
\ba
&&\s_{12} := \sum_{\ell = 1}^{n-1} h^\ell_1 h_{\ell 2} =
\sum_{j=1}^n (e_j^{~j})_1 (e_j^{~j})_2 - \frac{1}{n}\,\id_{12}
\quad\Rightarrow
\lb{Cn-sigma}\\
&&C_{12} = \s_{12} + \sum_{i\ne j} (e_i^{~j})_1 (e_j^{~i})_2 = P_{12} - \frac{1}{n}\,\id_{12}\ ,\quad
P_{12} = \sum_{i,j=1}^n (e_i^{~j})_1 (e_j^{~i})_2
\nn
\ea
($(P_{12})^{ij}_{i'j'} = \d^i_{j'} \d^j_{i'}$ is the permutation matrix) which is a well known formula for the
polarized Casimir operator in the tensor square of the defining $n$-dimensional representation of $s\ell(n)\,.$

\smallskip

Proceeding to the general (deformed, $SU(n)\,,\ n\ge 2$) case,
we shall view ${\mathcal M}_q$ as a submanifold of co-dimension $2$ of the $n(n+1)$ dimensional phase
space ${\mathcal M}_q^{\rm ex}$ of all $\{ a_{\alpha}^j \,,\, p_i\}\,.$ The constraint $P\approx 0$ in (\ref{AWn1}) will be supplemented by
a gauge condition which is a $q$-deformed version of (\ref{DaDp1}),
\begin{equation}
\label{Dpq}
D(a) \approx {\mathcal D}_q (p) := \prod_{i < j} \ [p_{ij}]\ , \qquad [p] = \frac{q^p - q^{-p}}{q-q^{-1}}\qquad
{\rm for}\qquad q = e^{-i\frac{\pi}{k}}
\end{equation}
(cf. (\ref{qcl})). The determinant $D(a)$ may be defined by either one of the relations
\begin{equation}\label{Da}
\epsilon_{i_n \dots i_1} a^{i_n}_{\alpha_n} \dots a^{i_1}_{\alpha_1} = D(a)\, \epsilon_{\alpha_n \dots \alpha_1}\ ,\quad
a^{i_n}_{\alpha_n} \dots a^{i_1}_{\alpha_1} \epsilon^{\alpha_n \dots \alpha_1} = \epsilon^{i_n \dots i_1} D(a)
\end{equation}
(we assume summation over repeated upper and lower indices and normalize the totally skew symmetric tensors by
$\epsilon_{n\dots 1} = 1 = \epsilon^{n\dots 1}$).
The corresponding adjugate matrix $A=(A^\alpha_j )$ such that
\be
\lb{aA}
a^i_\alpha A^\alpha_j = D(a)\, \delta^i_j\ ,\quad A^\alpha_i a^i_\beta = D(a)\, \delta^\alpha_\beta\qquad {\rm i.e.,}\quad
(a^{-1})^\alpha_i = \frac{A^\alpha_i}{D(a)}
\ee
can be determined from either one of the following equivalent equations:
\ba
&&a^{i_n}_{\alpha_n} \dots \, \widehat{a^{i_\ell}_{\alpha_\ell}}\, \dots a^{i_1}_{\alpha_1}
\epsilon^{\alpha_n \dots \alpha_\ell\dots\alpha_1} = \epsilon^{i_n \dots i_\ell\dots i_1} A_{i_\ell}^{\alpha_\ell}\ ,\nonumber\\
&&\epsilon_{i_n \dots i_\ell\dots i_1} a^{i_n}_{\alpha_n} \dots \, \widehat{a^{i_\ell}_{\alpha_\ell}}\, \dots a^{i_1}_{\alpha_1} =
A_{i_\ell}^{\alpha_\ell}\, \epsilon_{\alpha_n \dots \alpha_\ell\dots \alpha_1}\ ,
\label{AA}
\ea
the hat meaning omission (note that missing indices in the left hand side, e.g. $\alpha_\ell$ in the second equation,
correspond to summed up ones in the right hand side).

\smallskip

The choice (\ref{Dpq}) will lead to PB relations expressed in terms of a standard classical dynamical $r$-matrix
\cite{GN, BDF, Felder}. Upon quantization it will reproduce for $n=2$ the Pusz-Woronowicz $q$-deformed oscillators \cite{PW} (see Section 5.1
below). For the time being we only note that the expression ${\cal D}_q(p)$ (\ref{Dpq}) (just as ${\cal D}_1(p) = {\cal D}(p)$ (\ref{DaDp1})) is a
pseudoinvariant with respect to the $su(n)$ Weyl group. As $[p_{ij}] > 0$ for $0< p_{ij} < k$ ($i<j$), ${\cal D}_q(p)$ and hence, $D(a)$ are
positive if and only if $\sp$ is an internal point of the positive Weyl alcove, (\ref{AWn1}).

\smallskip

One can verify, using $\sum_{s=1}^n e_s^{~s} = \id\,,$ that the following equality holds:
\be
\lb{eq-pP}
p:=\sum_{s=1}^n p_s e_s^{~s} = \left( \frac{1}{n} \sum_{s=1}^n p_s \right) \id \,+\, \sum_{\ell = 1}^{n-1} p_{\ell \ell +1} h^\ell\qquad{\rm for}\quad
h^\ell = \sum_{s=1}^\ell e_s^{~s} - \frac{\ell}{n}\, \sum_{s=1}^n e_s^{~s}\ .
\ee
We shall assume that the {\em extended} diagonal monodromy matrix is given by
\be
M_p = e^{\frac{2\pi i}{k} p} = \bq^{2\, \left(\frac{1}{n} P + \sp \right)}\ ,\qquad \sp \in A_W\ ,
\lb{monex}
\ee
cf. (\ref{eq-pP}), (\ref{uuMp}), (\ref{AWn1}). Further, it is convenient to expand the form $\delta a a^{-1}$
(having non-zero trace in the {\em extended}, non-unimodular zero mode case)
into $n^2$ basic right-invariant forms $\Theta_k^j\,$ using the $n \times n$ Weyl matrices (\ref{eij}):
\begin{equation}
\label{eq90}
-i \, \delta \, a \, a^{-1} = e_j^{~\ell}\, \Theta_\ell^j \ \ ( \equiv \sum_{j , \ell = 1}^n e_j^{~\ell} \, \Theta_\ell^j \, )\quad
\Leftrightarrow \quad \Theta_\ell^j = -i \, {\rm tr}\, (e_\ell^{~j} \, \delta \, a \, a^{-1}) \ .
\end{equation}
Taking into account the Maurer-Cartan equations
\begin{equation}
\label{eq91}
\delta ( \delta\, a \, a^{-1} ) = ( \delta\, a \, a^{-1} )^2\qquad\Rightarrow\qquad \delta \, \Theta_\ell^j = i \, \Theta_s^j \, \Theta_\ell^s\ ,
\end{equation}
we can thus write the extension of the form $\Omega_q (a,M_p)$ (\ref{Ofvar}) (for $G=SU(n)$) as
\begin{equation}
\label{eq92}
\Omega_q^{\rm ex} = \sum_{s=1}^n \delta \, p_s \, \Theta_s^s - \frac{k}{4\pi} (q - q^{-1})
\sum_{j < \ell} \ [2 p_{j\ell}] \,\Theta_{\ell}^j \, \Theta_j^{\ell}\ .
\end{equation}
So the second term in the right hand side is not sensitive to the extension, while
the first ($k$-independent) one can be rewritten singling out the "total momentum" $P$ (\ref{bary}) as
\begin{equation}
\label{eq93}
\sum_{s=1}^n \delta \, p_s \, \Theta_s^s = \sum_{j=1}^{n-1} \, \delta \, p_{jj+1} \, \Theta^j + \delta \, P \, \Theta^n\ ,
\end{equation}
where
\ba
&&\Theta^j = (1- \frac{j}{n})  \sum_{s=1}^j \Theta_s^s - \frac{j}{n} \sum_{s=j+1}^n \Theta_s^s  \ , \qquad j=1,\dots, n-1\ ,\nn\\
&&\Theta^n = \frac{1}{n} \sum_{s=1}^n \Theta_s^s = - \frac{i}{n} \frac{\delta D(a)}{D(a)} \ .
\label{eq94}
\ea
Hence (cf. (\ref{Ofvar})),
\begin{equation}
\label{Oqex}
\Omega_q^{\rm ex} = \Omega_q (a,M_p) - \frac{i}{n}\, \delta P \frac{\delta D(a)}{D(a)}\ .
\end{equation}
As the $2$-form $\rho(M)$ is only restricted by (\ref{drho}), and $\theta(M)$ does not change upon extension (this is easy
to check using $M^{-1} \d M    \ \to \ M^{-1} \d M + \frac{2\pi i}{kn}\, \d P$), we can assume that
$\rho^{\rm ex} = \rho\,,$
and shall look for a closed, Weyl invariant $2$-form $\omega_q^{\rm ex}(p)$ such that the extended version of (\ref{Oq}),
\be
\Omega^{\rm ex} = \Omega_q^{\rm ex} - \frac{k}{4\pi} \,\rho  - \omega_q^{\rm ex}(p)\ ,
\lb{Oex}
\ee
reduces to $\Omega (a,M_p)$ for $D(a) \approx {\mathcal D}_q (p)$ and $P \approx 0\,.$
More specifically, we shall demand that
\be
\Omega^{\rm ex} = \Omega(a,M_p) - i \, \delta P \, \delta \chi \ ,
\qquad \chi := \frac{1}{n} \log \frac{D(a)}{{\mathcal D}_q (p)}\ .
\label{eq97}
\ee
Taking into account the definition of ${\cal D}_q(p)$ (\ref{Dpq}) and (\ref{Oqex}), this means that
\be
\omega_q^{\rm ex}(p) - \omega_q (p)= \frac{i}{n}\,  \frac{\d\,{\mathcal D}_q (p)}{~{\mathcal D}_q (p)}\, \d P =
\frac{i}{n}\, \sum_{j < \ell} \, \frac{\d\, [p_{j\ell}]}{~[p_{j\ell}]}\, \d P
= \frac{i \pi}{k n} \sum_{j < \ell} \,{\rm cot} \left( \frac{\pi}{k} \, p_{j\ell} \right) \delta p_{j\ell}\, \d P \ .
\lb{oex-o}
\ee
The (closed) $2$-form $\omega_q(p)$ is by definition $P$-independent while, splitting the terms proportional
to $\delta P$ in the most general expression for $\omega_q^{\rm ex}(p)\,,$ we obtain
\begin{equation}
\label{oexqp}
\omega_q^{\rm ex}(p) : = \frac{1}{2}\,\sum_{j\ne \ell} f_{j\ell}(p)\, \delta p_j \,\delta p_\ell =
\sum_{j< \ell} c_{j\ell}(p) \,\delta p_{j\ell}\, \delta P +
\sum_{j<\ell< m} d_{j\ell m} (p)\, \delta p_{j\ell}\, \delta p_{\ell m}
\end{equation}
where  $f_{j\ell}(p) = - f_{\ell j} (p)$ and
\begin{equation}
\label{f-and-c}
n \sum_{j<\ell} c_{j\ell} (p)\, \delta p_{j\ell} = \sum_{j<\ell} f_{j\ell}(p)\, \delta p_{j\ell}\ ,\quad
n\, d_{j\ell m}(p) = f_{j\ell}(p) + f_{\ell m}(p) - f_{j m}(p)\ .
\end{equation}
To derive (\ref{f-and-c}), we have used the identities
\ba
&&n p_\ell = P + P_\ell\ ,\quad P_\ell := \sum_s p_{\ell s}\ ,\nn\\
&&\sum_{j<\ell} f_{j\ell} (p)\, \delta p_{j\ell}\, \delta P_\ell =
\sum_{j<\ell <m} (f_{j\ell}(p) + f_{\ell m}(p) - f_{j m}(p))\, \delta p_{j\ell}\, \delta p_{\ell m}\ .\qquad\qquad
\lb{Pp}
\ea
It follows from (\ref{oex-o}) -- (\ref{f-and-c}) that the corresponding unextended $p$-dependent $2$-form is
\be
\o_q(p) = \frac{1}{n}\,\sum_{j<\ell< m} (f_{j\ell}(p) + f_{\ell m}(p) - f_{j m}(p))\,  \delta p_{j\ell}\, \delta p_{\ell m}\ .
\lb{unextoq}
\ee
Note that the expression (\ref{unextoq}) vanishes for $n=2$ as it should, due to the restrictions on the summation indices.

\bigskip

\noindent
{\bf Remark 3.4~} One could write a more general Weyl invariant second constraint $\chi \approx 0\,$ replacing ${\cal D}_q (p)$
(\ref{Dpq}) in the definition of $\chi$ (\ref{eq97}) by
\be
\Phi(p) = \prod_{j<\ell} F (p_{j\ell}) \qquad{\rm for} \quad F(p) = - F(-p)\ .
\lb{Phigen}
\ee
(It requires a suitable change in Eq.(\ref{oex-o}) where the logarithmic derivative of ${\cal D}_q (p)$ has to
be replaced by that of $\Phi(p)\,.$) Assuming that $\Phi(p)$ is {\em proportional} to
${\cal D}_q(p)$  gives rise to a $\o_q^{\rm ex}$ of type (\ref{oexqp}) with
\be
f_{j\ell}(p) = i \,\frac{F'(p_{j\ell})}{F(p_{j\ell})} = i \, \frac{\pi}{k}\, \left ( {\rm cot} ( \frac{\pi}{k}\, p_{j\ell} )
- \b (\frac{\pi}{k}\, p_{j\ell})\right)\ ,\ j\ne \ell \quad (\, \b (p) = - \b (-p))\ .
\label{f01}
\ee
This freedom fits the quasi-classical limit of the general solution of the quantum dynamical Yang-Baxter equation found in \cite{I2}.
Identifying $F(p)$ with the "quantum dimension" $[p]\,$ is equivalent to making the Ansatz
\begin{equation}
\label{2Max-ex}
f_{j\ell} (p) = i \left(\frac{\partial V^\ell}{\partial p_j} - \frac{\partial V^j}{\partial p_\ell}\right) \ ,\quad V^\ell(p) := \sum_{r<\ell} \log\, [p_{r\ell} ]\qquad
(\, \omega^{\rm ex}_q(p) = i\,\d V^\ell (p)\, \d p_\ell\,)\ .
\end{equation}
As one can see from (\ref{dynr}) below, this choice (which amounts to setting $\b(p)=0$ in (\ref{f01})) simplifies the expression for the classical dynamical $r$-matrix $r_{12}(p)\,.$

\medskip

\noindent
{\bf Remark 3.5~} We observe that Eqs. (\ref{oexqp}), (\ref{f01}) define a {\em non-trivial} cohomology class of closed meromorphic $2$-forms.
(The Ansatz (\ref{2Max-ex}) does not contradict this since the logarithm is not meromorphic.
We can still use Eq.(\ref{2Max-ex}) locally, say inside the positive Weyl alcove, in verifying that the form $\omega^{\rm ex}_q(p)$ is closed.)
The same remark holds for the change of variables $a \to a' = {\cal D}_q(p)^{\frac{1}{n}}\, a\,$
(formally relating $D(a') = {\cal D}_q(p)$ with $D(a) = 1$) which is not a legitimate "gauge transformation" in the class of meromorphic functions.

\subsubsection{Computing zero modes' Poisson and Dirac brackets}

\noindent
Our next task is to derive the PB relations among $a_{\alpha}^i$ and $p_j$
inverting the symplectic form (\ref{Oex}), (\ref{eq92}), (\ref{oexqp})
and taking into account the second class constraint (in Dirac's terminology \cite{Dir})
\begin{equation}
\label{eq11new.20}
P \left( = \sum_{j=1}^n p_j \right) \approx 0 \ , \qquad
\chi \left(= \frac{1}{n} \log \frac{D(a)}{\Phi(p)} \right) \approx 0 \ .
\end{equation}
If we regard $P \approx 0$ as a natural constraint, then $\chi \approx 0$ plays the role as associated (Weyl invariant)
gauge condition.

We recall (cf. (\ref{defOX}), (\ref{PBdef})) that given a symplectic form $\Omega$ and a Hamiltonian vector field
$X_f$ obeying the defining relation $\hat X_f \, \Omega = \delta f$, we can compute the PB $\{ f,g \}$ by setting
$\{ f,g \} = X_f \, g \equiv {\hat X}_f\, \d g\,.$
As the dependence of $\Omega^{\rm ex}$ (\ref{Oex}) on $P$ and $\chi$ is split (cf. (\ref{eq97})), the corresponding Hamiltonian vector fields are
\begin{equation}
\label{eq11new.22}
X_{\chi} = i \, \frac{\delta}{\delta P} \  , \quad X_P = -i \, \frac{\delta}{\delta \chi} \quad\Rightarrow\quad \{ \chi , P \} = i \ .
\end{equation}
The PB on ${\mathcal M}_q$ is reproduced by the Dirac bracket on ${\mathcal M}_q^{\rm ex}$:
\begin{equation}
\label{PBD}
\{ f,g \}_D = \{ f,g \} + \frac{1}{\{ P,\chi \}} \left( \{ f , P \} \{ \chi , g \} - \{f , \chi \} \{ P,g \} \right)\qquad
\left( \frac{1}{\{ P , \chi \}} = i \right)\ .
\end{equation}
In fact, the second term in the right-hand side of (\ref{PBD}) vanishes in most cases of interest since,
as we shall verify it by a direct computation below, $\chi$ is central for the zero modes' Poisson algebra
restricted to the hypersurface of the first constraint $P= 0$:
\begin{equation}
\label{chi-center}
\{ \chi , a_{\alpha}^j \} = 0 = \{\chi , p_{j\ell} \}\ .
\end{equation}
To obtain the PB on ${\mathcal M}_q^{\rm ex}\,,$ we have to invert the symplectic form (\ref{Oex})
\be
\Omega^{\rm ex} =\frac{k}{2\pi}\, {\rm tr}\, \d a a^{-1} \d M_p M_p^{-1}  - \,\omega_q^{\rm ex} (p) \,+\,
\frac{k}{4\pi} \left({\rm tr}\, \d a a^{-1} Ad_{M_p} \d a a^{-1} - \rho (a^{-1} M_p \, a ) \right)\,.
\lb{Oex-var}
\ee
In order to write it down in a manageable form, we use Eq.(\ref{defrhoK}) for $\rho (a^{-1} M_p \, a )$
noting that $K_M$ (\ref{KofM}) can be recast as
\be
K_M = \left( (1 + Ad_M) \fr + 1 - Ad_M \right) \left( (1 - Ad_M) \fr + 1 + Ad_M \right)^{-1}\ ,
\lb{KofM2}
\ee
and introduce the notation
\ba
&&\d p = \sum_{s=1}^n \d p_s\, e^{~s}_s = \frac{k}{2\pi i}\, \d M_p M_p^{-1}\ ,\qquad
\Theta := \sum_{j\ne \ell} \Theta^j_\ell\, e_j^{~\ell}\ ,\nn\\
&&A_\pm := 1\pm Ad_{M_p}\ ,\qquad \fr^a := Ad_a \, \fr\, Ad_a^{-1}\ ,\nn\\
&&K^a := Ad_a\, K_{a^{-1} M_p\, a}\, Ad_a^{-1} = (A_+ \fr^a + A_- ) (A_- \fr^a + A_+ )^{-1}\ .\qquad\qquad
\lb{notO}
\ea
(To derive the last equality in (\ref{notO}) from (\ref{KofM2}), we use that
$Ad_{a^{-1}M_pa} = Ad_a^{-1} Ad_{M_p}\, Ad_a\,.$)
It is easy to show that the operators $K^a\,$ and $\,\fr^a\,$ are skew symmetric together with $K_M\,$ and $\fr\,.$
We obtain
\ba
&&\frac{k}{4\pi}\, \rho (a^{-1} M_p \, a ) = \lb{rK}\\
&&= \frac{k}{8\pi}\, {\rm tr}\, \{( \d M_p M_p^{-1} - A_- (\d a a^{-1}))\, K^a\,( \d M_p M_p^{-1} - A_- (\d a a^{-1})) \} = \nn\\
&&= - \frac{k}{8\pi}\,{\rm tr}\, \{ (\frac{2\pi}{k}\, \d p - A_- \Theta )\,K^a\, (\frac{2\pi}{k}\, \d p - A_- \Theta) \} =\nn\\
&&= - \frac{1}{2}\,{\rm tr}\,\d p\, \frac{\pi}{k}\, K^a \,\d p + \frac{1}{2} \, {\rm tr}\, \d p \,K^a A_- \Theta -
\frac{k}{8\pi}\,{\rm tr}\, A_- \Theta\, K^a A_- \Theta\ ,
\nn
\ea
while the other term in (\ref{Oex-var}) containing $\Theta^j_\ell$ with $j\ne \ell$ can be rewritten as
\be
{\rm tr}\, \d a a^{-1} Ad_{M_p} \d a a^{-1} \ \
(\, = (\bar q-q) \sum_{j < \ell} \, [2 \, p_{j\ell}] \, \Theta_{\ell}^j \, \Theta_j^{\ell}\,\, )\,
= - \frac{1}{2}\, {\rm tr}\, A_- \Theta\, A_+ \Theta\ .
\lb{other}
\ee
Summing up the two terms pairing the off-diagonal forms and taking into account that
\ba
&&K^a A_- - A_+ = (A_+ \fr^a + A_-) (A_- \fr^a + A_+ )^{-1} A_- - A_+ =\nn\\
&&=(A_+ \fr^a + A_-) (\fr^a + \frac{A_+}{A_-})^{-1} - A_+ =\nn\\
&&= \left(A_+ \fr^a + A_- - A_+ (\fr^a + \frac{A_+ }{A_-} ) \right)\, (\fr^a + \frac{A_+ }{A_-})^{-1}= \nn\\
&&= \frac{A_-^2 - A_+^2}{A_-}\, (\fr^a + \frac{A_+ }{A_-} )^{-1} = - 4 \, \frac{Ad_{M_p}}{A_-} \, (\fr^a + \frac{A_+}{A_-} )^{-1}\ ,
\lb{KA}
\ea
we obtain
\ba
&&\frac{k}{8\pi} \left( \, {\rm tr}\, A_- \Theta\, K^a A_- \Theta\, -\, {\rm tr}\, A_- \Theta\, A_+ \Theta \right) =\nn\\
&&=- \frac{k}{2\pi}\, {\rm tr}\, A_- \Theta\,\frac{Ad_{M_p}}{A_-} \, (\fr^a + \frac{A_+}{A_-} )^{-1} \Theta
\equiv \frac{1}{2}\, {\rm tr}\, \Theta\,\frac{k}{\pi}\,(\fr^a + \frac{A_+}{A_-} )^{-1} \Theta\ .
\nn
\ea
The last equality follows from the fact that $A\equiv Ad_{M_p}$ is orthogonal with respect to ${\rm tr}\,$
(i.e. $^t\! A = A^{-1}$), hence $^t\!(1-A)\, A = (1 - A^{-1}) A = A-1\,$ so that, for $1-A$ is invertible, one has
\be
{\rm tr}\, (1-A) X\, \frac{A}{1-A}\, Y = {\rm tr}\, X\, \frac{A-1}{1-A}\, Y =-\, {\rm tr}\,X \, Y\ .
\lb{AXY}
\ee
Hence, in the basis of vector fields
$\{ \frac{\d}{\d p_s} , V^i_i , V^\ell_j \}\,$ dual to the $1$-forms
$\{ \d p_s , \Theta^i_i , \Theta^j_\ell \}\,,$ respectively
(all the indices running from $1$ to $n\,,$ and $j\ne \ell$),
the Poisson bivector matrix we obtain for (\ref{Oex-var}) has the following block form (in which $B$ is an $n\times n$ square matrix
and the block $D^{-1}$ is $n(n-1)\times n(n-1)$ while $C$ is an $n\times n(n-1)$ rectangular matrix, and
$f e^{~j} _j := \sum_\ell f_{\ell j}\, e^{~\ell}_\ell$):
\ba
&&{\begin{pmatrix}B&\id&C\cr -\id&0&0\cr - ^tC& 0& D^{-1}\end{pmatrix}}^{-1} =
\begin{pmatrix}0&-\id&0\cr \id& B+C D\, ^tC& - C D\cr 0 & - D\, ^t C & D\end{pmatrix}\ ,\nn\\
&&B = - f + \frac{\pi}{k}\, K^a\ ,\quad C = - \frac{1}{2}\, K^a A_-\ ,\quad D = \frac{\pi}{k}\, (\fr^a + \frac{A_+}{A_-} )\ .\qquad\quad
\lb{BCD}
\ea
Equivalently, the Poisson bivector is just
\be
{\cal P} = {\rm tr}\, \left( V \wedge \frac{\d}{\d p}  + \frac{1}{2}\, V \wedge F\, V \right)\ ,\quad
V:=\sum_{j,\ell} V^\ell_j e^{~j}_\ell \equiv \sum_i V^i_i e^{~i}_i + \sum_{j\ne\ell} V^\ell_j e^{~j}_\ell\ ,
\lb{PbV_def}
\ee
where the skew symmetric square matrix
\be
\lb{Pn2}
F\,:= \,\begin{pmatrix}B+C D\, ^tC& - C D\cr  - D\, ^t C & D\end{pmatrix}
\ee
is the $n^2\times n^2$ block in the lower right corner of (\ref{BCD}).

We shall show that, by using repeatedly the equality
$K^a (A_- \fr^a + A_+) = A_+ \fr^a + A_-\,$ following from (\ref{notO}) and the fact that
\ba
&&Ad_{M_p}\, e_i^{~j} = \sum_{r,s} e^{\frac{2\pi i}{k} p_{rs}} e_r^{~r}\, e_i^{~j} e_s^{~s} = \bq^{2 p_{ij}} e_i^{~j} \qquad\Rightarrow\nn\\
&&A_+ e^{~s}_s =\, ^t\!\!A_+ e^{~s}_s = 2\, e^{~s}_s\ ,\quad  A_- e^{~s}_s = \,^t\!\!A_- e^{~s}_s =0
\lb{AdMe1}
\ea
(cf. (\ref{eij})), the action of ${\cal P}$ (\ref{PbV_def}) can be actually simplified. We find that for $j\ne \ell\,,$
\ba
&&-\, {\rm tr}\,e^{~i}_i C D e^{~j}_\ell =
\frac{\pi}{2 k}\, {\rm tr}\,e^{~i}_i K^a\,A_- (\fr^a + \frac{A_+}{A_-} ) e^{~j}_\ell =\nn\\
&&= \frac{\pi}{2 k}\, {\rm tr}\,e^{~i}_i (A_+ \fr^a + A_- )\, e^{~j}_\ell = \frac{\pi}{k}\,
{\rm tr}\,e^{~i}_i\, \fr^a e^{~j}_\ell\ ,
\lb{CDr}
\ea
and, due to the skew symmetry of $K^a$ and $\fr^a\,,$
\ba
&&{\rm tr}\,e^{~i}_i (B+C D\, ^tC ) e^{~j}_j = {\rm tr}\,e^{~i}_i (- f + \frac{\pi}{k}\, K^a +
\frac{\pi}{4k} (A_+ \fr^a + A_-) {^t\!A_-} {^t\!K^a} ) e^{~j}_j =\nn\\
&&= - f_{ij} - \frac{\pi}{2k} {\rm tr}\,e^{~i}_i\,\, {^t[ K^a (A_- \fr^a + A_+ ) ]}\, e^{~j}_j =
- f_{ij} + \frac{\pi}{k} \, {\rm tr}\,e^{~i}_i \fr^a e^{~j}_j \ .
\lb{BCDC}
\ea
It follows further from (\ref{AdMe1}) that
\be
\frac{A_+}{A_-} \, e_j^{~\ell} = \frac{1+\bq^{2 p_{j\ell}}}{1-\bq^{2 p_{j\ell}}}\, e_j^{~\ell} =
\frac{e^{-i\frac{\pi}{k} p_{j\ell}}+e^{i\frac{\pi}{k} p_{j\ell}}}{e^{-i\frac{\pi}{k} p_{j\ell}}-e^{i\frac{\pi}{k} p_{j\ell}}}\, e_j^{~\ell}=
i \cot ( \frac{\pi}{k}\, p_{j\ell} ) e_j^{~\ell}\quad  {\rm for} \ j\ne\ell\ .
\lb{AdMe}
\ee
On the other hand, as $Ad_{a_1}^{-1} C_{12} = Ad_{a_2} C_{12}\,,$ we conclude that
\be
r^a_{12}\, a_1 a_2 = (\fr^a_1 C_{12}) \, a_1 a_2 = (Ad_{a_1} \fr_1 Ad_{a_1}^{-1} C_{12})\, a_1 a_2 = (Ad_{a_1 a_2} r_{12})\, a_1 a_2 = a_1 a_2\, r_{12}\ .
\lb{PBex-aa}
\ee
Combining these results and using ${\hat V}^\ell_j\, \d a^i_\a = i\, \d^i_j\, a^\ell_\a\,$ (cf. Eq.(\ref{eq90})) we finally obtain the PB
on ${\mathcal M}_q^{\rm ex}$:
\ba
&&\{ p_j , p_{\ell} \}= 0\ ,\qquad  \{ a_{\alpha}^j , p_{\ell} \}= i \, a_{\alpha}^j \,\delta_{\ell}^j\ ,\nn\\
&&\{ a_1 , a_2 \} = \left( r_{12} (p) - \frac{\pi}{k} \, r^a_{12} \right) a_1 \, a_2 \equiv
r_{12} (p) \, a_1 \, a_2 - \frac{\pi}{k} \, a_1 \, a_2 \, r_{12}\ .\qquad\quad
\lb{PBex}
\ea
Here the (standard) {\em constant} classical $r$-matrix (\ref{rstandard}) which corresponds to the operator $\fr$ acting as
\be
\fr\, e_s^{~s} = 0\ ,\quad \fr\, e_i^{~j} = e_i^{~j}\ ,\ i<j\ ,\quad \fr\, e_i^{~j} = - e_i^{~j}\ ,\ i>j
\lb{stand-r-op}
\ee
(cf. (\ref{re1})) has the form
\be
r^{\a\,\b}_{~{\a}' {\b}'} = -\, \epsilon_{\a\b}\, \d^\a_{{\b}'}\, \d_{{\a}'}^\b\ ,\qquad \epsilon_{\a\b} =
\left\{
\begin{array}{ll}
\, {~~}1&, \quad \a > \b\\
\, {~~}0&, \quad\a = \b\\
\, - 1&,\quad \a < \b
\end{array}
\right.\quad ,
\lb{stand-r-matr}
\ee
while the matrix
\be
r_{12} (p) = \sum_{j\ne \ell} \left( f_{j\ell}(p) (e_j^{~j})_1 (e_\ell^{~\ell})_2 -
i\frac{\pi}{k}\,{\rm cot} \left(\frac{\pi}{k} p_{j\ell}\right) (e^{~\ell}_j )_1 (e^{~j}_\ell )_2 \right)
\qquad (\,f_{j\ell}(p) = - f_{\ell j}(p) \,)
\lb{dynr}
\ee
(where $f_{j\ell}(p)\,$ is given in (\ref{f01})), with entries
\be
r (p)^{j~\ell}_{~j' {\ell}'} =
\left\{
\begin{array}{ll}
f_{j\ell}(p)\, \d^j_{j'} \d^\ell_{{\ell}'} -  i \frac{\pi}{k}\,{\rm cot} \left(\frac{\pi}{k} p_{j\ell}\right) \d^j_{{\ell}'} \d^\ell_{j'}
&{\rm for}\ j\ne\ell\ \ {\rm and}\ \ j'\ne\ell'\\
\, 0& {\rm for}\ j=\ell\ \ {\rm or}\ \ j'=\ell'
\end{array}
\right.
\lb{dyn-r-matr}
\ee
is the classical {\em dynamical} $r$-matrix solving the (modified) classical dynamical YBE
\begin{eqnarray}
&&[r_{12}(p) , r_{13}(p)] + [r_{12}(p) , r_{23}(p)] + [r_{13}(p) , r_{23}(p)]
+ {\rm Alt}\, ( d r ({p})) = \nn\\
&&=\frac{\pi^2}{k^2}\, [ C_{12} , C_{23} ] \ ,\lb{CDYBE}\\
&&{\rm Alt}\, ( d r ({p})) := -i \sum_{s=1}^n \frac{\partial}{\partial p_s}
\left(\, ({e^s_s})_1\, r_{23} (p) - ({e^s_s})_2\, r_{13} (p) + ({e^s_s})_3\, r_{12} (p) \,\right)
\nn
\end{eqnarray}
(cf. \cite{EV}). The difference between (\ref{CDYBE}) and the modified classical YBE (\ref{mCYBE}) satisfied by $r_{12}$
is in the term ${\rm Alt}\, (d r(p))$ containing derivatives in the dynamical variables $p_s\,.$
It is easy to see that (\ref{mCYBE}) and its dynamical counterpart (\ref{CDYBE}) guarantee
the Jacobi identity for the PB (\ref{PBex}).

Comparing (\ref{dyn-r-matr}) with (\ref{PBa1a2G}), we see that $\frac{\pi}{k}\,{\rm cot} \left( \frac{\pi}{k}\,p_{j\ell} \right)\,$
substitutes its undeformed ($k\to 0$) limit, $\frac{1}{p_{j\ell}}\,;$
the diagonal term reflects the gauge freedom in choosing $\omega_q^{\rm ex}(p)$ (\ref{oexqp}) and the determinant condition.
On the contrary, the presence of the constant $r$-matrix term is purely a deformation phenomenon.

\smallskip

In order to prove that the constraint $\chi$ is central on the hypersurface $P=0\,,$ i.e. that Eqs. (\ref{chi-center}) take place, one first derives
\ba
&&\{ a^j_\beta ,\, a^n_{\alpha_n} \dots a^1_{\alpha_1} \} =\sum_{\ell\ne j} f_{j\ell}(p)\,a^j_\beta \,
a^n_{\alpha_n} \dots \,a^\ell_{\alpha_\ell} \, \dots a^1_{\alpha_1} - \nonumber\\
&&- i\frac{\pi}{k}\,\sum_{\ell\ne j} \cot ( \frac{\pi}{k}p_{j\ell} )\,
a^\ell_\beta a^j_{\alpha_\ell}\, a^n_{\alpha_n} \dots \, \widehat{a^{\ell}_{\alpha_\ell}}\, \dots a^1_{\alpha_1} -\nonumber\\
&&- \frac{\pi}{k}\, \sum_\ell \epsilon_{\beta \alpha_\ell}\, a^\ell_\beta a^j_{\alpha_\ell} \,
a^n_{\alpha_n} \dots \, \widehat{a^{\ell}_{\alpha_\ell}}\, \dots a^1_{\alpha_1} \ .
\lb{ana}
\ea
The second and the third terms in (\ref{ana}) vanish when multiplied by $\epsilon^{\alpha_n \dots \alpha_\ell\dots\alpha_1}$
and summed over repeated indices, due to
\ba
&&\sum_{\ell\ne j} \cot ( \frac{\pi}{k}p_{j\ell})\,
a^\ell_\beta\, a^j_{\alpha_\ell}\, a^n_{\alpha_n} \dots \, \widehat{a^{\ell}_{\alpha_\ell}}\, \dots a^1_{\alpha_1}\,
\epsilon^{\alpha_n \dots \alpha_\ell\dots\alpha_1} = \nonumber\\
&&= \sum_{\ell\ne j} \cot ( \frac{\pi}{k}p_{j\ell})\,
a^\ell_\beta\, a^j_{\alpha_\ell} A^{\alpha_\ell}_\ell =
\sum_{\ell\ne j} \cot ( \frac{\pi}{k}p_{j\ell}) a^\ell_\beta\,D(a) \, \delta^j_\ell = 0\qquad\qquad
\label{8van}\\{\rm and}\nn\\
&&\sum_\ell \epsilon_{\beta \alpha_\ell}\, a^\ell_\beta \, a^j_{\alpha_\ell} \,
a^n_{\alpha_n} \dots \, \widehat{a^{\ell}_{\alpha_\ell}}\, \dots a^1_{\alpha_1} \,\epsilon^{\alpha_n \dots \alpha_\ell\dots\alpha_1}
= \nonumber\\
&&= \sum_\ell \epsilon_{\beta \alpha_\ell}\, a^j_{\alpha_\ell}\, A^{\alpha_\ell}_\ell \, a^\ell_\beta =
\epsilon_{\beta \alpha_\ell}\, a^j_{\alpha_\ell}\,D(a)\,
\delta_\beta^{\alpha_\ell} = 0
\lb{9van}
\ea
(cf. (\ref{Da}) -- (\ref{AA})). Hence,
\begin{equation}
\{ a^j_\beta , \log D(a) \} = \frac{1}{D(a)}\, \{ a^j_\beta , D(a) \} = \sum_{\ell\ne j} f_{j\ell}(p)\,a^j_\beta\ .
\label{aDa}
\end{equation}
On the other hand, the PB (\ref{PBex}) imply
\begin{equation}
\label{pavar}
\{ D(a) , p_{\ell} \} = i \,D(a)\quad\Rightarrow\quad \{ D(a) , p_{ j \ell} \} = 0\quad\Rightarrow\quad \{ \chi , p_{ j \ell} \} = 0\ ,
\end{equation}
as well as
\begin{equation}\label{aUp}
\{ a_\a^j , U(p) \} = \{ a_\a^j , p_{\ell} \} \frac{\partial U}{\partial p_{\ell}} (p) =
i\, \frac{\partial U}{\partial p_j} (p)\, a^j_\a   \ .
\end{equation}
In particular, the calculation of the PB (\ref{aUp}) for $U(p) = \log \Phi(p)\,,$
see (\ref{Phigen}), (\ref{f01}), gives the same result as (\ref{aDa}),
\be
\label{11}
\{ a_{\alpha}^j ,\, \log \Phi(p) \} = \sum_{i<\ell} f_{i\ell}(p) \left( \frac{\partial}{\partial p_j}\, p_{i\ell} \right)\, a^j_\a
= \sum_{\ell \ne j} f_{j\ell}(p)\, a^j_\alpha\ .
\ee
As $\chi = \frac{1}{n}\log\,\frac{D(a)}{\Phi(p)}\,,$ it follows from (\ref{aDa}) and (\ref{11}) that
\be
\{ \chi , a_{\alpha}^j \} = 0\quad \Rightarrow\quad\{ \frac{D(a)}{\Phi(p)} \ ,\, a_{\alpha}^j \} = 0\ .
\lb{DPa}
\ee
The first of these equations together with the last one in (\ref{pavar}) confirm the centrality of the constraint $\chi$ for $P=0$ (\ref{chi-center}).

\smallskip

The passage to the $(n+2)(n-1)$-dimensional (unextended) phase space ${\mathcal M}_q$ is straightforward;
using (\ref{PBD}), we see that of the three PB (\ref{PBex}) only the second one is changing and, as
\be
\{ a^j_\a , P \} = i\,  a^j_\a\ ,\qquad \{ \chi , p_{\ell} \} = \frac{1}{n}\, \{ \log D(a) , p_{\ell} \} = \frac{i}{n}
\lb{Dirap}
\ee
(cf. (\ref{pavar})), it follows that
\be
\label{PBapD}
\{ a_{\alpha}^j , p_{\ell} \}_D = i \left(\delta_{\ell}^j - \frac{1}{n} \right) a_{\alpha}^j \quad \Rightarrow\quad
\{ a_{\alpha}^j , p_{\ell m} \}_D =\{ a_{\alpha}^j , p_{\ell m} \} = i (\delta_{\ell}^j - \delta_m^j) \, a_{\alpha}^j\ .
\ee
On the other hand, $D(a)$ and $p_\ell\,$ have a vanishing Dirac bracket:
\be
\{ D(a) , p_{\ell} \}_D = \{ D(a) , p_{\ell} \} + i \,\{ D(a) , P \} \{ \chi , p_{\ell} \} = i D(a) + i . i\, n\, D(a)\, . \frac{i}{n} = 0\ .
\lb{Dap}
\ee
From now on we shall assume that all the brackets are the Dirac ones, skipping the subscript {\footnotesize{$D$}}.

\smallskip

We now proceed to computing the PB of the monodromy matrix $M=a^{-1} M_p\, a\,,$ cf. (\ref{aintertw}), and its Gauss components $M_\pm\,.$

\bigskip

\noindent
{\bf Remark 3.6~} As we shall see, in the quantized theory $p_{i\, i+1}\,$ become operators
whose eigenvalues label the representations of the current algebra,
while the entries of the quantum monodromy matrix $M\,$
are functions of the $U_qs\ell(n)\,$ generators which commute with the currents.
We should therefore expect, in particular, that in the classical case $M\,$ Poisson commutes
with $p_{ij}\,$ and hence, with the diagonal monodromy $M_p\,.$
Another implication of this fact would be that the PB of $M\,$
with the zero modes, as well as the PB between the matrix elements
of $M\,$ itself, do not contain the dynamical $r$-matrix.
All this is confirmed by the results of the explicit calculations carried below.

\bigskip

It follows from (\ref{PBapD}) and (\ref{AWn1}) that
\be
\{ a_{\alpha}^j , p_{\ell \ell+1} \} = i (h_\ell \, a)^j_\a\quad \Leftrightarrow\quad \{ \sp_1 , a_2 \} = - i\, \s_{12}\, a_2
\lb{asp}
\ee
($\s_{12}= h^\ell_1 h_{\ell 2}\,$ is the diagonal part of the polarized Casimir operator $C_{12}\,,$ see (\ref{Cn-sigma})) and hence,
\be
\{ M_{p 1} , a_2 \} = \frac{2\pi}{k}\, \s_{12}\, M_{p 1}\, a_2
\qquad (\,\{ M_{p 1} ,  M_{p 2} \} = 0\,) \ .
\lb{Mpa0}
\ee
From (\ref{PBex}) and (\ref{Mpa0}) one gets
\ba
&&\{ M_1 , a_2 \} = \{ a^{-1}_1 M_{p 1} a_1 , a_2 \} = \nn\\
&&= -  a_1^{-1} \{ a_1 , a_2 \}\, a_1^{-1} M_{p 1}\, a_1 + a^{-1}_1 \{ M_{p 1} , a_2 \}\, a_1 + a_1^{-1} M_{p 1} \{ a_1 , a_2 \} =\nn\\
&&= \frac{\pi}{k}\, a_2 (r_{12} M_1 - M_1 r_{12} ) +
\nn\\
&&+ a_1^{-1} ( M_{p 1} r_{12}(p) - r_{12}(p) M_{p 1} + \frac{2 \pi}{k}\, \s_{12}  M_{p 1} )\, a_1 a_2\ .
\lb{M1}
\ea
The classical dynamical $r$-matrix $r_{12} (p)\,$ (\ref{dyn-r-matr}) obeys the relation
\be
(\id - Ad_{M_{p 1}})\, r_{12}(p) = - \frac{\pi}{k}\, (\id + Ad_{M_{p 1}})\, (C_{12} - \s_{12} )\ ,
\lb{rp-sat}
\ee
cf. (\ref{AdMe}) (only the off-diagonal part of $r_{12}(p)\,$ survives after applying $\id - Ad_{M_{p 1}}$), which can be rewritten as
\be
M_{p 1} r_{12}(p) - r_{12}(p) M_{p 1} + \frac{2 \pi}{k}\, \s_{12}  M_{p 1}
= \frac{\pi}{k}\, ( M_{p 1} C_{12} + C_{12} M_{p 1} )\ .
\lb{adMreq}
\ee
(the $n^2\times n^2$ matrices $M_{p1}$ and $\s_{12}$ are diagonal and hence, commute with each other). We have, therefore,
\ba
&&\{ M_1 , a_2 \} = \frac{\pi}{k}\, a_2 (r_{12} M_1 - M_1 r_{12} )
+ \frac{\pi}{k}\,a_1^{-1} ( M_{p 1} C_{12} + C_{12} M_{p 1})\, a_1 a_2  =\nn\\
&&= \frac{\pi}{k}\, a_2 (r_{12} M_1 - M_1 r_{12} ) +\frac{\pi}{k}\,
a_1^{-1} (M_{p 1} a_1 a_2\, C_{12} + a_1 a_2\, C_{12}\, a_1^{-1} M_{p 1} a_1 ) = \lb{Mgen}\\
&&= \frac{\pi}{k}\, a_2 (r_{12} M_1 - M_1 r_{12} ) +
\frac{\pi}{k}\, a_2 (M_1 C_{12} + C_{12} M_1 ) = \frac{\pi}{k}\, a_2 (r^+_{12} M_1 - M_1 r^-_{12} )\qquad
\quad\nn
\ea
where $r^\pm_{12}\, = \, r_{12}\, \pm \, C_{12}$ are the $r$-matrices satisfying the CYBE (\ref{CYBE}).
The matrix elements of the monodromy $M\,$ Poisson commute with those of the diagonal one $M_p:$
\ba
&&\{ M_{p 1} , M_2 \} = \{ M_{p 1} , a_2^{-1} M_{p2}\, a_2 \} = \nn\\
&&=\frac{2 \pi}{k}\, a_2^{-1} (M_{p2}\, \s_{12}\, M_{p1} - \s_{12} \, M_{p1}\,  M_{p2} )\, a_2 = 0
\lb{PBMMp}
\ea
(we have used (\ref{Mpa0})).
Finally, from (\ref{Mgen}) and (\ref{PBMMp}) we obtain the PB of two monodromy matrices $M$:
\ba
&&\{ M_1 , M_2 \} = \{ M_1 , a_2^{-1} M_{p 2}\, a_2 \} = \nn\\
&&= a_2^{-1} M_{p 2} \{ M_1 , a_2 \} -\, a_2^{-1} \{ M_1 , a_2 \} a_2^{-1} M_{p 2}\, a_2 =
\nn\\
&&= M_2\, a_2^{-1} \{ M_1 , a_2 \} - a_2^{-1} \{ M_1 , a_2 \} M_2
= \frac{\pi}{k}\, [ M_2 \,, r^+_{12} M_1 - M_1 r^-_{12}\, ] \equiv\nn\\
&&\equiv  \frac{\pi}{k}\ (M_1 r_{12}^-\, M_2 + M_2\, r_{12}^+ M_1\, - M_1 M_2\, r_{12} - r_{12} M_1 M_2 )\,.
\lb{PBMM}
\ea

\smallskip

As already mentioned (at the end of Section 2), a basic property of the PB listed above is their
Poisson-Lie symmetry \cite{D1, S-T-S, D} with respect to constant right shifts of $a\,,$
\be
a\ \to\ a\, T\ ,\quad M\ \to \ T^{-1}\, M\, T \qquad (\, T\in G\,)\ ,
\lb{PLleft}
\ee
provided that the PB of the transformation group (are non-trivial and)
are given by the Sklyanin bracket (\ref{PBSkl}) $\{ T_1 , T_2 \} = \frac{\pi}{k}\, [r_{12}\,, T_1 T_2  ]$
(assuming that $\{ a_1 , T_2 \} = 0 = \{ M_1 , T_2 \}$).
It follows from (\ref{aintertw}) that the diagonal monodromy matrix $M_p = a M a^{-1}\,$ is
invariant with respect to (\ref{PLleft}), cf. Remark 3.6.
The PL symmetry of the chiral classical WZNW model, leading to {\em quantum group} \cite{D} symmetry
of the quantized theory, has been first explored in \cite{AS, G}.

\bigskip

To derive the PB of the Gauss components $M_\pm\,$ from those of the monodromy
matrix $M = M_+ M_-^{-1}$ in a systematic way, we can use the fact that, by (\ref{dMM+-}) and (\ref{KMMM}),
\be
\frac{1}{2}\, (K_M+\id )\, \d M M^{-1} = \d M_+ M_+^{-1}
\lb{KM+1}
\ee
and hence, for any (matrix) function $F\,$ on the phase space,
\be
\{ M_{+1} , F_2 \} =  \frac{1}{2} \left(\, (K_{M1}+\id )\,\{ M_1 , F_2 \} \, \right) M_{-1}\ .
\lb{rules}
\ee
The corresponding PB for $M_-\,$ can be now found from
\be
\{ M_{-1} , F_2 \} =  M_1^{-1} (\, \{ M_{+1} , F_2 \}  - \{ M_1 , F_2 \} M_{-1}\, )\ .
\lb{M->Mpm}
\ee
Combining (\ref{rules}) and (\ref{M->Mpm}) with (\ref{Mgen}) or (\ref{PBMM})
and using (\ref{KofM}), from which it follows that
\be
\frac{1}{2}\, (K_{M1}+\id )\, (r^+_{12} - Ad_{M_1} r^-_{12} ) = r^+_{12}
\lb{KofM+1}
\ee
we get, respectively,
\be
\lb{Mpma}
\{ M_{\pm 1} , a_2\} = \frac{\pi}{k}\, a_2\, r^\pm_{12}\, M_{\pm 1}\ ,\quad
\{ M_{\pm 1} , M_2 \} = \frac{\pi}{k}\, [ M_2 , r^\pm_{12}\, ]\, M_{\pm 1}\ .
\ee
As $M$ Poisson commutes with $p_\ell\,,$ (\ref{rules}), (\ref{M->Mpm}) imply the same for $M_\pm$:
\be
\{ M_\pm , p_\ell \} = \{ M , p_\ell \} = 0\ .
\lb{Mpmpl}
\ee
Note that the PB of $M_\pm\,$ displayed above are simpler than the
analogous brackets for $M\,.$ Applying once more (\ref{rules}),
we can obtain the PB among the Gauss components themselves. For example,
\ba
&&\{ M_{+ 1} , M_{+ 2} \} = \frac{1}{2} (\, ( K_{M1}+\id )\,\{ M_1 , M_{+ 2} \}\,) M_{- 1}= \nn\\
&&= - \frac{\pi}{2k}\,( (K_{M1}+ \id ) \,( r_{12}^- - Ad_{M_1} r_{12}^- )\,)\, M_1\, M_{+2}\, M_{-1}=\nn\\
&&= - \frac{\pi}{2k}\, (\,(K_{M1}+ \id ) \,
(r_{12}^+ - Ad_{M_1} r_{12}^- - 2 \, C_{12} )\, )\, M_{+ 1} M_{+2}=\nn\\
&&= \frac{\pi}{k}\, [ M_{+ 1} M_{+ 2} , r^+_{12}\, ] \,
=\, \frac{\pi}{k}\, [ M_{+ 1} M_{+ 2} , r_{12}\, ]\,.\quad\
\lb{MpmfromM}
\ea
To evaluate $(K_{M1}+\id) \, C_{12}\,$ in (\ref{MpmfromM}), we have used (\ref{altKM}), from which it follows that
\ba
&&(K_{M1}+\id) \, C_{12} = Ad_{M_{+1}} \, (r_1+\id)\, Ad^{-1}_{M_{+1}}\, C_{12} = \nn\\
&&= Ad_{M_{+1}}\, (r_1+\id)\, Ad_{M_{+2}}\, C_{12} = M_{+1} M_{+2}\, r^+_{12}\, M^{-1}_{+2} M^{-1}_{+1} \ .
\lb{KMC}
\ea
Here is the complete list of PB among $M_\pm\,:$
\be
\{ M_{\pm 1} , M_{\pm 2} \} = \frac{\pi}{k}\, [ M_{\pm 1} M_{\pm 2} , r_{12}\, ] \ ,\quad
\{ M_{\pm 1} , M_{\mp 2} \} = \frac{\pi}{k}\, [ M_{\pm 1} M_{\mp 2} , r^\pm_{12}\, ]\ .
\lb{Mpmmp}
\ee

\subsection{PB for the Bloch waves}

\noindent
The requirement that the covariant group valued chiral field $g (x)\,$ (\ref{gua}) is
unimodular implies that the determinants of the zero mode's matrix $(a^j_\a )\,$
and of the Bloch waves $( u^A_j (x))\,$ have inverse values
(after identifying ${\fp}\,$ and $p\,,$ cf. Remark 3.1).
We shall denote the determinant of the extended Bloch wave matrix by $ {\tilde D} (x) := {\det}\, u(x)\,$ so that
the analog of (\ref{Da}) holds,
\ba
&&u^{A_1}_{j_1}(x) u^{A_2}_{j_2} (x)\dots u^{A_n}_{j_n}(x)\,
\varepsilon^{j_1 j_2 \dots j_n} =
{\tilde D}(x)\, \varepsilon^{A_1 A_2 \dots A_n}
\quad\Rightarrow\nn\\
&&{\tilde D}(x) = \frac{1}{n!}\,\varepsilon_{A_1 A_2 \dots A_n}
u^{A_1}_{j_1}(x) u^{A_2}_{j_2}(x) \dots u^{A_n}_{j_n}(x)\,
\varepsilon^{j_1 j_2 \dots j_n} \ .
\lb{Dtilde}
\ea
Here again $\varepsilon_{A_1 A_2 \dots A_n} = \varepsilon^{A_1 A_2 \dots A_n}\,$
is the fully antisymmetric Levi-Civita tensor of rank $n\,,$ for which
\be
\varepsilon_{A_1 A_2 \dots A_n} \,
\varepsilon^{B_1 A_2 \dots A_n} \,
= \, (n-1)! \, \d^{B_1}_{A_1}\ .
\lb{normal}
\ee
In the extended Bloch waves' phase space ${\tilde D}(x)$ is {\em necessarily} $x$-dependent;
indeed, we set, in complete analogy with the zero mode case (\ref{monex}),
\be
M_p = u(-\pi )^{-1} u(\pi ) = \sum_{s=1}^n \bq^{2{p_s}} e^s_s\ ,\quad P := \sum_{s=1}^n p_s\ne 0\quad
\Rightarrow\quad\det M_p = e^{\frac{2\pi i}{k} P}
\lb{extMpBW}
\ee
and hence, ${\tilde D}(\pi ) = {\tilde D}(-\pi )\, e^{\frac{2\pi i}{k} P}\,$ where ${\tilde D} (x)\,$
is an abelian group valued field. To study its $x$-dependence, we take the derivative in $x$ of both sides of
the second equation (\ref{Dtilde}). Using the "classical KZ equation" (\ref{clKZ})
written in terms of $u (x)\,,$ the first equation in (\ref{Dtilde}) and (\ref{normal}), we obtain
\ba
&&\frac{d}{d x}\, {\tilde D}(x) = - \frac{i}{k}\, \frac{1}{n!}\,
\varepsilon_{A_1 A_2 \dots A_n} \,
\{ j^{A_1}_{B_1} u^{B_1}_{j_1} u^{A_2}_{j_2} \dots u^{A_n}_{j_n}\, +\nn\\
&&+ u^{A_1}_{j_1} j^{A_2}_{B_2} u^{B_2}_{j_2} \dots u^{A_n}_{j_n}\, +\dots
+ u^{A_1}_{j_1} u^{A_2}_{j_2} \dots j^{A_n}_{B_n} u^{B_n}_{j_n} \}\,
\varepsilon^{j_1 j_2 \dots j_n} =\nn\\
&&= - \frac{i}{k}\, \frac{1}{n!}\,
\varepsilon_{A_1 A_2 \dots A_n} \,
\{ j^{A_1}_{B_1} \, {\tilde D}(x) \varepsilon^{B_1 A_2 \dots A_n} \, +
j^{A_2}_{B_2} \, {\tilde D}(x) \varepsilon^{A_1 B_2 \dots A_n} \, +\dots
+\nn\\
&&+ j^{A_n}_{B_n} \, {\tilde D}(x) \varepsilon^{A_1 A_2 \dots B_n} \, \}
=  - \frac{i}{k}\, \frac{1}{n}\,
{\tilde D}(x) \{\, j^{A_1}_{A_1} \, + j^{A_2}_{A_2} \, +\dots\,
+ j^{A_n}_{A_n} \, \} =\nn\\
&&= - \frac{i}{k}\,
( {\rm tr}\, j (x)) {\tilde D}(x)\,\equiv
- \frac{i}{k}\, J(x)\, {\tilde D}(x)\ ,
\quad\quad J(x) := {\rm tr}\,j(x)\ .
\lb{Dtindep}
\ea
We shall parametrize ${\tilde D} (x)\,,$ setting accordingly
\be
{\tilde D} (x) = {\tilde D}\, e^{- \frac{i}{k} t(x)} \ ,\qquad
t(x) = J_0\, x + i \sum_{r\ne 0} \frac{J_r}{r} e^{-irx}\ ,
\lb{tildeDabel}
\ee
so that
\ba
&&t'(x) = J(x) = \sum_{r\in {\mathbb Z}} J_r e^{-i r x}\ ,\qquad J_r =\int_{-\pi}^\pi \, J(x) \, e^{irx}\,\frac{dx}{2\pi}\ ,\nn\\
&&t(\pi) = t(-\pi) + 2\pi J_0\quad\Rightarrow\quad J_0 = - P\ .
\lb{JrDabel}
\ea
Thus, the extension amounts to adding the modes of ${\tilde D} (x)\,$ which form
a denumerable (countably infinite) set of degrees of freedom. Denoting
\be
{\tilde\chi} := \frac{1}{n}\, {\rm log}\, ({\tilde D}\, {\cal D}_q (p) )\ ,
\lb{chitilde}
\ee
the reduction from the extended Bloch waves' phase space to the unextended one (in which
$u(x)$ has inverse determinant ${\tilde D}^{-1} = {\cal D}_q (p)$ !)
is performed, accordingly, by imposing the infinite set of constraints
\be
{\tilde\chi} \approx 0 \approx J_r\ ,\quad r \in {\Z}\ .
\lb{cu}
\ee
Writing $u(x)\,$ as a multiple of an (unimodular) element $u_0 (x)\in SU(n)\,,$
\be
u(x) = u_0 (x)\,{\tilde D}(x)^{\frac{1}{n}}
\lb{uu1}
\ee
and denoting the corresponding (Lie algebra valued) left invariant $1$-forms by
\be
U(x) := -i u^{-1} (x)\, \d u (x)\,,\qquad U_0(x) := -i u_0^{-1} (x)\, \d u_0 (x)\,,
\lb{LL1}
\ee
we obtain from (\ref{uu1}) and (\ref{tildeDabel}) the following
expressions for $U(x)\,$ and its derivative $U'(x)\,:$
\be
U(x) = U_0 (x) - \frac{i}{n} \frac{\d {\tilde D} (x)}{{\tilde D} (x)}\ ,\quad
\frac{\d {\tilde D} (x)}{{\tilde D} (x)} = \frac{\d {\tilde D}}{{\tilde D}} - \frac{i}{k}\, \d\, t(x)\ ,\quad
U'(x) = {U_0}'(x) - \frac{1}{n k} \,\d J(x)\ .
\lb{defL}
\ee
In terms of $U_0(x)$ ({\ref{LL1}), the symplectic form for the Bloch waves $\O_B = \O + \o_q\,$ (\ref{OB}) becomes
\be
\O_B  (u_0, \bq^{2\sp} )\, = {\rm tr}\, \left( \frac{k}{4\pi}\, \int_{-\pi}^{\pi} dx \, U_0'(x) U_0(x)
- \frac{1}{2}\, U_0 (-\pi )\, \d \sp\, \right) + \o_q (p)\ ,
\lb{OBunext}
\ee
and the extended symplectic form given by
\be
\O^{\rm ex}_B (u, M_p) = {\rm tr}\, \left( \frac{k}{4\pi}\, \int_{-\pi}^{\pi} dx\, U'(x) U(x)
- \frac{1}{2}\, U (-\pi )\, \d p\, \right) + \o_q^{\rm ex} (p)
\lb{OBWext}
\ee
reduces again (as it happens in the zero modes case, cf. (\ref{eq97}))
to the sum of $\O_B\,$ (\ref{OBunext}) and a part representing the (second class) constraints:
\be
\O^{\rm ex}_B (u, M_p) = \O_B (u_0 , \bq^{2\sp} ) - i\, \d P \, \d {\tilde \chi}
+ \frac{i}{nk}\,\sum_{r=1}^\infty\,\frac{\d J_{-r}\,\d J_r}{r}\ .
\lb{OPchi}
\ee
Deriving (\ref{OPchi}), we have assumed that $\o_q^{\rm ex} (p)\,$ given by (\ref{oexqp})
is related to $\o_q (p)\,$ by (\ref{oex-o}) and have
used (\ref{chitilde}) and (\ref{JrDabel}), the latter implying, in particular,
\ba
&&\int_{-\pi}^\pi\, dx \,x\, \d J (x)\, \d J_0 \,
=-  \sum_{r\ne 0}\,\int_{-\pi}^\pi\, dx\, x\, e^{-irx}\, \d J_r\, \d P\,  =\nn\\
&&=-  2\pi i\, \sum_{r\ne 0} \,\frac{(-1)^r}{r} \, \d J_r \, \d P \,
= - 2\pi\, \d  t(-\pi )\, \d P\ .
\lb{int-djdp}
\ea

To find the PB for the Bloch waves $u(x)$, we need to invert the symplectic form (\ref{OBWext}).
To this end, we shall introduce loop group (periodic) variables
\begin{equation}
\ell (x) = u(x)\, e^{-i\frac{p}{k}x} \ ,\qquad \ell (x+2\pi) = \ell (x)
\label{l-u}
\end{equation}
(the exponential factor compensating the non-trivial diagonal monodromy $M_p = \bq^{2p}$ of $u(x)$), in terms of which
the left invariant, matrix valued Bloch waves' $1$-forms are expressed as
\begin{equation}
i\, U(x) \equiv u^{-1} (x)\,\delta u (x)\,= \, e^{-i\frac{p}{k}x} \,\ell^{-1} (x)\,\delta \ell (x)\, e^{i\frac{p}{k}x} + i\, \frac{\delta p}{k}\, x\ .
\label{u-l}
\end{equation}
The mode expansion of the periodic matrix valued $1$-forms
\begin{equation}
- i k\, \ell^{-1} (x)\delta \ell (x) = \sum_{m\in \mathbb Z} {\Xi}_m\, e^{-imx}\ ,
\qquad  {\Xi}_m = \sum_{j,\ell =1}^n ({\Xi}_m)^j_\ell \, e^{~\ell}_j
\label{lmodes}
\end{equation}
allows to write the extended symplectic form simply as
\begin{eqnarray}
&&\Omega^{\rm ex}_B (u, M_{p}) - \o^{\rm ex}_q (p) = \frac{1}{k}\,{\rm tr}\, \{\delta (p\, \Xi_0 ) + i \sum_{m=1}^\infty m\, \Xi_{-m} \Xi_m \}=\nonumber\\
&&= \frac{1}{k}\,\sum_{\ell=1}^n \delta {p}_\ell\, (\Xi_0)^\ell_\ell
+ \frac{i}{2k}\, \sum_{m=-\infty}^\infty\, \sum_{j,\ell=1}^n (m+\frac{{p}_{j\ell}}{k} )\, (\Xi_{-m})^\ell_j (\Xi_m)^j_\ell
\ .\qquad\qquad
\label{OLinTheta}
\end{eqnarray}
(Note that $\mid\!\frac{{p}_{ij}}{k}\!\mid\, <1$ for $\sp\in A_W\,,$ cf. (\ref{AWn1}).) To derive (\ref{OLinTheta}), we deduce from
$\delta (\ell^{-1} \delta \ell ) = - (\ell^{-1} \delta \ell )^2$ that
\begin{equation}
\delta \, \Xi_n = \frac{1}{ik} \sum_{m} \Xi_{n-m} \Xi_m\quad\Rightarrow\quad
\delta \, \Xi_0 = \frac{1}{ik} \sum_{m} \Xi_{-m} \Xi_m
\label{dThetan}
\end{equation}
and use
\be
[p \,, e^{~\ell}_j\, ] = {p}_{j\ell}\, e^{~\ell}_j\ ,\qquad
e^{i\frac{p}{k} x}\, e_j^{~\ell} = e^{i\frac{{p}_{j\ell}}{k} x}\, e_j^{~\ell} \, e^{i\frac{p}{k} x}
\lb{pexp}
\ee
as well as the relations
\be
\ell^{-1} (-\pi) \delta \ell (-\pi) - \int_{-\pi}^{\pi} \frac{d x}{2\pi}\, x\,( \ell^{-1}(x)\delta \ell(x))'
= \int_{-\pi}^{\pi} \frac{dx}{2\pi}\, \ell^{-1}(x)\,\delta \ell(x) = \frac{i}{k}\,\Xi_0\ .\qquad
\label{intermed}
\ee

\bigskip

The form  $\Omega^{\rm ex}_B (u, M_{p}) $ (\ref{OLinTheta}) can be readily inverted in terms of the vector fields
$(V^m)^j_i\,,\, \frac{\delta}{\delta {p}_\ell}$ dual to the $1$-forms $(\Xi_m )^i_j\,,\, \delta {p}_\ell\,,$
respectively, to obtain the corresponding Poisson bivector:
\begin{eqnarray}
&&{\cal P} = k \sum_{\ell} (V^0)^\ell_\ell \wedge \frac{\d}{\d p_\ell}
+ \frac{k^2}{2}\,\sum_{j \ne \ell} f_{j\ell}(p)\, (V^0)^j_j\wedge (V^0)^\ell_\ell\, +
\label{Poisson-bi}\\
&&+ \, \frac{i k}{2}\,\left( \sum_{m\ne 0} \,\sum_{\ell}\frac{1}{m} (V^{-m})^\ell_\ell\wedge (V^m)^\ell_\ell \,
+ \sum_{m} \sum_{j\ne \ell}\, \frac{1}{m+\frac{ {p}_{j\ell}}{k}}
(V^{-m})^j_\ell \wedge (V^m)^\ell_j \right)\ .
\nn
\end{eqnarray}
From Eq.(\ref{u-l}) we obtain the contractions with $\delta u (x)$:
\begin{equation}
({\hat V}^m)^\ell_j\, \delta u (x) = \frac{i}{k}\, u (x)\, e^{~\ell}_j\, e^{-i (m+\frac{{p}_{j\ell}}{k}) x}\ ,\quad
\frac{\hat\delta}{\delta {p}_\ell} \, \delta u (x) = \frac{i}{k} \, x\, u(x) e^\ell_\ell\ .
\label{basic-on-v}
\end{equation}
This gives (trivially) $\{ p_j \,,\, p_\ell  \} = 0$ and
\begin{equation}
\{ u^A_j (x)\, ,\, p_\ell \} = i\, u^A_j (x) \delta_{j\ell}\quad\Rightarrow\quad
\{ (M_p)_\ell^\ell\,,\, u^A_j (x)\} = \frac{2\pi}{k}\, u^A_j(x) (M_p)_\ell^\ell\, \delta_{j\ell}\ .
\label{pPBex}
\end{equation}
The PB of two Bloch wave fields, on the other hand, is quadratic,
\begin{eqnarray}
&&\{ u_1(x_1)\,,\,u_2(x_2) \} \equiv {\cal P}\, (u (x_1) , u (x_2) ) =
- \, u_1(x_1) u_2(x_2)\, \sum_{j\ne \ell} f_{j\ell}(p) (e_j^{~j} )_1 (e_\ell^{~\ell} )_2 +
\nonumber\\
&&+\, u_1(x_1) u_2(x_2)
\left( \frac{\pi}{k}\, \varepsilon (x_{12}) \sum_{\ell} (e_\ell^\ell )_1 (e_\ell^\ell )_2 +
\frac{1}{i k}\, \sum_{j\ne\ell} \sum_{m\in{\mathbb Z}}
\frac{e^{i(m+ \frac{p_{j\ell}}{k})x_{12}}}{m+ \frac{p_{j\ell}}{k}}(e^{~j}_\ell )_1 (e^{~\ell}_j )_2  \right) = \nonumber \\
&&= \frac{\pi}{k}\, u_1(x_1) u_2(x_2)\,\left( \varepsilon (x_{12} ) \sum_{\ell} (e_\ell^{~\ell} )_1 (e_\ell^{~\ell} )_2 +
\sum_{j\ne\ell} \,\varepsilon_{\frac{p_{j\ell}}{k}} (x_{12})\, (e^{~j}_\ell )_1 (e^{~\ell}_j )_2 \right) - \nonumber\\
&&-\,  u_1(x_1) u_2(x_2)\, r_{12} (p) \ .
\label{uuPBex}
\end{eqnarray}
Here the classical dynamical $r$-matrix $r_{12} (p)$ coincides with (\ref{dynr}),
and the discontinuous functions $\varepsilon (x)$ and $\varepsilon_z (x)$
(it is appropriate to consider them as distributions) are given by the series
\begin{eqnarray}
&&\varepsilon (x) := \frac{1}{i\pi}\,\sum_{m\ne 0} \frac{e^{i m x}}{m} + \frac{x}{\pi}
= \frac{2}{\pi}\,\sum_{m=1}^\infty\, \frac{{\sin}\, m x }{m} + \frac{x}{\pi} \ ,\label{eps}\\
&&\varepsilon_z (x) := \frac{1}{i\pi}\, \sum_m\, \frac{e^{i(m+z)x} - 1}{m+z}\qquad(\, z\notin{\mathbb Z}\, )\ ,
\label{epsz}
\end{eqnarray}
respectively. The first one is just a twisted periodic generalization of the sign function $sgn (x)$,
\begin{eqnarray}
&&{\varepsilon} (x + 2\pi N) = {\varepsilon} (x) + 2 N \quad (\, N \in {\mathbb Z}\,)\ ,\qquad {\varepsilon} (0) = 0\ ,\nonumber\\
&&\varepsilon (x) = sgn (x) \quad{\rm for}\quad -2\pi < x < 2 \pi\ ,
\label{eps-sgn}
\end{eqnarray}
and its derivative is twice the {\it periodic} $\delta$-function
\begin{equation}
\delta_{per} (x) := \frac{1}{2\pi} \sum_m e^{imx} \equiv \sum_m \delta (x+2\pi m)\ .
\label{eps-perd}
\end{equation}
The properties of the second one, $\varepsilon_z (x)$ defined by (\ref{epsz}), follow from the
Euler formula\footnote{See e.g. \cite{Weil}. An integrated version of (\ref{epsz-cot}) appeared in \cite{BL};
we thank L. Feh\'er for indicating this reference to us.} for $\cot (\pi z)$
yielding (for $x \in {\mathbb R}\,,\ z \notin{\mathbb Z}$)
\begin{equation}
\lim_{N\to\infty} \,\frac{1}{\pi}\,\sum_{m=-N}^{N}  \, \frac{e^{i(m+z)x}}{m+z}\, =\,
{\cot} (\pi z) + i \,\varepsilon_z (x) \ ,\qquad \varepsilon_z (0) = 0\ .
\label{epsz-cot}
\end{equation}
The derivative of $\varepsilon_z (x)$ in $x$ is proportional to a
twisted version of the periodic $\delta$-function,
\begin{equation}
\frac{1}{2}\, \frac{\partial}{\partial x}\, \varepsilon_z (x) = e^{izx}\,\delta_{per} (x)
\label{dtwisted}
\end{equation}
which implies that, for $-2\pi < x < 2 \pi\,,\ \varepsilon_z (x) = sgn (x) = \varepsilon (x)$.
One concludes that for $-2 \pi < x_{12} <2 \pi$ the two terms in (\ref{uuPBex}) containing
$\varepsilon (x)$ and $\varepsilon_z (x)$ combine to produce the sign function times the permutation matrix
$P_{12} = \sum_{i,j} e^{~j}_i e^{~i}_j\,:$
\begin{eqnarray}
&&\label{uuPBsgn}
\{ u_1(x_1)\,,\,u_2(x_2) \} = u_1(x_1) u_2(x_2)\,( \frac{\pi}{k}\, sgn (x_{12}) P_{12}
- r_{12} (p)\,) \nonumber\\
&&{\rm for}\quad -2\pi < x_{12} <2\pi\ .
\end{eqnarray}
By the twisted periodicity of $u(x)\,$ and with the help of (\ref{pPBex}), one can reconstruct
the PB $\{ u_1(x_1)\,,\,u_2(x_2) \}$ for general $x_1$ and $x_2$ from the one in which the values of both arguments are restricted
to intervals of length $2\pi$ (as e.g. in (\ref{uuPBsgn})).
On the other hand, using the twisted periodicity of $\varepsilon (x)$ (\ref{eps-sgn})
and the twisted periodicity property
\be
\lb{tw-per}
\sum_m\, \frac{e^{i(m+z)(x+2\pi)}}{m+z} = e^{2\pi i z} \sum_m\, \frac{e^{i(m+z)x}}{m+z}\qquad (\,{\rm for}\ \  z\notin {\mathbb Z}\,)\ ,
\ee
one can show that the relation
\be
\lb{2pi}
\{ u_1(x_1+2\pi)\,,\,u_2(x_2) \} = \{ (u (x_1) M_p )_1\,,\,u_2(x_2) \}
\ee
holds, which provides a consistency check for (\ref{pPBex}) and (\ref{uuPBex}).

\smallskip

Proceeding to the Dirac brackets we first note that, as it follows from (\ref{OPchi}),
the infinite matrix of PB between the independent constraints
\be
\Phi = \{ P\,,\ {\tilde\chi}\,, \ J_r\,,\ r\ne 0\, \}\qquad
(\, P\equiv - J_0\,,\ {\tilde\chi} = \frac{1}{n}\,{\rm log} ({\tilde D} {\cal D}_q(p))\,)
\lb{BWconstr}
\ee
consists of $2\times 2\,$ non-degenerate (canonical) blocks
\be
( \{ \Phi_\ell , \Phi_{{\ell}\,^\prime} \} ) =
\begin{pmatrix}0& \{ P ,{\tilde\chi} \}&\dots&\dots&\dots&\dots\cr
\{ {\tilde\chi} , P \}& 0 &\dots&\dots&\dots&\dots\cr
\dots&\dots&\dots&\dots&\dots&\dots\cr
\dots&\dots&\dots&0&\{ J_r , J_{-r} \}&\dots\cr
\dots &\dots &\dots&\{ J_{-r} , J_r \}&0&\dots\cr
\dots&\dots&\dots&\dots&\dots&\dots\end{pmatrix}
\ .
\lb{PhiPB}
\ee
Hence, the Dirac bracket of any two phase variables $b(x_1)\,,\, c(x_2)\,$ from the Bloch waves sector is
\ba
&&\{ b(x_1 ) , c (x_2 ) \}_D = \{ b(x_1 ) , c(x_2 ) \} +\nn\\
&&+ \{ P ,\, {\tilde\chi} \}^{-1}\,
\left( \{ b(x_1 ) ,P \}\, \{ {\tilde\chi} , c (x_2 ) \} - \{ b(x_1 ) , {\tilde\chi} \}\, \{ P , c (x_2 ) \} \right) +
\lb{Dbr}\\
&&+ \sum_{r=1}^\infty \{ J_r , \, J_{-r} \}^{-1}\,
\left( \{ b(x_1 ) , J_r \}\,\{ J_{-r} , c (x_2 ) \} - \{ b(x_1 ) , J_{-r} \}\,\{ J_r , c (x_2 ) \} \right) \nn
\ea
i.e., to compute it we need to find the PB $\{ P , {\tilde\chi} \}\,,\ \{ J_r , \, J_{-r} \}$
as well as those of $b(x_1)$ and $c(x_2)\,$ with the constraints (\ref{BWconstr}).

As it follows directly from (\ref{OPchi}), the Hamiltonian vector field corresponding to $J_r\,,\, r\ne 0$
is $X_{J_r} = - i k n r \frac{\d}{\d J_{-r}}\,$ and that for $P\equiv - J_0$ is $X_P = - i \frac{\d}{\d {\tilde\chi}}\,,$ hence
\be
\{ J_r ,\, {\tilde\chi} \} = i\,\d_{r 0}\quad
(  \{ P ,\, {\tilde\chi} \} = - i \, )\ ,\qquad
\{ J_r ,\, J_s \} = - i k n r\,\d_{r+s ,\, 0}
\lb{PB-P-c}
\ee
and
\be
\{ P ,\, {\tilde\chi} \}^{-1} = i\ \ ,\qquad \{ J_r , J_{-r} \}^{-1} = \frac{i}{k n r}\ ,\quad r=1,2,\dots\ .
\lb{invjj1}
\ee
The PB of $P$ with the basic variables follow immediately from (\ref{pPBex}):
\be
\{ P ,\, u(x)\} \, = - i\, u(x)\ ,\qquad \{ P ,\, p_\ell \} = 0\ .
\lb{PB-P}
\ee
The PB of the modes $J_r$ of the abelian current $J(x)$ can be computed, by taking the trace,
from those for $j(x) = ik\, u'(x)u^{-1}(x)$ (cf. (\ref{clKZ})) which
follow, in turn, from those for $u(x)\,,$ (\ref{uuPBsgn}):
\be
\{ j_1(x_1) , u_2(x_2) \} \, =\, 2\pi i \, P_{12}\, u_2 (x_2)\,\d_{per} (x_{12})\ ,\qquad
\{ j (x_1) , p_\ell \} = 0\ .
\lb{jx-PB}
\ee
(Due to the periodicity of the current, $j(x+2\pi) = j(x)\,,$ the first PB
including the periodic $\d$-function (\ref{eps-perd}) is valid for arbitrary real $x_1 , x_2 .$)
Taking the trace in the first space and using ${\rm tr}_1 P_{12} = \sum_{i,j} \d_j^i (e^{~j}_i)_2 = \id_2\,,$ we obtain
\be
\{ J (x_1) , u (x_2) \} \, =\, 2\pi i \, u (x_2)\,\d_{per} (x_{12})\ ,\qquad
\{ J (x) , p_\ell \} = 0
\lb{Jx-PB}
\ee
or, in terms of modes (\ref{JrDabel}),
\be
\{ J_r , u(x) \} = i\, e^{irx} u(x)\ ,\qquad \{ J_r , p_\ell \} = 0 \ .
\lb{Jr-PB}
\ee
We finally note that the only non-trivial PB of ${\tilde\chi}$ (\ref{chitilde}) with the variables in (\ref{OPchi})
is the one with $P$; in particular, ${\tilde\chi}\,$ Poisson commutes with the differences $p_{j\ell}\,.$
Eqs. (\ref{uu1}), (\ref{tildeDabel}) (implying $\frac{\partial}{\partial P}\, u(x) = \frac{ix}{k n}\, u(x)$)
and the equality $p_\ell = \frac{1}{n} (P - \sum_{j=1}^n p_{j\ell})$ give
\be
\{ {\tilde\chi} , u(x) \} = \{{\tilde\chi} , P \} \frac{ix}{k n}\, u(x) = - \frac{x}{k n}\, u(x)\ , \qquad
\{ {\tilde\chi} , p_\ell \} = \frac{1}{n}\,\{ {\tilde\chi} , P \} = \frac{i}{n} \ .
\lb{chit-PB}
\ee
Hence, the terms that have to be added to $\{ u_1 (x_1) , u_2 (x_2) \}$ to obtain the corresponding Dirac bracket (\ref{Dbr}) are
\ba
&&\{ P , {\tilde\chi} \}^{-1}\, \left(\, \{ u_1 (x_1) , P \}\,
\{ {\tilde\chi} , u_2 (x_2 ) \} - \{ u_1 (x_1), {\tilde\chi} \} \,\{ P , u_2 (x_2 ) \}\,\right) =  \nn\\
&&= - \frac{x_{12}}{kn}\,u_1 (x_1)\,u_2 (x_2 ) \ ,\lb{Pchitterm}\\
&&\sum_{r=1}^\infty \,\{ J_r , J_{-r} \}^{-1}\, \left(\, \{ u_1 (x_1) , J_r \}\,
\{ J_{-r} , u_2 (x_2 ) \} - \{ u_1 (x_1), J_{-r} \} \,\{ J_r , u_2 (x_2 ) \}\,\right) = \nn\\
&&= \frac{i}{kn}\sum_{r=1}^\infty \, \frac{e^{irx_{12}} - e^{-irx_{12}}}{r} \,u_1 (x_1)\,u_2 (x_2 ) =
- \frac{2}{kn}\sum_{r=1}^\infty \, \frac{\sin r x_{12}}{r}\,u_1 (x_1)\,u_2 (x_2 )\ .\nn
\ea
Combining (\ref{uuPBsgn}) and (\ref{Pchitterm}), we obtain, for $-2\pi < x_{12} <2\pi$
\ba
&&\{ u_1 (x_1) , u_2 (x_2) \}_D  = \{ u_1 (x_1) , u_2 (x_2) \} -
\frac{\pi}{n k}\, u_1 (x_1) \, u_2 (x_2)\, sgn (x_{12}) =\nn\\
&&= u_1 (x_1) \, u_2 (x_2)\, ( \frac{\pi}{k}\, sgn (x_{12} )\, C_{12} - r_{12} (p) )
\lb{uuDir}
\ea
where $C_{12} = P_{12} - \frac{1}{n} \id_{12}\,,$ see (\ref{Cn-sigma}) and we have made use of the expansion
(\ref{eps}) for the twisted periodic $\varepsilon (x)$ as well of (\ref{eps-sgn}).
The Dirac bracket of $u^A_j (x)\,$ with $p_\ell\,$ is
\be
\{ u^A_j (x) , p_\ell \}_D = \{ u^A_j (x) , p_\ell\} + i\,\{ u^A_j (x) , P \} \{ \tilde{\chi} , p_\ell \} =
i\, u^A_j (x)\, (\d_{j\ell} - \frac{1}{n} )
\lb{DPBdiffer2}
\ee
implying
\be
\{ u^A_j(x),\, p_{\ell \ell +1} \}_D = i\, (u (x)\, h_\ell)^A_j\ ,\qquad
\{ u_1(x) ,\, M_{p 2}  \}_D = - \frac{2\pi}{k}\, u_1(x) M_{p 2}\, \s_{12}\ .
\lb{uMp-PB}
\ee
Due to the twisted periodicity of $u(x)\,,$ (\ref{uuDir}) and (\ref{uMp-PB}) allow to calculate
$\{ u_1(x_1) \,,\, u_2(x_2) \}_D$ for arbitrary values of $x_1$ and $x_2\,.$

\smallskip

The Dirac PB involving the $su(n)\,$ current $j(x)$ can be
obtained either directly from (\ref{uuDir}) and (\ref{clKZ}) or by
applying the Dirac reduction to (\ref{jx-PB}). One gets \ba
&&\{ j_1(x_1) , u_2(x_2) \}_D \, =\, 2\pi i \, C_{12}\, u_2 (x_2) \,\d_{per} (x_{12})\quad\Leftrightarrow\nn\\
&&\{ j_a (x_1) , u (x_2) \}_D \, =\, 2\pi i \, T_a \, u (x_2)\,\d_{per} (x_{12})\ ,\quad{\rm or}\lb{curf1}\\
&&\{ j_m^a , u (x) \}_D \, =\, i \, t^a \, u (x)\, e^{imx}\nn\\
&&{\rm for}\quad j(x) = j^a (x)\, T_a \ (\, \equiv j_a (x)\, t^a\, )\, = \sum_m j_m^a\, T_a\, e^{-im x} \nn
\ea
and further (from now on we shall skip the
subscript {\scriptsize{$D$}} for the Dirac brackets),
\ba &&\{ j_1 (x_1) , j_2 (x_2) \} = 2\pi i\, [ C_{12} , j_2 (x_2) ] \,\d_{per} (x_{12}) \,
+ 2 \pi k \, C_{12}\, {\d}_{per}'(x_{12})\quad\Leftrightarrow\nn\\
&&\{ j_a(x_1) , j_b (x_2) \} = 2\pi\, f_{ab}^{~~c} j_c (x_2)\,\d_{per} (x_{12}) + 2 \pi k \, \eta_{ab}\, {\d}_{per}'(x_{12})\ ,\quad{\rm or}\qquad\nn\\
&&\{ j_m^a , j_n^b \} \, =\, f^{ab}_{~~c}\, j_{m + n}^c - i\, k\, m\, \eta^{ab}\, \d_{m+n , 0}\qquad (\, [ t^a , t^b ] = i f^{ab}_{~~c}\, t^c \,) \ .
\lb{KacM1}
\ea
Eq.(\ref{KacM1}) is the classical (PB) counterpart of the defining relations of the {\em affine (current) algebra}
$\widehat{\cal G}\,$ at level $k\,$ while (\ref{curf1}), whose form could be actually anticipated from the fact that $j(x)$ is the Noether
current generating left translations, shows that $u(x)$ is a {\em primary field} corresponding to the fundamental representation of ${\cal G} = su(n)\,.$

\smallskip

The PB of the chiral component of the Sugawara stress energy
tensor (\ref{Tchir}), $T(x) = \frac{1}{2k}\,{\rm tr}\, j^2(x) = \frac{1}{2k}\,\eta^{ab} j_a(x) j_b (x)$
are easy to compute from those of the current (\ref{KacM1}).
Making use of the total antisymmetry of the structure constants $f_{abc}$ (\ref{fabc}), we obtain
\ba
&&\{ j_a (x_1) , \,{\rm tr}\, j^2 (x_2) \} = \eta^{bc} \{ j_a(x_1) , j_b(x_2) j_c(x_2) \} = 4 \pi k\, j_a (x_2)\, \d_{per}'(x_{12})\ ,\quad {\rm or}\nn\\
&&\{ j_m^a ,\, \eta_{bc}\sum_\ell j_{-\ell}^b\, j_{n+\ell}^c\, \} = - 2\,i\,k\, m\, j_m^a
\lb{j-j2}
\ea
and hence,
\be
\{ j (x_1) , T (x_2) \} = 2 \pi \, j (x_2) \d_{per}'(x_{12})\ .
\lb{jT}
\ee
On the other hand, the current-field PB (\ref{curf1}), together with (\ref{clKZ}), imply
\be
\{ T (x_1) , u (x_2) \} = \frac{2\pi i}{k}\, j(x_1)\, u(x_2)\, \d_{per}(x_{12}) = - 2 \pi\, u'(x_2)\, \d_{per}(x_{12})\ .
\lb{stressf1}
\ee
Introducing the mode expansion $T(x) = \sum_m L_m e^{-imx}\,,$ one derives from Eqs. (\ref{jT}) and (\ref{stressf1}), respectively,
the following PB characterizing the chiral stress energy tensor modes as generators of local diffeomorphisms:
\ba
&&\{ j (x) , L_n \} = \frac{d}{dx} \left(\, j(x) e^{inx}\right)\qquad\Leftrightarrow\qquad
\{ j_m^a , L_n \} = - i\, m \, j_{m+n}^a\ ,\nn\\
&&\{ u(x) , L_n \} = e^{inx}\, \frac{d u}{d x} (x) \ .
\lb{LPB}
\ea
Eq.(\ref{jT}) also implies
\be
\{T (x_1) , T (x_2) \} = \frac{2\pi}{k} \, {\rm tr}\, (j(x_1) j (x_2)) \, \d_{per}'(x_{12})\ .
\lb{TT}
\ee

\smallskip

Clearly, Eqs. (\ref{clKZ}) and (\ref{DPBdiffer2}) imply that the current $j(x)$ (and hence, the stress energy tensor $T(x)$)
commute with $p_\ell\,,$ i.e.
\be
\{ j^a_m , p_\ell \} = 0\ ,\qquad \{ L_n , p_\ell \} = 0\ .
\lb{jTpl}
\ee

\smallskip

We shall finalize this section by showing how the basic properties of a classical dynamical $r$-matrix (see \cite{EV}) arise
as consistency conditions for the Poisson structure of the Bloch waves, i.e. how the mere existence of
(\ref{uuDir}) and (\ref{uMp-PB}) restricts $r_{12}(p)\,.$
The most important among them, that $r_{12}(p)\,$ solves the classical dynamical Yang-Baxter equation (\ref{CDYBE}),
follows from the Jacobi identity for the PB (\ref{uuDir}).
Indeed, performing the calculation, one gets the triple tensor product $u_1 (x_1)\, u_2 (x_2)\, u_3 (x_3) \,$
multiplied from the right by an expression containing three
different kinds of commutators, of $C$-$C\,,\ C$-$r\,,$ and $r$-$r\,$ type, respectively.
The first group of terms produces the right-hand side of (\ref{CDYBE}),
$\frac{\pi^2}{k^2}\, [ C_{12} , C_{23} ]\,.$
To see this, one uses (\ref{CCrel}) and the following quadratic identity satisfied by the sign function,
invariant with respect to point permutations:
\be
{sgn}\, (x_{13})\, {sgn}\, (x_{32})\, +\,
{sgn}\, (x_{21})\, {sgn}\, (x_{13})\, +\,
{sgn}\, (x_{32})\, {sgn}\, (x_{21})\, =\, - 1\ .
\lb{eps2}
\ee
The second group containing mixed commutators is actually zero, due to the invariance of $C_{12}\,$
with respect to the $ad\,{\cal G}\,$ action (\ref{ad-inv12}) implying, for example,
$[ r_{13}(p) + r_{23}(p) , C_{12} ] = 0\,.$
The third group (of $r$-$r$ terms) multiplying $u_1 (x_1)\, u_2 (x_2)\, u_3 (x_3) \,$
gives rise to the left hand side of the modified classical  dynamical YBE (\ref{CDYBE}).

The skew-symmetry of (\ref{uuDir}) implies {\em "unitarity"}, $r_{12} (p) + \, r_{21} (p) = 0\,.$
Finally, Eqs. (\ref{DPBdiffer2}) or (\ref{uMp-PB}) and the Jacobi identity involving
$u_1 (x_1) , u_2 (x_2)\,$ and $p_{\ell}$ (or $p_{\ell \ell+1}\,,$ respectively)
impose the {\em zero weight} condition on $r_{12}(p)\,,$
\ba
&&\qquad\, [ ( e_\ell^\ell)_1 + ( e_\ell^\ell )_2 \,,\, r_{12} (p) ] = 0\ ,\qquad \ell = 1,\dots , n\nn\\
&&\Rightarrow\quad [ h_{\ell 1} +  h_{\ell 2}\, ,\, r_{12}(p) ] = 0\ ,\qquad \ell = 1,\dots , n-1\ .
\lb{neutrality}
\ea
One can explicitly check that $r_{12}(p)$ given by (\ref{dynr}), (\ref{dyn-r-matr})
indeed satisfies all the three conditions specified above. Note that our classical dynamical
YBE (\ref{CDYBE}) is written in a form that keeps track (in the term ${\rm Alt}\, (d r (p))$) of the
extension of the phase space. Also, $r_{12}(p)$ (\ref{dynr}) only depends on the differences $p_{j\ell}$ (cf. (\ref{f01})),
but its diagonal part does {\em not} belong to $su(n)\wedge su(n)\,.$

\smallskip

The first expression for the dynamical $r$-matrix appeared already in the early studies of the
chiral WZNW model \cite{BDF} (see also \cite{BFP} for further generalization in a direction
different from ours). Classification theorems for classical dynamical $r$-matrices in various cases
(for Kac-Moody algebras, simple Lie algebras etc. as well such with a spectral parameter) can be found in \cite{EV}.

\subsection{PB for the chiral field $g(x)$. Recovering the $2D$ field}

We have described so far (in full details, for $G=SU(n)$) the two basic canonical versions of the chiral WZNW model,
the first one described in terms of the Bloch wave field $u(x)\,$ with diagonal monodromy matrix $M_p\,,$
whose quadratic PB (\ref{uuDir}) involve the classical {\em dynamical} $r$-matrix $r_{12}(p)\,$
and the second, in terms of chiral field $g(x)\,$ with general ($G$-valued) monodromy matrix $M$.
These two pictures are intertwined by the zero modes $a\,$ obeying (\ref{aintertw}).

\subsubsection{The Poisson brackets of the chiral field $g(x)$}

We shall now use the PB for the zero modes $a^j_\a\,$ and the Bloch waves $u(x)^A_j\,$ to find the PB for the chiral field
$g(x)^A_\a\,$ (\ref{gua}). As explained in Section 3.1, the two constituents of $g(x)^A_\a\,$ can be treated as independent
(and therefore, Poisson commuting), only at the end we should identify the variables
${\mathfrak p}\,$ (for the Bloch waves) and $p\,$ (for the zero modes) and hence,
the corresponding diagonal monodromies. This prescription is equivalent to introducing an additional set of {\em first class constraints}:
\be
{\cal C}_p := {\mathfrak p} - p \approx 0\quad\Rightarrow\quad
M_{\mathfrak p} \ ( \, = u(x)^{-1} u(x+2\pi ) \, ) \ \approx M_p\ .
\lb{MpMp}
\ee
So the PB of the covariant group valued field $g(x)= u(x)\, a$ are obtained by combining (\ref{uuDir}) and (\ref{PBex}):
\ba
&&\{ g_1(x_1) , g_2(x_2) \} =
\left( \{ u_1(x_1) , u_2(x_2) \} a_1 a_2 + u_1(x_1) u_2(x_2) \{ a_1 , a_2 \} \right)_{\mid{\cal C}_p\approx 0} =\nn\\
&&= u_1(x_1) u_2(x_2)\, \left( ( \frac{\pi}{k}\, C_{12}\, {sgn} (x_{12}) - r_{12} (p))\, a_1 a_2 +
r_{12} (p)\, a_1 a_2 - \frac{\pi}{k}\, a_1 a_2 \, r_{12} \right) =\nn\\
&&= \frac{\pi}{k}\, g_1(x_1) g_2(x_2) \, (C_{12}\, {sgn} (x_{12}) - r_{12} ) \equiv \lb{gpb}\\
&&\equiv - \frac{\pi}{k}\, g_1(x_1) g_2(x_2) \,(r^-_{12}\, \theta (x_{12}) + r^+_{12}\, \theta (x_{21}) )\ ,
\quad -2\pi < x_{12} <2\pi
\nn
\ea
where $r_{12}$ is given by (\ref{stand-r-matr}) and $\theta (x)$ is the Heaviside step function,
\be
\theta (x) = \left\{\begin{array}{ll} 0\ ,\ x\le 0\\ 1\ ,\ x > 0 \end{array}\right.\ ,\qquad \theta (x) - \theta (- x) = {sgn} (x) \ .
\lb{heavi}
\ee
Identifying the monodromy matrix $M\,$ with that of the zero modes, one trivially obtains, from (\ref{Mgen})
and  (\ref{Mpma})
\be
\lb{Mgeng}
\{ M_1 , g_2 (x)\} = \frac{\pi}{k}\, g_2 (x)\, (r^+_{12} M_1 - M_1 r^-_{12} )\ ,\quad
\{ M_{\pm 1} , g_2 (x)\} = \frac{\pi}{k}\, g_2 (x)\, r^\pm_{12}\, M_{\pm 1}\ .
\ee
The compatibility of the PB (\ref{gpb}) and (\ref{Mgeng}) can be easily checked, e.g.
\ba
&&\{ g_1(x_1) , g_2(x_2) \} =  - \frac{\pi}{k}\, g_1(x_1)\, g_2(x_2) \,r^+_{12}\qquad{\rm for}\quad -2\pi < x_{12} <0\quad \Rightarrow\nn\\
&&\{ g_1(x_1 + 2\pi) , g_2(x_2) \} = \{ g_1(x_1) , g_2(x_2) \} M_1 + g_1(x_1) \{ M_1 , g_2 (x_2)\} = \nn\\
&&=- \frac{\pi}{k}\, g_1(x_1)\, g_2(x_2) \,r^+_{12} M_1 + \frac{\pi}{k}\,g_1(x_1)\, g_2 (x_2)\, (r^+_{12} M_1 - M_1 r^-_{12} ) =\nn\\
&&=- \frac{\pi}{k}\, g_1(x_1 +2\pi)\, g_2(x_2) \,r^-_{12} \qquad{\rm for}\quad g_1(x_1 + 2\pi) = g_1(x_1) M_1\ .
\lb{gpb-comp}
\ea
The current and hence, the stress energy tensor, Poisson commute with the zero modes, so that their
PB with the chiral field $g(x)$ are analogous to those given in (\ref{curf1}) and (\ref{LPB}), respectively.
We have, in particular,
\be
\{ j_m^a , g (x) \}  = i\,  t^a  g (x)\, e^{imx}\ ,\qquad \{ g(x) , L_n \} = e^{inx}\, \frac{d g}{d x} (x)\ .
\lb{jTg}
\ee

\subsubsection{Symmetries of the chiral PB}

A guiding principle in quantization is to retain the invariance of the classical system replacing,
if needed, the classical notions of symmetry by appropriate quantum analogs.
The set of chiral PB is preserved by the following transformations (the first two of them are inherited from the corresponding properties of the
Bloch waves, while the third is shared with the zero modes):

(1) {\em $G$-valued periodic left shifts}
\be
g(x)\ \to\ h(x) \, g(x)\ ,\qquad h(x)\in G\ , \quad h(x+2\pi) = h(x)
\lb{Gleft}
\ee
are generated by the chiral current $j(x)\,$ (cf. Section 2.4). This transformation does not affect the zero modes; accordingly,
the PB of $j(x)\,$ with the left chiral field $g(x)\,$ is the same as its bracket with the Bloch wave, (\ref{curf1}):
\be
\{ j_1(x_1) , g_2(x_2) \} \, =\, 2 \pi i \, C_{12}\, g_2 (x_2)\,\d_{per} (x_{12})\ .
\lb{curg}
\ee
To prove that the PB (\ref{curg})is also invariant with respect to (\ref{Gleft}) (the current itself transforming as $j(x)\to\ h(x)\, j(x)\, h(x)^{-1}$),
we use the fact that the tensor product $h_1(x_1) h_2(x_2)$ commutes with $C_{12}$ when multiplied with the periodic delta function.

\smallskip

(2) {\em Chiral conformal symmetry} with respect to smooth monotonic coordinate transformations of the type
\be
x\ \to \ f(x)\ ,\quad f'(x) > 0 \qquad (\, f(\pm\pi) = \pm\pi \ ,\ -\,\pi < x < \pi\,)\ .
\lb{chiralconf}
\ee
Checking the invariance of Eq.(\ref{gpb}) with respect to (\ref{chiralconf}), one uses the following
obvious property of the step function under such mappings:
\be
\theta (f(x_1) - f(x_2)) = \theta (x_{12})\,.
\lb{tf}
\ee
Alternatively, using (\ref{jTg}), one can validate the infinitesimal conformal invariance
of (\ref{gpb}) generated by the modes $L_n$ of the stress energy tensor.
The invariance of (\ref{KacM1}) and (\ref{curg}) is equivalent to the following easily verifiable relations:
\ba
&&\{ \{ j^a_m , L_r \} , j^b_n \} + \{ j^a_m , \{ j^b_n , L_r \} \} = f^{ab}_{~~c} \{ j^c_{m+n} , L_r \}\ ,\nn\\
&&\{ \{ j^a_m , L_n \} , g(x) \} + \{ j^a_m , \{ g(x) , L_n \} \} = i\, t^a \{ g(x) , L_n \} e^{imx}\ .\qquad
\lb{clVir}
\ea
This is the classical prerequisite of the invariance of the quantized chiral model with respect to infinitesimal diffeomorphisms
(implemented by the {\em Virasoro algebra}).

\smallskip

(3) {\em Poisson-Lie symmetry } with respect to constant {\em right} shifts of the chiral field $g(x)\,.$
The left sector PB are invariant with respect to the transformations
\be
g_L(x)\ \to\ g_L(x)\, T_L\ ,\quad M_L\ \to \ T_L^{-1}\, M_L\, T_L\qquad (\, T_L\in G\,)\ ,
\lb{PLleftg}
\ee
provided that
\be
\{ g_{L1} , T_{L2} \} = 0\ ,\qquad \{ T_{L1} , T_{L2} \} = \frac{\pi}{k}\, [r_{12} \,, T_{L1} T_{L2} ]\ ,
\lb{PLdefg}
\ee
cf. (\ref{PBSkl}). It was proposed already in the early papers on the subject \cite{MR, F1, AS, G} that the PL symmetry is to be
replaced, in the quantized chiral WZNW theory, by quantum group invariance of the corresponding exchange relations.

\subsubsection{The classical right movers' sector; the "bar" variables}

As already noted in Section 2.3, transferring the PB structure from the left to the right movers' sector
(written in terms of chiral fields $g_L\,$ and $g_R\,$ such that $g(x^+, x^-) = g_L (x^+ )\, g_R^{-1} (x^- )\,,$ cf. (\ref{LR}))
amounts to a mere change of sign, see (\ref{O-O}), (\ref{Oc}) and (\ref{O2alt}), (\ref{O}). The extreme simplicity of this
"rule of thumb" makes it quite suitable for practical applications concerning the classical model. This will be exemplified in the
following Section 3.5.4 where the locality and monodromy invariance of the $2D\,$ field will be examined.

It is easy to foresee, however, that the pair of chiral variables $g_L\,,\,g_R\,$ will not be convenient in the quantum case
when the interpretation of the matrix inverse would lead to considerable difficulties. In addition, being formally equivalent to replacing
the level $k\,$ by its opposite $- k\,,$ the thumb rule forces us to use $q^{-1}$ rather than $q$ (\ref{qcl}) as a classical
deformation parameter for the right sector, and this fact will persist in the quantum case as well.
Both problems are trivially overcome by just setting
\be
\bar g (\bar x) = g^{-1}_R (\bar x)\ ,\quad \bar g (\bar x + 2\pi ) = \bar M\, \bar g (\bar x )\quad
(\, \bar M = M_R^{-1}\,)\ ,\quad\bar g (\bar x) = \bar a\, \bar u (\bar x)
\lb{ggbar}
\ee
for $x = x^+\,,\ \bar x = x^-\,$ so that now $g^A_B(x,\bar x) = g^A_\a (x)\, {\bar g}^\a_B (\bar x)\,.$
With the "bar" variables the left and the right sector are put on equal footing;
we shall also have, eventually, the same deformation parameter $q\,$ for both sectors.

As the chiral Poisson brackets provide the basis for the canonical quantization performed in the following Chapter 4, we shall collect below
those already obtained for the left sector and also derive the corresponding ones for the right sector in the bar variables
by changing the sign in (\ref{gpb}), (\ref{uuDir}), (\ref{PBex}) and (\ref{Mgen}) and then substituting (\ref{ggbar}). We thus get
\ba
&&\{ g_1(x_1) , g_2(x_2) \} = \frac{\pi}{k}\, g_1(x_1)\, g_2(x_2) \, (C_{12}\, {sgn} (x_{12}) - r_{12} ) = \nn\\
&&= - \frac{\pi}{k}\, g_1(x_1)\, g_2(x_2) \,(r^-_{12}\, \theta (x_{12}) + r^+_{12}\, \theta (x_{21}) )\ ,
\quad -2\pi < x_{12} <2\pi\ ,\nn\\\nn\\
&&\{ {\bar g}_1({\bar x}_1) , {\bar g}_2({\bar x}_2) \}= \frac{\pi}{k}\, (r_{12} - C_{12}\, {sgn} ({\bar x}_{12}) ) \,
{\bar g}_1({\bar x}_1)\, {\bar g}_2({\bar x}_2) \,= \nn\\
&&= \frac{\pi}{k}\,(r^-_{12}\,\theta ({\bar x}_{12}) + r^+_{12}\, \theta ({\bar x}_{21}) )\,\, {\bar g}_1({\bar x}_1)\, {\bar g}_2({\bar x}_2)  \ ,
\quad -2\pi < {\bar x}_{12} <2\pi\ ;\qquad\qquad
\lb{gpbbar}
\ea
\ba
&&\{ u_1 (x_1) , u_2 (x_2) \}= u_1 (x_1) \, u_2 (x_2)\, ( \frac{\pi}{k}\, C_{12}\, sgn (x_{12} ) - r_{12} (p) ) = \nn\\
&&= - u_1 (x_1) \, u_2 (x_2)\, (r^-_{12}(p)\, \theta (x_{12}) + r^+_{12}(p)\, \theta (x_{21}) )\ ,
\quad -2\pi < x_{12} <2\pi\ ,\qquad\quad
\nn\\\nn\\
&&\{ {\bar u}_1({\bar x}_1) , {\bar u}_2({\bar x}_2) \}= (\bar r_{12} (\bar p) -  \frac{\pi}{k}\, C_{12}\, {sgn} ({\bar x}_{12}) ) \,
{\bar u}_1({\bar x}_1)\, {\bar u}_2({\bar x}_2) \,= \nn\\
&&= (\bar r^-_{12} (\bar p) \,\theta ({\bar x}_{12}) + \bar r^+_{12} (\bar p)\, \theta ({\bar x}_{21}) )\, {\bar u}_1({\bar x}_1)\, {\bar u}_2({\bar x}_2)  \ ,
\quad -2\pi < {\bar x}_{12} <2\pi\ \qquad\qquad\qquad
\lb{uuDirbar}
\ea
(for $r^\pm_{12}= r_{12} \pm  C_{12}\,,\ r^\pm_{12} (p) = r_{12}(p) \pm \frac{\pi}{k}\, C_{12}\,$
and $\ \bar r^\pm_{12}(\bar p) = \bar r_{12}(\bar p) \pm \frac{\pi}{k}\, C_{12}$ with $\bar p = p_R$), as well as
\ba
&&\{ a_1 , a_2 \} = r_{12}(p) \, a_1 \, a_2 - \frac{\pi}{k} \, a_1 \, a_2 \, r_{12}
= r_{12}^{(\pm)} (p) \, a_1 \, a_2 - \frac{\pi}{k} \, a_1 \, a_2 \, r_{12}^{(\pm)}\ ,\qquad\quad
\nn\\
&&\{ {\bar a}_1 , {\bar a}_2 \}
= \frac{\pi}{k}\, r_{12}\, {\bar a}_1 \, {\bar a}_2 - {\bar a}_1 \, {\bar a}_2 \,\bar r_{12}(\bar p)
= \frac{\pi}{k}\, r_{12}^{(\pm)} {\bar a}_1 \, {\bar a}_2 - {\bar a}_1 \, {\bar a}_2 \,\bar r^{(\pm)}_{12}(\bar p)
\lb{aabar}
\ea
for $\bar a = a_R^{-1}\,.$  The PB involving $\bar p\,$ follow from (\ref{DPBdiffer2}) and (\ref{PBapD}), so we have
\ba
&&\{ u^A_j (x) , p_\ell \} = i\, (\d_{j\ell} - \frac{1}{n} )\, u^A_j (x) \ ,\qquad\,
\{ a_{\alpha}^j , p_{\ell} \} = i\, (\delta_{\ell}^j - \frac{1}{n} )\, a_{\alpha}^j \ ,\nn\\
\lb{PBaupDbar}
&&\{ {\bar u}_A^j (\bar x) , {\bar p}_\ell \} = i\, (\d^j_\ell - \frac{1}{n} )\, {\bar u}_A^j (\bar x)\ ,\qquad\
\{ {\bar a}^{\alpha}_j , {\bar p}_{\ell} \} = i\, (\delta_{j \ell} - \frac{1}{n} )\, {\bar a}^{\alpha}_j\ .\qquad\quad
\ea
The PB of the general monodromy matrices (recall that $\bar M = M_R^{-1}$ (\ref{ggbar})) are
\ba
&&\{ M_1 , g_2(x) \} = \frac{\pi}{k}\, g_2(x) (r^+_{12} M_1 - M_1 r^-_{12} )\ ,\nn\\
&&\{ {\bar M}_1 , {\bar g}_2 (\bar x) \} =
\frac{\pi}{k}\, ( r_{12}^- {\bar M}_1 - {\bar M}_1 r^+_{12} )\, {\bar g}_2 (\bar x) \ ,\lb{Mgenbar}\\
&&\{ M_1 , a_2 \} = \frac{\pi}{k}\, a_2 (r^+_{12} M_1 - M_1 r^-_{12} )\ ,\quad
\{ {\bar M}_1 , {\bar a}_2 \} = \frac{\pi}{k}\, ( r_{12}^- {\bar M}_1 - {\bar M}_1 r^+_{12} )\, {\bar a}_2 \ ,
\nn
\ea
cf. (\ref{Mgeng}), (\ref{Mgen}), (\ref{PBMM}), and
\ba
&&\{ M_1 , M_2 \} = \frac{\pi}{k}\ (M_1 r_{12}^-\, M_2 + M_2\, r_{12}^+ M_1\, - M_1 M_2\, r_{12} - r_{12} M_1 M_2 )\ ,\nn\\
&&\{ \bar M_1 , \bar M_2 \} = \frac{\pi}{k}\
(\bar M_1 \bar M_2\, r_{12} + r_{12} \bar M_1 \bar M_2 - \bar M_1\, r_{12}^+ \bar M_2\, - \bar M_2\, r_{12}^-\,\bar M_1   )\ .\qquad\qquad
\lb{PBMMbar}
\ea
Finally, the PB of the Gauss components of the monodromy matrices
(such that $M = M_+ M_-^{-1}\,$ and $\bar M = \bar M_-^{-1} \bar M_+\,,\ \bar M_\pm = M_{R\pm}^{-1}$) with the chiral fields or zero modes read
\ba
&&\{ M_{\pm 1} , g_2(x)\} = \frac{\pi}{k}\, g_2(x)\, r^\pm_{12}\, M_{\pm 1}\ ,\quad
\{ \bar M_{\pm 1} , \bar g_2(\bar x)\} = - \frac{\pi}{k}\, \bar M_{\pm 1} \, r^\pm_{12}\,\bar g_2(\bar x) \ ,\nn\\
&&\{ M_{\pm 1} , a_2\} = \frac{\pi}{k}\, a_2\, r^\pm_{12}\, M_{\pm 1}\ ,\quad
\{ \bar M_{\pm 1} , \bar a_2\} = - \frac{\pi}{k}\, \bar M_{\pm 1}\, r^\pm_{12}\,\bar a_2 \qquad\qquad
\lb{Mpmga-bar}
\ea
(cf. (\ref{Mgeng}), (\ref{Mpma})).
It is remarkable that the PB of $\bar M_\pm\,$ with themselves are {\em identical} to those of $M_\pm\,$ (\ref{Mpmmp}):
\ba
&&\{ M_{\pm 1} , M_{\pm 2} \} = \frac{\pi}{k}\, [ M_{\pm 1} M_{\pm 2} , r_{12}\, ] \ ,\quad \{ M_{\pm 1} , M_{\mp 2} \} = \frac{\pi}{k}\, [ M_{\pm 1} M_{\mp 2} , r^\pm_{12}\, ]\ ,\nn\\
&&\{ \bar M_{\pm 1} , \bar M_{\pm 2} \} = \frac{\pi}{k}\, [ \bar M_{\pm 1} \bar M_{\pm 2} , r_{12}\, ] \ ,\quad
\{ \bar M_{\pm 1} , \bar M_{\mp 2} \} = \frac{\pi}{k}\, [ \bar M_{\pm 1} \bar M_{\mp 2} , r^\pm_{12}\, ]\ .\qquad\qquad
\lb{Mpmmp-bar}
\ea

\subsubsection{Back to the $2D$ WZNW model}

To complete the "classical part" of this review, we shall show that expressing the $2D\,$ field
$g(x^+ , x^- )\,$ in terms of its chiral components (\ref{LR}) is selfconsistent. This is not obvious since we have allowed
the left and right monodromy matrices $M_L\,,\, M_R\,$ to be independent, cf. (\ref{PC}),
whereas the single-valuedness of $g(x^0 , x^1)$
(strict periodicity in the compact space variable $x^1\,$  or, equivalently,
condition (\ref{gper}) for $g(x^+ , x^- )$) requires $M_L\,$ and $M_R\,$ to be equal, see Eq.(\ref{cM}).
The latter relation cannot be imposed "in the strong sense" since the PB of left and
right chiral variables differ in sign, but it is perfectly sound {\em as a constraint}.
Indeed, to obtain the $2D\,$ field from its (independent) chiral components,
one has to project the phase space ${\cal S}_L \times {\cal S}_R\,$ on $\tilde{\cal S}\,$ (\ref{extPh}),
and this amounts to imposing the (matrix valued) gauge condition
\be
M_L \approx M_R\ ,
\lb{constrC}
\ee
cf. (\ref{O2alt}). Now the fact that left and right PB only differ in sign is exactly what is needed for
the constraints ${\cal C}:= M_L - M_R\,$ to be first class \cite{BFP}:
\be
\{ {\cal C}_1 , {\cal C}_2 \} = \{ M_{L1} - M_{R1} , M_{L2} - M_{R2} \} =
\{ M_{L1}  , M_{L2} \} + \{ M_{R1}  , M_{R2} \} \approx 0\ .
\lb{C1class}
\ee

The "observable" field $g (x^+ , x^- )= g_L (x^+ ) \, g_R^{-1} (x^- )\,$ (\ref{LR}) has to be gauge invariant.
Indeed, using (\ref{Mgeng}) and its right sector analog, we obtain
\ba
&&\{ {\cal C}_1 , g_{L2}\, g_{R2}^{-1}  \} = \{ M_{L1} , g_{L2} \}\, g_{R2}^{-1} + g_{L2}\, g_{R2}^{-1} \{ M_{R1} , g_{R2} \}\, g_{R2}^{-1} = \nn\\
&&= \frac{\pi}{k}\, g_{L2} (r_{12}^+ M_{L1} - M_{L1} r_{12}^- )\, g_{R2}^{-1} -
\frac{\pi}{k}\, g_{L2} (r_{12}^+ M_{R1} - M_{R1} r_{12}^- )\, g_{R2}^{-1} =\nn\\
&&= \frac{\pi}{k}\, g_{L2} \,
(r_{12}^+\, {\cal C}_1 - {\cal C}_1\, r_{12}^- )\, g_{R2}^{-1}\,\approx \,0\ .
\lb{gginv}
\ea
The $2D$ field is also local (already "in the strong sense") since,
according to (\ref{gpb}), for $- 2\pi < x_{12}^\pm < 2\pi\,$ we have
\ba
&&\{ g_1 (x_1^+ , x_1^- ) , g_2 (x_2^+ , x_2^- ) \} =
\{ g_{L1} (x_1^+ ) , g_{L2} (x_2^+ ) \}\,
g_{R2}^{-1} (x_2^-)\, g_{R2}^{-1} (x_2^-) +\nn\\
&&+\, g_{L1} (x_1^+ )\, g_{L2} (x_2^+ )\, g_{R1}^{-1} (x_1^-)\, g_{R2}^{-1} (x_2^-)\,
\{ g_{R1} (x_1^- ) , g_{R2} (x_2^- ) \}\, g_{R1}^{-1} (x_1^-)\, g_{R2}^{-1} (x_2^-)
=\nn\\
&&= \frac{\pi}{k}\,( {sgn}\, (x_{12}^+ ) - {sgn}\, (x_{12}^- ) )\,
g_{L1} (x_1^+ )\, g_{L2} (x_2^+ )\,
C_{12}\, g_{R1}^{-1} (x_1^-)\, g_{R2}^{-1} (x_2^-)\ ,
\lb{gloc}
\ea
and ${sgn}\, (x_{12}^+ ) = {sgn}\, (x_{12}^- )\,$ for $x_{12}$ {\em spacelike}
(i.e., $x_{12}^+\, x_{12}^- > 0\,,$ see (\ref{conev})).

\medskip

\noindent
{\bf Remark 3.7~} The reason for Eqs. (\ref{C1class}) -- (\ref{gloc}) to hold,
i.e. the fact that the left and right sector PB only differ in sign,
presupposes the equality of the classical constant $r$-matrices appearing in both.
If we restrict ourselves to chiral fields with {\em diagonal} monodromy matrices, cf. Remark 2.4
(and hence, do {\em not} introduce zero modes), we should replace (\ref{constrC}) by the constraint
$M_{p_L} \approx M_{p_R}\,.$ To ensure the locality of the $2D\,$ field $u(x)\, \bar u (\bar x)\,$
as in (\ref{gloc}), we should choose in this case equal classical {\em dynamical} $r$-matrices
for the left and right sectors. In the presence of the chiral zero modes, however, the dynamical $r$-matrices
in the two sectors can be given by {\em different} functions of the respective arguments.
(This amounts to choosing different $\b(p)\,$ in (\ref{f01}); we shall make use of the quantum counterpart of
this fact to impose, in Section 4.6.2 below, identical exchange relations for the left and right zero mode operators.)
What is needed, on top of the mentioned equality of the left and right {\em constant} $r$-matrices,
is to choose identical dynamical $r$-matrices for the Bloch waves and zero modes of {\em same} chirality
(i.e., $r_{12}(p)\,$ in (\ref{uuDirbar}) and (\ref{aabar}) should be the same, as well as $\bar r_{12} (\bar p)$).
This requirement stems from the decomposition (\ref{gua}) of the chiral fields into Bloch waves and zero modes,
cf. Remark 3.1.

\medskip

Assuming that the left and right sector constant $r$-matrices coincide, we can also prove that
the matrix elements of the $2D\,$ field $g(x^+ , x^- )\,$ Poisson commute with those of
$M^{-1}_{L\pm } M^{~}_{R\pm }\,,$ again "in the strong sense".
Indeed, using (\ref{Mgeng}) and its right sector counterpart, we obtain
\ba
&&\{ (M^{-1}_{L\pm\,})_1  (M^{~}_{R\pm})_1 \, , \, g_2 (x^+ , x^- ) \} =\nn\\
&&= -\, (M^{-1}_{L\pm\,})_1 \, \{ (M^{~}_{L\pm\,})_1 \,,\, g_{L2} (x^+) \} \, (M^{-1}_{L\pm\,})_1\,  (M^{~}_{R\pm})_1 \, g^{-1}_{R2} (x^-) -\nn\\
&&\quad -\, (M^{-1}_{L\pm\,})_1\,  g_{L2}(x^+)\,  g^{-1}_{R2}(x^-) \, \{ (M^{~}_{R\pm\,})_1 \,,\, g_{R2}(x^-) \}\,  g^{-1}_{R2}(x^-) =\nn\\
&&= -\, \frac{\pi}{k}\, (M^{-1}_{L\pm})_1\,  g_{L2} (x^+ ) \,  r^\pm_{12}\,  (M^{~}_{R \pm})_1\,  g^{-1}_{R2} (x^- ) +\nn\\
&&\ \ \ +\, \frac{\pi}{k}\, (M^{-1}_{L\pm})_1\,  g_{L2} (x^+ )\,   r^\pm_{12}\,  (M^{~}_{R \pm})_1\,
g^{-1}_{R2} (x^- ) = 0\ .
\lb{Mpm2dg}
\ea
Clearly, the zero mode analog of (\ref{Mpm2dg}) (which we shall write using the inverse product
$(M^{-1}_{R\pm})_1 (M^{~}_{L\pm\,})_1$) is also valid, cf. (\ref{Mpmga-bar}):
\be
\{ (M^{-1}_{R\pm})_1(M^{~}_{L\pm\,})_1   \,,\, Q_2 \} = 0\ ,\qquad Q:= a_L a^{-1}_R\ .
\lb{Mpm2a}
\ee
In the quantized theory, where the factors $M_\pm\,$ of the monodromy matrix (\ref{M+-})
(satisfying $R$-matrix quadratic equations) can be
conveniently parametrized in terms of the generators of the Hopf algebra $U_q(s\ell (n))\,$
(see \cite{FRT} and Section 4.3 below),
the vanishing of the {\em commutators} of $(M^{-1}_{R\pm})_1(M^{~}_{L\pm\,})_1 \,$
with $g(x^+,x^-)\,$ and $Q = a_L a^{-1}_R$ implies the "gauge invariance" of the latter with
respect to the (inverse) coproduct action of the quantum group.
In this sense the quantum group symmetry remains "hidden" in the $2D$ WZNW theory, see e.g. \cite{GS}.

\section{Quantization}

\setcounter{equation}{0}
\renewcommand\theequation{\thesection.\arabic{equation}}

{\em Quantization} of a classical system involves two steps:
\\
(i) a {\em deformation of the algebra of dynamical variables} such that the commutator of any two of them,
$f$ and $g\,,$ is given by a power series in the Planck constant $\hbar$ with leading term proportional to their PB:
\be
[ f , g ] = i \hbar\, \{ f , g \} + {\cal O} (\hbar^2 )\ .
\lb{commPB}
\ee
\noindent
(ii) {\em constructing a state space}, i.e. an inner product vector space which
carries a positive energy representation of the above quantum algebra.\footnote{Any positive
linear functional on a $C^*$-algebra of norm $1$ defines a state via the Gelfand-Naimark-Segal construction.
For a review and applications of the GNS construction to axiomatic QFT, see \cite{BLOT}.}

\smallskip

The first step is rather straightforward for a classical observable algebra of conserved currents
(like the chiral currents $j_L(x^+) \equiv j (x^+)$ and $j_R(x^-)$) that span a Lie algebra under
Poisson brackets. It is more involved when dealing with group-like objects like $g(x^+ , x^-)\,,$ and especially
with their gauge dependent chiral components. We shall start with the quantization of the chiral current algebra
reviewing, in particular, the change in the level in the Sugawara formula and then proceed to our main task,
the $R$-matrix quantization of the group valued chiral fields $g(x)$ and of the zero modes in the case of $G=SU(n)\,$
and the quantum group symmetry of their exchange relations.
The chiral state space will be then constructed as a representation of the chiral fields' algebra built on
a non-degenerate (cyclic) lowest energy vector, the {\em vacuum} $\vac\,,$ satisfying  $L_0 \mid 0 \rangle = 0\,.$
The inner product on such a space is defined by introducing a left ("bra"-) vacuum such that $\lvac L_0 = 0\,.$
(We expect that the reader is familiar with the basic notions of $2D$ CFT -- see e.g. \cite{DFMS, FSoT}.)

\subsection{The chiral conformal current algebra}

The quantum counterpart of the classical current PB (\ref{KacM1}) are the standard relations for the affine Kac-Moody
(current) algebra $\widehat{\cal G}\,$ at level $k$:
\be
[ j_m^a , j_n^b ] = i f^{ab}_{~~c} \, j_{m+n}^c + k\, m\, \eta^{ab} \,\d_{m+n, 0}\ .
\lb{KM}
\ee
The Planck constant $\hbar\,$ is hidden here in a rescaling of the current, $j \to \hbar\, j\,$ and of the level, $k \to \hbar\, k =: \bar k\,,$
cf. Remark 4.1 below, so that the right-hand side of (\ref{KM}) written in terms of the new variables is proportional to $\hbar\,.$

The local diffeomorphism invariance (\ref{LPB}) can also be extended to the quantum theory:
\be
[ j (x) , L_n ] = i\,\frac{d}{dx} \left(\, j(x) e^{inx}\right)\ .
\lb{jLcomm}
\ee
As (\ref{jLcomm}) implies
\be
[ j_m^a , L_n ] = m \, j_{m+n}^a \qquad\Rightarrow\qquad L_0\, j_m^a \mid 0 \rangle = j_m^a (L_0 - m) \vac\ ,
\lb{Ljvac}
\ee
it follows from the positive energy requirement that
\be
j_m^a \vac = 0 \qquad{\rm for}\qquad m\ge 0\ .
\lb{jonvac}
\ee

Keeping with tradition in the quantum CFT, we shall introduce at this point the {\em analytic $z$-picture} using the complex variables
\be
z := e^{i x^+}\ ,\quad \bz := e^{- i x^-}
\lb{zzbar}
\ee
in which a chiral field $\varphi (x)\,$ of dimension $\Delta$ is substituted by a field $\phi (z)$ such that
\be
\varphi (x) = z^\Delta \, \phi (z)\ .
\lb{xz}
\ee
Note that in {\em Euclidean} space-time (defined as the set of {\em real}\, Wick-rotated points
$(i x^0 , x^1) \to (x^0 , x^1) \in {\mathbb R}^2 \subset \mathbb C^2$)\ the variables $z$ and $\bz$ are complex conjugate,
\be
\lb{Eucl}
x^0\ \to\ - i\, x^0 \quad \Rightarrow \quad z \ \to\ e^{x^0 + i x^1}\ ,\quad \bz \ \to\ e^{x^0 - i x^1}
\ee
and that the infinite future/past limits $x^0\to \infty$ and $x^0\to - \infty$
correspond to $|z|\to\infty$ and $|z|\to 0\,,$ respectively.

The counterpart of (\ref{jLcomm}) for an arbitrary {\em primary} (with respect to the Virasoro algebra)
chiral field $\phi$ of dimension $\D$ reads
\be
[ L_n , \phi (z)] = z^n ( z \frac{d}{d z} + (n + 1)\,\Delta ) \,\phi (z)\ .
\lb{phiLcomm}
\ee
The deviation of $\Delta$ from its canonical (integer or half integer) value signals a field strength renormalization.

We shall have, as a consequence of energy positivity, analyticity of the vacuum expansion in both $z$ and $\bz\,;$
for example, for a primary chiral field it only involves non-negative integer powers of $z\,,$
\be
\phi (z) \vac = \sum_{m=0}^\infty \phi_{-m-\Delta}\, z^m \vac\ .
\lb{phionvac}
\ee
Calculating the norm square of (\ref{phionvac}) provides a power series convergent for $|z| < 1\,,$ by the following
general argument. Conformal (M\"obius) invariance implies
\be
L_n \vac = 0 = \lvac L_n\quad {\rm for} \quad n=0,\pm 1\ .
\lb{Moebius}
\ee
The notion of $z$-picture conjugate of a complex chiral field $\phi(z)$ of dimension $\D$ \cite{DFMS}
and the $2$-point function (determined from (\ref{phiLcomm}) and (\ref{Moebius})),
\be
\lb{phistar}
\phi(z)^* = {\bar z}^{-2\Delta} \, \phi^* ( {\bar z}^{-1} ) \ ,\qquad
\lvac \phi^* (z_1)\, \phi(z_2) \vac = N_\phi\,z_{12}^{-2\D}
\ee
yield the following expression for the norm square of the vector (\ref{phionvac}):
\be
\lb{normphionvac}
\Vert\, \phi (z) \vac\, \Vert^2 = {\bar z}^{-2\Delta} \,\lvac \phi^* ( {\bar z}^{-1} )\, \phi (z) \vac =
N_\phi\, (1 - |z|^2 )^{-2\D}\ .
\ee

\smallskip

For the $z$-picture current (which, abusing notation, we again denote by $j$), Eq.(\ref{jLcomm}) takes the form
\be
[ L_n , j^a (z) ] = \frac{d}{dz}\, (z^{n+1} j^a (z))\qquad (\, j^a(z) = \sum_m j^a_m\, z^{-m-1}\ ,\ \ \Delta (j) =1 )\ .
\lb{jzLcomm}
\ee
Proceeding to the quantum version of the Sugawara formula, we shall use the following
definition (cf. \cite{FSoT}) for an infinite sum of {\em normal products} of current modes,
\be
{\rm tr}\, \sum_\ell : j_{-\ell}\, j_{n+\ell} :\ =
{\rm tr}\left( \sum_{\ell=1}^\infty + \sum_{\ell=-n}^\infty\right) \, j_{-\ell}\, j_{n+\ell} \equiv
\eta_{ab} \left( \sum_{\ell=1}^\infty + \sum_{\ell=-n}^\infty\right) \, j_{-\ell}^a\, j_{n+\ell}^b\qquad
\lb{npcm}
\ee
where $j_m := j_m^a T_a\,.$ It has the virtue that, applied to a finite energy state, only a finite number of terms survive.
We shall prove (comparing the resulting commutator with the mode expansion of $T(x)$ in the PB relations (\ref{jT})) that
the sum (\ref{npcm}) is proportional to $L_n$ and will compute the proportionality coefficient:
\ba
&&[ j_m^a\,,\,{\rm tr}\,\sum_\ell : j_{-\ell} j_{n+\ell} : ] =
\eta_{bc}\, \left( \sum_{\ell=1}^\infty + \sum_{\ell=-n}^\infty\right) \, [ j_m^a\,,\, j_{-\ell}^b\, j_{n+\ell}^c\, ] =\nn\\
&&= k\, m\, j_{m+n}^a \left( \sum_{\ell = 1}^\infty (\d_{m-\ell, 0} + \d_{m+n+\ell , 0} ) +
\sum_{\ell = -n}^\infty (\d_{m-\ell , 0} + \d_{m+n+\ell , 0}) \right) + \nn\\
&&+\, i\, \eta_{bc} f^{ab}_{~~d} \left( \sum_{\ell=1}^\infty ( j_{m-\ell}^d j_{n+\ell}^c - j_{-\ell}^d j_{m+n+\ell}^c) +
\sum_{\ell = -n}^\infty (j_{m-\ell}^d j_{n+\ell}^c - j_{-\ell}^d j_{m+n+\ell}^c ) \right) = \nn\\
&&= k\, m\, j_{m+n}^a \left( \left( \sum_{\ell = 1}^\infty + \sum_{\ell = - \infty}^0 \right) \d_{m\ell} +
\left( \sum_{\ell = 1}^\infty + \sum_{\ell = - \infty}^0 \right) \d_{m+n+\ell , 0} \right) + \nn\\
&&+\, i\, \eta_{bc} f^{ab}_{~~d} \times
\left\{
\begin{array}{ll}
0\ , \quad m= 0\\
\left( \sum_{\ell=1}^m + \sum_{\ell=-n}^{m-n-1} \right) \, \frac{1}{2}\, [j_{m-\ell}^d\, ,\, j_{n+\ell}^c\, ] \ , \quad m > 0\\
- \left( \sum_{\ell=1}^{|m|} + \sum_{\ell=-n}^{|m|-n-1} \right)\,  \frac{1}{2}\, [j_{-\ell}^d\, ,\, j_{m+n+\ell}^c\, ]\ , \quad m < 0\\
\end{array}
\right.\ = \nn\\
&&= 2\, k\, m\, j_{m+n}^a + i^2\,m\, f^{ab}_{~~d} f^d_{~bs}\, j_{m+n}^s = 2\, h\, m\, j_{m+n}^a\ ,\qquad h:= k+ g^\vee\ .
\lb{commjnpcm}
\ea
(In the last equality we have used (\ref{adff}).) As anticipated, only finite sums are involved at the final step
of the computation (\ref{commjnpcm}). The {\em quantum shift} of the level $k$ to the {\em height} $\, h\,$ affects the
normalization of the WZNW stress energy tensor so that, to comply with the standard commutation relations of
the Virasoro algebra (see e.g. \cite{Kac, KR}),
\be
[ L_m , L_n ] = (m-n) \, L_{m +n} + \frac{c}{12}\, m (m^2 -1)\, \d_{m+n ,\,0}\ ,
\lb{Vir}
\ee
one should set
\be
L_n = \frac{1}{2h}\,{\rm tr}\left( \sum_{\ell=1}^\infty + \sum_{\ell=-n}^\infty\right) \, j_{-\ell}\, j_{n+\ell}\qquad
\Rightarrow\qquad c = \frac{k}{h}\, {\rm dim}\, {\cal G}
\lb{Ln}
\ee
(cf. \cite{GO86} where one can find a list of the authors who have contributed to deriving the correct result).
The Sugawara formula (\ref{Ln}) and (\ref{jonvac}) imply
\be
L_n\vac = 0 \qquad{\rm for}\qquad n\ge -1\ .
\lb{Lonvac}
\ee

\smallskip

The local diffeomorphisms in $z$ and $\bz$ are generated by the mutually commuting
modes $L_n$ and ${\bar L}_n$ of the left and right component of the stress energy tensor
\be
T(z) = \sum_m \frac{L_m}{z^{m+2}}\ ,\qquad {\bar T}(\bz) = \sum_m \frac{{\bar L}_m}{\bz^{m+2}}\ ,\qquad
[L_m , {\bar L}_n ] = 0\ .
\lb{Tz}
\ee
We shall write the quantum analog of the $2D$ group valued field (\ref{LR}) as
\be
g (z, \bz ) = g(z)\, {\bar g}(\bz ) \equiv ( g^A_\a(z)\, {\bar g}^\a_B(\bz ))\ ,
\lb{LRq}
\ee
where ${\bar g}\,$ replaces $g_R^{-1}\,.$ Then the current-field PB in (\ref{jTg}) yields the commutation relation
\be
[ j_m^a , g (z, \bz) ] = - z^m\, t^a \, g (z, \bz) \ .
\lb{c-f}
\ee
Requiring that $g (z, \bz)$ also satisfies (\ref{phiLcomm}) for $n=0$ and $L_0$ given by (\ref{Ln}),
\be
L_0 = \frac{1}{h}\,{\rm tr}\, (\frac{1}{2}\,j_0^2 + \sum_{m=1}^\infty j_{-m} j_m )
\lb{L0}
\ee
is equivalent to imposing the Knizhnik-Zamolodchikov equation \cite{KZ, T} in an operator form,
\ba
&&h \,\frac{\partial}{\partial z}\, g (z, \bz) =\,- :\! j(z)\, g(z, \bz)\! :\  =
- T_a\, (j^a_{(+)}(z)\,g(z, \bz) + g(z, \bz)\,j^a_{(-)}(z))\ ,\nn\\
&&j^a_{(+)}(z) := \sum_{m = 0}^\infty j^a_{-m-1}\, z^m\ ,\quad j^a_{(-)}(z) := \sum_{m= 0}^\infty j^a_m\, z^{-m-1}
\lb{KZ}
\ea
and fixes the conformal dimension $\Delta$ of $g$ to
\be
\Delta = \frac{C_2(\pi_f)}{2h} = \frac{n^2-1}{2nh}\ .
\lb{conf-dim-g}
\ee
A similar equation involving the right current dictates the same value for ${\bar \Delta}\,.$
Here $C_2(\pi_f) = n-\frac{1}{n}$ is the value (\ref{c2piL}) of the quadratic Casimir operator (\ref{CasCW})
in the defining $n$-dimensional representation $\pi_f$ of $su(n)\,.$ These two operator KZ equations are the quantum
counterparts of the definitions (\ref{jLR}) of the classical chiral currents.

\smallskip

More generally, if $\phi_\L(z)$ is a ${\widehat{\cal G}}$-primary chiral field transforming under an IR of weight $\L$ of the simple
compact Lie algebra ${\cal G}\,,$ i.e. if
\ba
&&[ j^a_{(-)} (z_1)\,, \phi_\L(z_2) ] = - \pi_\L (t^a) \frac{1}{z_{12}}\, \phi_\L(z_2) \ ,\nn\\
&&[ \phi_\L(z_1)\,, j^a_{(+)} (z_2) ] =  \pi_\L (t^a) \frac{1}{z_{12}}\, \phi_\L(z_1) \ ,
\lb{Ward}
\ea
then $\phi_\L(z)$ has conformal dimension
\be
\Delta (\L) = \frac{C_2(\pi_\L)}{2h}
\lb{conf-dim-L}
\ee
and satisfies the KZ equation
\be
h \,\frac{d}{d z}\,\phi_\L(z) = - \pi_\L(T_a)\, (j^a_{(+)}(z)\,\phi_\L(z) + \phi_\L(z)\,j^a_{(-)}(z))\ .
\lb{KZL}
\ee
Here $\pi_\L(T_a)$ and $\pi_\L (t^b)$ are dual bases in the (finite dimensional) representation space of ${\cal G}$
of highest weight $\L$ and $\frac{1}{z_{12}}$ in (\ref{Ward}}) is understood as the power series
$\frac{1}{z_1}\,\sum_{m=0}^\infty\left( \frac{z_2}{z_1} \right)^m$ for $|z_1| > |z_2|\,$
(therefore it is {\em not} strictly antisymmetric but satisfies $\frac{1}{z_{12}} + \frac{1}{z_{21}} = \d (z_{12})$ \cite{FSoT, Kac98}).
The KZ equation (\ref{KZL}), the operator {\em Ward identity} (\ref{Ward}) and Eq.(\ref{jonvac})  allow to write
a system of partial differential equations for the vacuum expectation value
\be
W_N = \langle 0 \mid \phi_{\L^{(1)}} (z_1)\dots \phi_{\L^{(N)}}(z_N) \vac
\lb{W-N}
\ee
in its primitive domain of analyticity in which $|z_1| > |z_2| \dots > |z_N|$:
\ba
&&\left(h \frac{\partial}{\partial z_i} + \sum_{j=1}^{i-1} \frac{C_{ij}\,(\L^{(i)} , \L^{(j)})}{z_{j\, i}} -
\sum_{j=i+1}^N \frac{C_{ij}\,(\L^{(i)} , \L^{(j)})}{z_{ij}}\right) W_N = 0\ ,\nn\\
&&i=1,\dots ,N\ ,\qquad C_{ij}\,(\L^{(i)} , \L^{(j)}) := \eta^{ab}\, \pi_{\L^{(i)}} ( T_a ) \otimes \pi_{\L^{(j)}} (T_b )\ .\qquad
\lb{KZW-N}
\ea

\smallskip

To summarize: the infinite chiral symmetry of the WZNW model, which involves both a local chiral internal symmetry
expressed by the current-field commutation relations (CR) (\ref{Ward}) and (infinitesimal) diffeomorphism invariance
of primary fields (\ref{phiLcomm}), allows to compute the {\em anomalous dimension} $\Delta$
(\ref{conf-dim-g}) of the primary field $\phi_{\L}$ deriving on the way the operator KZ equation (\ref{KZL}).
This is a remarkable non-perturbative result and deserves recalling its main ingredients.

\noindent
(i) The requirement of infinite chiral invariance at the classical level led to the addition of the multivalued
Wess-Zumino term to the classical action $S[g]$ (\ref{Swznw0}).

\noindent
(ii) Demanding the path integral measure involving the factor $e^{i S[g]}$ to be single valued
yields the quantization of the coupling constant $k$ (ultimately identified with the affine Kac-Moody level).

\noindent
(iii) The quantum Sugawara formula (\ref{Ln}), which gives rise to a (non-perturba-tive) renormalization of $k\,,$ relates
the internal symmetry with the conformal properties. The non-integer anomalous dimension $\Delta$
(\ref{conf-dim-L}) implies, in particular, the presence of a non-trivial monodromy in the chiral theory.

\noindent
(iv) The non-perturbative character of the outcome is displayed by the fact that the renormalized coupling constant $h$
appears in the denominator of the anomalous dimension $\Delta\,.$

\noindent
(v) The operator equation (\ref{KZL}) along with the Ward identity (\ref{Ward}) allows to write down the system of partial
differential equations (\ref{KZW-N}) for the correlation functions. The operator in the left hand side of (\ref{KZW-N}) has
a nice geometric interpretation as a flat connection (see e.g. \cite{Ka}). The system admits an explicit solution in terms of a
multiple integral representation \cite{KZ, DF, ZF, CF, STH, FGP}.

\subsection{The exchange algebra of the chiral field $g(x)$}

The naive idea of just replacing PB by commutators fits the cases of free or Lie-algebra valued fields but
is no longer applicable to group-like quantities which have quadratic PB relations. The simplest example
is provided by the Weyl form of the canonical CR involving the groups of unitary operators $e^{i \alpha p}$ and $e^{i \beta x}\,,$
\be
e^{i \alpha p} \, e^{i \beta x} = e^{i \hbar \alpha \beta} \, e^{i \beta x} \, e^{i \alpha p}\, .
\lb{CCR}
\ee
We can recover the PB as a quasi-classical limit of the quantum {\em exchange relations} setting
\be
\{ e^{i \alpha p} , e^{i \beta x} \} = \lim_{\hbar \to 0} \frac{1}{i\hbar} \, [ e^{i \alpha p} , e^{i \beta x}] =
\a\b \, e^{i \beta x} \, e^{i\alpha p}\ .
\lb{WCR}
\ee

To quantize the classical {\em chiral WZNW\ } PB relations (\ref{gpb}), we shall look for
quadratic exchange relations for $g(x)$ \cite{B, MR, F1, AS, G}, setting in the real ($x$-) picture
\be
\lb{ggR}
g_1(x_1)\, g_2 (x_2) = g_2(x_2) \, g_1(x_1)\, R_{12} (x_{12})\ ,\quad -2\pi < x_{12} <2\pi
\ee
where
\be
\lb{Rx}
R_{12} (x) = R_{12}^-\, \theta (x) + R_{12}^+\, \theta (-x)\ ,\qquad R_{12}^- = R_{12}\ ,\quad R_{12}^+ = R_{21}^{-1}\ ,
\ee
the {\em quantum $R$-matrix} $R_{12}\,$ being an invertible matrix satisfying the {\em quantum Yang-Baxter equation} (QYBE)
\be
R_{12}\, R_{13}\, R_{23}\, = \,R_{23}\, R_{13}\, R_{12}
\lb{QYBE}
\ee
and reproducing the classical $r$-matrix $r^-_{12}$ in the quasi-classical limit.
The relation between $R^-$ and $R^+$ in (\ref{Rx}) ensures the compatibility between the exchange relations
for $x_1 < x_2$ and $x_1 > x_2$ while the QYBE is a consistency condition for the associativity of triple products
of chiral field operators.

The properties of the quantum exchange relations are revealed by studying their {\em quantum group symmetry},
the quantum counterpart of the Poisson-Lie structure (discussed in Section 2.4). A key to understanding
quantum groups $\fU$, in particular {\em quantum universal enveloping algebras} (QUEA) $U_q({\cal G})$
is provided by the notion of {\em coproduct} $\Delta:\,\fU\to\fU\,,$ which teaches us  how to "add" quantum
numbers passing from a single particle to a many particle system and has a bearing on the quantum statistics.
The crucial property which distinguishes the
QUEA coproduct from that of the standard undeformed universal enveloping algebra $U({\cal G}) = U_1 ({\cal G})$
is the possibility $\Delta$ to be non-symmetric, i.e. (using the convenient Sweedler's notation \cite{Sweedler})
\be
{\Delta}(X) := \sum_{(X)} X_1\otimes X_2 \ne \sum_{(X)} X_2\otimes X_1 =: {\Delta}' (X) \ .
\lb{DDp}
\ee
The breaking of {\em cocommutativity}, i.e. of the symmetry of the coproduct,
implies that quantum mechanical multiparticle wave functions (or
correlation functions, in QFT) cannot transform covariantly under
the group of permutations. The exchange symmetry that replaces it should commute with
the coproduct $\Delta (X)\,.$ One can construct such a substitute of permutation
for {\em almost cocommutative} Hopf algebras (see Appendix B where this and related notions are recalled and illustrated on examples)
for which a special element ${\cal R} \in \fU\otimes\fU\,,$ called the {\em universal $R$-matrix}, exists
that intertwines between the coproduct $\Delta (X)$ and its opposite $\Delta'(X)$:
\be
{\cal R}\,\Delta (X) = \Delta' (X) \, {\cal R}\ .
\lb{intR}
\ee
This notion will be applicable to the above exchange relations if the matrix $R=R_{12}$ in
(\ref{Rx}) can be obtained from ${\cal R}$ when applied to the tensor square of the defining
representation of $U_q({\cal G})\,.$ The object of main interest for us is the {\em braid
operator} that combines $R$ with the permutation operator $P = P_{12}$ so that it commutes with the
coproduct
\be
{\hat R} := P R\ ,\quad \Delta' (X) = P\, \Delta (X)\, P\quad\Rightarrow\quad
\Delta(X)\,{\hat R} = {\hat R}\,\Delta (X)
\lb{PRco}
\ee
and satisfies the {\em braid group relations} (for $\hR_i := \hR_{i\,i+1}$)
\be
\quad\hR_i\, \hR_{i+1} \hR_i\, =\, \hR_{i+1} \hR_i\, \hR_{i+1}\ ,\qquad
\hR_i\, \hR_j\, =\, \hR_j\, \hR_i\quad{\rm for}\quad |i-j|>1\ ,
\lb{braidR}
\ee
the first of which follows from the Yang-Baxter equality (\ref{QYBE}) for $R_{ij}\,.$

The analytic ($z$-) picture exchange relations are then expressed in terms of the corresponding matrix $\hat R$:
\ba
&&g^A_\a(z_1)\, g^B_\b (z_2) =\,
\stackrel{\curvearrowright}{g^B_\rho (z_2)\, g^A_\s (z_1)}\!{\hat R}^{\rho\s}_{~\a\b}\qquad\quad
(\, {\hat R}^{\rho\s}_{~\a\b} \equiv R^{\s\rho}_{~\a\b}\,)\ ,\lb{ggRa}\\
&&(\, z_{12} \stackrel{\curvearrowright}{\rightarrow} z_{21} = e^{-i\pi} z_{12}\quad
{\rm for} \quad |z_1| > |z_2|\ ,\ \pi > {\rm arg} (z_1) > {\rm arg} (z_2) > - \pi\,)\ .
\nn
\ea
They involve analytic continuation along a path that exchanges two neighbouring arguments of the multivalued
{\em chiral (conformal) blocks}. (Analyticity in the domain indicated
in the last equation (\ref{ggRa}), cf. e.g. \cite{FHIOPT}, is a consequence of energy positivity.)

The multivaluedness of chiral blocks reflects the fact that the (complex) configuration space is
not simply connected. The quantum group symmetry and the braid group statistics generalize in a sense
the {\em Schur-Weyl duality} between an internal unitary symmetry group and the permutation group\footnote{
See \cite{TH} for a pedagogical survey of Schur-Weyl duality and references to the pioneer work of Arnold \cite{Ar}
that links the braid group with the topology of configuration space. The similarity between
Schur-Weyl duality and Doplicher-Roberts theory of superselection sectors \cite{DR} is commented
in \cite{HF}.} to the case of correlation functions with non-trivial monodromy.
There is a {\em gauge freedom} in the choice of the braid operator related to the ambiguity in the definition
of the chiral components $g(z)$ and $\bar g(\bz)$ of $g(z ,\bar z )$ (\ref{LRq}).
We shall opt for the simple, {\em numerical} $SL_q(n)$ $R$-matrix of \cite{FRT} for the $SU(n)$ WZNW model under consideration
ensuring the simple covariance and braiding properties of the matrix chiral fields at the expense of dropping
chiral covariance under the (antilinear) complex conjugation and the related unitarity property, which will be only satisfied by the $2D$ field
$g (z, \bz )$ (\ref{LRq}). We shall only require that the {\em regularized quantum determinant} of $g(z)$
\be
D_q (g\,; z_1,\dots ,z_n) := \frac{1}{[n]!}\,\prod_{1\le i<j\le n} z_{ij}^{\frac{n+1}{nh}}\,
\epsilon_{A_1 \dots A_n} \,g^{A_1}_{\a_1}(z_1)\dots g^{A_n}_{\a_n}(z_n)\, \varepsilon^{\a_1 \dots \a_n}
\lb{D(g)}
\ee
belongs to the conformal class of the unit operator. The necessity to use the deformed ("quantum") $\varepsilon$-tensor
$\varepsilon^{\a_1 \dots \a_n}$ will be explained  in Section 4.4 below where we introduce the similar notion of quantum determinant for
the zero modes\footnote{The "quantum factorial" $[n]!\,$ is defined in (\ref{antis-l}).}.
Here we shall only provide the argument for the $z$-depending prefactor.

Let $G=SU(n)\,$ and denote by $w_n$ the $n$-point conformal block
\be
w_n = w_n (z_1,\dots ,z_n)^{A_1\dots A_n}_{~\a_1\dots\a_n} = \langle 0 \mid g^{A_1}_{\a_1}(z_1)\dots g^{A_n}_{\a_n}(z_n) \vac\ .
\lb{Wn}
\ee
It satisfies the KZ equation (\ref{KZW-N}) for $N=n$ and all $\pi_{\L^{(i)}} = \pi_f$ so that
\be
C_{ij}(\L^{(i)} , \L^{(j)}) = C_{ij} = P_{ij} - \frac{1}{n} \id_{ij} = C_{j\,i}\ ,\quad i,j=1,\dots ,n\ ,
\lb{nL1}
\ee
cf. (\ref{Cn-sigma}). As the full antisymmetry of $\epsilon_{A_1\dots A_n}$ implies
\be
\epsilon_{A_1\dots A_i\dots A_j \dots A_n} P^{A_i A_j}_{B_i B_j}
= \epsilon_{A_1\dots B_j\dots B_i \dots  A_n} = - \epsilon_{A_1\dots B_i\dots B_j \dots  A_n}\ ,
\lb{epsP}
\ee
the KZ linear system (\ref{KZW-N}) reduces to
\be
\left\{\frac{\partial}{\partial z_i} - \frac{n+1}{n h} \left( \sum_{j=1}^{i-1} \frac{1}{z_{j\,i}} -
\sum_{j=i+1}^n \frac{1}{z_{ij}} \right) \right\} p_n (z_1,\dots ,z_n) = 0\ ,\quad i=1,\dots ,n
\lb{KZp_n}
\ee
for
\be
p_n (z_1,\dots ,z_n) := \frac{1}{[n]!}\,\epsilon_{A_1\dots A_n}\, w_n (z_1,\dots ,z_n)^{A_1\dots A_n}_{~\a_1\dots\a_n}\, \varepsilon^{\a_1\dots \a_n}
\lb{Wn1}
\ee
and hence,
\be
p_n (z_1,\dots ,z_n) = c\,\prod_{1\le i<j\le n} z_{ij}^{- \frac{n+1}{n h}}\ ,\quad c = const\ .
\lb{p_n}
\ee
For $c=1$ and $D_q (g\,; z_1,\dots ,z_n)$ given by (\ref{D(g)}), Eq.(\ref{p_n}) is equivalent to
\be
\langle 0 \mid D_q (g\,; z_1,\dots ,z_n)\vac = 1\ .
\lb{Dg1}
\ee
The prefactor can also be deduced from (\ref{conf-dim-L}) and the identity
\be
2\, \Delta (\L^1) - \Delta (\L^2) = \frac{n+1}{nh} \quad (\, = \Delta (2 \L^1) - 2 \Delta (\L^1)\,)
\lb{id-pre}
\ee
and then verified by the KZ equation (the values of the quadratic Casimir in the symmetrized and antisymmetrized square,
$\pi_{2\L^1} \equiv \pi_s$ and $\pi_{\L^2} \equiv \pi_a\,,$ of the defining representation $\pi_{\L^1} \equiv \pi_f$ are, respectively
\be
C_2(\pi_s) = 2 \frac{(n-1)(n+2)}{n} \ ,\qquad C_2(\pi_a) = 2 \frac{(n+1)(n-2)}{n}\ ,
\lb{C-as}
\ee
cf. (\ref{llab})). Note that $\left({n}\atop{2} \right) \frac{n+1}{nh} = n \, \Delta\,\,$
for $\Delta\,$ the dimension (\ref{conf-dim-g}) of the primary field $g(z)\,.$

\medskip

Eq.(\ref{ggR}) is also invariant with respect to {\em $G$-valued periodic left shifts}
and {\em chiral conformal transformations} (the quantum version of (\ref{Gleft}), (\ref{chiralconf})).
The invariance of the exchange relations (\ref{ggR}) with respect to constant right shifts
\be
g(x)\to g(x)\, T\ ,
\lb{gT}
\ee
the counterpart of the Poisson-Lie symmetry of the corresponding PB,
implies the {\em RTT relations} \cite{D, FRT}
\be
R_{12}\, T_1\, T_2 \,=\, T_2\, T_1\, R_{12} \qquad\Leftrightarrow \qquad
R_{21}^{-1}\, T_1\, T_2\, = \,T_2\, T_1\, R_{21}^{-1}\ .
\lb{RTT}
\ee
So a natural choice for the quantum $R$-matrix is the Drinfeld-Jimbo \cite{D, J} $n^2\times n^2$ matrix used in \cite{FRT} to define
the quantum group $SL_q(n)\,,$
\be
R_{12} = ( R^{\a\b}_{~\a'\b'} )\ ,\qquad
R^{\a\b}_{~\a'\b'} = q^\frac{1}{n}\left( \d^\a_{\a'}\d^\b_{\b'} + (q^{-1}- q^{\epsilon_{\a\b}} )\, \d^\a_{\b'}\d^\b_{\a'} \right)
\lb{R}
\ee
(all indices running from $1$ to $n$ and the sign convention on the skew-symmetric
$\epsilon_{\a\b}$ being fixed in (\ref{stand-r-matr})), where $q$ is the corresponding {\em quantum} deformation parameter.

\smallskip

The value of $q$ in ({\ref{R}) may not coincide with the "classical" one (\ref{qcl}) but the quasi-classical expansion of (\ref{R}) with
\be
q=1-i\frac{\pi}{k} + {\cal O} (\frac{1}{k^2})
\lb{qq-k}
\ee
has to reproduce the standard $s\ell(n)$ $r$-matrix (\ref{rstandard}), (\ref{stand-r-matr}).
To this end, it is convenient to rewrite $R_{12}$ and $r_{12}$ in the following compact form using the diagonal $n^2\times n^2$ matrix
$\epsilon_{12} = {\rm diag} (\epsilon_{\a\b})\,$ (i.e., $\epsilon^{\a \b}_{{\a}' {\b}'} = \epsilon_{\a\b}\, \d^\a_{{\a}'} \d^\b_{{\b}'}$)
satisfying $\epsilon_{12}\, P_{12} = - P_{12}\, \epsilon_{12}\,$:
\be
R_{12} = q^\frac{1}{n}\, (\id_{12} + (q^{-1} - q^{\epsilon_{12}} )\, P_{12} )\ ,\qquad r_{12} = - \, \epsilon_{12}\, P_{12}\ .
\lb{Rr-compactly}
\ee

\smallskip

\noindent
{\bf Remark 4.1~} To show that the quantum exchange relations reproduce the WZNW model PB in the quasi-classical limit
we can introduce at an intermediate step the Planck constant $\hbar\,$ and the
dimensionful overall coefficient ${\bar k}\,$ to the action (\ref{Swznw0}) setting $k = \frac{\bar k}{\hbar}\,$
so that, effectively, $\hbar \to 0 \ \Leftrightarrow\ \frac{1}{k} \to 0\,.$
If one considers angular momentum type variables $\bar p_{ij}$ which also have the dimension
of an action, then the corresponding dimensionless quantities are given by $p_{ij} = \frac{\bar p_{ij}}{\hbar}\,$
so that the quasi-classical limit can be recovered from their scaling behaviour:
\be
\lb{quasicl}
\hbar \to 0 \quad \Leftrightarrow\quad
\frac{1}{k} \to 0 \,,\quad p_{ij} \to \infty \,, \quad \frac{p_{ij}}{k} \quad{\rm finite}\ .
\ee
The {\em undeformed quantum} limit, on the other hand, corresponds to finite $p_{ij}\,,$ neglecting all terms
of the type $\frac{p_{ij}}{k}$ in the expansion in powers of $\frac{1}{k}\,.$

\smallskip

Using (\ref{Rr-compactly}), it is straightforward to show that right-hand side of the PB (\ref{gpb}) is reproduced, up to an $i$-factor,
by the leading term in the expansion in powers of $\frac{1}{k}$ of the commutator following from (\ref{ggR}).
In particular, the classical $r$-matrices $r^\pm$ appear in the expansion of the quantum $R$-matrix,
\ba
&&R_{12} = \id_{12} - i\frac{\pi}{k}\, r^-_{12} +{\cal O}(\frac{1}{k^2})\ ,\quad
R_{21} = \id_{12} + i\frac{\pi}{k}\, r^+_{12} +{\cal O}(\frac{1}{k^2})\ ,\quad{\rm or}\nn\\
&&R^\pm_{12} = \id_{12} - i\frac{\pi}{k}\, r^\pm_{12} +{\cal O}(\frac{1}{k^2})\qquad (\, R^-_{12} := R_{12}\,,\ R^+_{12} := R_{21}^{-1}\, )\ .
\lb{Rr}
\ea
To verify the compatibility of (\ref{Rr}) for $r_{12}^\pm = r_{12} \pm C_{12}\,,$ we take into account that
$r_{12} = - r_{21}$ and $C_{12} = C_{21}\,.$ (The overall coefficient $q^\frac{1}{n}$ of $R_{12}$ is important:
the first non-trivial term in its expansion contributes to the polarized Casimir operator $C_{12}=P_{12}-\frac{1}{n}\, \id_{12}$ (\ref{Cn-sigma}).)
These expansions also ensure that the Sklyanin bracket (\ref{PBSkl}) emerges as the quasi-classical limit of the RTT relations (\ref{RTT}).
(In both cases one has to take into account the fact that the matrix elements of $g(x)$, as well
as those of $T$, commute in this limit so that $g_1(x_1)\, g_2(x_2) = g_2(x_2)\, g_1(x_1)$ and $T_1 T_2 = T_2 T_1$.)

\smallskip

Demanding that the eigenvalues of the braid matrix $\hat R$ agrees with the conformal dimensions implies
that the correct value of the {\em quantum} deformation parameter $q$ (satisfying (\ref{qq-k})) is
\be
q = e^{- i\frac{\pi}{h}}\ ,\qquad h := k + g^\vee
\lb{height-h}
\ee
i.e., the level $k\,$ of the classical expression (\ref{qcl}) has to be replaced again by the {\em height} $h\,.$
To begin with, we note that for $R$ given by (\ref{R}), (\ref{Rr-compactly}),
$\hat R = P R$ (\ref{PRco}) satisfies the Hecke algebra relation
\be
(q^{-\frac{1}{n}}\,\hat R - q^{-1} ) (q^{-\frac{1}{n}}\,\hat R + q) = 0
\lb{Hecke}
\ee
and hence, has only two different eigenvalues\footnote{
This is the main reason for constraining ourselves to the case of $G=SU(n)$. The braid operators obtained from
the $R$-matrices for the deformations of other simple (compact) classical groups have {\em three} different eigenvalues
\cite{FRT} and are more difficult to handle.}, $q^{-1+\frac{1}{n}}$ and $- q^{1+\frac{1}{n}}\,.$
These have to be compared with the braiding properties following from the exchange relations (\ref{ggRa}).
Conformal invariance fixes the $3$-point functions of primary fields up to normalization (see e.g.
\cite{DFMS, FSoT}) so that we have
\ba
&&\langle \Delta_s\! \mid g_1(z_1)\, g_2(z_2) \vac =
N^{(s)}_{12} z_{12}^{\Delta_s - 2 \Delta}\ ,\quad (P_{12} - 1)\, N^{(s)}_{12} = 0\ ,\nn\\
&&\langle \Delta_a\! \mid g_1(z_1)\, g_2(z_2) \vac =
N^{(a)}_{12} z_{12}^{\Delta_a - 2 \Delta}\ ,\quad (P_{12} + 1)\, N^{(a)}_{12} = 0\ ,\qquad
\lb{Nsa}
\ea
where the normalization matrices $N^{(s,\, a)} = (\, {N^{(s,\, a)}}^{AB}_{\a\b} )$ have both $SU(n)$
and quantum group indices inherited from the chiral fields.
The conformal dimension $\Delta$ in (\ref{Nsa}) is given by (\ref{conf-dim-g}), while $\Delta_{s,\,a}= \frac{C_2(\pi_{{\bar s},\,{\bar a}})}{2h}
= \frac{C_2(\pi_{s,\,a})}{2h}$ (cf. (\ref{C-as})) are the conformal dimensions of the WZWN primary fields conjugate to the
symmetric and antisymmetric $SU(n)$ tensors, respectively. Applying now (\ref{ggRa}) to (\ref{Nsa}), we obtain
\ba
&&N^{(s)} \hat R = e^{-i\frac{\pi}{h} (C_2(\pi_f)-\frac{1}{2}C_2(\pi_s))} P\, N^{(s)} =
e^{-i\frac{\pi}{h}(-1+\frac{1}{n})}\, N^{(s)}\  ,\nn\\
&&N^{(a)} \hat R = e^{-i\frac{\pi}{h} (C_2(\pi_f)-\frac{1}{2}C_2(\pi_a))} P\, N^{(a)} =
- e^{-i\frac{\pi}{h}(1+\frac{1}{n})}\, N^{(a)}\  .
\lb{Ex}
\ea
Hence, the matrices $N^{(s,\, a)}$ intertwine the corresponding symmetric and antisymmetric eigenspaces
of the permutation $P$ and the braid operator $\hat R$ which have the same dimensions $\left({n+1}\atop{2}\right)$ and
$\left({n}\atop{2}\right)$, respectively. Comparing the eigenvalues of $\hat R$ following from (\ref{Ex})
with those predicted by (\ref{Hecke}), we fix the value of the quantum deformation parameter $q$ (\ref{height-h}) for $G=SU(n)$:
\be
q = e^{- i\frac{\pi}{h}}\ ,\quad h = k + n\qquad (\, q^{\pm \frac{1}{n}} = e^{\mp i\frac{\pi}{n h}}\, )\ .
\lb{h-SUn}
\ee

\subsection{Monodromy, its factorization and the QUE algebra}

Noting that $L_0 - {\bar L_0}$ is the generator of translation in $x^1$ and that the spin (or, rather, the helicity)
$\Delta-\bar\Delta$ vanishes (i.e., $g(z,\bar z)$ is a Lorentz scalar field), we deduce that the
periodicity of $g(x^0,x^1)$ in $x^1$ (cf. (\ref{gper}), (\ref{xz}) and (\ref{zzbar})) is equivalent to the univalence property of $g(z,\bz)$:
\be
e^{2\pi i (L_0 - \bar L_0)} \, g(z,\bar z) \, e^{2\pi i (\bar L_0 - L_0)} =
g(e^{2\pi i} \, z, e^{-2\pi i} \, \bar z) = g(z,\bar z) \ .
\lb{gzzbar-per}
\ee
Eq.(\ref{gzzbar-per}) would be satisfied if the monodromy matrices $M \ ( = M_L )\,$ and $\bar M \ ( = M^{-1}_R )\,$ of the chiral components of $g(z,\bar z)\,,$ defined by
\ba
&&e^{2\pi iL_0} \, g^A_\a(z) \, e^{-2\pi i L_0} = e^{2\pi i \Delta} \, g^A_\a(e^{2\pi i}\,z)
= g_{\sigma}^A(z) \, M^{\sigma}_{~\alpha}\ ,\nn\\
&&e^{- 2\pi i \bar L_0} \, \bar g^\a_B (\bar z) \, e^{2\pi i \bar L_0} =
e^{-2\pi i {\bar\Delta}} \, \bar g^\a_B (e^{-2\pi i}\, \bar z ) = {\bar M}^\a_{~\rho}\, \bar g^\rho_B (\bar z)
\lb{gzM}
\ea
were inverses of each other. (The classical counterpart of this property
of the chiral splitting is spelled out in Proposition 2.1, see further Eq.(\ref{O2alt}).
As already mentioned, it requires a gauge theory framework which, in the quantum case, involves singling an
appropriate physical space of states. This problem is approached, for $n=2\,,$ in Section 5.4.2.)

Applying the first relation (\ref{gzM}) to the vacuum vector $\vac$ and using (\ref{conf-dim-g}), we obtain that
\be
M^\a_{~\b} \vac = q^{- C_2(\pi_f)} \d^\a_\b \vac = q^{\frac{1}{n} - n} \d^\a_\b \vac
\lb{M0}
\ee
i.e., the vacuum is annihilated by the off-diagonal elements of $M\,$ and is a common eigenvector of the diagonal ones,
corresponding to the (common) eigenvalue $q^{\frac{1}{n} - n}\,.$ This suggests a modification of the factorization
(\ref{M+-}) of the quantum monodromy matrix $M$ in upper and lower triangular Gauss components:
\be
M=q^{\frac{1}{n} - n} M_+ M_-^{-1} \qquad (\, {\rm diag}\, M_+ = {\rm diag}\, M_-^{-1}\,)\ .
\lb{M+-q}
\ee

We postulate the following quantum exchange relations for $M_\pm\,$:
\ba
&&g_1(x)\, R_{12}^\mp M_{\pm 2} = M_{\pm 2}\, g_1(x) \qquad\ (\, R^-_{12} = R_{12}\,,\ R^+_{12} = R_{21}^{-1}\, )\ ,\lb{Mg}\\ \nn\\
&&R_{12} M_{\pm 2} M_{\pm 1} = M_{\pm 1} M_{\pm 2} R_{12}\ ,\quad R_{12} M_{+2} M_{-1} = M_{-1} M_{+2} R_{12}\ .\qquad\quad
\lb{Mpmq}
\ea
Using the quasi-classical asymptotics (\ref{Rr}) of the quantum $R$-matrix, it is not hard to check that the $\frac{1}{k}$
expansions of the commutators following from (\ref{Mg}) and (\ref{Mpmq}) reproduce the corresponding PB in the second
relation (\ref{Mgeng}) and (\ref{Mpmmp}), respectively. The resulting exchange relation between $M$ (\ref{M+-q}) and $g(x)$ is
\be
g_1 (x)\, R_{12}^- M_2 = M_2\, g_1(x)\, R_{12}^+\ .
\lb{Mgq}
\ee
It guarantees the compatibility of Eq.(\ref{ggR}) for $x_2 < x_1 < x_2 + 2\pi\,$ when we have
\ba
&&g_1(x_1)\, g_2 (x_2) = g_2(x_2)\,  g_1(x_1) R_{12}^- \ ,\nn\\
&&g_1(x_1)\, g_2 (x_2+2\pi) = g_2(x_2+2\pi)\, g_1(x_1) R_{12}^+\ ,\lb{ggR-M}\\
&&g_2 (x_2+2\pi) \equiv g_2 (x_2) M_2\ .\nn
\ea
The exchange relations for the matrix elements of $M$ following from (\ref{Mpmq}) can be written as a
{\em reflection equation} \cite{Ch84, Sk88} that is quadratic in the $R$-matrix:
\be
M_1\, R_{12}\, M_2\, R_{21} = R_{12}\, M_2\, R_{21}\, M_1\quad\Leftrightarrow\quad
{\hat R}_{12}\, M_2\, {\hat R}_{12}\, M_2 = M_2\, {\hat R}_{12}\, M_2\, {\hat R}_{12}\ .
\lb{Mexch}
\ee
The quasi-classical limits of (\ref{Mgq}) and (\ref{Mexch}) agree with the first PB relation (\ref{Mgeng}) and with (\ref{PBMM}), respectively.

\smallskip

Using the explicit form (\ref{R}) of the quantum $R$-matrix, one can write the RMM relations (\ref{Mpmq}) for $M_\pm$ in components:
\ba
&&[ (M_\pm)^\a_{~\rho} , (M_\pm)^\b_{~\s} ] =
(q^{\epsilon_{\s\rho}} - q^{\epsilon_{\a\b}} )\, (M_\pm)^\a_{~\s} (M_\pm)^\b_{~\rho}\ ,
\nn\\
&&[ (M_-)^\a_{~\rho} , (M_+)^\b_{~\s} ] =\lb{RM}\\
&&= (q^{-1} - q^{\epsilon_{\a\b}})\, (M_+)^\a_{~\s} (M_-)^\b_{~\rho} -
(q^{-1} - q^{\epsilon_{\s\rho}})\, (M_-)^\a_{~\s} (M_+)^\b_{~\rho}\ .
\nn
\ea
We shall denote
\be
{\rm diag}\, M_+  = {\rm diag}\, M_-^{-1}\ =: D = (d_\a \d^\a_\b)\ ,\quad \det D := \prod_{\a=1}^n d_\a = 1
\lb{MpmD1}
\ee
(cf. (\ref{diagMM})). From (\ref{RM}) we obtain, in particular,
\ba
&&d_\a\, d_\b = d_\b\, d_\a\qquad\quad (\, (M_+)^\a_{~\a} = d_\a \ , \ (M_-)^\a_{~\a} = d^{-1}_\a\,)\ ,\lb{dMpm}\\
&&d_\a (M_+)^\b_{~\a} = q^{-1}\,  (M_+)^\b_{~\a}\, d_\a\ , \quad d_\b (M_+)^\b_{~\a} = q\, (M_+)^\b_{~\a}\, d_\b\ ,\qquad \a > \b\ ,\nn\\
&&d_\a (M_-)^\a_{~\b} = q\,  (M_-)^\a_{~\b}\, d_\a\ , \quad d_\b (M_-)^\a_{~\b} = q^{-1}\, (M_-)^\a_{~\b}\, d_\b\ ,\qquad \a > \b\ , \nn\\
&&[ (M_-)^\a_{~\b} , (M_+)^\b_{~\a} ] = \l\, ( d_\a^{-1} d_\b  - d_\a d_\b^{-1} )\ ,\qquad \a > \b\qquad (\,\l = q-q^{-1}\,)\ .
\nn
\ea
(Using the triangularity of $M_+$ and $M_-$ in deriving (\ref{dMpm}) is crucial; as $d_\a$ commute,
their order in the product defining $\,\det D$ is not important.)

\smallskip

A natural coalgebra structure on the algebra generated by the entries of $M_\pm$ is given by
\ba
&&\Delta ((M_\pm)^\a_{~\b}) = ( M_\pm )^\a_{~\s}\otimes (M_\pm)^\s_{~\b}\ ,\nn\\
&&\varepsilon ((M_\pm)^\a_{~\b}) = \d^\a_\b\ ,\quad S ((M_\pm)^\a_{~\b}) =  (M_\pm^{-1})^\a_{~\b}\ .
\lb{Hopf-FRT}
\ea
(In computing $M^{-1}_\pm$ one should take into account the non-commutativity of the matrix elements.)
Following \cite{FRT}, we are going to show that the Hopf algebra determined by (\ref{RM}), (\ref{MpmD1}) and (\ref{Hopf-FRT})
is a cover of the QUEA $U_q (s\ell(n))\,$ defined in Appendix B.

\smallskip

Due to the triangularity, the coproduct (\ref{Hopf-FRT}) of a matrix element of $M_+$ or $M_-$
belonging to the corresponding "$m$-th diagonal" (for $m=1,\dots ,n$) contains exactly $m$ summands.
Thus, the diagonal elements $d_\a\,,\ \a=1,2,\dots ,n\,$ ($m=1$) are {\em group-like}
($\Delta (d_\a) = d_\a\otimes d_\a\,,\ \varepsilon (d_\a)=1\,,\ S(d_\a) = d_\a^{-1}$), while
\ba
&&\Delta ((M_+)^i_{~i+1}) = d_i\otimes (M_+)^i_{~i+1} + (M_+)^i_{~i+1} \otimes d_{i+1}\ ,\nn\\
&&\Delta ((M_-)^{i+1}_{~i}) = (M_-)^{i+1}_{~i}\otimes d_i^{-1} + d_{i+1}^{-1} \otimes (M_-)^{i+1}_{~i}
\lb{DeltaMpm}
\ea
for $1\le i\le n-1$ (here $m=2$). The comparison with (\ref{copr}) suggests that
\be
(M_+)^i_{~i+1} = x_i\, F_i\, d_{i+1}\ ,\quad (M_-)^{i+1}_{~i} = y_i\, d_{i+1}^{-1}\, E_i\ , \quad d_i^{-1} d_{i+1} = K_i
\lb{MpmFE}
\ee
where $x_i$ and $y_i$ are some yet unknown $q$-dependent coefficients.
The second and third relation (\ref{dMpm}) (for $\a=i+1\,,\ \b = i$) are satisfied if
\be
d_\a = k_{\a-1} k^{-1}_\a\quad (\,k_0 = k_n = 1\,)\quad\Rightarrow\quad \prod_{\a=1}^n d_\a = 1\ ,
\lb{dkk}
\ee
the new set of independent Cartan generators $k_1, \dots , k_{n-1}$ obeying
\ba
&&k_i := \prod_{\ell =1}^i d_\ell^{-1}\ ,\quad K_i = k_{i-1}^{-1} k_i^2 k_{i+1}^{-1}\ ,\quad i=1,2,\dots,n-1\ ,\nn\\
&&k_i k_j = k_j k_i\ ,\quad k_i\, E_j = q^{\d_{ij}} E_j\, k_i\ ,\quad k_i\, F_j = q^{-\d_{ij}} F_j\, k_i\ ,\nn\\
&&\Delta (k_i) = k_i\otimes k_i\ ,\quad \varepsilon (k_i)=1\ ,\quad S(k_i) = k_i^{-1}\ .\lb{dk}
\ea
Inserting (\ref{MpmFE}) into the last Eq.(\ref{dMpm}) and using the second and third relation (\ref{dMpm})
from which it follows that $[ d_{i+1} , (M_-)^{i+1}_{~i} (M_+)^i_{~i+1} ] = 0$, we obtain
\be
x_i\, y_i = - \l^2\ ,\qquad i=1,\dots , n-1\ .
\lb{xiyi}
\ee
We note further that the commutation relation (\ref{RM}) of $(M_+)^i_{~i+2}$ with $d_\a$ (\ref{dkk})
suggests that $(M_+)^i_{~i+2}$ contains the step operators $F_i$ and $F_{i+1}\,$ only. Assuming that it is proportional to
$(F_{i+1} F_i - z F_i F_{i+1} ) D_{i+2}$ where $D_{i+2}$ is group-like and $z$ is another unknown $q$-dependent coefficient,
taking the corresponding coproduct (\ref{Hopf-FRT}) and using (\ref{MpmFE}), (\ref{copr}) gives
\be
(M_+)^i_{~i+2} = - \frac{x_i x_{i+1}}{\l}\,[F_{i+1},  F_i ]_q\, d_{i+2}\ ,\quad (\, [ A , B ]_q := A B - q B A )\ .
\lb{M+i2}
\ee
A similar calculation shows that
\be
(M_-)^{i+2}_{~i} = \frac{y_i y_{i+1}}{\l}\, d_{i+2}^{-1}\,[ E_i , E_{i+1} ]_{q^{-1}}\ .
\lb{M-i2}
\ee
From now on we shall fix the coefficients $x_i$ and $y_i$ satisfying (\ref{xiyi}) in a symmetric way:
\be
x_i = - \l\ , \quad y_i = \l\ .
\lb{fix-xiyi}
\ee
Computing from (\ref{RM}) the commutators of $(M_+)^i_{~i+2}$ with $(M_+)^i_{~i+1}$ and $(M_+)^{i+1}_{~i+2}\,,$
and of $(M_-)^{i+2}_{~i}$ with $(M_-)^{i+1}_{~i}$ and $(M_-)^{i+2}_{~i+1}\,,$ we obtain relations equivalent to
\ba
&&[ (M_+)^i_{~i+1} , (M_+)^i_{~i+2} ]_q = 0\ ,\quad\ [(M_+)^i_{~i+2} , (M_+)^{i+1}_{~i+2} ]_q = 0\ ,\nn\\
&&[ (M_-)^{i+1}_{~i} , (M_-)^{i+2}_{~i} ]_q = 0\ , \qquad [ (M_-)^{i+2}_{~i} , (M_-)^{i+2}_{~i+1} ]_q = 0
\lb{MpmMpmq}
\ea
which are in fact the non-trivial $q$-Serre relations (\ref{Sq}) written in the form
\ba
&&[ F_i , [ F_i , F_{i+1} ]_{q^{-1}} ]_q \, = \, 0  = \  [ F_{i+1} , [ F_{i+1} , F_i ]_q \, ]_{q^{-1}} \ ,\nn\\
&&[ E_i , [ E_i , E_{i+1} ]_{q^{-1}} ]_q = \, 0 \, = \, [ E_{i+1} , [ E_{i+1} , E_i ]_q \, ]_{q^{-1}} \ .
\lb{Sq-alt}
\ea
Proceeding in a similar way, one can obtain the higher off-diagonal terms of the matrices $M_\pm\,$
(for example, $(M_+)^1_{~4} = - \l\, [ F_3 , [ F_2 , F_1 ]_q ]_q\, d_4$).

The result can be summarized in
\be
M_+ = (\id - \l\, N_+ ) \, D\ ,\qquad M_- = D^{-1} \, (\id + \l\, N_- )
\lb{MpmNpmD}
\ee
where the {\em nilpotent} matrices $N_+$ and $N_-$ are upper and lower triangular, respectively, with matrix
elements given by the corresponding (lowering and raising) {\em Cartan-Weyl} generators of $U_q(s\ell (n))\,$ (see e.g. \cite{R, KhT}),
while the non-trivial entries $d_\a\,,\ \a=1,\dots ,n\,$ of the diagonal matrix $D$ are determined by (\ref{dkk}), (\ref{dk}).
Writing $K_i = q^{H_i}$ would allow us to present the Cartan elements $k_i$ as $k_i = q^{H^i}$ where
$H_i = \sum_{j=1}^{n-1} c_{ij} H^j = 2 H^i - H^{i-1} - H^{i+1}\,$ so that an inverse formula
expressing $k_i$ in terms of $K_i$ would involve "$n$-th roots" of the latter (as $\det (c_{ij}) = n\,;$ cf. also (\ref{al-d})).
In this sense the Hopf algebra $U_q^{(n)}(s\ell (n))\,$ generated by $E_i , F_i , k_i ,\ i=1,\dots , n-1$
(called the "simply-connected rational form" in \cite{CP}) is an {\em $n$-fold cover} of $U_q(s\ell (n))\,.$

\smallskip

Taking into account (\ref{M+-q}), the condition (\ref{M0}) turns out to be consistent with the QUEA invariance of the vacuum vector,
\be
X \vac = \varepsilon (X) \vac
\lb{Uqvac}
\ee
where $\varepsilon (X)$ is the counit (\ref{Hopf-FRT}); in accord with the above we may assume that $X\in U_q^{(n)}(s\ell (n))\,.$

\smallskip

We shall display below the matrices $N_\pm$ and $D$ (\ref{MpmNpmD}) in the cases $n=2$ and $n=3\,:$
\ba
{\bf n=2:}\quad D = \begin{pmatrix}k^{-1}&0\cr 0&k\end{pmatrix}\ (\, K = k^2\,) \ ,\ \
N_+ = \begin{pmatrix}0&F\cr0&0\end{pmatrix}\ ,\  N_- = \begin{pmatrix}0&0\cr E&0\end{pmatrix}\,;\quad
\lb{MD2}\\
\nn\\
{\bf n=3:}\quad D = \begin{pmatrix}k_1^{-1}&0&0\cr 0&k_1 k_2^{-1}&0\cr 0&0& k_2\end{pmatrix}\qquad\
(\, K_1 = k_1^2 k_2^{-1}\,,\ K_2 = k_1^{-1} k_2^2 \,) \ ,\quad\quad\quad\, \nn\\
N_+ = \begin{pmatrix}0&F_1&[F_2 , F_1 ]_q\cr0&0&F_2\cr 0&0&0\end{pmatrix}\ ,\ \ \,
N_- = \begin{pmatrix}0&0&0\cr E_1&0&0\cr [E_1 , E_2 ]_{q^{-1}}& E_2&0\end{pmatrix}\ ,\qquad\,\
\lb{MD3}\\
(\id + \l N_- )^{-1} = \id - \l \begin{pmatrix}0&0&0\cr E_1&0&0\cr [E_1 , E_2 ]_q& E_2&0\end{pmatrix}\ .\qquad\qquad\qquad\qquad\quad\quad\
\nn
\ea

\smallskip

The symmetric choice (\ref{fix-xiyi}) of the normalization is singled out, up to a sign, by the following additional requirement.
There exists a {\em transposition} $X \to X '\,,$ an {\em involutive} linear algebra antihomomorhism (and coalgebra
homomorphism,
$('\,\otimes\ ')\circ \Delta\, (X) = \Delta\, (X')\,,\ \e (X') = \e (X)$), acting on the generators as
\ba
&&{k_i}' = k_i \quad (\,\Rightarrow\ \, {K_i}' = K_i\ ,\quad  {d_\a}' = d_\a\,)\ ,\nn\\
&&{E_i}' = d_i^{-1} F_i\, d_{i+1} = q^{-1} F_i K_i\ ,\quad {F_i}' = d_{i+1}^{-1} E_i\, d_i = q\, K_i^{-1} E_i
\lb{'}
\ea
(cf. (\ref{CRq}), (\ref{Sq}) and (\ref{copr}), respectively).
We observe that demanding $x_i = - y_i$ (cf. (\ref{MpmFE}) and (\ref{xiyi})) is equivalent to requiring the standard
{\em matrix transposed} $\,^t\!M_\pm$ to coincide with the {\em algebraic transposition} of $M_\mp^{-1}$ determined by (\ref{'})
(so that these two different transformations give the same result when applied to the monodromy matrix $M\,;$ see Eq.(\ref{Mpr}) below):
\be
(M_\pm )^\b_{~\a} = (( M^{-1}_\mp )^\a_{~\b} )'\qquad\Rightarrow\qquad M^\b_{~\a} = (M^\a_{~\b})'\ .
\lb{Mtr}
\ee

The parametrization (\ref{MpmNpmD}) of the matrix elements of $M_\pm$ in terms of the QUEA generators relates
two Hopf algebras that seem very different. As it has been already mentioned, the deep result that the Hopf algebra defined by
(\ref{Mpmq}), (\ref{MpmD1}) and (\ref{Hopf-FRT}) is a cover of the QUEA $U_q (s\ell(n))$ has been obtained by Faddeev, Reshetikhin and Takhtajan
in \cite{FRT} (in fact it is more general, applying, for suitably chosen numerical $R$-matrices, to the quantum deformations
introduced by Drinfeld \cite{D} and Jimbo \cite{J} of all classical simple Lie algebras ${\cal G}$).

The main idea in \cite{FRT} is that an appropriately defined deformation ${\rm Fun} (G_q)$ of the algebra of functions
on a matrix Lie group $G$ should be dual to a certain cover of the QUEA $U_q({\cal G})$ where ${\cal G}$
is the Lie algebra of $G\,.$ The "classical" counterpart of this duality is the realization, due to L. Schwartz,
of $U({\cal G})$ as the (non-commutative) algebra of distributions on $G$ supported by its unit element,
$U({\cal G})\simeq C^{-\infty}_e(G)$ (see Theorem 3.7.1 in \cite{C06}).

In \cite{FRT} the Hopf algebra covering $U_q({\cal G})$ (generated, in our notation, by the matrix elements of $M_\pm$)
was constructed as the dual of a quotient of the RTT algebra (\ref{RTT}) defining ${\rm Fun} (G_q)$.
In particular, the Hopf algebra (\ref{Mpmq}), (\ref{MpmD1}), (\ref{Hopf-FRT}) is dual to ${\rm Fun}\, (SL_q(n))\,,$
the ${\rm det}_q(T)= 1\,$ quotient of the {\em RTT}\, algebra (\ref{RTT})
(for an appropriate definition of the quantum determinant) with coalgebra relations written in matrix form as
\be
\Delta (1) = 1\otimes 1\ ,\quad \Delta (T) = T\otimes T\ ,\quad \varepsilon (T) = \id\ ,\quad S(T) = T^{-1}\ .
\lb{coRTT}
\ee
Moreover, it has been shown that relations (\ref{Mpmq}), (\ref{MpmD1}), (\ref{Hopf-FRT})
can be derived from an explicitly given pairing $\langle M_\pm , T \rangle$ expressed in terms of $R^\mp\,.$

\subsection{The zero modes' exchange algebra}

Our next step will be to find appropriate quantum relations corresponding to the PB of the zero modes. We shall first
postulate the exponentiated quantum version of (\ref{PBapD}),
\be
\lb{ExRap}
q^{p_j}\, a_{\alpha}^i = a^i_\a\, q^{p_j + v^{(i)}_j} \ ,\quad v^{(i)}_j = \d^i_j - \frac{1}{n}\quad\Rightarrow\quad
q^{p_{j\ell}} a_{\alpha}^i = a_{\alpha}^i \, q^{p_{j\ell} + \delta_j^i - \delta_{\ell}^i}
\ee
where the operators $q^{p_j}\,,\ i=1,\dots ,n\ $ are mutually commuting and their product is equal to the unit operator:
\be
q^{p_i} q^{p_j} = q^{p_j} q^{p_i}\ ,\qquad \prod_{j=1}^n q^{p_j} = 1\ .
\lb{prod-p=1}
\ee
As the quantum matrix $a$ is a group-like quantity, it is natural to assume that it obeys quadratic exchange relations of the form
\cite{AF, BF}
\be
R_{12} (p) \, a_1 \, a_2 = a_2 \, a_1 \, R_{12}
\lb{ExRaa1}
\ee
involving the {\em quantum dynamical $R$-matrix} $R_{12}(p)\,$ as well as the constant $R$-matrix $R_{12}\,$ (\ref{R}),
that reproduce the PB $\{ a_1 , a_2 \}$ (\ref{PBex}) in the quasi-classical limit.
Eqs. (\ref{ExRap}), (\ref{prod-p=1}) and (\ref{ExRaa1}) determine the {\em quantum matrix algebra} ${\mathcal M}_q (R(p) , R)\,.$

As one may expect from (\ref{ggR}), (\ref{Rx}), Eq.(\ref{ExRaa1}) has two equivalent forms,
\be
R^\pm_{12} (p) \, a_1 \, a_2 = a_2 \, a_1 \, R^\pm_{12}\ ,\quad
R^-_{12}(p) := R_{12}(p)\ ,\quad R^+_{12}(p) := R_{21}^{-1}(p)
\lb{ExRaa}
\ee
which can be also written as a braid relation (note that ${\hat R}_{12}= P R_{12}^-$ implies ${\hat R}^{-1}_{12} = P R_{12}^+$):
\be
\hat R_{12}(p) \, a_1 \, a_2 = a_1 \, a_2 \, \hat R_{12}\ , \quad
\hat R_{12}(p) := P R_{12}^- (p)\quad\Rightarrow\quad {\hat R}^{-1}_{12}(p) = P R_{12}^+ (p)\ .
\lb{ExRaa2}
\ee
Using (\ref{quasicl}) to determine the leading terms in $\hbar\,$ in the quasi-classical expansion of (\ref{ExRaa}), we conclude that
$R^\pm_{12}(p)$ have to reproduce in the large $k$ limit the classical dynamical $r$-matrices $r^\pm_{12} (p)\,,$
\be
R^\pm_{12}(p) = \id - i\, r^\pm_{12} (p) + {\cal{O}} ( \frac{1}{k^2} )\ ,\qquad r^\pm_{12} (p) = r_{12}(p) \pm \frac{\pi}{k}\, C_{12}
\lb{Rp-cond}
\ee
with $r_{12}(p)$ given by (\ref{dynr}), (\ref{f01}). Indeed, assuming (\ref{Rp-cond}) and (\ref{Rr}) and taking into account that
the entries of $a$ classically commute (so that $a_1\, a_2 = a_2\, a_1$),
we conclude that the leading terms in $\frac{1}{k}$ of (\ref{ExRaa}) exactly match the PB (\ref{PBex}).

\smallskip

Applying the two sides of Eq.(\ref{QYBE}) to the right of the triple tensor product $a_3\, a_2\, a_1$ and using (\ref{ExRaa})
and the CR (\ref{ExRap}), we obtain, as consistency condition, the {\em quantum dynamical YBE}
\ba
&&R_{12} (p - v_{(3)}) \, R_{13} (p) \, R_{23} (p - v_{(1)}) = R_{23} (p) \, R_{13} (p - v_{(2)}) \, R_{12} (p) \quad\Leftrightarrow\nn\\
&&\hat R_{12} (p) \, \hat R_{23} (p-v_{(1)}) \, \hat R_{12} (p) = \hat R_{23} (p-v_{(1)}) \, \hat R_{12} (p) \, \hat R_{23} (p-v_{(1)})\ .\qquad\qquad
\lb{QDYBE}
\ea
The following example explains the above short-hand notation:
\begin{equation}
\label{pv1}
{\hat R}_{23} (p-v_{(1)})_{~j_1 j_2 j_3}^{i_1 i_2 i_3} = \delta_{j_1}^{i_1} \, R (p-v^{(i_1)})_{~j_2 j_3}^{i_3 i_2} \ .
\end{equation}
Eq.(\ref{QDYBE}) appeared in the early days of the $2D$ CFT in the paper \cite{GN} by Gervais and Neveu on the Liouville model
and attracted wide interest ten years later due to the work of Felder \cite{Felder}.

Following Etingof and Varchenko \cite{EV2}, we shall call {\em quantum dynamical $R$-matrix} an
invertible solution $R_{12}(p)$ of (\ref{QDYBE}) satisfying, in addition, the {\em zero weight condition}
\be
[ h_{\ell 1} + h_{\ell 2} \,,\, R_{12} (p) ] = 0\ ,\quad \ell = 1,\dots , n-1\ .
\lb{nRp}
\ee
Eq.(\ref{nRp}) looks natural as it implements at the quantum level the classical condition (\ref{neutrality}) for $r_{12}(p)\,.$
It strongly restricts the off-diagonal elements of the $n^2\times n^2$ matrix $R_{12} (p)\,,$ implying the {\it ice condition}
\be
\label{ice}
R_{~i' j'}^{ij} (p)= 0 \qquad \mbox{unless} \quad i = i' \, , \ j = j' \quad \mbox{or} \quad i = j' \, , \ j = i'
\ee
which is in turn equivalent to
\be
\lb{Rp-ice}
q^{- \frac{1}{n}}\,{\hat R}^{ij}_{~i'j'} (p)\, =\,a_{ij}(p)\, \delta^i_{j'} \delta^j_{i'} + b_{ij}(p)\, \delta^i_{i'} \delta^j_{j'}\qquad
(\,b_{ii}(p) = 0\,)\ .
\ee
(The last convention makes the representation (\ref{Rp-ice}) unambiguous.)

The Hecke relation (\ref{Hecke}) for $\hat R$ implies a similar equation for $\hat R(p)$:
\be
(q^{-\frac{1}{n}}\,\hat R(p) - q^{-1} ) (q^{-\frac{1}{n}}\,\hat R(p) + q) = 0\ .
\lb{HeckeRp}
\ee

Finally, the property of the operators $\hat R_{i\, i+1}(p)\,$ to generate a representation of the braid group
(namely, the commutativity of distant braid group generators (\ref{braidR})) is ensured by the additional requirement
\be
\hat R_{12} (p+v_{(1)}+v_{(2)} )\, =\, \hat R_{12}(p)\quad \Leftrightarrow\quad
\hat R^{ij}_{~k\ell} (p)\, a^k_\a a^\ell_\b \, =\, a^k_\a a^\ell_\b \,\hat R^{ij}_{~k\ell} (p) \ .
\lb{Rpvv}
\ee

The general solution for ${\hat R} (p)$ of the type (\ref{Rp-ice}) satisfying (\ref{QDYBE}), (\ref{HeckeRp}) and (\ref{Rpvv})
has been found in \cite{HIOPT} (based on the paper \cite{I2}; see also \cite{EV2}). It can be brought to the following canonical form:
\ba
&&a_{ij}(p) = \a_{ij}(p_{ij}) \,\frac{[p_{ij}-1]}{[p_{ij}]} \ , \quad b_{ij}(p) = \frac{q^{-p_{ij}}}{[p_{ij}]}\ ,\quad i\ne j\nn\\
&&(\,\a_{ji}(p_{ji}) = \frac{1}{\a_{ij}(p_{ij})} \,)\ ,\quad a_{ii}(p) = q^{-1}\ ,\quad b_{ii}(p) = 0\ .
\lb{canRp}
\ea
For any given pair $(i,j) \ (i\ne j)$, the ice condition provides a convenient representation of the $(i,j)$ block of ${\hat R}(p)$ as a $4\times 4$ matrix
which, assuming the ordering $(ii) , (ij) , (ji) , (jj)$ of the rows and columns, takes thus the form
\be
\lb{RRp2}
{\hat R}^{(ij)}(p)\, =\,
q^{\frac{1}{n}}\begin{pmatrix}q^{-1}&0&0&0\cr
0&\frac{q^{-p_{ij}}}{[p_{ij}]}&\a_{ij}(p_{ij})\,\frac{[p_{ij} -1]}{[p_{ij}]}&0\cr
0&(\a_{ij}(p_{ij}))^{-1} \frac{[p_{ij} +1]}{[p_{ij}]}&-\frac{q^{p_{ij}}}{[p_{ij}]}&0\cr
0&0&0&q^{-1}\end{pmatrix}\ .
\ee
Using the expansions
\be
\lb{exppk}
\frac{[p\pm 1]}{[p]} = 1\pm \frac{\pi}{k} \cot ( \frac{\pi}{k} p) + O(\frac{1}{k^2})\ ,\quad
\frac{q^{\pm p}}{[p]} = \frac{\pi}{k} \left( \cot ( \frac{\pi}{k} p) \mp\, i \right) + O(\frac{1}{k^2})\ ,
\ee
one recovers in the quasi-classical limit (\ref{Rp-cond}) the classical dynamical $r$-matrix $r_{12}(p)\,$ (\ref{dyn-r-matr}) for
\ba
&&\a_{ij}(p_{ij}) = 1 + \frac{\pi}{k}\, \b (\frac{\pi}{k}\,p_{ij}) + O(\frac{1}{k^2}) \qquad (\,\b(p) = - \b(-p)\,)\ ,\nn\\
&&f_{j\ell}(p) =  i \frac{\pi}{k} \left(\cot ( \frac{\pi}{k}\, p_{j\ell}) - \b (\frac{\pi}{k}\, p_{j\ell})\right) \ ,
\lb{f-alpha}
\ea
cf. (\ref{f01})\footnote{In (\ref{f01}), the condition $\b_{j\ell} (p_{j\ell}) = \b(p_{j\ell})$ has been imposed to ensure the Weyl invariance of the constraint $\chi\,.$}.
Here again, the expansion of the coefficient $q^{\frac{1}{n}}$ provides the $\frac{1}{n}$ term for $C_{12}$ (\ref{Cn-sigma}).

\smallskip

In contrast with the constant $\hat R$ case, the representation of the braid group generated by ${\hat R}(p)$ is "nonlocal".
The second equation (\ref{QDYBE}) suggests that the braid operators corresponding to the dynamical $R$-matrix should be defined by
${\hat R}_1(p) = {\hat R}_{12} (p)\,,\ {\hat R}_2(p) = {\hat R}_{23} (p - v_{(1)})\,.$ In general, we shall define
the (renormalized) $i$-th braid operator as
\be
b_i(p) = q^{-\frac{1}{n}} {\hat R}_i(p) := q^{-\frac{1}{n}} {\hat R}_{i i+1} (p - \sum_{\ell = 1}^{i-1} v_{(\ell)} )
\lb{dyn-braid}
\ee
which guarantees that the braid group relations (\ref{braidR}) are satisfied.

The Hecke condition for the renormalized braid operators $b_i := q^{-\frac{1}{n}} {\hat R}_i$ and $b_i(p)$ (\ref{dyn-braid})
(Eqs. (\ref{Hecke}) and (\ref{HeckeRp}), respectively) can be equivalently expressed in their spectral decomposition
in terms of two orthogonal idempotents $\frac{q^{\pm 1}\! \idd \pm b_i}{[2]}$ with coefficients $q^{-1}$ and $- q\,,$ respectively.
A renormalized version of this, more suitable for the root of unity case, is to set
\be
b_i = q^{-1} \id -  A_i\ ,\qquad b_i(p) = q^{-1} \id - A_i(p)\ ,
\lb{biAi}
\ee
where $A_i \equiv A_{i i+1}$ and $A_i(p)$ are the constant and dynamical $q$-{\em antisymmetrizers}, respectively.
Now the full set of relations (\ref{braidR}) and (\ref{Hecke}) satisfied by the braid operators,
\ba
&&b_i^2 = (q^{-1} - q)\, b_i + \id\ ,\nn\\
&&b_i \, b_j \, b_i = b_j \, b_i \, b_j \quad \mbox{for}\quad | i -j |  = 1\ ,\nn\\
&&b_i\, b_j = b_j \, b_i = 0 \quad \mbox{for}\,\quad | i -j |  \ge 2
\lb{b-Hecke}
\ea
can be rewritten equivalently as
\ba
&&A_i^2 = [2]\, A_i\qquad (\, [2] = \, q + q^{-1}\,)\ ,\nn\\
&&A_i \, A_j \, A_i - A_i = A_j \, A_i \, A_j - A_j \quad \mbox{for}\quad | i -j\, |  = 1\ ,\nn\\
&&[A_i , A_j] = 0 \quad \mbox{for}\quad | i -j\, |  \ge 2
\lb{q-antisymm}
\ea
(identical relations exist for $b_i(p)\,$ and $A_i(p)$).

\medskip

\noindent
{\bf Remark 4.2~} The abstract algebra generated by $\id\, ,\, b_1 \,,\, \dots \,,\, b_{m-1}\,,$ subject to relations (\ref{b-Hecke})
(or by $\id , A_1 , \dots , A_{m-1}\,$ and (\ref{q-antisymm}), respectively), is known as the {\em Hecke algebra} $H_m (q^{-1})\,$
(see e.g. \cite{CP, GdlHJ}). Regarded as an one-parameter deformation of the group algebra of a Coxeter group (here of the symmetric
group of $m\,$ elements, see (\ref{Wrels})), it is also called the {\em Iwahori-Hecke algebra of type $A\,$}. Its quotient defined
by imposing the stronger condition
\be
A_i \, A_j \, A_i = A_i\quad {\rm for}\quad| i -j\, |  = 1
\lb{TL}
\ee
is the well known {\em Temperley-Lieb} algebra ${\cal T}{\cal L}_m (\beta)\,$ \cite{TL}
(for $\b = [2]^2$) that has numerous applications in lattice models of statistical mechanics\footnote
{An infinite "tower" of such algebras defined in terms of {\em projectors} satisfying
(\,$E_i^2 = E_i\,$ and) $\,\b\, E_i \, E_j \, E_i = E_i\,$ for $\,| i -j\, |  = 1\,$ has been used by V.F.R. Jones in the
classification of inclusions of von Neumann subfactors \cite{Jones83} and in the construction of a new polynomial invariant of links \cite{Jones85}.}.
Note that the second set of relations in (\ref{b-Hecke}) and (\ref{q-antisymm})
are only relevant for $m > 2\,$ (and the third set, even for $m > 3$).

\medskip

The operators $A_i$ and $A_i(p)$ provide two different deformations of the projector on the skewsymmetric part of the tensor square
of an $n$-dimensional vector space. We shall proceed, following the paper \cite{HIOPT} (in which ideas, techniques and
results from \cite{Gur, GPS} and \cite{I2} have been further developed), with the definitions of the corresponding
higher order antisymmetrizers acting on the (tensor products of the) auxiliary index spaces and the Levi-Civita
($\varepsilon$-)tensors related to them. This will allow us to introduce the notion of {\em quantum determinant}
$D_q(a)$ of the zero modes matrix (with non-commuting entries) $(a^i_\a)\,$ and find the appropriate quantum counterpart
of the determinant condition (\ref{DaDp1}).

\smallskip

The constant solution of the YBE (\ref{R}) gives rise to (\ref{biAi}) with
\be
A_1 \equiv A_{12} = q^{-\epsilon} \id_{12} - P_{12} = ( A_{~\alpha' \beta'}^{\alpha \beta} )\ ,\qquad
A_{~\alpha' \beta'}^{\alpha \beta} = q^{\epsilon_{\beta \alpha}} \, \delta_{\alpha'}^{\alpha} \, \delta_{\beta'}^{\beta} -
\delta_{\beta'}^{\alpha} \, \delta _{\alpha'}^{\beta}\ .
\lb{A1const}
\ee
Following \cite{HIOPT}, we shall introduce inductively higher order antisymmetrizers $A_{\ell m}$ projecting on the $q$-skewsymmetric
tensor product of $n$-dimensional spaces with labels $\ell , \ell + 1,\dots , m\,,\ 1\le \ell\le m\,$ by
\ba
&&A_{\ell\, m+1}= q^{-m+\ell-1}\, A_{\ell\, m} - \frac{1}{[m-\ell]!}\,A_{\ell\, m}\, b_m \, A_{\ell\, m}
\ ,\quad A_{\ell \ell} = \id\ ,\qquad\nn\\
&&[m]! = [m][m-1]!\ ,\quad [0]! = 1\ .
\lb{antis-l}
\ea
The operators $A_{\ell\, m}$ (for $\ell < m$) are thus multilinear functions of $b_\ell, b_{\ell +1} ,\dots , b_{m-1}\,.$
Their projector properties follow from the general relation
\be
A_{\ell\, m}\, A_{1 j} = A_{1 j}\, A_{\ell \, m} = [m - \ell +1]!\, A_{1 j}\qquad{\rm for}\qquad 1\le \ell \le m \le j\ ;
\lb{unP}
\ee
in particular, $A_{1 j}^2 = [j]!\, A_{1j}\,.$ In the non-trivial case when $\ell < m\,,$ Eq.(\ref{unP}) can be proved by induction, starting with
\be
A_{\ell\,\ell +1}\, A_{1 j} = A_{1 j}\, A_{\ell\,\ell+1} =[2]\, A_{1 j}
\quad\Leftrightarrow\quad b_\ell\, A_{1 j} = A_{1 j}\, b_\ell  = - q\, A_{1 j}
\lb{bA}
\ee
for $1 \le \ell \le j-1\,.$ Indeed, suppose that (\ref{unP}) is correct for $1\le \ell < m \le j-1\,.$ Then, from (\ref{antis-l}) one obtains
\ba
&&A_{\ell\, m+1}\, A_{1 j} = A_{1 j}\, A_{\ell\, m+1}  =\left( q^{-m+\ell -1} [m-\ell +1]! + q\, \frac{[m-\ell +1]!^2}{[m-\ell ]!}\right) A_{1j} = \nn\\
&&= [m-\ell +1 ]!\, ( q^{-m+\ell -1} + q \, [m-\ell +1] ) A_{1 j} = [m-\ell +2 ]!\, A_{1 j}\ .\qquad\qquad
\lb{unP1}
\ea
One can verify that the definition of $A_{1 j+1}\,,\ j=1,2,\dots$ implied by (\ref{antis-l}),
\be
A_{1 j+1} = q^{-j} A_{1j} - \frac{1}{[j-1]!}\,A_{1j}\, b_j\, A_{1j}
\equiv \frac{1}{[j-1]!}\,A_{1j} A_{j\, j+1} A_{1j} - [j-1] A_{1j}
\lb{antis-j}
\ee
is equivalent also to
\be
A_{1 j+1} = \frac{1}{[j-1]!}\,A_{2\, j+1}\, A_{12} \, A_{2\, j+1} - [j-1]\, A_{2\, j+1}\ ,
\lb{alt-antis}
\ee
the equality of (\ref{antis-j}) and (\ref{alt-antis}) generalizing the first relation (\ref{q-antisymm}).

\smallskip

As already mentioned, the unusual normalization of the antisymmetrizers adopted here is suitable for the case when $q^h = -1\,.$
Indeed, as $h = n+k >n\,,$ all $A_{1j}$ are well defined for $1\le j\le n+1\,.$ Further, one can show that the dimension
of the image of $A_{1j}$ (i.e., its {\em rank}) is equal, for any $j$ in this range, to the dimension $\left( {n\atop j}\right)$
of the fully skew-symmetric IR of the symmetric group ${\cal S}_j$ corresponding to the single column Young diagram with $j\,$ boxes
so that, in particular,
\be
\lb{A1n}
A_{1\,n+1} = 0\ ,\qquad {\rm rank}\, A_{1\, n} = 1\qquad\Rightarrow\qquad
A_{1\,n} = (\varepsilon^{\alpha_1 \ldots \alpha_n} \, \varepsilon_{\beta_1 \ldots \beta_n})\ .
\ee
The Levi-Civita tensors $\varepsilon\,$ with upper indices belong to the eigenspaces corresponding to the eigenvalue $[2]$ of all
$A_j\,,\ j=1,\dots , n-1\,$ and those with lower indices, to the corresponding eigenspaces of the transposed $A_j\,,$ i.e.
\ba
&&A^{\alpha_i \alpha_{i+1}}_{~\s_i \s_{i+1}} \,\varepsilon^{\alpha_1\dots \s_i \s_{i+1}\dots \alpha_n} =
[2]\,\varepsilon^{\alpha_1\dots \alpha_i \alpha_{i+1}\dots \alpha_n}\ , \nonumber\\
&&\varepsilon_{\alpha_1\dots \s_i \s_{i+1}\dots \alpha_n} A_{~\alpha_i \alpha_{i+1}}^{\s_i \s_{i+1}} =
[2]\,\varepsilon_{\alpha_1\dots \alpha_i \alpha_{i+1}\dots \alpha_n}
\lb{eqs-eps}
\ea
(see the first relation (\ref{bA})). By (\ref{A1const}), this implies e.g. that
\ba
&&\varepsilon_{\alpha_1\dots \a_{i+1} \a_i \dots \alpha_n} =
- q\, \varepsilon_{\alpha_1\dots \a_i \a_{i+1}\dots \alpha_n}\quad{\rm for}\quad \a_{i+1} < \a_i\ ,\qquad
\varepsilon_{\alpha_1\dots \a \a \dots \alpha_n} = 0\ ,\nn\\
&&{\rm i.e.}\qquad \varepsilon_{\alpha_1\dots \a_{i+1} \a_i \dots \alpha_n} =
- q^{\epsilon_{\a_i \a_{i+1}}}\, \varepsilon_{\alpha_1\dots \a_i \a_{i+1}\dots \alpha_n}\ ,
\lb{ee}
\ea
see (\ref{stand-r-matr}). As the matrix of the operator $A_{i i+1}$ is symmetric,
$A^{\alpha' \beta'}_{~\alpha \beta} = A_{~\alpha' \beta'}^{\alpha \beta}\,,$
the solutions of (\ref{eqs-eps}) with identical ordered sets of upper and lower indices only differ by a proportionality factor
and, in particular, can be chosen to be equal. Then the normalization condition implied by (\ref{unP}), (\ref{A1n})
\be
A_{1n}^2 = [n]!\, A_{1n}\quad \Rightarrow\quad \varepsilon_{\alpha_1 \ldots \alpha_n} \varepsilon^{\alpha_1 \ldots \alpha_n} = [n]!
\lb{een!}
\ee
fixes them up to a sign. Thus, the constant Levi-Civita tensors vanish whenever some of their indices coincide while, in our conventions,
\begin{equation}
\label{q-eps}
\varepsilon^{\alpha_1 \ldots \alpha_n} = \varepsilon_{\alpha_1 \ldots \alpha_n} = q^{- \frac{n(n-1)}{4}} \, (-q)^{\ell (\sigma)}
\qquad \mbox{for} \quad \sigma = \left( {n\ \ldots\ 1}\atop{~\alpha_1 \ldots ~\alpha_n} \right)
\in {\mathcal S}_n\ ,
\end{equation}
where ${\mathcal S}_n$ is the symmetric group of $n$ objects and $\ell (\sigma)$ is the {\em length} of the permutation\footnote{
The length $\ell(\s)$ of a permutation $\s$ (\ref{q-eps}) is equal to $inv(\s)\,,$ the number of {\em inversions} which, in our notation, are the
pairs $(\a_i, \a_j)$ such that $\a_i < \a_j$ for $i < j\,.$ Let $Z(n, \ell)$ be the number of permutations in ${\cal S}_n$
of length $\ell\,.$ The normalization factor in Eq.(\ref{q-eps}) is derived using the well known formula for the generating
function of $Z(n, \ell)\,$
$$\sum_{\s\in{\mathcal S}_n} t^{\ell(\s )} = \sum_{\s\in{\mathcal S}_n} t^{inv(\s )} =
\sum_{\ell =0}^{\left( n\atop 2\right)} Z(n,\ell )\, t^\ell = (1+t)(1+t+t^2)\dots (1+t+\dots + t^{n-1})
\eqno{(*)}$$
and the relation $1 + q^2 +\dots + q^{2(n-1)} = q^{n-1} \, [n]\,,$ implying $\,\sum_{\s\in{\mathcal S}_n} q^{2\ell(\s )} = q^{\frac{n(n-1)}{2}}\, [n]!\,.$
The discovery (in 1970!) of the fact that formula $(*)$ has been actually found by
Benjamin Olinde Rodrigues \cite{Rod} in 1839 (see e.g. \cite{CSZ}) is attributed to Leonard Carlitz.}
$\sigma\,.$ The $q \to 1$ limit of (\ref{q-eps}) reproduces the ordinary (undeformed) Levi-Civita tensor
$\epsilon_{\a_1 \ldots \a_n}$ normalized by $\epsilon_{n \ldots 1} = 1$ whose non-zero components are simply $(-1)^{\ell (\sigma)}$.
We also have \cite{HIOPT}
\be
\varepsilon^{\a \s_1 \ldots \s_{n-1}} \, \varepsilon_{\s_1 \ldots \s_{n-1} \b} = (-1)^{n-1}\, [n-1]! \, \d^\a_\b =
\varepsilon_{\b \s_1 \ldots \s_{n-1}} \, \varepsilon^{\s_1 \ldots \s_{n-1} \a } \ .
\lb{NK}
\ee

\smallskip

The dynamical antisymmetrizer $A_1(p) \equiv A_{12}(p) = ( A(p)_{~i'j'}^{ij} )$
deduced from (\ref{biAi}), (\ref{dyn-braid}), (\ref{Rp-ice}) and (\ref{canRp}) has the form
\ba
&&A(p)_{~i'j'}^{ij} = \frac{[p_{ij} - 1]}{[p_{ij}]}\,
(\delta_{i'}^i \, \delta_{j'}^j - \alpha_{ij}(p_{ij}) \,\delta_{j'}^i \, \delta_{i'}^j )
\quad{\rm for}\quad i\ne j\quad {\rm and}\quad i' \ne j'\ ,\nn\\
&&A(p)_{~i'j'}^{ij} = 0\quad{\rm for}\quad i=j\quad {\rm or}\quad i' = j'\ .
\lb{A1dyn}
\ea
Higher order dynamical antisymmetrizers $A_{1j}(p)$ can be found by a procedure similar to the one used for the constant ones \cite{HIOPT}.
In particular, $A_{1n}(p)\,$ is of rank $1$ and hence,
\be
A_{1n}(p) = (\epsilon^{i_1 \ldots i_n} \, (p)\,\epsilon_{j_1 \ldots j_n} \, (p) ) = \frac{1}{[n]!}\, A_{1n}^2(p)\quad
\Rightarrow \quad \epsilon^{i_1 \ldots i_n} \, (p)\,\epsilon_{i_1 \ldots i_n} \, (p) = [n]!\ .
\lb{een!dyn}
\ee
The choice $\alpha_{ij}(p_{ij}) = 1\,$ simplifies considerably the above expressions
and we shall assume it in what follows, unless explicitly stated otherwise.
In this case the dynamical analogs of Eqs. (\ref{eqs-eps}), (\ref{ee}) for the $\epsilon$-tensors read
\ba
&&\epsilon_{i_1 \dots\, i\, i\, \dots i_n} \, (p) =\epsilon^{i_1 \dots\, i\, i\, \dots i_n} \, (p) = 0\ ,\nn\\
&&[p_{i_{\mu+1}i_\mu } +1]\,\epsilon^{i_1 \dots i_{\mu+1} i_\mu \dots i_n} \, (p) =
[p_{i_\mu i_{\mu+1}}+1]\,\epsilon^{i_1 \dots i_{\mu} i_{\mu+1} \dots i_n} \, (p)\ ,\nn\\
\lb{eqs-eps-p}
&&\epsilon_{i_1 \dots i_{\mu+1} i_\mu \dots i_n} \, (p) =
- \,\epsilon_{i_1 \dots i_{\mu} i_{\mu+1} \dots i_n} \, (p) \quad {\rm for}\quad i_\mu \ne i_{\mu+1}\ .
\ea
Fixing the remaining ambiguity by choosing the $\epsilon$-tensor with lower indices to be equal to
the ($p$-independent) undeformed Levi-Civita tensor $\epsilon_{i_1 \ldots i_n} = \epsilon^{i_1 \ldots i_n}$ eventually leads to
the following solution satisfying the normalization condition in (\ref{een!dyn}):
\be
\epsilon_{i_1 \ldots i_n} \, (p) = \epsilon_{i_1 \ldots i_n} \ ,\qquad
\epsilon^{i_1 \ldots i_n} \, (p) = \epsilon^{i_1 \ldots i_n}
\prod_{1 \leq \mu < \nu \leq n} \frac{[p_{i_{\mu} i_{\nu}} - 1]}{[p_{i_{\mu} i_{\nu}}]}\ \ .
\lb{epsilon-p}
\ee
The {\em non-zero components} of the dynamical $\epsilon$-tensor with upper indices (which should be therefore all different) can be also written as
\be
\epsilon^{i_1 \ldots i_n} \, (p) = \frac{(-1)^{\frac{n(n-1)}{2}}}{{\cal D}_q(p)}\,\prod_{1 \leq \mu < \nu \leq n} [p_{i_{\mu} i_{\nu}} - 1]\ ,\qquad
{\cal D}_q(p) := \prod_{i<j} [p_{ij}]\ .\qquad\qquad
\lb{eps-Dqp}
\ee

\smallskip

In order to complete the study of the quantum matrix algebra ${\mathcal M}_q\,,$ we define the {\it quantum determinant}
\be
\lb{Dqa}
\det (a) = D_q (a) := \frac{1}{[n]!} \, \epsilon_{i_1 \ldots i_n} (p) \, a_{\alpha_1}^{i_1} \ldots a_{\alpha_n}^{i_n} \,
\varepsilon^{\alpha_1 \ldots \alpha_n}\ .
\ee
The definition (\ref{Dqa}) of the quantum determinant is justified by the following statement
(see Proposition~6.1 of \cite{HIOPT}).

\medskip

\noindent {\bf Proposition 4.1~} {\it The product $\, a_{\alpha_1}^{i_1} \ldots a_{\alpha_n}^{i_n}\, $ intertwines between the constant and dynamical
Levi-Civita tensors:}
\be
\epsilon_{i_1 \ldots i_n} (p) \, a_{\alpha_1}^{i_1} \ldots a_{\alpha_n}^{i_n} =
D_q (a) \, \varepsilon_{\alpha_1 \ldots \alpha_n}\ ,\quad
a_{\alpha_1}^{i_1} \ldots a_{\alpha_n}^{i_n} \,
\varepsilon^{\alpha_1 \ldots \alpha_n} = \epsilon^{i_1 \ldots i_n} \, (p) \, D_q (a)\ .
\lb{det-intertw}
\ee

\smallskip

\noindent {\bf Proof~}
Denote $a_{\alpha_1}^{i_1} \ldots a_{\alpha_n}^{i_n} =: a_1 \dots a_n \,;$ then (\ref{ExRaa}) implies
\ba
&&a_1 \dots a_n \, {\hat R}_{i\,i+1} = a_1\dots a_{i-1} a_i a_{i+1} {\hat R}_{i\,i+1} a_{i+2}\dots a_n =
\lb{aaR-Rpaa}\\
&&= a_1\dots a_{i-1}\, {\hat R}_{i\,i+1} (p)\, a_i\, a_{i+1}\, a_{i+2}\dots a_n =
{\hat R}_{i\,i+1} (p - \sum_{\ell = 1}^{i-1} v_{(\ell)} )\, a_1 \dots a_n
\nn
\ea
for $1\le i\le n-1$ which, due to (\ref{dyn-braid}), (\ref{biAi}), is equivalent to
\be
a_1 \dots a_n\, A_i = A_i(p)\, a_1 \dots a_n \qquad\Rightarrow\qquad a_1 \dots a_n\, A_{1n} = A_{1n}(p)\, a_1 \dots a_n \ .
\lb{ApA}
\ee
Multiplying the last equality (\ref{ApA}) by $A_{1n}(p)$ from the left, or by $A_{1n}$ from the right,
we obtain the following two relations,
\ba
A_{1n}(p)\, a_1 \dots a_n = \frac{1}{[n]!}\, A_{1n}(p)\, a_1 \dots a_n\, A_{1n} = a_1 \dots a_n \, A_{1n}
\lb{ApA2}
\ea
which are equivalent to (\ref{det-intertw}) (to prove this we use the rank $1$ projector properties
of the constant and dynamical antisymmetrizers $A_{1n}$ and $A_{1n}(p)$ (\ref{A1n}), (\ref{een!}) and (\ref{een!dyn})). \eod

\smallskip

The quantum counterpart of the vanishing PB (\ref{Dap}) is the commutativity of $D_q(a)$ with $q^{p_j}\,,$
an immediate corollary of the commutation relations (\ref{ExRap}) and the definition (\ref{Dqa}) of the quantum determinant:
\be
q^{p_j} D_q(a) = D_q(a)\, q^{p_j + \sum_{i=1}^n v^{(i)}_j} = D_q(a)\, q^{p_j}\ .
\lb{Dqap}
\ee
On the other hand, the exchange of $D_q(a)$ and $a^i_\a$ produces a $p$-dependent coefficient,
\be
\lb{Dqa-K}
D_q(a)\, a^i_\a = K_i(p)\, a^i_\a\, D_q(a)\ ,\qquad i =1,\dots ,n \ ,
\ee
where the function $K_i(p)$ is given explicitly by
\be
K_i(p) := \frac{(-1)^{n-1}}{[n-1]!}\,\epsilon_{i j_1 \ldots j_{n-1}}\, \epsilon^{j_1 \ldots j_{n-1} i} (p-v^{(i)}) =
\prod_{j\ne i} \frac{[p_{ij}]}{[p_{ij}-1]}
\lb{Dqaa}
\ee
(cf. \cite{HIOPT}, Proposition 6.2). So the centrality of a function of the type
$\frac{D_q(a)}{\Phi_q(p)} \in {\cal M}_q$ which reduces, effectively, to the quantum analog of (\ref{DPa}),
\be
[\, \frac{D_q(a)}{\Phi_q(p)}\,,\, a^i_\a ] = 0
\lb{qcent}
\ee
will be guaranteed if $\Phi_q(p)$ satisfies an equation analogous to (\ref{Dqa-K}),
\be
\lb{Fpa}
\Phi_q(p)\, a^i_\a = K_i(p)\, a^i_\a\, \Phi_q(p)\ .
\ee
It is easy to prove that (\ref{Fpa}) takes place for
\be
\Phi_q(p) = {\cal D}_q(p)
\lb{Fqp}
\ee
(note that ${\cal D}_q(p)$ introduced in (\ref{eps-Dqp}) coincides with its classical expression (\ref{Dap}), only the value of the
deformation parameter is different). The quasi-classical expansions of these relations agree with (\ref{aDa}), (\ref{11})
and (\ref{f01}) (for $\b(p) = 0$).

\smallskip

It is thus consistent to impose the {\em determinant condition}
\be
\lb{Dqa=Dqp}
\det (a) \equiv D_q (a) = {\mathcal D}_q (p)
\ee
as an additional constraint on the quantum matrix $a$ and define the zero modes' quantum algebra as the quotient
of ${\mathcal M}_q (R(p) , R)$ with respect to the two-sided ideal generated by (\ref{Dqa=Dqp}); we shall denote
this quotient henceforth simply as ${\cal M}_q\,.$ Note that the determinant condition is {\em $n$-linear} whereas the
exchange relations (\ref{ExRaa}) are quadratic so they are only mixing in the degenerate case $n=2\,.$

\smallskip

Quantizing (\ref{Mgen}), we obtain the zero modes exchange relations with the monodromy matrix $M$ which are essentially the same as
those for $g(z)$ (\ref{Mgq}):
\be
a_1\, R_{12}^- M_2 = M_2\, a_1\, R_{12}^+ \qquad\ (\, R^-_{12} = R_{12}\,,\ R^+_{12} = R_{21}^{-1}\, )\ .
\lb{Maq}
\ee
We shall assume that the classical relation (\ref{aintertw}) is retained at the quantum level:
\be
\lb{aMMpa}
M_p \, a \, = \, a \, M\ .
\ee
It allows to compare (\ref{Maq}) with the first relation (\ref{ExRap}) which can be written in the form
\be
a_1\, M_{p2}\, =\, q^{2\s_{12}}\, M_{p2}\, a_1\ ,\qquad (q^{2 \s_{12}})^{ij}_{\ell m} =
q^{2(\d_{ij}-\frac{1}{n})}\, \d^i_\ell \, \d^j_m
\lb{ExRap2}
\ee
where $\s_{12}$ is the diagonal part of the polarized Casimir operator (\ref{Cn-sigma}).
Using the exchange relations (\ref{ExRaa1}), we derive a compatibility condition between
the last three equalities expressing the inverse of the dynamical $R$-matrix in terms of $R_{12}(p)\,$
and the diagonal monodromy matrix $M_p\,:$
\ba
R_{12}(p)\, q^{2 \s_{12}} \,M_{p2}\, R_{21}(p) \,M^{-1}_{p2} = \id_{12}\quad \Leftrightarrow\quad
({\hat R}_{12}(p) )^{-1} \, = \,q^{2 \s_{12}} M_{p2}\, {\hat R}_{12}(p)\, M^{-1}_{p 1}\ .\qquad\quad
\lb{Rpinv}
\ea
One can verify that Eq.(\ref{Rpinv}) holds for ${\hat R}_{12}(p)$ given by (\ref{Rp-ice}), (\ref{canRp})
and $M_p$ {\em proportional} to ${\rm diag}\, ( q^{ - 2 p_1} ,\dots , q^{ - 2 p_n})$
(see the next subsection).

It should be also mentioned that the PB (\ref{Mpmpl}) quantize trivially to
\be
[ (M_\pm)^\a_{~\b} \,,\, p_\ell\, ]\, =\, 0\,=\, [ M^\a_{~\b} \,,\, p_\ell\, ] \quad \Rightarrow\quad [M_{\pm 1} , M_{p2} ] = 0 = [ M_1 , M_{p2} ]\ .
\lb{Mpmplq}
\ee

\smallskip

We shall conclude this subsection with the quantum group transformation properties of the quantum zero mode's matrix.
The exchange relations between the Gauss components of the monodromy $M_\pm$ and $a$
(the quantization of the first relation (\ref{Mpma})) read
\be
M_{\pm 2}\, a_1 = a_1\, R_{12}^\mp M_{\pm 2}\ ;
\lb{aMpm}
\ee
of course, Eq.(\ref{Maq}) follows from here as it should. Recasting (\ref{aMpm}) in a form involving the antipode $S$ (\ref{Hopf-FRT}),
\be
M_{\pm 2}\, a_1\, S(M_\pm )_2 = a_1\, R_{12}^\mp\qquad (\,{\rm i.e.,}\quad
(M_\pm)^\b_{~\rho}\, a^i_\a\, S((M_\pm)^\rho_{~\gamma}) = a^i_\s\, (R^\mp)^{\s\b}_{~\a\g}\,)
\lb{aMpm-comp}
\ee
defines the quantum group action on the zero modes.
Writing down explicitly equations (\ref{aMpm-comp}) that only include the diagonal and next-to-diagonal elements of $M_\pm$ (i.e.,
fixing $\g = \b$ or $\g = \b\pm 1\,,$ respectively), using the parametrization of $M_\pm$ from the previous Section 4.3, as well as the formula
\be
R_{12}^+ = R_{21}^{-1} = q^{-\frac{1}{n}}\, (\id_{12} + (q - q^{\epsilon_{12}} )\, P_{12} )
\lb{R+compactly}
\ee
(cf. (\ref{Mg}) and (\ref{Rr-compactly})), we obtain
\ba
&&d_\b\, a^i_\a d_\b^{-1} = q^{\frac{1}{n}-\d_{\a\b}}\, a^i_\a\ ,\quad\quad\quad k_a\,a^i_\a\, k_a^{-1}= q^{\theta_{a \a}- \frac{a}{n}}\, a^i_\a\nn\\
&&\mbox{for}\quad \theta_{a\a} = \left\{\begin{array}{ll}\, 1\,, \ &a\ge \a\\\, 0\,, \ &a < \a\\ \end{array}\right.\ ,\quad\,
K_a\,a^i_\a\, K_a^{-1}= q^{\d_{a\,\a}-\d_{a+1\,\a}} a^i_\a\ ,\nn\\
&&[E_a,\,a^i_\a ]\,=\,\d_{a+1\,\a}\,a^i_{\a-1}\,K_a\ ,\qquad
[K_a F_a ,\,a^i_\a ] \,=\,\d_{a\,\a}\, K_a \, a^i_{\a+1}\nn\\
&&(\mbox{or, equivalently,}\quad F_a \, a^i_\a \,=\, q^{\d_{a+1\, \a}- \d_{a \a}}\, a^i_\a\, F_a \,+\, \d_{a\, \a}\, a^i_{\a +1}\,)\ ,\nn\\
&&a=1,\ldots ,n-1\ ,\quad \a,\b = 1,\dots ,n
\lb{AdXa}
\ea
(note that $\theta_{ij} - \theta_{i-1\, j} = \d_{ij}$).
Remarkably, relations (\ref{AdXa}) imply that the rows of the zero modes matrix $a^i = (a^i_\a)_{\a=1}^n\,,\ i=1,\dots , n\,$
form {\em $\,U_q$-vector operators}\footnote{$U_q$-tensor operators have been introduced in \cite{RS92, S93}.}
for the $n$-fold cover $U_q^{(n)}(s\ell(n))\,$ of $U_q(s\ell(n))\,,$ i.e.
\be
Ad_X (a^i_\a ) = a^i_\s (X^f)^\s_\a\ \ ,\qquad{\rm where} \qquad Ad_X(z) := \sum_{(X)} X_1 z\, S(X_2)\ .
\lb{tens-op}
\ee
In (\ref{tens-op}) $X\mapsto X^f$ is the defining $n\times n$ matrix representation so that
\be
(K^f_a)^\s_\a = q^{\d_{a\,\a}-\d_{a+1\,\a}} \d^\s_\a\ ,\quad (E^f_a)^\s_\a = \d^\s_{\a -1} \d_{a\,\s}\ ,\quad
(F^f_a)^\s_\a = \d^\s_{\a +1}\d_{a\,\a}\quad
\lb{Xf}
\ee
($k_a^f$ and $d_\b^f$ are defined accordingly, see (\ref{AdXa})),
and $X_1$ and $X_2$ are the factors appearing in the $U_q$ coproduct written as $\Delta(X) = \sum_{(X)} X_1\otimes X_2\,,$
see (\ref{copr}) in Appendix B. Hence, albeit quite differently looking, relations
(\ref{aMpm}), (\ref{AdXa}) and (\ref{tens-op}) express the same property of the zero modes' matrix, namely its
covariance with respect to $U_q$. As the initial formulae (\ref{aMpm}) and (\ref{Mg}) for the transformation of
the zero modes' matrix $a$ and of the chiral field $g(x)$ are identical, the same applies to $g(x)$ as well.

\smallskip

One can show further that, as devised by Pusz and Woronowicz \cite{PW} back in the late 1980's,
the zero modes' exchange relations (\ref{ExRaa1}) transform covariantly
with respect to the quantum group action (\ref{aMpm}), in the following sense:
\be
M_{\pm 3}\, (R_{12}(p)\, a_1\, a_2 - a_2\, a_1 R_{12})\,M_{\pm 3}^{-1} =
(R_{12}(p)\, a_1\, a_2 - a_2\, a_1 R_{12})\, R_{13}^\mp\, R_{23}^{\mp} \ .
\lb{Mpm-aex}
\ee
To verify (\ref{Mpm-aex}), one uses the relation $[M_{\pm 3} , R_{12}(p)]\,$
(see (\ref{Mpmplq})), Eq.(\ref{aMpm-comp}) and the quantum YBE (\ref{QYBE}) in the form
\be
R_{12}\, R_{13}^\mp\, R_{23}^{\mp}\, =\, R_{23}^\mp\, R_{13}^{\mp}\, R_{12}\ .
\lb{QYBE1}
\ee
In the spirit of the discussion at the end of Section 4.3, (\ref{Mpm-aex}) has to be considered as dual to
the obvious invariance of the exchange relations (\ref{ExRaa1}) with respect to the action $a \to a \, T\,$
where $T$ obey the RTT relations (\ref{RTT}).

All this applies to the exchange relations (\ref{Mg}) for $g(x)\,$ as well.

\subsection{The WZNW chiral state space}

Our next task will be to construct the state space of the quantized WZNW model as
a vacuum representation of the quantum exchange relations.

\smallskip

We shall assume that the quantized chiral field $g(z)$ splits as in (\ref{gua}),
\be
g^A_\a(z) = u^A_j(z) \otimes a^j_\a
\lb{guaq}
\ee
where the field $u(z) = (u^A_i(z))$ has diagonal monodromy,
\be
e^{2\pi i L_0} u^A_j(z)\, e^{- 2\pi i L_0} = e^{2\pi i \Delta} \, u^A_j(e^{2\pi i}\,z) = (M_{\mathfrak p})^i_j\,u^A_i(z)
\lb{uuMpq}
\ee
and further, that the zero modes "inherit" the diagonal monodromy matrix $M_{\mathfrak p}$ of $u(z)$ in (\ref{uuMpq}), in the sense that
\be
(M_{\mathfrak p})^i_j\,u^A_i(z) \otimes a^j_\a = u^A_i(z) \otimes (M_p)^i_j\, a^j_\a = u^A_i(z) \otimes a^i_\s M^\s_\a
\lb{inhMp}
\ee
(cf. (\ref{gzM}) and (\ref{aMMpa})). To ensure that (\ref{inhMp}) takes place, we shall require that
$( {\hat {\mathfrak p}}_i - {\hat p}_i )\,  {\cal H} =\, 0\,$
as a constraint characterizing the WZNW chiral state space (cf. Remark 3.1; we shall put temporarily hats on the
operators to distinguish them from their eigenvalues).
Clearly, this will take place if the chiral field (\ref{guaq}) acts on
\be
\lb{space}
{\cal H}\, =\, \bigoplus_p\,{\cal H}_p\otimes {\cal F}_p
\ee
where both ${\cal H}_p$ and ${\cal F}_p\,$ are eigenspaces corresponding to the same eigenvalues of the
collections of commuting operators ${\hat{\mathfrak p}} \ = ({\hat {\mathfrak p}}_1 ,\dots , {\hat{\mathfrak p}}_n)\,$
and ${\hat p} \ = ({\hat p}_1 ,\dots , {\hat p}_n)\,,$ respectively, so that
\be
( {\hat{\mathfrak p}}_i\otimes \id - \id \otimes {\hat p}_i )\, {\cal H}_p \otimes {\cal F}_p = 0\ ,\quad i=1,\dots , n\ .
\lb{1pp1}
\ee
Assuming that ${\cal H}$ is generated from the vacuum vector by polynomials in $g(z)$ (and its derivatives)
automatically provides this structure.

The quantum counterparts of the PB (\ref{jTpl}) and (\ref{curf1}),
\be
[ j^a_m , p_\ell\, ] = 0 = [ L_n , p_\ell \, ] \ ,\qquad [ j^a_m , u^A_i(z) ] = -\, z^m \, (t^a)^A_B\, u_i^B(z)
\lb{curfq}
\ee
show that ${\cal H}_p$ are representation spaces of both the current algebra $\widehat{su}(n)_k$ (\ref{KM}) and the Virasoro algebra (\ref{Vir}), while
$u(z)$ is an {\em affine primary field}. On the other hand, the quantum analog of (\ref{DPBdiffer2}), written as
\be
p_\ell \, u^A_i (z)  \, =\, u^A_i (z)\, ( p_\ell + v^{(i)}_\ell )\ ,\qquad v^{(i)}_\ell = \d^i_\ell - \frac{1}{n}
\lb{gCVO}
\ee
implies that the operators $u_i(z) = (u_i^A(z))$ intertwine ${\cal H}_p$ and ${\cal H}_{p+v^{(i)}}\,$ i.e.,
are generalized {\em chiral vertex operators} (CVO) \cite{TK, DFMS}.

Likewise, the PB (\ref{PBapD}) is quantized to
\be
p_\ell \, a^i_\a  \, =\, a^i_\a\, ( p_\ell + v^{(i)}_\ell ) \equiv a^i_\a\, ( p_\ell + \d^i_\ell -\frac{1}{n})
\quad\Rightarrow\quad [ p_{j\ell} \,, a^i_\a ] = (\d_j^i - \d_\ell^i)\, a^i_\a\quad
\lb{pacomm}
\ee
which implies the first equation (\ref{ExRap}).
According to (\ref{Mpmplq}), every ${\cal F}_p$ is invariant with respect to the action of
(the $n$-fold cover $U_q$ of) $U_q(s\ell(n))\,,$ the rows $a^i = (a^i_\a)$ of the zero modes' matrix acting as
"$q$-vertex operators" (cf. (\ref{ExRap})). The reducibility properties of the corresponding representations
will be studied in detail in what follows.

\smallskip

Having in mind (\ref{uuMpq}) and (\ref{inhMp}), one should expect that
\be
\lb{detaM}
\det (M_p\, a) = \det (a) = \det (a M)
\ee
for appropriately defined $\det (M_p\, a)$ and $\det (a M)\,.$ The first relation (\ref{detaM}) suggests that the quantum
diagonal monodromy matrix $M_p$ also gets a "quantum correction" to its classical expression (\ref{uuMp})
(as the general monodromy $M$ does, cf. (\ref{M+-q})):
\be
\lb{Mpq}
(M_p)_j^i = q^{ - 2 p_i + 1-\frac{1}{n}} \, \delta_j^i\ .
\ee
Indeed, the non-commutativity of $q^{{p}_j}\,$ and $a^i\,,$ see (\ref{ExRap}),
exactly compensates the additional factors $q^{1-\frac{1}{n}}\,$ when computing
\be
\det (M_p\, a) :=
\frac{1}{[n]!}\, \epsilon_{i_1 \ldots i_n}\, (M_p\, a)_{\a_1}^{i_1} \ldots (M_p\, a)_{\a_n}^{i_n}\, \e^{\a_1 \dots \a_n}\ .
\ee
To prove this, assume that $i_\mu \ne i_\nu$ for $\mu\ne\nu$
(so that, in particular, $\prod_{\mu=1}^n\, q^{-2p_{i_\mu}} = \prod_{i=1}^n\, q^{-2p_i} = \id\,$); we then have
\be
q^{-2{p}_{i_1} +1-{1\over n}}\, a^{i_1}_{\a_1}\,
q^{-2{p}_{i_2} +1-{1\over n}}\, a^{i_2}_{\a_2}\,\ldots
q^{-2{p}_{i_n} +1-{1\over n} }\, a^{i_n}_{\a_n} \, =\,
a^{i_1}_{\a_1} a^{i_2}_{\a_2}\ldots a^{i_n}_{\a_n}
\lb{qsum}
\ee
since, moving all $q^{-2{p}_{i_\mu} +1-{1\over n}}$ terms either to the leftmost or to the rightmost position,
we get trivial overall numerical factors:
\be
\lb{qsum1}
q^{n(1-{1\over n}) - {2\over n} (1+2+\dots + n-1)}\, =\, 1\ \,= q^{n(1-{1\over n}) - 2n + {2\over n} (1+2+\dots + n)}\ .
\ee
Hence, defining simply
\be
\det (M_p) := \prod_{i=1}^n q^{-2p_i} \ (\, = 1\, ) \ ,
\lb{detMp}
\ee
we also obtain
\be
\det (M_p\, a) = \det (M_p)\, \det (a) = \det (a)\, \det (M_p)\ .
\lb{DaDMp}
\ee
Understanding the second relation (\ref{detaM}) turns out to be more intriguing \cite{FH2};
it is relegated to Appendix C where we also justify the appropriate definition of $\det (M)\,.$

In accord with (\ref{Mpmplq}), it follows from (\ref{aMMpa}) that the elements of $M$ commute with
$q^{p_i}$ and hence, with $M_p\,$ (\ref{Mpq}).

\smallskip
Eq.(\ref{gCVO}) implies that the exchange relations between $q^{p_j}$ and $u^A_i(z)$ are
identical to those for the zero modes (\ref{ExRap}):
\be
q^{p_j}\, u^A_i(z) = u^A_i(z)\, q^{p_j + \d^i_j - \frac{1}{n}} \qquad\Rightarrow\qquad
q^{p_{j\ell}} u^A_i(z) = u^A_i(z) \, q^{p_{j\ell} + \delta_j^i - \delta_{\ell}^i}\ .
\label{ExRup}
\ee
(Together with (\ref{1pp1}), this is the reason why $M_p$ should multiply $u(z)$ from the {\em left} in (\ref{uuMpq}).)
As expected, in the quantum theory the spectrum of the commuting operators $p_i\,,\ i=1,\dots ,n\,$ acting on ${\cal H}$ (\ref{space})
will be {\em discrete}\,; to determine it we only need, in addition to (\ref{ExRup}), the corresponding eigenvalues on the vacuum.
Combining (\ref{ExRup}) with (\ref{uuMpq}) and (\ref{Mpq}), we obtain
\be
q^{\frac{1}{n}-n}\, u^A_i(0)\, {\mid 0 \rangle} = u^A_i(0)\, q^{-2p_i - 1 + \frac{1}{n}}\, {\mid 0 \rangle}\quad\Leftrightarrow\quad
u^A_i(0)\, q^{-2p_i}\, {\mid 0 \rangle} = q^{1-n}\, u^A_i(0)\, {\mid 0 \rangle}\ .
\lb{uqp-vac}
\ee
Eq.(\ref{uqp-vac}) admits the following interpretation. The vacuum eigenvalues $p^{(0)}_i$ on ${\mid 0 \rangle}$ are equal to
the barycentric coordinates of the Weyl vector $\rho$ (\ref{llab}),
\be
p_i\, \vac = p^{(0)}_i\,\vac\ ,\quad p^{(0)}_i = \ell_i(\rho) = \frac{n+1}{2} - i\ ,\quad i=1,\dots, n
\lb{vac-Weyl}
\ee
(so that, in particular, $q^{-2p^{(0)}_1} = q^{1-n}$),  and
\be
u^A_i (z) \,{\mid 0 \rangle} = 0\quad{\rm for}\quad i\ge 2\ .
\lb{u2.n}
\ee
A similar condition appears for the zero modes due to (\ref{aMMpa}) and (\ref{M0}):
\be
(M_p)^i_j\, a^j_\a \,{\mid 0 \rangle} = a^i_\s\, M^\s_\a \,{\mid 0 \rangle} \qquad\Leftrightarrow\qquad
a^i_\a \, q^{-2p_i}\, {\mid 0 \rangle} = q^{1-n}\, a^i_\a\, {\mid 0 \rangle}\ .
\lb{ap-vac}
\ee
Hence, the assumption that (\ref{vac-Weyl}) holds leads us to the counterpart of (\ref{u2.n}) for the zero modes:
\be
(q^{p_i} - q^{\frac{n+1}{2} - i}) \,{\mid 0 \rangle} = 0\ ,\quad i=1,\dots, n \quad\Rightarrow\quad
a^i_\a \,{\mid 0 \rangle} = 0\quad{\rm for}\quad i\ge 2\ .
\lb{a2.n}
\ee

\smallskip

As the exchange relations (\ref{ExRup}) (or (\ref{ExRap})) imply
\be
u^A_i(z)\,:\ {\cal H}_p \ \rightarrow\ {\cal H}_{p+v^{(i)}}\ ,\qquad a^i_\a\,:\ {\cal F}_p \ \rightarrow\ {\cal F}_{p+v^{(i)}}\ ,
\lb{cqvo}
\ee
they completely determine, together with (\ref{vac-Weyl}), the spectrum of $p$ on the chiral state space (\ref{space})
under the assumption that ${\cal H}$ is generated from the vacuum by polynomials in $g(z)$ (\ref{guaq}).
(The uniqueness of the vacuum requires the spaces ${\cal H}_{p^{(0)}}$ and ${\cal F}_{p^{(0)}}$ to be one dimensional, so that
${\cal H}_{p^{(0)}}\otimes \, {\cal F}_{p^{(0)}} = {\mathbb C}\vac\,.$)
The first thing to say about the spectrum is that it is certainly a subset of the lattice of {\em shifted} integral $s\ell(n)$ weights
\be
p = \L + \rho\quad \Leftrightarrow\quad p_{i\,i+1} = \l_i + 1\qquad{\rm for}\quad \L = \sum_{i=1}^{n-1} \l_i\, \L^i\ ,\quad \l_i \in {\mathbb Z}\ ,
\lb{sp-p-r}
\ee
see (\ref{lambda-ell}) and (\ref{Wv}) (it follows from (\ref{sp-p-r}) that all $p_{ij}$ have integer eigenvalues).
The shifted weight interpretation is also supported by the observation that,
according to (\ref{Dqa=Dqp}), the quantum determinant $\det (a) = {\cal D}_q(p)\,$ of the zero modes' matrix is strictly positive
for $q^h = -1$ for integer values of $p_{i \,i+1}$ satisfying $p_{i \,i+1} \ge 1\ ,\ p_{1n} \le h-1\,.$ By (\ref{sp-p-r}), these coincide with
the shifted {\em dominant} weights lying in the level $k$ positive Weyl alcove, with Dynkin labels characterized by
$\l_i \ge 0\ ,\ \sum_{i=1}^{n-1} \l_i \le k\,,$ a fact that might be anticipated by the classical correspondence, see (\ref{AWn}).

It is natural to start the study of the WZNW space of states with the representation of the chiral zero modes' algebra ${\cal M}_q\,.$
Being $z$-independent, it is a quantum system with a {\em finite} number of degrees of freedom and state space
\be
{\cal F} = {\cal F} ({\cal M}_q) := {\cal M}_q\vac \ .
\lb{F}
\ee
The dynamical $R$-matrix (\ref{RRp2}) is singular for $[p_{ij}] = 0\,,$ so that the exchange relations (\ref{ExRaa1}) are ill defined on ${\cal F}\,$
for $q$ given by (\ref{height-h}) ($q^h=-1$), as $[nh]=0$ for any integer $n\,.$ This problem has however a simple solution; indeed,
getting rid of the denominators in (\ref{RRp2}) (for $\alpha_{ij}(p_{ij}) = 1$) and using the identity $[p-1]-q^{\pm 1} [p] = -\, q^{\pm p}\,,$ we obtain the set of relations
\ba
&&a^j_\b a^i_\a\, [p_{ij}-1] = a^i_\a a^j_\b\, [p_{ij}] -\,a^i_\b\, a^j_\a \, q^{{\epsilon}_{\a\b}p_{ij}} \quad
(\,{\rm for}\quad i\ne j \quad {\rm and}\quad\alpha\ne\beta )\ ,\nn\\
&&[a^j_\alpha , a^i_\alpha ] = 0\ ,\qquad a^i_\alpha a^i_\beta = q^{{\epsilon}_{\alpha\beta}}\, a^i_\beta a^i_\alpha\ ,\qquad
\a , \b , i, j = 1, \dots, n\ ,
\lb{aa2}
\ea
with ${\epsilon}_{\alpha\beta}$ as defined in (\ref{stand-r-matr}).
We shall replace from now on the relations (\ref{ExRaa1}) by their "regular form" (\ref{aa2}).
Thus the algebra ${\cal M}_q$ is defined by (\ref{aa2}), (\ref{ExRap}), (\ref{prod-p=1})
and the determinant condition (\ref{Dqa=Dqp}). We assume that ${\cal M}_q$ contains {\em polynomials}
in $a^i_\a$ and {\em rational functions} of $q^{p_j}\,.$

\smallskip

To avoid confusion between the operators and their eigenvalues we shall put, when needed, hats on the {\em operators}
$\hp_{ij}\,.$Note that, evaluated on a given ${\cal F}_p\,,$ the operators $p_{ij}\,$ in the first relation (\ref{aa2})
can be replaced by their (integer) eigenvalues so that the coefficients of the three (bilinear in $a^i_\a$) terms
become just ordinary ($q$-) numbers:
\be
(\hp_{ij} - p_{ij})\, {\cal F}_p = 0\qquad\Rightarrow\qquad (q^{\hp_{ij}} - q^{p_{ij}})\, {\cal F}_p = 0\ .
\lb{Fpdef}
\ee

\subsubsection{Fock representation of ${\cal M}_q$ for generic $q$}

We shall call the vacuum representation (\ref{F}) of the algebra ${\cal M}_q$ determined by
(\ref{a2.n}) and (\ref{vac-Weyl}) "Fock representation".
Due to (\ref{Uqvac}) (with the counit defined in (\ref{coun}), (\ref{dk})) and (\ref{AdXa}), it is clear that ${\cal F}$ is an $U_q$-invariant space.
The two questions of prime importance for us will be its $U_q$-module structure and the construction of convenient bases.
We shall first explore both of them in the case of generic $q$ for which we have a satisfactory theory and consider the root of
unity case (\ref{height-h}) only at the end.

The following result (also valid for $q=1$) was first established, for general $n$, in \cite{FHIOPT} (for $n=2$, cf. \cite{BF}).

\medskip

\noindent
{\bf Proposition 4.2~} {\em For generic $q\,$ the Fock space ${\cal F}\,$ (\ref{F}) is a direct sum of irreducible $U_q(s\ell(n))$ modules ${\cal F}_p\,:$
\be
{\cal F} = \bigoplus_p\, {\cal F}_p\qquad\quad (\, {\cal F}_{p^{(0)}} = {\mathbb C}\vac )\ .
\lb{Fock-n}
\ee
Here $p$ runs over all shifted dominant weights of $s\ell (n)\,$ and each ${\cal F}_p$ enters into the direct sum with multiplicity one.
In other words, ${\cal F}$ provides a model \cite{BGG} for the finite dimensional representations of $U_q(s\ell(n))\,.$}

\medskip

To prove this statement, we shall introduce bases of vectors in ${\cal F}_p$ labeled by {\em semistandard Young tableaux},
see e.g. \cite{FulH} and \cite{FM}. The key point is to realize that Eqs. (\ref{cqvo}) and (\ref{sp-p-r}) imply that, in
the Young tableaux language, the multiplication by $a^i_\a$ is equivalent to adding a box (labeled by $\a$) to the $i$-th row; in particular,
\be
a^i_\a \,:\ Y_{\l_1 ,\dots ,\l_{n-1} }  \ \rightarrow\ \ Y_{\l_1 ,\dots , \l_{i-1} -1,\, \l_i +1 ,\dots ,\, \l_{n-1} } \ ,\quad i = 1,\dots , n
\lb{aY}
\ee
where $Y_{\l_1 ,\dots ,\l_{n-1} }$ is the {\em Young diagram} corresponding to ${\cal F}_p\,$
(here $Y_{0 ,\dots , 0 }$ is identified with ${\cal F}_{p^{(0)}}\,,$ the one dimensional vacuum subspace).
Thus, the entries of the zero modes' matrix appear as natural variables for a {\em non-commutative} polynomial
realization of the finite dimensional representations of $U_q(s\ell(n))$.\footnote{Note that this realization has a
non-trivial $q=1$ counterpart. The proof given below goes essentially without any modification in the undeformed case
as well since, for generic $q\,,$ $[n]$ vanishes only for $n=0\,.$}

The correspondence between the labels of ${\cal F}_p\,$ and $Y_{\l_1 ,\dots ,\l_{n-1} }\,$ is made explicit by the following

\medskip

\noindent
{\bf Theorem 4.1~} (cf. Lemma 3.1 of \cite{FHIOPT}) {\em For generic $q$ the space ${\cal F}$
(\ref{F}) is spanned by "antinormal ordered" polynomials applied to the vacuum vector
\ba
&&P_{m_{n-1}} (a^{n-1})\, \dots\, P_{m_2} (a^2)\, P_{m_1} (a^1) \vac\nn\\
&&{\rm with} \qquad m_1 \ge m_2 \ge\dots \ge m_{n-1}
\lb{PolF}
\ea
where each $P_{m_i}(a^i)$ is a homogeneous polynomial of degree $m_i\,$ in $\, a^i_1 ,\dots , a^i_n\,$
or, alternatively, by vectors of the type
\ba
&&P_{\l_1} ({\Delta}^{(1)})\,P_{\l_2} ({\Delta}^{(2)}) \dots \, P_{\l_{n-1}} ({\Delta}^{(n-1)}) \vac\nn\\
&&{\rm with} \qquad \l_i = m_i - m_{i+1}\ge 0\quad (m_n\equiv 0)
\lb{PolF-alt}
\ea
where $\Delta^{(i)}_{\a_i \dots \a_1} := a^i_{\a_i} \dots a^1_{\a_1}\,,\ i=1,\dots, n-1$ are "strings" of
antinormal ordered operators of length $i\,.$}

\medskip

One can check that a vector of the type (\ref{PolF}) belongs to the space ${\cal F}_p\,$ which
is a common eigenspace of the commuting operators $\hp = (\hp_1 , \dots , \hp_n)$ with eigenvalues satisfying
$p_{i i+1} = \l_i +1\,.$ If the total number of zero mode operators acting on the vacuum is $N\,,$ then the
inequalities in (\ref{PolF}) and (\ref{PolF-alt}) correspond to the partition $N = \sum_{i=1}^{n-1} m_i = \sum_{j=1}^{n-1} j\,\l_j\,$
or, in other words, to the Young diagram $Y_{\l_1 ,\dots ,\,\l_{n-1} }\,;$ in (\ref{PolF}) the diagram
is built row by row while (\ref{PolF-alt}) corresponds to a construction column by column.

\medskip

\noindent
{\bf Proof of Theorem 4.1~} We shall start by assuming that $n\ge 3\,;$ the case $n=2$ is special
(and simpler) and will be considered separately at the end. The proof is based on the following three Lemmas.

\smallskip

\noindent
{\bf Lemma 4.1~} {\em If $P(a^i ,\dots , a^1)$ is a (unordered) polynomial in $a^\ell_\a$ for $1 \le \ell \le i\,$
(and arbitrary $1\le \a \le n$), then}
\be
a^j_\b\, P (a^i , \dots, a^1) \vac = 0 \qquad {\rm for}\qquad  3\le i+2\le j \le n\ .
\lb{L1}
\ee

\noindent
{\bf Lemma 4.2~} {\em The "string vectors" of length $i\ge 2$}
\be
v^{(i)}_{\a_i \dots \a_1} := a^i_{\a_i} a^{i-1}_{\a_{i-1}} \dots a^1_{\a_1} \vac \ ,\qquad 2\le i \le n
\lb{string-v}
\ee
{\em  are $q$-antisymmetric, i.e.}
\be
v^{(i)}_{\a_i \dots \a_{\ell +1} \a_\ell \dots \a_1} =
-\, q^{\epsilon_{\a_\ell \a_{\ell +1}}}\, v^{(i)}_{\a_i \dots \a_\ell \a_{\ell +1} \dots \a_1}\ .
\lb{vi-q-anti}
\ee
{\em String vectors of length $n$ are proportional to the vacuum vector} $\vac\,.$

\smallskip

\noindent
{\bf Lemma 4.3~} {\em The product of two operators of type $a^{i+1}$ annihilates a string vector of length $i\,$
for an arbitrary combination of their lower indices:}
\be
a^{i+1}_\a \, a^{i+1}_\b\, v^{(i)}_{\g_i \dots \g_1} = 0 \qquad {\rm for}\qquad 1\le i \le n-1\ .
\lb{L3}
\ee

\medskip

\noindent
{\bf Proof of Lemma 4.1~} To show that Eq.(\ref{L1}) takes place, we first note that
\be
\hp_{\ell j} \vac = p^{(0)}_{\ell j} \vac = (j - \ell) \vac\ ,\qquad 1\le \ell , j \le n
\lb{pjl-on-vac}
\ee
(see (\ref{vac-Weyl})) and hence, by (\ref{ExRap}),
\ba
&&[\hp_{\ell j} - 1]\, P_{m_n \dots \, m_1} (a^n, a^{n-1}, \dots, a^1) \vac =\nn\\
&&= [m_\ell - m_j + j - \ell - 1 ]\, P_{m_n\dots \, m_1} (a^n, a^{n-1}, \dots, a^1) \vac\qquad
\lb{evs-plj}
\ea
for any {\em homogeneous} polynomial of order $m_r\, (\ge 0)\,$ in $a^r\,, \ 1\le r \le n\,.$
Eq.(\ref{L1}) follows from the consecutive application of the equality
\ba
&&a^j_\b\, a^\ell_\a \, P_{m_i \dots  m_1} (a^i ,\dots, a^1) \vac =
%nn\\ &&= \frac{1}{[p_{\ell j} - 1 ]}\,a^j_\b\, a^\ell_\a \,[\hp_{\ell j} - 1]\, P_{m_i \dots  m_1} (a^i , \dots, a^1) \vac =\qquad
\lb{L1a}\\
&&= \frac{1}{[p_{\ell j} - 1 ]}\,
( a^\ell_\a \,a^j_\b\, [p_{\ell j}]\,-\,a^\ell_\b \,a^j_\a\, q^{{\epsilon}_{\a\b}p_{\ell j}})\,
P_{m_i  \dots  m_1} (a^i ,\dots, a^1) \vac
\nn
\ea
for $\a\ne \b\,,$ with
\be
p_{\ell j} = m_\ell + j - \ell \ge 2\qquad {\rm for} \qquad 1 \le l \le i\ ,\quad i +2 \le j \le n
\lb{plj}
\ee
(it is essential that $p_{\ell j} - 1 \ne 0$); for $\a = \b$ the operators
$a^j_\a$ and $a^\ell_\a$ simply commute, see (\ref{aa2}). As $j\ge 3\,,$ moving the operators $a^j$
to the right until they reach the vacuum and using (\ref{a2.n}), we prove that expressions of the
type (\ref{L1a}) (and hence, (\ref{L1})) vanish.

\medskip

\noindent
{\bf Proof of Lemma 4.2~} It is clear in the first place that a string vector vanishes if any two neighbouring indices $\a_{\ell +1}$ and
$\a_\ell\,,$ for $\ell = 1,\dots, i-1\,,$ coincide (if this is the case, we can exchange the corresponding operators $a^{\ell +1}_{\a_{\ell +1}}$
and $a^\ell_{\a_\ell}\,$ and then apply Lemma 4.1). If $\a_{\ell +1} \ne \a_\ell\,,$ we can use the first exchange relation (\ref{aa2}) in the form
\be
a^{\ell +1}_{\a_{\ell +1}}\, a^\ell_{\a_\ell}\, [\hp_{\ell \ell+1}] =
a^\ell_{\a_\ell}\, a^{\ell +1}_{\a_{\ell +1}}\, [\hp_{\ell \ell+1} + 1]
- a^{\ell +1}_{\a_\ell}\, a^\ell_{\a_{\ell +1}}\,q^{\epsilon_{\a_\ell \a_{\ell +1}} \hp_{\ell \ell + 1}}
\lb{aaP}
\ee
and, as the first term in the right hand side vanishes when evaluated on $v^{(\ell -1)}_{\a_{\ell -1} \dots \a_1}\,$
$\,( v^{(0)} \equiv\, \vac )\,$ while the eigenvalue $p_{\ell \ell+1} =1\,,$ deduce relation (\ref{vi-q-anti}).
For $i=n\,$ it complies with the properties of the $\varepsilon$-tensor (\ref{ee}) since
\ba
&&v^{(n)}_{\a_n \dots \a_1} \equiv \epsilon_{i_n \dots i_1} a^{i_n}_{\a_n} \dots a^{i_1}_{\a_1} \vac =
\e_{\a_n \dots \a_1}\,D_q(a) \vac = \e_{\a_n \dots \a_1}\,{\cal D}_q(p^{(0)}) \vac \ , \nn\\
&&{\cal D}_q(p^{(0)}) = \prod_{1\le \ell <j\le n} [j-\ell\,] = \prod_{\ell =1}^{n-1} [\ell ]!
\lb{vn-q-anti}
\ea
(the first equality (\ref{vn-q-anti}) follows from Lemma 4.1; we then use (\ref{det-intertw}),
(\ref{Dqa=Dqp}) and (\ref{pjl-on-vac})).

\medskip

\noindent
{\bf Proof of Lemma 4.3~} Eq.(\ref{L3}) is a simple consequence of the $q$-symmetry of the product
$a^{i+1}_\a a^{i+1}_\b$ and the $q$-antisymmetry of the string vectors (Lemma 4.2).
Denote a vector of the type (\ref{L3}) by
\be
w_{\a\b\g} \equiv w_{\a \b \g {\{ \s \}}} := a^{i+1}_\a \, a^{i+1}_\b \,v^{(i)}_{\g\, \s_{i-1}\dots\, \s_1}
= a^{i+1}_\a \, v^{(i+1)}_{\b\, \g\, \s_{i-1}\dots\, \s_1} \ , \quad 1\le i\le  n-1
\lb{wabg}
\ee
(the indices $\s_{i-1} , \dots , \s_1$ are irrelevant for the argument that follows).
The point is that the ensuing symmetry of the tensor $w_{\a\b\g}$ is contradictory,
i.e. incompatible with its non-triviality. Indeed, exchanging the indices arranged as $\g , \b , \a$
back to $\a , \b , \g$  in the two possible ways and using the last equality (\ref{aa2}) and (\ref{vi-q-anti})
we obtain, respectively
\ba
&&w_{\g\b\a} = q^{\epsilon_{\g\b}}\, w_{\b\g\a} = -\, q^{\epsilon_{\g\b} + \epsilon_{\a\g}}\, w_{\b\a\g} =
-\,q^{\epsilon_{\g\b} + \epsilon_{\a\g} + \epsilon_{\b\a}}\, w_{\a\b\g}\ \ \quad{\rm or}\nn\\
&&w_{\g\b\a} = -\, q^{\epsilon_{\a\b}}\, w_{\g\a\b} = - \, q^{\epsilon_{\a\b} + \epsilon_{\g\a}}\, w_{\a\g\b} =
q^{\epsilon_{\a\b} + \epsilon_{\g\a} + \epsilon_{\b\g}}\, w_{\a\b\g}\ ,\quad{\rm i.e.}\nn\\
&&w_{\a\b\g} = -\,q^{2(\epsilon_{\a\b} + \epsilon_{\b\g} + \epsilon_{\g\a})}\,w_{\a\b\g}\qquad
\Rightarrow\qquad w_{\a\b\g} = 0\ .
\lb{wgba}
\ea

\medskip

Returning to the {\em proof of Theorem 4.1}, we shall first show that a weaker form of (\ref{PolF}) takes place, namely
all vectors in ${\cal F}$ are linear combinations of vectors
\be
P_{m_n} (a^n)\,P_{m_{n-1}} (a^{n-1})\, \dots\, P_{m_2} (a^2)\, P_{m_1} (a^1) \vac\ ,\quad m_i \ge m_j\quad{\rm for} \quad i<j\ .
\lb{PolFn}
\ee
By making use of Lemmas 4.1 and 4.3, one can easily exhaust the list of vectors created from the vacuum by a small number
(say, $N\le 3$) operators $a^i_\a$:
\ba
&&N=1:\quad a^1_\a \vac \ ;\nn\\
&&N=2:\quad a^1_\a\, a^1_\b\vac\ ,\quad a^2_\a\, a^1_\b \vac = v^{(2)}_{\a\b}\ ;\nn\\
&&N=3:\quad a^1_\a\, a^1_\b\, a^1_\g \vac\ ,\quad a^2_\a\, a^1_\b\, a^1_\g \vac\ ,\quad a^3_\a\, a^2_\b\, a^1_\g \vac = v^{(3)}_{\a\b\g}\nn\\
&&{\phantom{N=3:\quad}} (\, a^2_\a\, a^1_\b\, a^1_\g \vac = [2]\, a^1_\b\, v^{(2)}_{\a\g} - q^{2\epsilon_{\b\a}}\, a^1_\a \, v^{(2)}_{\b\g}\,)\ ;\nn\\
&&\dots\qquad
\lb{F123}
\ea
Due to the $q$-(anti)symmetry in the lower indices, not all combinations (\ref{F123}) are linearly independent.
Obviously, all vectors in the list (\ref{F123}) are of the form (\ref{PolFn}).
We shall assume that the arrangement (\ref{PolFn}) can be made for any number of zero modes' operators
not larger than certain $N\,$ and then perform the induction in $N$.
To this end we shall prove that the action of $a^j_\b\,$ on a vector
\be
P_{m_i}(a^i) \dots \dots P_{m_1} (a^1) \vac \qquad{\rm for}\quad N=m_1 + \dots + m_i\ ,\quad 1\le i\le n
\lb{PolN}
\ee
either produces again vectors of the form (\ref{PolFn}), or gives zero. The former is certainly correct for $j=i+1\,$
and the latter for $n\ge j\ge i+2\,,$ by Lemma 4.1. So it is necessary to show that an operator of type
$a^j_\b\,,\ 1\le j\le n-1$ acting on (\ref{PolN}) can be moved to the right through $P_{m_i}(a^i)$ for any $j< i\le n$
and $m_i > 0$. This amounts to proving that the corresponding eigenvalue of $[{\hp}_{ij} - 1]\,,\ i>j$
is different from zero; to this end we could write
\ba
&&a^j_\b\, P_{m_i}(a^i)\dots P_{m_j} (a^j) \dots P_{m_1} (a^1) \vac = \nn\\
&&= \frac{1}{[p_{ij}-1]}\, a^j_\b\, a^i_\a\, [{\hp}_{ij}-1]\, P_{m_i-1}(a^i)\dots
P_{m_j} (a^j) \dots P_{m_1} (a^1) \vac\qquad\quad
\lb{aiPmj}
\ea
and apply the first relation (\ref{aa2}) if $\a \ne \b\,,$ or just use the
second relation (\ref{aa2}) if $\a=\b\,.$ By the general formula (\ref{evs-plj})
\be
p_{ij} = m_i - 1 - m_j + j - i \qquad (\, \le - 2\ \ {\rm for}\ \ m_i \le m_j\ \ {\rm and}\ \ j < i\,)\ ,
\lb{evs-pij}
\ee
hence the quantum brackets in the right-hand side of (\ref{aiPmj}) do not vanish.
As a result, the operator $a^j\,$ can always join its companions of the same type.
Our next step will be to show that this will not violate the inequalities
among $m_i$ in (\ref{PolFn}) i.e., if $m_j = m_{j-1}\,,$
\be
a^j_\a\, P_{m_{j-1}} (a^j)\, P_{m_{j-1}} (a^{j-1}) \dots P_{m_1} (a^1) \vac = 0\ ,\qquad 2\le j \le n\ .
\lb{mi=mi-1}
\ee
Eq.(\ref{mi=mi-1}) can be proved by pulling consecutively the rightmost operators of type $a^2 , a^3, \dots , a^j$
until they form a string of length $j$ with the rightmost "free" $a^1\,.$ Using the property of strings
\be
[ \hp_{rs} , \Delta^{(j)} ] = 0 \qquad \mbox{for} \qquad 1\le r < s\le j \le n\ ,
\lb{prop-str}
\ee
we can proceed in the same way, eventually expressing (\ref{mi=mi-1}) as a linear combination of vectors of the kind
\be
P_{m_{j-2} - m_{j-1}} (a^{j-2}) \dots P_{m_1 - m_{j-1}}(a^1) \, a^j_\b\, P_{m_{j-1}} (\Delta^{(j)}) \vac\ ,
\qquad 2\le j \le n-1\qquad
\lb{last-i1}
\ee
(strings of length $n\,$ that would appear for $j = n\,$ are eliminated by (\ref{vn-q-anti})). To confirm (\ref{mi=mi-1})
-- and hence, (\ref{PolF}), it remains to prove the following generalization of Lemma 4.3:
\be
a^j_\b\, P_{m} (\Delta^{(j)}) \vac = 0\qquad{\rm for}\qquad 2\le j \le n-1\ ,\quad m\ge 0 \ .
\lb{genL3}
\ee
The proof of (\ref{genL3}) can be done by induction in $m\,.$ The case $m=0\,$ is covered by (\ref{a2.n}) and $m=1\,,$ by (\ref{L3}).
For $m\ge 2\,$ we shall use (\ref{aaP}) to extract a $q$-antisymmetric term from $P_{m} (\Delta^{(j)}) \vac\,$ which vanishes when
acted upon by $a^j_\b\,,$ due to an immediate generalization of (\ref{wabg}), (\ref{wgba}):
\ba
&&a^j_\b\, P_{m} (\Delta^{(j)}) \vac = a^j_\b\,a^j_{\a_j} a^{j-1}_{\a_{j-1}} \dots a^1_{\a_1} \, P_{m -1} (\Delta^{(j)}) \vac = \nn\\
&&= a^j_\b \left( \frac{1}{2}\, ( a^j_{\a_j} a^{j-1}_{\a_{j-1}} - a^j_{\a_{j-1}} a^{j-1}_{\a_j} q^{\epsilon_{\a_{j-1} \a_j}} ) +
\frac{[2]}{2}\, a^{j-1}_{\a_{j-1}} a^j_{\a_j} \right) \times\nn\\
&&\times \ a^{j-2}_{\a_{j-2}} \dots a^1_{\a_1} \, P_{m -1} (\Delta^{(j)}) \vac \ ,\qquad 2\le j \le n-1\ .
\lb{gen-string-v}
\ea
Further, the operator $a^j_{\a_j}\,$ from the remaining last term in the big parentheses of (\ref{gen-string-v}) can be moved to the right until
one gets a linear combination of terms of the type $P_{1} (\Delta^{(j)})\, a^j_\rho \, P_{m-1} (\Delta^{(j)}) \vac\,.$
Thus Eq.(\ref{genL3}) follows from the same assumption for $m-1\,.$

\smallskip

A similar procedure (grouping the operators in strings of decreasing length) leads to (\ref{PolF-alt}).
By the technique used in (\ref{gen-string-v}), based on Eq.(\ref{aaP}), one can prove that any of the
strings is $q$-antisymmetric on its lower indices; this generalizes Lemma 4.2.

\smallskip

To complete the proof of Theorem 4.1, we shall consider separately the special case $n=2$ when the determinant condition is also bilinear
as the exchange relations (\ref{aa2}). Denoting $p := p_{12}\,,$ we have (for $\a_{12}(p_{12}) = 1$ in (\ref{A1dyn}))
\be
{\cal D}_q(\hp) = [\hp\, ]\ ,\qquad
\epsilon^{12} (\hp) = - \,\frac{[\hp - 1]}{[\hp]}\ ,\qquad
\epsilon^{21} (\hp) = \frac{[\hp+1]}{[\hp]}
\lb{eps-p-n2}
\ee
(cf. (\ref{eps-Dqp})) so that, combining (\ref{det-intertw}) and (\ref{Dqa=Dqp}), we obtain
\ba
&&\epsilon_{ij}\, a^i_\a a^j_\b\ (\,\equiv a^2_\a a^1_\b - a^1_\a a^2_\b\,)\,
= [{\hat p}\,]\, \varepsilon_{\a\b}\ ,\quad\a,\b=1,2\nn\\
&&(\,\e_{12} = - q^{\frac{1}{2}} = \e^{12}\,,\ \e_{21} = q^{- \frac{1}{2}} = \e^{21}\,) \qquad
\Rightarrow\qquad a^1_\a a^2_\a = a^2_\a a^1_\a\ ,\nn\\
&&a^2_\a a^1_\b \, \varepsilon^{\a\b} = [\hp+1]\ ,\qquad a^1_\a\, a^2_\b \, \varepsilon^{\a\b} = - \,[\hp-1]\ ,
\lb{detc-n2-1}\\
&&a^i_\alpha a^i_\beta \, \varepsilon^{\alpha\beta} = 0\quad (\,{\rm i.e.,}\quad a_2^i\, a_1^i = q\, a_1^i a_2^i\, )\ ,\quad i=1,2\ .
\lb{detc-n2-2}
\ea
It is not difficult to see that Eqs.(\ref{detc-n2-1})
(which are {\em inhomogeneous} in $a^i_\a$) and (\ref{detc-n2-2})
imply the homogeneous exchange relations (\ref{aa2}) for $n=2\,.$ An important consequence of (\ref{detc-n2-1}) is that
the exchange of operators with different upper indices (in particular, their "antinormal ordering")
can be performed already at the algebraic level, which directly implies Theorem 4.1.\eod

\medskip

\noindent
{\bf Proof of Proposition 4.2~}

By Theorem 4.1, for generic $q\,$ any vector in ${\cal F}\,$ is a linear combination of vectors belonging to the spaces ${\cal F}_p\,$
where the (barycentric shifted weight) labels $p = (p_1,\dots , p_n)$ are related to the Dynkin labels of
Young diagrams $Y_{\l_1,\dots, \l_{n-1}}\,$ of $s\ell(n)\,$ type by $p_{i i+1} = \l_i + 1\,,\ i = 1,\dots , n-1\,.$

As the $U_q(s\ell(n))\,$ generators only change the lower indices of the zero mode operators,
it follows that each ${\cal F}_p$ is a $U_q(s\ell(n))\,$ invariant space. In particular, all vectors generated from the vacuum by
homogeneous polynomials are weight vectors (eigenvectors of all $K_i\,,\ i=1,\dots , n-1$),
the weights depending solely on the set of $N$ lower indices. Both (\ref{PolF}) and (\ref{PolF-alt}) have an obvious interpretation as
filling in the boxes of the Young diagram $Y_{\l_1,\dots, \l_{n-1}}$ with numbers from $1$ to $n$ corresponding to the arrangement
of the lower indices along its rows or columns, respectively. One infers from the last equation (\ref{aa2}) the $q$-symmetry of the row
fillings, and from the generalization of Lemma 4.2, the $q$-antisymmetry of the column ones.
On the other hand, the exchange operations (\ref{aa2}) we use to express a vector of the form (\ref{PolF}) as a linear combination
of vectors (\ref{PolF-alt}) (and vice versa) leave the set of lower indices invariant.
We thus have the same situation as in the $s\ell(n)$ case where, for enumerational purposes, one introduces bases of vectors labeled by
semistandard Young tableaux, with indices "weakly increasing" (i.e., non-decreasing) along rows and strictly increasing along columns.

Each ${\cal F}_p\,$ contains a unique, up to normalization, highest (resp., lowest) weight vectors (HWV and LWV)
\be
| HWV \rangle_p \equiv |\l_1  \dots \l_{n-1} \rangle\qquad {\rm and} \qquad | LWV \rangle_p \equiv |-\l_{n-1}\,\dots\,-\l_1 \rangle
\lb{HLWV1}
\ee
satisfying
\ba
&&(K_i -q^{\lambda_i})\, | \l_1\dots \l_{n-1} \rangle = 0 = (K_i - q^{-\lambda_{n-i}} )\, | -\l_{n-1}\,\dots\, -\l_1  \rangle\ ,\nn\\
&&E_i\, | \l_1\dots \l_{n-1} \rangle = 0 = F_i\, | -\l_{n-1}\,\dots\, -\l_1  \rangle\ ,\quad 1\le i\le n-1 \ .\quad
\lb{CartEF}
\ea
These are given by
\ba
&&| \lambda_1 \dots \lambda_{n-1} \rangle =
(\Delta^{(1)}_{11})^{\lambda_1} (\Delta^{(2)}_{2 1})^{\lambda_2}\dots
(\Delta^{(n-2)}_{n-2\, 1})^{\lambda_{n-2}} (\Delta^{(n-1)}_{n-1\, 1})^{\lambda_{n-1}} \vac \sim\nn\\
&&\sim (a^{n-1}_{n-1})^{m_{n-1}} (a^{n-2}_{n-2})^{m_{n-2}} \dots (a^2_2)^{m_2}\, (a^1_1)^{m_1} \vac \ ,\nn\\ \nn\\
&&| - \lambda_{n-1}\,\dots \,-\lambda_1 \rangle =
(\Delta^{(1)}_{nn})^{\lambda_1} (\Delta^{(2)}_{n n-1})^{\lambda_2}\dots
(\Delta^{(n-2)}_{n3})^{\lambda_{n-2}} (\Delta^{(n-1)}_{n2})^{\lambda_{n-1}}\vac \sim\nn\\
&&\sim (a^{n-1}_2)^{m_{n-1}} (a^{n-2}_3)^{m_{n-2}} \dots (a^2_{n-1})^{m_2}\, (a^1_n)^{m_1} \vac  \ ,\nn\\ \nn\\
&&\Delta^{(i)}_{\a + i-1\, \a} := a^i_{\a +i-1}\, a^{i-1}_{\a +i-2}\dots a^1_\a\ ,\nn\\
&&\l_i = m_i - m_{i+1} = p_{i i+1} - 1\ ,\quad i=1,\dots , n-1\ .
\lb{HLWV}
\ea
As for generic $q\,$ the $U_q(sl(n))$ (finite-dimensional) representation theory (including weight space decomposition and dimensions)
is essentially the same as that for $s\ell(n)$ \cite{CP}, we conclude that the spaces ${\cal F}_p$ for $p_{i i+1} = \l_i +1\,,\ \l_i \ge 0$
exhaust the list of $U_q(sl(n))$ IR. The dimension (\ref{Weyldim}) and the {\em quantum dimension} of ${\cal F}_p$ are given by
\ba
&&{\rm dim}\,{\cal F}_p = \prod_{1\le i < j\le n} \frac{p_{ij}}{p_{ij}^{(0)}} =
\frac{{\cal D}_1(p)}{{\cal D}_1(p^{(0)})} = \frac{1}{\prod_{\ell =1}^{n-1} \ell !}\,{\cal D}_1(p) =: d(p)\ ,\lb{qdimFp}\\
&&{\rm qdim}\,{\cal F}_p := {\rm Tr}_{{\cal F}_p} \prod_{i=1}^{n-1} K_i = \prod_{1\le i < j\le n} \frac{[p_{ij}]}{[p_{ij}^{(0)}]}
= \frac{{\cal D}_q(p)}{{\cal D}_q(p^{(0)})}
= \frac{1}{\prod_{\ell =1}^{n-1} [\ell ]!}\,{\cal D}_q(p) =: d_q(p)\nn
\ea
(cf. \cite{CP}, Example 11.3.10).
According to Theorem 4.1, every vector in ${\cal F}$ has a finite number of components belonging to different ${\cal F}_p\,.$
It is obvious from the definition that vectors belonging to ${\cal F}_p$ and ${\cal F}_{p^\prime}$ for $p\ne p^\prime\,$ are linearly independent.
It follows that the Fock space ${\cal F}$ (\ref{F}), originally defined as a vacuum representation space of the zero modes algebra ${\cal M}_q\,,$ is
equal to the direct sum (\ref{Fock-n}). This completes the proof of Proposition 4.2 (for generic $q$).
\hfill\eod

\smallskip

\noindent
{\bf Remark 4.3~} Note that (\ref{detc-n2-1}) takes place also for $q$ a root of unity. Hence, for $n=2$ Theorem 4.1
applies to the Fock space ${\cal F} = \oplus_{p=1}^\infty\,{\cal F}_p$ of the WZNW chiral zero modes as well,
where the spaces ${\cal F}_p$ are generated from the vacuum by homogeneous monomials in $a^1$ of order
$( \l = )\ p-1 \,.$ In this case, however, ${\cal F}_p$ carry {\em indecomposable} representations of $U_q\,.$

\bigskip

We define next a {\em linear} antiinvolution ("transposition") on ${\cal M}_q$ \cite{FHIOPT} by
\ba
&&(X Y )' = Y' X'\quad  \forall X, Y \in {\cal M}_q\ ,\qquad (q^{\hp_i})' = q^{\hp_i}\ ,\nn\\
&&{\cal D}_q^{(i)}(\hp) (a^i_\a)' = {\tilde a}^\a_{i} := {1\over{[n-1]!}}\,
{\epsilon}_{i i_1 \dots i_{n-1}}\, a^{i_1}_{\a_1}\dots a^{i_{n-1}}_{\a_{n-1}}\,{\varepsilon}^{\a\a_1 \dots \a_{n-1}} \ ,\qquad
\lb{prim}
\ea
where ${\cal D}_q^{(i)}(p)\,$ is equal to $1\,$ for $n=2\,$ while, for $n\ge 3\,,$ is given by the product
\be
\lb{minor}
{\cal D}_q^{(i)} (p) = \prod_{j<l,\, j\ne i\ne l} [p_{jl}]\qquad
\left(\,\Rightarrow\ [{\cal D}_q^{(i)} (\hp)\,,\, a^i_\a ] = 0 = [{\cal D}_q^{(i)} (\hp)\,,\, {\tilde a}^\a_{i} ] \,\right)\,.
\ee
The matrix $({\tilde a}^\a_{i})$ is thus the {\em (left) adjugate matrix}\ of
$(a^i_{\a})\,:$
\ba
&&{\tilde a}^\a_{i} a^i_{\b} = {1\over{[n-1]!}}\,{\epsilon}_{i i_1 \dots i_{n-1}}\,
a^{i_1}_{\a_1}\dots a^{i_{n-1}}_{\a_{n-1}}a^i_{\b}\,\e^{\a\a_1 \dots \a_{n-1}} =\nn\\
&&= \frac{(-1)^{n-1}}{[n-1]!}\,\e^{\a\a_1 \dots \a_{n-1}} \e_{\a_1 \a_2\dots \a_{n-1} \b}\, D_q(a) = D_q(a)\,\d^\a_\b\quad
\lb{a-inv}
\ea
(we have used the antisymmetry of ${\epsilon}_{i i_1 \dots i_{n-1}}\,$ and further, (\ref{det-intertw}) and (\ref{NK})).
In other words,
\be
{\tilde a}^\a_i = D_q(a)\,(a^{-1})^\a_i = {\cal D}_q(\hp) \,(a^{-1})^\a_i \quad\mbox{where}\quad
(a^{-1})^\a_{i}\, a^i_{\b} = \d^\a_\b\ ,\quad a^i_\a (a^{-1})^\a_j = \d^i_j
\lb{a-1}
\ee
(the fact that the matrix $a^{-1}$ defined by (\ref{a-1}), (\ref{prim}) is also a {\em right} inverse of $a$ can be
demonstrated in a similar way as (\ref{a-inv}) by using the properties of the dynamical antisymmetrizers and $\epsilon$-tensors \cite{HIOPT}).
Note that, due to (\ref{a-inv}) (and in conformity with (\ref{Dqa=Dqp})), the determinant $D_q(a)$ of the zero modes' matrix is invariant
with respect to the transposition:
\ba
&&(D_q(a))'\,\d^\a_\b = (a^i_\b)'({\tilde a}^\a_{i})' = \frac{1}{{\cal D}_q^{(i)}(\hp)}\, {\tilde a}^\b_{i}\,
{\cal D}_q^{(i)}(\hp)\,a^i_\a = {\tilde a}^\b_{i} a^i_\a = D_q(a)\,\d^\b_\a\, \ ;\nn\\
&&(D_q(a))' = ({\cal D}_q(\hp))' = {\cal D}_q(\hp) = D_q(a)\ .
\lb{Dqatransp}
\ea
It also follows that the transposed elements $(a^i_\a)'$ obey
\be
\sum_{i=1}^n (a^i_\a)'\, {\cal D}_q^{(i)}(\hp)\, a^i_\b = {\cal D}_q (\hp)\,\d^\a_\b\ ,\qquad
\sum_{\a=1}^n a^i_\a \frac{1}{{\cal D}_q (\hp)}\, (a^j_\a)'  = \frac{1}{{\cal D}_q^{(j)}(\hp)}\,\d^i_j \ .\lb{ladjug}
\ee
The involutivity of the transposition derives from the fact that the last two equations are valid with $(a^i_{\a})''$ in place of $a^i_{\a}\,.$

\smallskip

To compute correlation functions (like in (\ref{Nsa})), we shall equip the chiral state space (\ref{space})
with a left ("bra") vacuum state ${\lvac}$, defining thus a linear functional on the chiral field algebra.
This will allow us to define, in particular, a {\em bilinear} form
$\langle\, .\! \mid\! .\, \rangle\,:\ {\cal F} \times {\cal F} \rightarrow \C\,$ on the zero modes' Fock space (\ref{F})
such that, for any two vectors in ${\cal F}\,$ of the form $|\Phi \rangle = A \vac\,,\
|\Psi \rangle = B \vac\,$ where $A, B \in {\cal M}_q\,,$
\be
\lb{dual}
\langle \Phi \,\mid \Psi \rangle := \lvac\! A'\, B\! \vac\ .
\ee
To this end, we shall require the left vacuum to be orthogonal to any ${\cal F}_p$ with $p\ne p^{(0)}\,,$
and normalized ($\lvac\! 0 \rangle = 1$):
\ba
&&\lvac\! C\! \vac = c_0 \qquad \forall\, C\in {\cal M}_q\ ,\qquad{\rm where}\nn\\
&&C\! \vac = c_0\! \vac + \sum_{p\ne p^{(0)}} \mid C_p \rangle\ ,
\qquad \mid C_p \rangle \in {\cal F}_p\ .
\lb{lvac}
\ea
It is clear that the only non-trivial monomials in $a^i$ contributing to the vacuum expectation value (\ref{lvac})
are those of the form (\ref{PolFn}) with $m_1 = \dots = m_n$ which could be further reduced by using (\ref{vn-q-anti}).
From the invariance of $D_q(a)$ and $q^{\hp_i}$ with respect to the transposition (\ref{Dqatransp})
and their commutativity, (\ref{Dqap}) we deduce that
\be
\lvac\! C\! \vac = \lvac\! C'\! \vac \qquad \forall\, C\in {\cal M}_q
\lb{C'vac}
\ee
and hence (with the same conventions as above),
\be
\langle \Phi \mid C \mid \Psi \rangle  = \lvac A' C B \vac = \lvac B' C' A \vac = \langle \Psi \mid C' \mid \Phi \rangle\qquad
\forall\, C\in {\cal M}_q
\lb{C'}
\ee
(by taking $C=\id\,$ in Eq.(\ref{C'}) we infer, in particular, that the bilinear form (\ref{dual}) is {\em symmetric}).
We thus have, for any $\mid \Psi \rangle \in {\cal F}\,,$
\ba
&&\lvac  a^j_\a \mid \Psi \rangle = \langle \Psi \mid (a^j_\a)' \vac = 0 \quad {\rm for}\quad j = 1,\dots , n-1\nn\\
&&{\rm i.e.}\qquad \lvac  a^j_\a  = 0\ ,\quad j\le n-1\ ,\lb{Dual1}\\
&&\langle \Phi \mid q^{\hp_{ij}}\! \mid \Psi \rangle = \langle \Psi \mid q^{\hp_{ij}}\! \mid \Phi \rangle
= q^{p_{ij}} \langle \Psi \mid \Phi \rangle = q^{p_{ij}} \langle \Phi \mid \Psi \rangle\nn\\
&&{\rm i.e.}\qquad \langle \Phi \mid q^{\hp_{ij}} = q^{p_{ij}} \langle \Phi \mid\qquad
\forall\, \mid \Phi \rangle \in {\cal F}_p
\lb{Dual2}
\ea
(cf. (\ref{a2.n}), (\ref{prim}), and (\ref{Fpdef}), respectively). It easily follows from (\ref{Dual2}) that all the
irreducible $U_q(s\ell(n))$ modules ${\cal F}_p$ and ${\cal F}_{p'}$ (\ref{Fock-n}) with $p\ne p'$ are orthogonal to each other.

Eqs. (\ref{prim}), (\ref{a-1}), (\ref{Dqa=Dqp}) and the relation $a\, M \,= \, M_p \, a\,$ (which can be considered, for a given $M_p\,,$
as a {\em definition} of the monodromy matrix $M\,$ for the zero mode sector) imply
\be
(M^\a_{~\b})'\, (a^{-1})^\a_{i}\, =\, (a^{-1})^\b_{j}\, (M_p)^j_i\qquad \Rightarrow\quad
( M^\a_{~\b} )'\, = (a^{-1} M_p\, a )^\b_{~\a} = \,M^\b_{~\a}
\lb{Mpr}
\ee
i.e., the transposition of an entry of $M\,$ coincides with the corresponding entry of its transposed,
in the usual matrix sense, $M' = \,^t\!M\,.$ In agreement with the opposite triangularity of the
Gauss components $M_\pm\,$ (\ref{M+-q}), this is compatible with Eq.(\ref{Mtr}), $(M_\pm)' =\, ^t(M_\mp^{-1})\,$
which implies, in turn, Eq.(\ref{'}) for the transposed of the Chevalley generators of $U_q (s\ell(n))\,.$

It follows trivially from the definition (\ref{dual}) that, for any
$\mid \Phi \rangle\,,\mid \Psi \rangle \in {\cal F}_p\,$ and any $X\in U_q(s\ell(n))\,,$
\be
\langle X \Phi \mid \Psi \rangle = \langle \Phi \mid X' \mid \Psi \rangle\ ,
\lb{bfinv}
\ee
i.e. the bilinear form is $U_q(s\ell(n))${\em -invariant} (see Section 9.20 of \cite{Ja} for a proof that, for generic $q\,,$
a form with this property is essentially unique and {\em non-degenerate}). It is equally simple to derive,
by analogy with (\ref{Dual1}) and using (\ref{Uqvac}) and $\e (X') = \e (X)\,,$ the invariance of the left vacuum:
\be
0 = \lvac (X - \varepsilon (X) )\qquad \forall\, X\in U_q(s\ell(n))\ .
\lb{Uqlvac}
\ee
It has been proven in \cite{FHIOPT} for $n=2, 3\,$ (and conjectured to hold in general)
that the scalar squares of the highest and lowest weight vectors (\ref{HLWV}) are
\be
\lb{scsq}
\langle HWV \mid HWV \rangle_p = \prod_{i<j} [p_{ij}-1]! = \langle LWV \mid LWV \rangle_p \ .
\ee

\subsubsection{Fock representation of ${\cal M}_q$ for $q = e^{-i\frac{\pi}{h}}$}

After having studied the structure of the Fock representation of the algebra ${\cal M}_q$ for generic $q\,,$
we now return to our genuine problem, assuming that the deformation parameter is an (even) root of unity,
$q=e^{-i\frac{\pi}{h}}\,,\ h=k+n$ (\ref{h-SUn}).
The fact that in this case $[N h]=0$ for any $N \in {\Z}\,$ changes drastically the picture.
We shall point out and comment on the main differences below.
Albeit the full combinatorial description of the Fock space
${\cal F}\,$ (\ref{F}) for $n\ge 3\,$ still remains a challenge,
the observations and the technical tools described below could help us understand better its structure.

The classification of Fock states for $q\,$ generic
and $N\ge 3\,$ was based on the three lemmas derived in the previous subsection.
It is not difficult to see that Lemma 4.2 and Lemma 4.3 hold in the root of unity case as well.
The proof of Lemma 4.1 however fails since in this case $[p_{ij}-1]\,$ can vanish which makes
impossible the exchange of $a^j_\b\,$ and $a^i_\a\,$ for $\a \ne \b\,;$ indeed, in this case
\be
[\hat p_{ij}-1] \, v = 0\quad\Leftrightarrow\quad \hat p_{ij}\,v = (Mh+1)\,v\ ,\quad M \in {\mathbb Z}
\quad\Rightarrow\quad q^{\epsilon\hat p_{ij}}\,v = (-1)^M q^\epsilon \,v
\lb{pij1}
\ee
(for $\epsilon = \pm 1$) so (\ref{aa2}) reduces in this case to just the $q$-symmetry of $a^i_\a a^j_\b \, v\,:$
\be
[\hat p_{ij}-1] \, v = 0\quad\Rightarrow\quad a^i_\a a^j_\b \, v = q^{\epsilon_{\a\b}} a^i_\b \, a^j_\a \, v\ .
\lb{pij-symm}
\ee
Note that the vanishing of the other $p$-dependent coefficient in (\ref{aa2}) implies, on the other hand,
the symmetry of $a^j_\a\, a^i_\b \, v\,$ in the {\em upper} indices:
\be
[\hat p_{ij}]\, v = 0\quad\Leftrightarrow\quad \hat p_{ij}\, v = Mh\,v\ , \quad M\in {\mathbb Z}
\quad\Rightarrow\quad
 a^i_\a  a^j_\b \, v = a^j_\a  a^i_\b \, v \ .
\lb{pij0}
\ee

The proof of Lemma 4.1 cannot be applied, for example, to the vector
\be
v_{\a \b_1 \b_2} := a^j_\a \, a^1_{\b_1} \,  a^1_{\b_2}\dots a^1_{\b_{h+3-j}}\, \vac \qquad {\rm for}\qquad j\ge 3
\lb{s}
\ee
which is of the form envisaged in (\ref{L1}). This is an important issue: if $v_{\a \b_1 \b_2} \ne 0\,,$
it would mean that, for $n\ge 3\,,$ the spectrum of $\hp = (\hp_1, \dots , \hp_n)\,$ on ${\cal F}\,$
includes non-dominant (shifted integral) $s\ell (n)$ weights. As mentioned above, when the index $\a$ is different
from all $\b_i\,,\ i=1,\dots , h+3-j\,,$ it is not possible to use (\ref{aa2}) to move $a^j$ to the right
until it reaches and annihilates the vacuum, since
\ba
&&[\hp_{1j} -1]\,a^1_{\b_2}\dots a^1_{\b_{h+3-j}}\!\vac =
a^1_{\b_2}\dots a^1_{\b_{h+3-j}}[\hp_{1j} + h+1-j]\!\vac =\nn\\
&&= [h]\, a^1_{\b_2}\dots a^1_{\b_{h+3-j}}\!\vac = 0\ .
\lb{ex2}
\ea
It turns out, however, that the vector (\ref{s}) is $q$-antisymmetric in the first pair of indices and
$q$-symmetric in the second,
\be
- q^{-\epsilon_{\a\b}}\, v_{\b\a \g} = v_{\a \b \g} = q^{\epsilon_{\b\g}} \, v_{\a \g \b}
\lb{b}
\ee
and, as a result, vanishes. Indeed, it follows from (\ref{b}) that
\be
v_{\a\b\g} = - q^{-\epsilon_{\a\b}}\, v_{\b \a \g} = -
q^{-\epsilon_{\a\b}+\epsilon_{\a\g}} \,v_{\b \g \a} =
q^{-\epsilon_{\a\b}+\epsilon_{\a\g} - \epsilon_{\b\g}} \,v_{\g \b \a}
\lb{vabg1}
\ee
but also
\be
v_{\a\b\g} = q^{\epsilon_{\b\g}}\, v_{\a \g \b} = - q^{\epsilon_{\b\g}-\epsilon_{\a\g}} \,v_{\g \a \b} =
- q^{\epsilon_{\b\g}+\epsilon_{\g\a}+\epsilon_{\a\b}} \,v_{\g \b \a}
\lb{vabg2}
\ee
or,
\be
v_{\a\b\g} = q^{-\epsilon_{\a\b} - \epsilon_{\b\g} -\epsilon_{\g\a}}\, v_{\g\b\a} =
- q^{\epsilon_{\a\b}+\epsilon_{\b\g}+\epsilon_{\g\a}} \, v_{\g\b\a}\ \ (\, = 0 \, )
\lb{vabg3}
\ee
since the relative factor is equal to $-1\ $ (for $\b = \g$) or to $\ - q^{\pm 2} \ne 1\,.$

We shall provide details of the proof of (\ref{b}) since they appear to be typical for the root of unity case.
The $q$-symmetry of $v_{\a\b\g}\,$ in $\b\,$ and $\g\,$ is implied directly by the second
Eq.(4.184). To prove its $q$-antisymmetry in the first two indices, we write
\be
v_{\a\b\g} = a^j_\a \, a^1_{\b} \, v_\g \quad{\rm where}\quad v_\g := a^1_\g \,a^1_{\b_3} \dots a^1_{\b_{h+3-j}}\, \vac\ .
\ee
There are $h+2-j\,$ operators $a^1\,$ applied to the vacuum in $v_\g\,$ so that, in particular,
by (\ref{ExRap}) and (\ref{pjl-on-vac}),
\be
\hp_{1j}\, v_\g = (h+1)\, v_\g\qquad {\rm and} \qquad a^j_\s \, v_\g = 0\quad \forall\, \s \ .
\lb{a}
\ee
The last equality follows since
$a^j_\s \, v_\g = a^j_\s\, a^1_\g \,v\,,\ \hp_{1j} \, v = h \, v \,$ etc.,
so one can apply repeatedly (\ref{aa2}), starting with
\be
a^j_\s\, v_\g = a^j_\s\, a^1_\g \,v = \frac{1}{[h-1]} \,\, a^j_\s\, a^1_\g\, [\hp_{1j}-1]\,v =\ \dots
\lb{av}
\ee
until $a^j\,$ reaches the vacuum. If $\a = \b\,,$ then
\be
v_{\a\a\g} = a^j_\a \, a^1_{\a} \, v_\g = a^1_\a \, a^j_{\a} \, v_\g = 0\ ,
\lb{va}
\ee
and this is equivalent to $- v_{\a\a\g} = v_{\a\a\g}\,,$ a particular case of the first Eq.(\ref{b}).
Assume now that $\a \ne \b\,;$ again by (\ref{aa2}) (with $i\leftrightarrow j\,,$ followed by $i=1$),
Eq.(\ref{a}) implies that
\be
[\hp_{j1}-1]\, a^1_\b\, a^j_\a\, v_\g = 0 = a^j_\a a^1_\b\, [\hp_{j1}] \, v_\g -
\,a^j_\b\, a^1_\a \, q^{{\epsilon}_{\a\b}\hp_{j1}}\, v_\g
\lb{paa}
\ee
and the first Eq.(\ref{b}) for $\a \ne \b\,$ follows since $\hp_{j1}\, v_\g = - (h+1) \, v_\g\,,$ cf. (\ref{a}):
\ba
&&- a^j_\a a^1_\b\, [h+1]\, v_\g -\,a^j_\b\, a^1_\a \, q^{-{\epsilon}_{\a\b} (h+1)} \,v_\g = 0\quad\Leftrightarrow\nn\\
&&a^j_\a a^1_\b \, v_\g \equiv v_{\a\b \g} = - q^{-{\epsilon}_{\a\b}} \,a^j_\b\, a^1_\a \, v_\g \equiv
- q^{-{\epsilon}_{\a\b}} \, v_{\b\a\g}\ .
\lb{aav}
\ea
Thus, $\, a^j_\a \, a^1_{\b_1} \,  a^1_{\b_2}\dots a^1_{\b_{h+3-j}} \vac = 0 \quad {\rm for}\quad j\ge 3\,.$

This result is easily generalized to vectors of the form
\be
w_{\a\b\g} = a^j_\a a^i_\b \,a^i_\g\,w\ ,\qquad p_{ij}\, w = N h \, w\ ,\qquad a^j_\s \,a^i_\g \,w = 0\ \ \forall\, \s
\lb{avgen}
\ee
for $n\ge j \ge i+2 \ge 3\,,$ i.e.,
\be
- q^{-\epsilon_{\a\b}}\, w_{\b\a \g} = w_{\a \b \g} = q^{\epsilon_{\b\g}} \,w_{\a \g \b}
\qquad\Rightarrow\qquad w_{\a\b\g} = 0 \ .
\lb{avgen1}
\ee
But even (\ref{avgen1}) is yet a consequence of a more universal technique we are going to describe now.

\smallskip

It turns out that the exchange relations (\ref{aa2}) for the zero modes
take a remarkably simple and transparent form when written in terms of the
$q$-symmetric and $q$-antisymmetric projections of the bilinear combination $a^i_\a a^j_\b\,,$
\be
a^i_\a a^j_\b = A^{ij}_{\a\b} + S^{ij}_{\a\b}\ ,\quad A^{ij}_{\a\b} = - q^{-\epsilon_{\a\b}} A^{ij}_{\b\a}\ ,
\quad S^{ij}_{\a\b} = q^{\epsilon_{\a\b}} S^{ij}_{\b\a}\ .
\lb{SA}
\ee
Here the $q$-antisymmetric part is defined by using (\ref{A1const}),
\be
\ \ \, [2]\, A^{ij}_{\a\b} := a^i_{\a'} a^j_{\b'} A^{\a' \b'}_{~\a\b} = \left\{
\begin{array}{ll}
&q^{-\epsilon_{\a\b}} a^i_\a a^j_\b - a^i_\b a^j_\a\ ,\quad \a\ne \b \\
&0\ ,\quad \a=\b
\end{array}
\right.
\lb{Adef}
\ee
and for the $q$-symmetric one,
\be
[2]\, S^{ij}_{\a\b} := a^i_{\a'} a^j_{\b'} S^{\a' \b'}_{~\a\b} = \left\{
\begin{array}{ll}
&q^{\epsilon_{\a\b}} a^i_\a a^j_\b + a^i_\b a^j_\a\ ,\quad \a\ne \b \\
&[2]\, a^i_\a a^j_\a\ \ ( \equiv [2]\, a^j_\a a^i_\a)\ ,\quad \a=\b
\end{array}
\right.\quad ,
\lb{Sdef}
\ee
we need to introduce the corresponding operator
\be
S^{\a' \b'}_{~\a\b} = [2]\, \d^{\a'}_\a \d^{\b'}_\b - A^{\a' \b'}_{~\a\b} =
s_{\a\b} \d^{\a'}_\a \d^{\b'}_\b + \d^{\a'}_\b \d^{\b'}_\a \ ,
\lb{q-symm}
\ee
where
\be
s_{\a\b} := \left\{ \begin{array}{ll}
&q^{\epsilon_{\a\b}}\ ,\quad \a\ne \b \\
&[2]-1 \equiv q+q^{-1}-1\ ,\quad \a=\b
\end{array}\right. \ .
\lb{sab}
\ee
Rewriting the first relation (\ref{aa2}) in terms of $S^{ij}_{\a\b}\,$ and $A^{ij}_{\a\b}\,$ using (\ref{SA}),
\ba
&&[\hp_{ij}-1]\, (S^{ji}_{\b\a} + A^{ji}_{\b\a}) = [\hp_{ij}]\, (q^{\epsilon_{\a\b}} S^{ij}_{\b\a} -
q^{- \epsilon_{\a\b}} A^{ij}_{\b\a}) - q^{\epsilon_{\a\b} \hp_{ij}} (S^{ij}_{\b\a} + A^{ij}_{\b\a}) = \nn\\
&&= (q^{\epsilon_{\a\b}}[\hp_{ij}] - q^{\epsilon_{\a\b}\hp_{ij}} )\, S^{ij}_{\b\a} -
(q^{- \epsilon_{\a\b}}[\hp_{ij}] + q^{\epsilon_{\a\b}\hp_{ij}} )\, A^{ij}_{\b\a}
\lb{S+A}
\ea
we obtain, with the help of the $q$-identities $q^{\pm \epsilon} [p] \mp q^{\epsilon p} = [p\mp 1]\,,$
the following relation between the matrices $S^{ij} := (S^{ij}_{\a\b})\,,\ A^{ij} := (A^{ij}_{\a\b})\,$:
\be
[\hp_{ij}-1]\,(S^{ij} - S^{ji} - A^{ji}) = [\hp_{ij}+1]\,A^{ij}\ .
\lb{rel1}
\ee
Exchanging $i\,$ and $j\,$ in (\ref{rel1}), we get
\be
[\hp_{ij}+1]\,(S^{ij} - S^{ji} + A^{ij}) = -\, [\hp_{ij}-1]\,A^{ji}\ .
\lb{rel2}
\ee
Now summing both sides of (\ref{rel1}) and (\ref{rel2}), we obtain
\ba
&&([\hp_{ij}-1] + [\hp_{ij}+1])\,( S^{ij} - S^{ji}) = [2]\,[\hp_{ij}] \,(S^{ij} - S^{ji}) = 0\nn\\
&&\quad\Rightarrow\quad S^{ij} = S^{ji}\quad\Leftrightarrow\quad
a^i_\a a^j_\b - a^j_\a a^i_\b = - q^{-\epsilon_{\a\b}} (a^i_\b a^j_\a - a^j_\b a^i_\a)\qquad\qquad
\lb{rel-add}
\ea
(we use $[p-1]+[p+1]=[2]\,[p\,]\,;$ the implication and the equivalence follow from (\ref{pij0})
and the definition (\ref{Sdef}), respectively).
Plugging (\ref{rel-add}) into (\ref{rel1}) or (\ref{rel2}),
we obtain the relation for the $q$-antisymmetric combinations:
\be
[\hp_{ij}+1]\,A^{ij} + [\hp_{ij}-1]\,A^{ji} = 0\ .
\lb{rel-A}
\ee
Albeit derived for ($i\ne j\,$ and) $\,\a\ne\b\,,$ these identities also hold for $\a = \b\,;$
in particular, the second relation (\ref{aa2}) is equivalent to $S^{ij}_{\a\a} = S^{ji}_{\a\a}\,.$
On the other hand, the last relation (\ref{aa2}) is the same as
\be
A^{ii} = 0\ .
\lb{rel-Aii}
\ee
So the simple operator identities (\ref{rel-add}), (\ref{rel-A}) and (\ref{rel-Aii}) completely replace (\ref{aa2}).

\smallskip

Having introduced the $q$-symmetrizers (\ref{q-symm}), we can rewrite the braid group relations
(\ref{b-Hecke})) (or, equivalently, (\ref{q-antisymm})) in various disguises.
Using for brevity the common notation $A_i \equiv A_{i i+1}\,,\ S_i \equiv S_{i i+1}\,,\ i = 1, 2,\dots $
we first deduce by $S_i = [2] - A_i\ ( = q + b_i)\,$ that
\be
S_i^2 = [2]\, S_i\ ,\quad A_i S_i = 0 = S_i A_i\ ,\quad [S_i , S_j ] = 0 = [S_i , A_j]\quad{\rm for}\quad |i-j|\ge 2\ .
\lb{SiAj}
\ee
We shall write down here just two of the whole variety of relations following from
$A_i \, A_{i+1} \, A_i - A_i = A_{i+1} \, A_i \, A_{i+1}- A_{i+1}\,$ (and (\ref{SiAj})):
\ba
&&A_i S_{i+1} A_i + S_{i+1} A_i S_{i+1} - [2] (A_i S_{i+1} + S_{i+1} A_i) + A_i + S_{i+1} = [2]\ ,\nn\\
&&S_i A_{i+1} S_i + A_{i+1} S_i A_{i+1} - [2] (A_{i+1} S_i + S_i A_{i+1}) + S_i + A_{i+1} = [2]\ .\qquad\quad
\lb{ASA2}
\ea
The presence of the free terms in (\ref{ASA2}) implies that a $3$-tensor $ v = v_{\a\b\g} \,$ that is $q$-symmetric
in the first pair of indices ($v\, A_1 = 0$) and $q$-antisymmetric in the second ($v\, S_2 = 0$)
or vice versa (i.e. $v\,S_1 = 0$ and $\, v\, A_2 = 0$), is zero --
something we have already proved, cf. (\ref{wgba}) and (\ref{avgen1}), respectively.

\medskip

We shall list below a few complications one has to confront when considering the zero modes' algebra and its Fock
representation at roots of unity.

\smallskip

\noindent{\bf (1)~} {\em The determinant $D_q(a)$ has zero eigenvalues on ${\cal F}$ so $a$ is not invertible.}

\smallskip

As the determinant $D_q(a)$ is equal, by definition, to ${\cal D}_q(\hp)\,,$ it vanishes on every subspace ${\cal F}_p$ characterized
by (\ref{Fpdef}) such that $p_{ij} \in \Z h$ for some pair $(i,j)\,,\ 1\le i < j \le n\,.$ Hence, the zero modes' operator matrix $a$
is not invertible, see (\ref{a-1}). For a similar reason (as ${\cal D}_q^{(i)}(p)$ (\ref{prim}) may vanish),
the bilinear form (\ref{dual}) is not well defined, except for $n=2\,.$

\smallskip

\noindent{\bf (2)~} {\em The zero modes' algebra ${\cal M}_q$ has a non-trivial (two-sided) ideal.}

\smallskip

The key to this property of ${\cal M}_q$ is the relation (valid for $i\ne j$ and $\a\ne\b$)
\be
\lb{genex}
[\hp_{ij}-1](a^j_\b)^m a^i_\a =
a^i_\a(a^j_\b)^m [\hp_{ij}]- [m] (a^j_\b)^{m-1}a^i_\b \,a^j_\a\, q^{\epsilon_{\a\b} \hp_{ij}}
\ee
generalizing the first Eq.(\ref{aa2}) for any positive integer $m\,.$\footnote{
Eq.(\ref{genex}) can be easily proven by induction, using the $q$-number relation
$$
[p+m] = [p][m+1] - [p-1][m]\ .
$$}
Therefore, assuming that $(a^j_\b)^m = 0\ \ \forall\ j,\b$ for {\em generic} $q\,$ would imply $(a^j_\b)^{m-1} = 0$
etc., leading eventually to trivialization. For $q^h=-1\,,$ however, putting in (\ref{genex}) (for $m=h$)
\be
(a^j_\b)^h = 0\ ,\qquad 1\le j\,,\b \le n
\lb{ah}
\ee
does not imply further relations for the lower powers. As we are mainly interested in the Fock representation
of ${\cal M}_q$ in which all the eigenvalues of $\p_{ij}$ are integers (cf. (\ref{sp-p-r})), we could also assume that
\be
q^{2 h \hp_{ij}} = 1\ ,\qquad 1\le i,j \le n\ .
\lb{qhpij}
\ee
Thus, if ${\cal J}^{(h)}_q \subset {\cal M}_q$ is the two-sided ideal generated by the $h$-th powers of all $a^i_\a\,$
and the $2h$-th powers of $q^{\hp_{ij}}\,,$ the quotient ${\cal M}^{(h)}_q := {\cal M}_q / {\cal J}^{(h)}_q$ is non-trivial.
For $n=2\,$ it is easy to deduce from Eqs. (\ref{detc-n2-1}), (\ref{detc-n2-2}), (\ref{ah}) and (\ref{qhpij})
that  ${\cal M}_q^{(h)}$ is finite ($2 h^5$-) dimensional; the corresponding Fock representation
\be
{\cal F}^{(h)} = {\cal M}^{(h)}_q\vac
\lb{Fock-h}
\ee
is $h^2$-dimensional \cite{FHT2}.

\smallskip

\noindent{\bf (3)~} {\em Indecomposable representations of $U_q(s\ell(n))$ appear.}

\smallskip

This issue will be discussed at length in the following section for $n=2\,.$ Here we shall only recall that the decomposition
of the Fock space ${\cal F} = \oplus_{p=1}^\infty\,{\cal F}_p\,$ (for $p\equiv p_{12}$) still takes place in this case (Remark 4.3).
Even so, the statement of Proposition 4.2 does not hold as it stays; it turns out \cite{FHT7} that only the $U_q(s\ell(2))$ representations on
${\cal F}_p$ with $p\le h$ are irreducible while those with $p > h$ are either indecomposable,
for $p \notin {\mathbb N} h\,,$ or fully reducible, for $p \in {\mathbb N} h\,.$
(As we shall see in the next Section, the true symmetry algebra in this case is in fact a finite dimensional quotient of $U_q(s\ell(2))\,.$)
The dimension and the quantum dimension of each ${\cal F}_p$ (\ref{qdimFp}) are equal to
\be
{\rm dim}\,{\cal F}_p = p\ ,\qquad {\rm qdim}\,{\cal F}_p = [p]\ ,
\lb{qdim-n2}
\ee
respectively; hence, the quantum dimension of ${\cal F}_p$ vanishes for $p \in {\mathbb N} h\,.$

\smallskip

As we do not have full control of the situation for $n\ge 3\,,$ we shall focus further our attention mainly on the $n=2$ case.
Before that, however, we shall complete this section with some general remarks on the role of the elementary CVO $u(z)$
and the quantum group covariant chiral field $g(z)\,,$ cf. (\ref{gCVO}) and (\ref{gT}).

\subsubsection{Braiding of the chiral quantum fields}

In analogy to (\ref{ggR}) (or (\ref{ggRa})) and (\ref{ExRaa2}), we shall postulate braiding relations for $u(x)$ of the type
\be
\lb{uuRp}
u_1(x_1)\, u_2 (x_2) = u_2(x_2) \, u_1(x_1)\, (R_{12}(p)\, \theta (x_{12}) + R_{21}^{-1}(p)\,\theta (x_{21}))
\ee
(for $-2\pi < x_{12} <2\pi$) or, equivalently, exchange relations for $u(z)$
\be
u^A_i(z_1)\, u^B_j (z_2) =\,\stackrel{\curvearrowright}{u^B_\ell (z_2)\, u^A_m (z_1)}\!{\hat R}(p)^{\ell m}_{~ij}\ ,
\qquad {\hat R}(p) = P R (p)
\lb{uuRp2}
\ee
in the analyticity domain specified in (\ref{ggRa}). Eq.(\ref{uuRp}) involving the dynamical quantum $R$-matrix (\ref{RRp2})
should serve as a quantum version of the PB (\ref{uuDir}). One may think that the singularity of $R(p)$ for $q$ a root of unity
could be resolved in the same way as it was done for the zero modes where we replaced the relations following from (\ref{ExRaa2})
by their regular counterparts (\ref{aa2}). The discussion in the beginning of Section 3.4 however shows that we should supplement
the exchange relations of $u(z)$ by a relation for its (regularized) determinant, and in the quantized theory this has to be
proportional to the {\em inverse} of the (operator) function ${\cal D}_q(p)\,$ -- which is ill defined too.

We can use analytical methods to tackle the problem by using the KZ equation (\ref{KZW-N}). To this end, we identify
the spaces ${\mathcal H}_p$ as infinite dimensional ${\widehat{su}}(n)_k$ current algebra modules (cf. (\ref{curfq})) characterized by highest weight
(which also means, due to (\ref{Ln}), also lowest energy) subspaces ${\mathcal V}_p\,:$
\be
j_n^a\, {\mathcal V}_p = 0\quad\Rightarrow\quad L_n \, {\mathcal V}_p = 0\quad \mbox{for}\quad n > 0\ .
\lb{jLnVp}
\ee
Further, ${\cal V}_{p^{(0)}}$ is $1$-dimensional and coincides with the vacuum subspace; in addition to (\ref{jLnVp}),
the vacuum vector $\mid\! 0\rangle$ is assumed to carry zero charge and, as a consequence of the Sugawara formula, is also conformal invariant,
see (\ref{jonvac}), (\ref{Lonvac}).

In general, any ${\cal V}_p$ is generated from the vacuum by a primary field $\phi_\L(z)$ satisfying (\ref{Ward}) (for
$p =\L + \rho$) so that
\be
{\cal V}_p = \phi_\L (0)\! \vac\quad\Rightarrow\quad
j_0^a\, {\mathcal V}_p = - \pi_\L(t^a)\,{\mathcal V}_p\ ,\quad L_0 \, {\cal V}_p = \Delta (\L )\,{\cal V}_p
\lb{jL0Vp}
\ee
where $\Delta (\L)$ is the conformal dimension (\ref{conf-dim-L}) of $\phi_\L(z)$ (the first implication follows from
(\ref{Ward})\footnote{Note that the minus sign ensures the compatibility between the commutation relations of $j^a_0$ and $t^a$ as
$\ [ j_0^a , j^b_0 ]\, {\mathcal V}_p = [ \pi_\L(t^b) , \pi_\L(t^a)]\, {\mathcal V}_p = - i f^{ab}_{~~~c} \pi_\L(t^c)\, {\mathcal V}_p =
i f^{ab}_{~~~c} j^c_0\, {\mathcal V}_p\ .$}
and the second, from (\ref{L0}) and (\ref{jLnVp})). In our context the primary fields can be constructed, in principle, as composite operators in
the elementary CVO $u(z)\,.$

Thus we can think of ${\cal H}_p$ as ${\widehat {su}}(n)_k$ current algebra highest weight modules defined by (\ref{jLnVp}) and (\ref{jL0Vp}).
Let us now consider a matrix element of the type
\be
\langle \Phi_{p'}\mid u^A_i(z_1) \, u^B_j(z_2 ) \mid \Phi_p \rangle \qquad{\rm for}\qquad \Phi_p \in {\cal H}_p\ ,\quad \Phi_{p'}\in {\cal H}_{p'}
\lb{KZuu}
\ee
The CVO $u_i(z)$ are assumed to intertwine between ${\cal H}_p$ and ${\cal H}_{p+ v^{(i)}}\,,$ see (\ref{gCVO}).
In order to avoid the difficulty of dealing with non-dominant weights, we assume
that all representations involved are integrable, i.e. all $p_{ij}$ satisfy $1 \le p_{ij} \le h-1\,$ for $i<j$
(or, which amounts to the same, that -- for fixed {\em dominant} $p$ and $p'$ -- the level $k$ is high enough).
Then we can expect that $(\ref{KZuu})$ is well defined unless $p_{ij}$ approaches $h\,.$

It is possible to {\em derive} the braiding relations (\ref{uuRp}) in this setting, and the following is a summary
of the corresponding computation performed in \cite{HST}. Due to the $SU(n)$ invariance, (\ref{KZuu}) could be only
non-zero for $p' = p + v^{(i)} + v^{(j)}\,$ so let us consider the $4$-point function
\be
W_4 := W_4 (z,  z_1 , z_2 , w)= \langle 0\mid \phi_{{\L^*}}(z)\, u^A_i(z_1) \, u^B_j(z_2)\, \phi_\L (w) \mid 0\rangle
\lb{W4}
\ee
where $\L^*$ is the $su(n)$ representation conjugate to $\L + \L^i + \L^j\,.$ Taking into account the M\"obius invariance
\cite{DFMS, FSoT}, (\ref{W4}) can be reduced, up to appropriate conformal factors, to a $4$-point function
$W_4 (\infty,  1 , \eta , 0)$ on a primary analyticity domain containing the real values of $\eta$ between $0$ and $1\,.$
For $i\ne j$ the two possible channels (with intermediate states belonging to ${\cal H}_{p+ v^{(i)}}$
and ${\cal H}_{p+ v^{(j)}}$, respectively) are identified by their analytic behaviour at $\eta \sim 0\,.$
For each of them the ensuing "reduced KZ equation" leads to an ordinary linear equation of hypergeometric type in $\eta\,.$
In the case $i=j$ there is a single first order equation.

The braiding of the corresponding solutions recovers exactly the quantum dynamical $R$-matrix ${\hat R}(p)$ (\ref{RRp2}).
The mutual normalization of the solutions to the reduced
KZ equation for $i\ne j$ has poles (or, conversely, zeroes) at $p_{ij} = N h\,$ for $i<j\,$ and  $N\,$ a positive integer.
As expected, (\ref{KZuu}) makes sense for {\em integrable} (shifted) dominant weights ($p_{i\,i+1} \ge 1\,,\ p_{1n} \le h-1$) which are the only ones
that appear when considering the model in the framework of rational CFT but are not sufficient for a consistent description
of the canonical quantization of the chiral theory.

By contrast, the solutions of the KZ equations for the analog of (\ref{KZuu})
\be
\langle \Phi_{p'}\mid g^A_\a (z_1) \, g^B_\b(z_2 ) \mid \Phi_p \rangle
\lb{KZgg}
\ee
involving the chiral field $g(x)$ (\ref{ggRa}) are well defined for any (dominant) $p\,$ and $p'\,.$
Their braiding reproduces the exchange relations (\ref{ggRa}) which do not depend on $p\,.$
What actually happens is that the meaningless matrix elements and exchange relations of the CVO are
"regularized" by the zeroes in the corresponding expressions for the zero modes.
A convenient basis of {\em regular} solutions of the KZ equations for a general $4$-point function has been introduced for $n=2$ in \cite{STH}.

As it has been already explained, a complete description of the $n\ge 3$ case would require studying more general representations
of both the zero modes' and the affine algebra corresponding to non-dominant $p\,.$ We shall restrict our
attention in the next Section to $n=2$ in which case this obstruction does not occur.

\section{Zero modes and braiding beyond the unitary limit for \boldmath$n=2$}

\setcounter{equation}{0}
\renewcommand\theequation{\thesection.\arabic{equation}}

We shall collect here, for reader's convenience, the necessary formulae for the $n=2$ case derived so far.
The $q$-antisymmetrizers of (\ref{biAi}) (Section 4.4) are rank one operators and in particular,
$A^{\rho\s}_{~\a\b} = \e^{\rho\s} \e_{\a\b}\,,$ cf. (\ref{A1const}).
The constant $R$-matrix (\ref{R}) gives then rise to the braid operator
\be
q^{- \frac{1}{2}} {\hat R}^{\rho\s}_{~\a\b} = q^{-1} \d^\rho_\a \d^\s_\b - \, \e^{\rho\s} \e_{\a\b}\qquad
(\,\e_{12} = \e^{12} = - q^{\frac{1}{2}}\ , \ \ \e_{21} = \e^{21} = q^{- \frac{1}{2}}\,)\ .\quad
\lb{braidR2}
\ee
In view of Remark 4.2 and Eq.(\ref{A1n}), this case is characterized by the fact that the Hecke representation
(\ref{b-Hecke}) factors through the Temperley-Lieb algebra. Using
$\,\e_{\a\s}\, \e^{\s\b} = - \d^\b_\a = \e^{\b\s} \e_{\s\a}\,,$ it is easy to verify indeed that
\ba
&&A_1\, A_2\, A_1 - A_1 = 0 = A_2\, A_1\, A_2 - A_2 \qquad{\rm with}\nn\\
&&(A_1)^{\a_1\a_2\a_3}_{~\b_1\b_2\b_3} = A^{\a_1\a_2}_{~\b_1\b_2} \,\d^{\a_3}_{\b_3}\qquad {\rm and} \qquad
(A_2)^{\a_1\a_2\a_3}_{~\b_1\b_2\b_3} = \d^{\a_1}_{\b_1}\,A^{\a_2\a_3}_{~\b_2\b_3} \ \, .\quad
\lb{TL2}
\ea
The corresponding dynamical $R$-matrix (\ref{RRp2}) reads
\be
\lb{Rpn=2}
{\hat R}_{12}(p)\, =\,
q^{\frac{1}{2}}\begin{pmatrix}q^{-1}&0&0&0\cr
0&q^{-p}\over{[p]}&\a(p)\,{{[p -1]}\over{[p]}}&0\cr
0&\a(p)^{-1}\,{{[p+1]}\over{[p]}}&-{{q^{p}}\over{[p]}}&0\cr
0&0&0&q^{-1}\end{pmatrix}\ ,\qquad p =p_{12}\ .
\ee
For $\a(p) = 1\,$ the quadratic $n=2$ determinant conditions (\ref{det-intertw}), (\ref{Dqa=Dqp})
(implying in this case the exchange relations (\ref{aa2})) can be written as
\be
a^j_\a a^i_\b - a^i_\a a^j_\b\, = [{\hat p}_{ij}\,]\,\varepsilon_{\a\b}\ ;\qquad
a^j_\a a^i_\b \, \varepsilon^{\a\b} = [\hp_{ij}+1]\quad (\,i\ne j\,)\ ,\qquad
a^i_\alpha a^i_\beta \, \varepsilon^{\alpha\beta} = 0
\lb{detc-n2}
\ee
(cf. (\ref{detc-n2-1}), (\ref{detc-n2-2})). Using (\ref{braidR2}), we can replace the first and/or the third relation (\ref{detc-n2}) by
\be
q^{\frac{1}{2}}\,a^i_\rho a^j_\s {\hat R}^{\rho\s}_{~\a\b} = a^j_\a a^i_\b - q^{1-\hp_{ij}} \e_{\a\b} \quad (\,i\ne j\,)\ ,\qquad
q^{\frac{1}{2}}\,a^i_\rho a^i_\s {\hat R}^{\rho\s}_{~\a\b} = a^i_\a a^i_\b \ ,\quad
\lb{altEx}
\ee
respectively \cite{FHT2, FHT3}. For $n=2\,$ Eq.(\ref{ExRap}) gives simply
\be
q^{\hp}\, a^1_\a = a^1_\a\, q^{\hp+1}\ ,\qquad q^{\hp}\, a^2_\a = a^2_\a\, q^{\hp-1}\ ,
\lb{ExRapn2}
\ee
and the relations (\ref{a2.n}) and (\ref{Dual1}) reduce to the standard creation and annihilation operator conditions
\be
a^2_\a \,{\mid 0 \rangle} = 0\ ,\qquad \lvac a^1_\a = 0\ .
\lb{a-vac}
\ee
The $U_q^{(2)}(s\ell(2))\,$ covariance properties (\ref{AdXa}) of the zero modes read
\ba
&&k\, a^i_1 k^{-1} = q^{\frac{1}{2}} a^i_1\ ,\quad k\, a^i_2 k^{-1} = q^{-\frac{1}{2}} a^i_1\qquad (\, k^2 = K \,)\ ,\nn\\
&&[E , a^i_1 ] = 0\ ,\quad [E , a^i_2 ] = a^i_1 \, K\ ,\nn\\
&&F \, a^i_1 = q^{-1} a^i_1 \, F + a^i_2\ ,\quad F \, a^i_2 = q\, a^i_2 \, F\ .
\lb{AdXa1}
\ea

\subsection{The Fock representation of the zero modes' algebra}

A basis
\be
\{\, |p,m{\cal i}\ ,\quad p=1,2,\dots\ ,\ \ 0\le m\le p-1\,\}
\lb{base2}
\ee
in the Fock space ${\cal F} = {\cal M}_q\vac\,$ is obtained by acting on the vacuum by homogeneous polynomials in
the creation operators $a^1_\a\,$ (of degree $p-1$):
\be
\lb{basis2}
|p,m\rangle := (a^1_1 )^m (a^1_2 )^{p-1-m} \vac \qquad (\,|1, 0\rangle \equiv \vac\ ,\quad (q^{\hat p} - q^p)\, |p,m\rangle = 0\, )\ .
\ee
For a given $p\,,$ all vectors $|p,m\rangle$ in the allowed range of $m$ form a basis in ${\cal F}_p$ so that
\be
{\cal F} = \oplus_{p=1}^\infty\,{\cal F}_p \qquad (\,{\rm dim}\,{\cal F}_p = p\ ,\quad {\rm qdim}\,{\cal F}_p = [p]\,)\ ,
\lb{FFp-dim}
\ee
see (\ref{qdim-n2}). By (\ref{detc-n2}) and (\ref{a-vac}), the operators $a^i_\a\,$ act on the basis vectors as
\ba
&&a^1_1 | p , m \rangle  =  | p+1 , m+1 \rangle  \ ,\nn\\
&&a^1_2 | p , m \rangle  =  q^m | p+1 , m \rangle \ ,\nn\\
&&a^2_1 | p , m \rangle  =  - q^{\frac{1}{2}} [p-m-1] | p-1 , m \rangle \ ,\nn\\
&&a^2_2 | p , m \rangle  =  q^{m-p+\frac{1}{2}} [m] | p-1 , m-1 \rangle  \ .
\lb{apmn2}
\ea
The $U_q(s\ell(2))\,$ transformation properties follow from (\ref{AdXa1}) and (\ref{Uqvac}),
\ba
&&K\, |p,m\rangle = q^{2m-p+1} |p,m\rangle \ ,\nn\\
&&E\, |p,m\rangle  = [p-m-1]\, |p,m+1\rangle \ ,\nn\\
&&F\, |p,m\rangle  = [m]\,\, |p,m-1\rangle
\lb{Uqprop2}
\ea
(in particular, all basis vectors (\ref{basis2}) are eigenvectors of $K$).
The transposition (\ref{prim}) is the linear transformation acting on the ${\cal M}_q\,$ generators as

\be
\lb{transp2}
(q^{\hat p})' = q^{\hat p}\ ,\quad (a^i_\a)' = {\epsilon}_{i\! j}\, {\e}^{\a\b} a^j_\b\ ,\quad{\rm i.e.}
\quad (a^1_1)' = q^{\frac{1}{2}} a^2_2\ ,\ \ (a^1_2)' = - q^{-\frac{1}{2}} a^2_1\ .
\ee
The $U_q(s\ell(2))\,$ generators $E\,$ and $K\,$ and their transposed (\ref{'}) are expressed as bilinear combinations in $a^j_\a$:
\ba
&&E = - q^{-\frac{1}{2}} a^1_1 a^2_1\ , \qquad q^{-1} F K = q^{\frac{1}{2}}  a^1_2 a^2_2 = E'\ ,\nn\\
&&K = q^{\frac{1}{2}} a^2_2 a^1_1 - q^{-\frac{1}{2}} a^1_1 a^2_2 =
q^{\frac{1}{2}} a^1_2 a^2_1 - q^{-\frac{1}{2}} a^2_1 a^1_2 = K'\ .
\lb{EFH}
\ea
The algebraic relations (\ref{EFH})
(derived in Appendix A of \cite{FHIOPT}) are valid in the Fock space representation, cf. (\ref{apmn2}) and (\ref{Uqprop2}).
Note that neither $F\,$ alone nor $K^{-1}$ appear; the generators $E, E', K\,$ obey the relation
$q\, E E' - q^{-1} E' E = \frac{K^2 - 1}{\l}\,.$

\medskip

To compute the inner product (\ref{dual}) of the basis vectors (\ref{basis2}), we first observe that
$\langle p', m' | p , m \rangle $ vanishes if either $p' \ne p$ or $m' \ne m$ (this follows easily
from (\ref{transp2}), (\ref{detc-n2}) and (\ref{a-vac})). Then we can apply directly (\ref{apmn2}) to obtain\footnote{
For generic $q\,,$ this result proves (\ref{scsq}) as $|p, p-1\rangle\,$ and $|p,0\rangle\,$
are the highest and lowest weight vector of ${\cal F}_p\,,$ respectively.}
\be
\lb{bilin2}
\langle p', m' | p , m \rangle = \d_{pp'}\, \d_{mm'}\, q^{m(m+1-p)} [m]! [p-m-1]!\ .
\ee
Thus all vectors $|p,m\rangle$ are mutually orthogonal, and the only ones that have non-zero scalar squares
are those for which
\be
1\le p \le h\ ,\ \ 0\le m\le p-1 \quad{\rm or}\quad h+1\le p\le 2h-1\ ,\ \ p-h\le m\le h-1\ \ .
\lb{nzscsq}
\ee
It is easy to see that conditions (\ref{nzscsq}) determine a $h^2$-dimensional subspace of ${\cal F}$ isomorphic to ${\cal F}^{(h)}$ (\ref{Fock-h}).

\subsection{The restricted quantum group}

\subsubsection{Action of $U_q (s\ell(2))\,$ on the zero modes' Fock space ${\cal F}$}

According to the general relations displayed in Appendix B.1, the QUEA $U_q\equiv U_q (s\ell(2))\,$
is a Hopf algebra with generators $E\,,\, F\,$ and $K^{\pm 1}\,$ satisfying
\ba
&&K E K^{-1} = q^2 E\ , \quad K F K^{-1} = q^{-2} F\ ,\quad K K^{-1} = K^{-1} K = \id\ ,\nn\\
&&[E , F ] = \frac{K-K^{-1}}{q-q^{-1}}
\lb{Uqsl2-alg}
\ea
and coalgebra structure defined by
\ba
&&\Delta (K) = K \otimes K \ , \ \ \Delta (E) = E \otimes K + \id\, \otimes E\ ,\ \ \Delta (F) = F \otimes \id + K^{-1} \otimes F \ ,\nn\\
&&\varepsilon (K) = 1 \ , \quad \varepsilon (E) = \varepsilon (F) = 0 \ ,\nn\\
&&S(K) = K^{-1} \ , \quad  S(E) = - E \, K^{-1} \ , \quad S(F) = - K \, F\ .
\lb{coalg2}
\ea
It is easy to see, however, that its representation on the Fock space ${\cal F}\,$ (\ref{Uqprop2}) is subject to the additional relations
\be
E^h = 0 = F^h \ ,\quad K^{2h} = \id\ .
\lb{Uq-res}
\ee
The quotient Hopf algebra defined by (\ref{Uqsl2-alg}), (\ref{Uq-res}) and (\ref{coalg2}) has been introduced in \cite{FGST1} under the name
of the {\em restricted quantum group} $\bU (s\ell (2))\,.$ As we only consider the $n=2$ case, we shall denote it for brevity as just $\bU\,.$

It is clear that $\bU$ is finite dimensional: the commutation relations (\ref{Uqsl2-alg}) allows any monomial in the generators
to be expressed in terms of ordered ones and (\ref{Uq-res}) restrict the maximal powers, so its dimension is $2h^3\,.$
A Poincar\'e-Birkhoff-Witt (PBW) basis is provided e.g. by the elements
\be
 E^\mu F^\nu K^n\quad {\rm for} \quad 0\le \mu ,\nu \le h-1\ ,\ \ 0\le n\le 2h-1 \ .
\lb{PBW-Uqres}
\ee
As $q^{2h} = 1\,,$ the element $K^h$ belongs to the {\em centre} ${\cal Z}$ of $\bU\,.$

It is customary (see e.g. \cite{CP}) to define, up to rescaling,  the Casimir operator in the deformed case as
\be
C = \l^2\, F E + q K + q^{-1} K^{-1} \, (\, = \l^2\, E F + q^{-1} K + q K^{-1}\,)\, \in {\cal Z}\ ,\quad \l = q-q^{-1}\ .
\lb{C}
\ee
Evaluating (\ref{C}) on the basis vectors $| p, m\rangle$ by using (\ref{Uqprop2})
and taking into account (\ref{basis2}) and (\ref{FFp-dim}), one obtains
\be
(C - q^p - q^{- p}) \, \, {\cal F}_p = 0 \qquad \Rightarrow \qquad (C - q^{\hp} - q^{- \hp}) \, \, {\cal F} = 0 \ .
\ee

\smallskip

The representation theory of $\bU$ has been thoroughly studied in \cite{FGST1, FGST2}.
It has a finite set of irreducible representations which is easy to describe.
It is clear from (\ref{Uq-res}) that the dimension of an IR cannot exceed $h\,$ (abusing notation, we shall denote it again by $p$).
Further, the spectrum of $K$ in a $p$-dimensional IR is non-degenerate and coincides with a set of the type
\be
S^{(p)}_\ell := \{ q^\ell , q^{\ell + 2} , \dots , q^{\ell + 2p-2} \}\quad
(\, \ell \in {\mathbb Z}\ ,\ -h+1 \le \ell \le h\ ,\ 1\le p \le h\, )\ ,
\lb{K-spec}
\ee
the first and the last eigenvalue corresponding to the lowest and highest weight vector, respectively
(the fact that the spectrum only contains integer powers of $q$ follows from the last equation in (\ref{Uq-res})).
Evaluating the Casimir operator (\ref{C}) on these two vectors imposes the following restriction on $\ell\,:$
\be
q^{\ell -1}+q^{-\ell +1 } = q^{\ell + 2p -1} + q^{-\ell - 2p +1} \quad\Rightarrow\quad \ell + p = 1\,{\rm mod}\, h\ .
\lb{CLH}
\ee
For a fixed dimension $p\,,$ (\ref{CLH}) has two solutions for $\ell\,$ in the allowed range, $\ell_+ = 1-p\,$ and $\ell_- = 1+h-p\,$
(the corresponding lowest weights, and therefore all weights, differ in sign: $q^{\ell_-} = - q^{\ell_+}$).
So there are $2h$ (equivalence classes of) irreducible representations $V_p^\pm\,$ of $\bU$ labeled by their highest weight $\pm q^{p-1}\,:$
\ba
&&V_p^\epsilon\, :\quad {\rm spec}\, K = \epsilon\, \{ q^{1-p} , q^{3-p} , \dots , q^{p-1} \}\ ,\quad p = 1,2,\dots , h\ ,\quad \epsilon = \pm\ ,\nn\\
&&\dim\, V_p^\epsilon = p\ ,\quad {\rm qdim}\, V_p^\epsilon := {\rm Tr}_{{\cal V}_p^\epsilon} K =
\epsilon\, [ p ]\ ,\quad (C - \epsilon (q^p + q^{-p})) \, V_p^\epsilon = 0\ .\qquad\qquad
\lb{specK-Vp}
\ea
We shall refer to the sign $\epsilon\,$ as to the {\em parity} of the IR $V_p^\epsilon\,.$ By (\ref{specK-Vp}) and (\ref{C}),
a characterization of a canonical basis $\{ v_{p, m}^\epsilon \}\,$ in $V_p^\epsilon\,$ invariant under a rescaling
$E \to \rho\, E\,,\ F \to \rho^{-1}\, F\ \,(\rho > 0)\,$ which preserves all defining relations (\ref{Uqsl2-alg}), (\ref{coalg2}),
is provided by the relations
\ba
&&(K - \epsilon\,q^{2m-p+1}) \, v_{p, m}^\epsilon  = 0 \qquad (\, 1\le p \le h\ ,\ \ 0\le m\le p-1 \, )\ ,\lb{EFK-eps}\\
&&(EF - \epsilon \,[m][p-m] )\, v_{p, m}^\epsilon  = 0 = (FE - \epsilon \,[m+1][p-m-1] )\, v_{p, m}^\epsilon \ .\nn
\ea

Returning to the Fock space representation of $\bU\,$ we see that ${\cal F}_p \simeq V^+_p\,$ for $1\le p \le h\,$ while the negative parity IR
first appear as subrepresentations of the spaces ${\cal F}_{h+p}\,,$ each of which contains {\em two} irreducible submodules isomorphic to
$V^-_p\,$ spanned by $\{ | h + p , m {\cal i} \}\,$ and $\{ | h + p , h+ m {\cal i} \}\,$ for $m=0,\dots , p - 1\,,$ respectively.
For $1\le p \le h-1\,$ the quotient of ${\cal F}_{h+p}\,$ by the direct sum of invariant subspaces is isomorphic to $V_{h-p}^+\,$ or, in terms of exact sequences,
\be
0\ \rightarrow\ V^-_p \oplus V^-_p\ \rightarrow\ {\cal F}_{h+p}\ \rightarrow V^+_{h-p}\ \rightarrow\ 0\ .
\lb{shexseq}
\ee
For $p = h\,$ the two negative parity submodules exhaust the content of ${\cal F}_{2h} = V^-_h \oplus V_h^-\,.$
More generally, the $\bU\,$ module structure of ${\cal F}_{Nh+p}\,$ for $N\in {\mathbb Z}_+\,$ and $1\le p \le h\,$ is described by the short exact sequence \cite{FHT7}
\ba
&&0\ \ \rightarrow\ \underbrace{V^{\epsilon (N)}_p \oplus V^{\epsilon (N)}_p\dots \oplus V^{\epsilon (N)}_p}
\ \ \rightarrow\ \ {\cal F}_{Nh+p}\ \
\rightarrow\ \  \underbrace{V^{-\epsilon (N)}_{h-p}\oplus \dots \oplus V^{-\epsilon (N)}_{h-p}}\ \rightarrow\ 0\ ,\nn\\
&& \hspace{22mm} \#\, (N+1) \hspace{58mm}\#\, N
\lb{shexseqN}
\ea
where $\epsilon (N)= (-1)^N$ is the parity of the integer $N\,$ and $V^\pm_0 := \{ 0 \} \,$
(we have $N+1\,$ submodules $V^{\epsilon (N)}_p\,$ and a quotient module which is a direct sum of $N\,$ copies of $V^{-\epsilon (N)}_{h-p}\,$).

The subquotient structure of ${\cal F}\,$ as a representation space of $\bU\,$ for $h=3\,$ is displayed on Figure 1 below.

\bigskip

%%%%%%%%%%%%
\begin{figure}[htb] \centering \includegraphics*[bb=10 190 580 620, width=340pt]{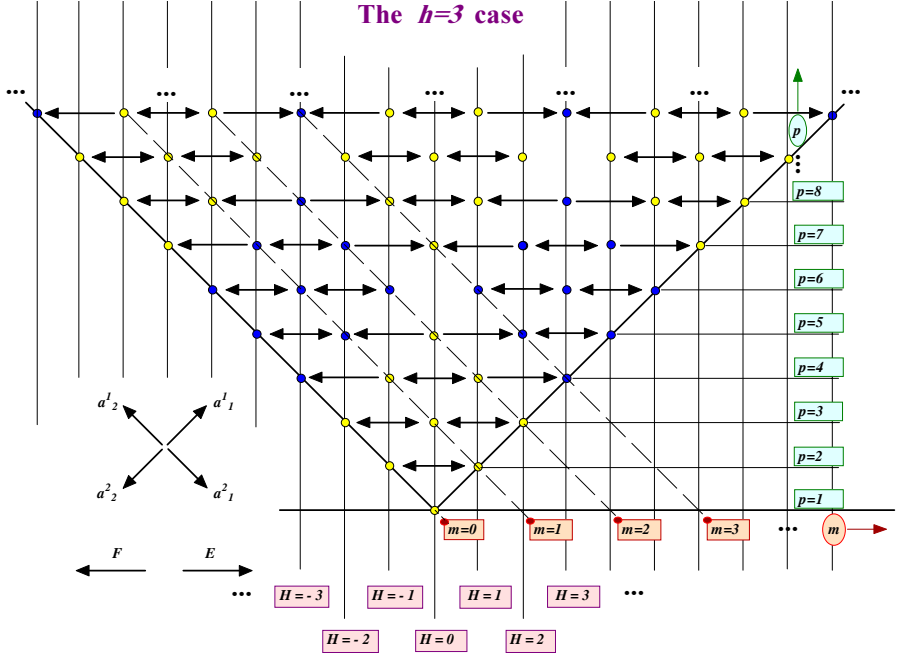}
\caption{\footnotesize{The $\bU$ representation on the Fock space ${\cal F}\,$ for $q= e^{\pm i\frac{\pi}{3}}\,.$
Vectors belonging to subquotients of type $V_p^+\,$ (for some $p\,$) are represented by yellow circles
($\circ\,$ in black and white print) and those belonging to $V_p^-\,,$ by blue ones (${\bullet}\,$ in BW).}
The eigenvalues of $K = q^H\,$ can be read off from those of $H\,.$} \end{figure}
%%%%%%%%%%%%%

\subsubsection{Quasitriangular  twofold cover $\bbU\,$ of $\bU$}

In accord with the consideration carried in Section 4.3, the Gauss components of the monodromy matrix $M_\pm\,$ for $n=2\,$ can be parametrized
in terms of the twofold cover $U_q^{(2)}(s\ell(2))\,$ of $U_q(s\ell(2))\,$ with Cartan element $k$ satisfying
\ba
&&k\,E = q E\, k\ ,\quad k\, F = q^{-1} F\, k\ ,\quad [E , F ] = \frac{k^2-k^{-2}}{q-q^{-1}}\quad (\, k^2 = K \,)\ ,\nn\\
&&\Delta (k) = k\otimes k\ ,\quad \varepsilon (k)=1\ ,\quad S(k) = k^{-1}\ .
\lb{dk2}
\ea
By (\ref{Uqvac}) and (\ref{tens-op}) we obtain the action of its generators on the basis (\ref{base2})
which are of course the same as in (\ref{Uqprop2}), except for
\be
k\, |p,m\rangle = q^{m - \frac{p-1}{2}} |p,m\rangle \ .
\lb{kprop2}
\ee
Restricting the Hopf algebra $U_q^{(2)}(s\ell(2))$ by the ensuing additional relations
\be
E^h = 0 = F^h \ ,\quad k^{4h} = \id
\lb{bUq-res}
\ee
one obtains the $4h$-dimensional double cover $\bbU$ of $\bU\,$ with a PBW basis provided by the elements
\be
E^\mu F^\nu k^n\ ,\quad  0\le \mu ,\nu \le h-1\ ,\ \ 0\le n\le 4h-1 \ .
\lb{PBW-Uqres2}
\ee
The important property of $\bbU\,$ is that it is {\em quasitriangular} i.e., there exists a universal $R$-matrix (\ref{intR})
${\cal R}\in \bbU \otimes \bbU$ satisfying (\ref{qtr}), while $\bU\,$ itself is not.

By contrast, $\bU\,$ (but not $\bbU$) is a {\em factorizable} Hopf algebra which means that the (universal)
monodromy matrix ${\cal M} = {\cal R}_{21} {\cal R}\,$ belongs to $\bU \otimes \bU$ and has maximal rank
($2 h^3 $), see Appendix B.3. A hint to this feature is provided by the following observation.
Using (\ref{M+-q}) for $n=2\,,$ as well as (\ref{MpmNpmD}), (\ref{MD2}) and (\ref{dk2}),
we deduce that the entries of the monodromy matrix $M\,$ only contain $K \in \bU\,$ and not its "square root"
$k \in \bbU\,:$
\ba
&&q^{\frac{3}{2}} M = M_+ M_-^{-1} =
\begin{pmatrix} k^{-1}& -\l\, F k \cr 0&k \end{pmatrix}
\begin{pmatrix}k^{-1}&0\cr - \l\, E k^{-1} & k \end{pmatrix} =\nn\\
&&= \begin{pmatrix}q\l^2 FE + K^{-1} & -\l\, F K \cr - q \l\, E& K \end{pmatrix}\ , \qquad \l = q - q^{-1}\ .
\lb{calcM2}
\ea

As the Hopf algebras under consideration are finite dimensional, all the constructions are purely algebraic.
An efficient way of finding the universal $R$-matrix is the {\em Drinfeld double} construction \cite{D, RS, Ka, Ma}
since the double of any Hopf algebra is canonically quasitriangular (and factorizable).
The quasitriangularity of $\bbU$ follows from the fact that it is a quotient of the ($16 h^4$-dimensional)
double of any of its Borel Hopf-subalgebras \cite{FGST1, FHT7}\footnote{The conventions in the journal paper
\cite{FHT7} are updated in its last arXiv version and coincide with those adopted here.}, see Appendix B.2.
We start e.g. with the $4h^2$-dimensional Hopf algebra $U_q(\bo_+)\,$ generated by $F\,$ and $k_+\,$ to find
$U_q(\bo_-)\,$ generated by $E\,$ and $k_-\,$ as its dual, and put at the end $k_+ = k_- =: k\,.$
In such a way we derive the (lower triangular) universal $R$-matrix of $\bbU$ given by the triple sum
\be
{\cal R} = \frac{1}{4h}\, \sum_{\nu = 0}^{h-1}
\frac{q^{-\frac{\nu (\nu -1)}{2}} (-\l)^\nu}{[\nu ]!}\,
F^\nu\otimes E^\nu \sum_{m ,\,n=0}^{4h-1} q^{\frac{mn}{2}} k^m \otimes k^n  \ \in\, \bbU\otimes\bbU\ .
\lb{RbD}
\ee
This expression allows to recover the $4\times 4$ matrix $R_{12}$ (\ref{R}), given explicitly in this case by
\be
R_{12} = q^{\frac{1}{2}} \begin{pmatrix}q^{-1}&0&0&0\cr 0&1&0&0\cr 0&-\l&1&0\cr 0&0&0&q^{-1}\end{pmatrix}\ ,
\lb{R2}
\ee
from the universal $R$-matrix (\ref{RbD}) by taking the generators of $\bbU$ in the $2$-dimensional representation $\pi_f$:
\be
E^f = \begin{pmatrix}0&1\cr0&0\end{pmatrix}\,,\quad
F^f = \begin{pmatrix}0&0\cr1&0\end{pmatrix}\,,\quad
k^f = \begin{pmatrix}q^{\frac{1}{2}}&0\cr0&q^{- \frac{1}{2}}\end{pmatrix}\ .
\lb{bUf}
\ee
Indeed, using $(E^f)^2 = 0 = (F^f)^2\,$ and the summation formula
\be
\sum_{m=0}^{4h-1} q^{\frac{mj}{2}} =
\left\{
\begin{array}{ll}
4h \ &{\rm for}\ j = 0\ {\rm mod}\ 4h\\
0 \ & {\rm otherwise}
\end{array}
\right.\ ,
\lb{sum-m}
\ee
one obtains from (\ref{RbD}) and (\ref{bUf})
\ba
&&(\pi_f\otimes \pi_f )\, {\cal R} =  \frac{1}{4h}\,
\left( \id_2 \otimes \id_2 - \l\, F^f \otimes E^f \right)\,
\sum_{m ,\, n=0}^{4h-1} q^{\frac{mn}{2}} (k^f)^m\otimes (k^f)^n =\nn\\
&&= \begin{pmatrix}1&0&0&0\cr
0&1&0&0\cr
0&-\l&1&0\cr
0&0&0&1\end{pmatrix}
\begin{pmatrix}
q^{-\frac{1}{2}}&0&0&0\cr
0&q^{\frac{1}{2}}&0&0\cr
0&0&q^{\frac{1}{2}}&0\cr
0&0&0&q^{-\frac{1}{2}}\end{pmatrix}
= R_{12}\ .
\ea
Remarkably, the expression for the universal monodromy matrix ${\cal M} = {\cal R}_{21} {\cal R}\,,$
\be
{\cal M} =\frac{1}{2h}\,\sum_{\mu ,\nu = 0}^{h-1}
\frac{(-\l)^{\mu+\nu}q^{\frac{\nu(\nu+1)-\mu(\mu-1)}{2}}}{[\mu ]![\nu ]!}\,
\sum_{m,\,n = 0}^{2h-1} q^{mn +\nu(n-m) }
E^\mu F^\nu k^{2m} \otimes F^\mu E^\nu k^{2n}
\lb{Mmatr}
\ee
only contains even powers of $k$ and hence, belongs to $\bU \otimes \bU \,.$ Moreover, ${\cal M}\,$ (\ref{Mmatr}) is of the type
(\ref{Mm}) so that $\bU\,$ is factorizable. This is the reason why we shall be interested mainly in $\bU\,$ in what follows, with
$\bbU\,$ playing an auxiliary role providing the universal $R$-matrix ${\cal R}$ in terms of which ${\cal M}\,$ is constructed.

\medskip

\noindent {\bf Remark 5.1~} The other admissible (upper triangular) universal $R$-matrix of $\bbU$ is found by
exchanging the places of $U_q(\bo_+)\,$ and $U_q(\bo_-)\,$ in the double and has the following form:
\be
{\cal R}_{21}^{-1} = \frac{1}{4h}\, \sum_{m ,\,n=0}^{4h-1} q^{-\frac{mn}{2}} k^m \otimes k^n \sum_{\nu = 0}^{h-1}
 \frac{q^{\frac{\nu (\nu -1)}{2}} \l^\nu}{[\nu ]!}\, E^\nu\otimes F^\nu \ .
\lb{RbD21}
\ee
It gives rise to the inverse of the monodromy matrix ${\cal M}^{-1} = {\cal R}^{-1} {\cal R}^{-1}_{21}\,.$

\medskip

It is instructive to note that the matrix (\ref{calcM2}) is equal to $(\pi_f\otimes id)\,{\cal M}\,.$ To verify this we observe that,
due to the nilpotency of $E^f$ and $F^f\,,$ one is left in the first sum in (\ref{Mmatr}) with the terms with $\mu ,\nu = 0,1$ only:
\ba
&&(\pi_f\otimes id)\,{\cal M} = \frac{1}{2h}\,\sum_{m,\, n=0}^{2h-1}
(q^{mn}\,\id_2 \otimes \id  - \l\, q^{mn+n-m+1} F^f\otimes E - \nn\\
&&- \l\, q^{mn} E^f\otimes F + \l^2 q^{mn+n-m+1} E^f F^f \otimes F E )\, (K^f)^m\otimes K^n=\nn\\
&&\nn\\
&&= \frac{1}{2h}\,\sum_{m,\,n=0}^{2h-1}\,
\begin{pmatrix}(q^{m(n+1)} + \l^2 q^{mn+n+1} FE)\, K^n & -\l\, q^{m(n-1)} F K^n \cr
-\l\, q^{mn+n+1} E K^n & q^{m(n-1)} K^n \end{pmatrix}\ .\qquad\qquad
\lb{calcM}
\ea
(We have applied (\ref{bUf}) from which it follows that
\be
\lb{EFKf}
E^f F^f = \begin{pmatrix} 1&0\cr0&0 \end{pmatrix}\ ,\qquad (K^f)^m = \begin{pmatrix}q^m&0\cr0&q^{-m}\end{pmatrix}
\ee
and evaluated the tensor product as a Kronecker product of matrices.) Proceeding with the summation in $m$
and using $\sum_{m=0}^{2h-1} q^{m j} = 2h\, \d_{j\,,\, 0\, {\rm mod}\, 2h}\,,$ we finally obtain
that (\ref{calcM}) indeed coincides with (\ref{calcM2}):
\be
(\pi_f\otimes id)\,{\cal M} = \begin{pmatrix}q\l^2 FE + K^{-1} & -\l\, F K \cr - q \l\, E& K \end{pmatrix} = q^{\frac{3}{2}} M \ .
\lb{piidM}
\ee

\subsubsection{The factorizable Hopf algebra $\bU$ and its Grothendieck ring}

A partial information about indecomposable representations is provided by their content
in terms of irreducible modules, independently of whether they appear as its submodules or subquotients.
It is captured by the concept of the {\em Grothendieck ring} (GR).
We write $\pi = \pi_1 + \pi_2\,$ if one of the representations in the right hand
side is a submodule of $\pi\,$ while the other is the corresponding quotient representation,
and complete the structure to that of an abelian group by introducing formal differences
(so that e.g. $\pi_1 = \pi - \pi_2\,$) and zero element, given by the vector $\{ 0 \}\,.$
To define the GR multiplication, we start with the tensor product of the IR $\pi_{V_1}\,$ and $\pi_{V_2}\,$
(with representation spaces $V_1$ and $V_2\,,$ respectively) defined by means of the coproduct,
\be
\pi_{V_1\otimes V_2}  := (\pi_{V_1}\otimes\, \pi_{V_2})\, \Delta
\lb{tens-ring}
\ee
and further, represent each of the (in general, indecomposable) summands in the expansion by the GR sum of its irreducible submodules
and subquotients (thus "forgetting" its indecomposable structure). By a construction due to Drinfeld \cite{D3}, the GR of the $\bU\,$
representations turns out to be equivalent to a subring of its centre generated by the Casimir operator $C\,$ (\ref{C}).

\smallskip

Let ${\mathfrak A}\,$ be a factorizable Hopf algebra with monodromy matrix ${\cal M}$; then there is an isomorphism between
the (commutative) algebra of the {\em ${\mathfrak A}$-characters}
\be
{\mathfrak C}{\mathfrak h}  := \{\, \phi \in {\mathfrak A}^*\, |\ \phi (x y) = \phi (S^2(y) x)
\ \ \forall\, x,y\in {\mathfrak A} \}
\lb{Ch-Ad*inv}
\ee
and the centre ${\cal Z} \in {\mathfrak A}\,,$ given by the {\em Drinfeld map} \cite{D3, FGST1}
\be
{\mathfrak A}^*\ \rightarrow\ {\mathfrak A}\,,\qquad \phi\ \mapsto\ (\phi\otimes id ) ({\cal M})
\lb{Dr-map}
\ee
(see Appendix B.3). Let further $g\,$ be a {\em balancing element}\footnote{The existence of a balancing element is not granted,
and it may be not unique. An element $g\in\fU$ satisfying the first relation (\ref{balance}) is called "group-like".}
of $\fU\,,$ i.e. an element satisfying
\be
g\in {\mathfrak A}\ ,\qquad \Delta (g) = g\otimes g\ ,\qquad S^2 (x) = g\, x\, g^{-1}\quad\forall x\in {\mathfrak A}\ .
\lb{balance}
\ee
Then any finite dimensional representation $\pi_V\,$ of ${\mathfrak A}$ (with representation space $V$) gives rise to a
${\mathfrak A}$-character $Ch_V^g\,$ defined by the {\em $q$-trace}
\be
Ch_V^g\, (x) := {\rm Tr}_{\pi_V} (g^{-1} x)\qquad \forall x\in {\mathfrak A}\ ;
\lb{canCh}
\ee
any $Ch_V^g$ belongs indeed to ${\mathfrak C}{\mathfrak h}$ (\ref{Ch-Ad*inv}) since
\be
Ch_V^g\, (S^2(y)\, x) = {\rm Tr}_{\pi_V} (g^{-1} S^2(y)\, x) = {\rm Tr}_{\pi_V} (y g^{-1} x) = Ch_V^g\, (xy)\ .
\lb{canch}
\ee
The corresponding {\em Drinfeld images}
\be
D (\pi_V) := (Ch_V^g\otimes id ) ({\cal M)} \in\, {\cal Z}\quad
\lb{D-im}
\ee
form a subring of the centre ${\cal Z}$ isomorphic to the GR.

\smallskip

We shall use the factorizability of $\bU\,$ to explore the GR ${\mathfrak S}_{2h}\,$ generated by its IR.
It is easy to see that both $K$ and $K^{h+1}$ satisfy the conditions (\ref{balance}); note that $K^h \in {\cal Z}\,.$
Choosing $K$ as balancing element for $\bU\,,$ the Drinfeld image of the
$2$-dimensional representation $\pi_f\,$ (\ref{bUf}) is just the Casimir operator (\ref{C}):
\be
\lb{ChKM}
(Ch_{\pi_f}^K\otimes id ) ({\cal M)} = C \qquad{\rm for}\qquad Ch_{\pi_f}^K (x) = {\rm Tr}_{\pi_f} (K^{-1} x)\ .
\ee
The computation of (\ref{ChKM}) amounts to applying (\ref{piidM}) and (\ref{EFKf}):
\ba
&&{\rm Tr} ((K^f)^{-1} (\pi_f\otimes id)\,{\cal M}) = {\rm Tr} \left\{ \begin{pmatrix}q^{-1}&0\cr0&q\end{pmatrix}
\begin{pmatrix}q\l^2 FE + K^{-1} & -\l\, F K \cr - q \l\, E& K \end{pmatrix} \right\} =\nn\\
&&\nn\\
&&= \l^2 FE + q K + q^{-1}K^{-1}  = C\ .
\lb{Tr1}
\ea
The alternative choice of  $K^{h+1}\,$ as balancing element (cf. Eqs. (3.3) and (4.7) of \cite{FGST1}) leads to the opposite sign in
(\ref{Tr1}) since $(K^f)^h = - \id_2\,.$

It follows from (\ref{specK-Vp}) that the $q$-dimension of an IR (and hence, of any representation) of $\bU\,$
is just its $q$-trace evaluated at the unit element:
\be
{\rm qdim}\, V = {\rm Tr}_V K = {\rm Tr}_V K^{-1} = Ch_{V}^K (\id )\ .
\lb{Ch-qdim}
\ee

\smallskip

The following Proposition shows that the commutative algebra generated by the Casimir operator $C\,$ (\ref{C})
is $2h$-dimensional and contains the central element $K^h\,.$
As a preliminary step, we note that the following relations can be easily proved by induction in $r\,:$
\ba
&&\l^{2r} E^r F^r = \prod_{s=0}^{r-1} (C - q^{-2s-1} K - q^{2s+1} K^{-1})\ ,\nn\\
&&\l^{2r} F^r E^r = \prod_{s=0}^{r-1} (C - q^{2s+1} K - q^{-2s-1} K^{-1} )\ .
\lb{ErFr}
\ea
Recall also that the Chebyshev polynomials of the first kind are defined by
\be
T_m\, (\cos t)  =  \cos m\, t\qquad (\,{\rm deg}\, T_m = m\, )\ .
\lb{Cheby1}
\ee

\medskip

\noindent
{\bf Proposition 5.1~} {\em

\noindent
(a) The central element $K^h\,$ (of order $2$) is related to $C\,$ by
\be
K^h = - T_h (\frac{C}{2})\ .
\lb{qhHTh1}
\ee
\noindent
(b)
The Casimir operator (\ref{C}) satisfies the equation
\be
P_{2h}(C) := \prod_{s=0}^{2h-1} (C - \b_s )= 0\ , \qquad \b_s = q^s + q^{-s} = 2\,\cos \frac{s\pi}{h}\ .
\lb{P2h=0}
\ee}
\medskip

\noindent
{\bf Proof~}
Writing the formula
\be
\cos N t - \cos N y = 2^{N-1} \prod_{s=0}^{N-1} ( \cos t - \cos (y+\frac{2\pi s}{N} ) )
\lb{flaRG1}
\ee
(see, e.g., $1.395$ in \cite{GR}) for $2\,\cos t =: C\,$ and $e^{iy} =: Z\,$ such that $Z^{2N} = 1 \,$
(and hence, $Z^N = \cos N y $), and applying it to the case when $C\,$ (given by (\ref{C})) and $Z\,$ are
commuting operators in a finite dimensional space, we find
\be
2\, (T_N(\frac{C}{2}) - Z^N ) = \prod_{s=0}^{N-1} ( C - e^{\frac{2\pi i s}{N}} Z - e^{-\frac{2\pi i s}{N}} Z^{-1}  ) \ .
\lb{TNZ}
\ee
Two special cases of (\ref{TNZ}): a)\ $N=h\,,\ Z = q^{-1} K\,$  and b)\ $N=2h\,,\ Z = 1\,$ give
\be
2\, (\, T_h (\frac{C}{2}) + K^h ) = \prod_{s=0}^{h-1} (C-q^{-2s-1} K -q^{2s+1} K^{-1})
\lb{qhHTh}
\ee
and
\be
2\, (\, T_{2h} (\frac{C}{2}) - 1 ) \equiv 4\, \bigl( (T_{h} (\frac{C}{2}) )^2 - 1 \bigr) = P_{2h} (C)\ ,
\lb{P2hT2h}
\ee
respectively. Setting in (\ref{ErFr}) $r=h\,$ and using (\ref{Uq-res}), we deduce that the product in (\ref{qhHTh}) vanishes, proving thus (a).
Further, (b) follows from (\ref{P2hT2h}), (\ref{qhHTh1}) and (\ref{Uq-res}):
\be
P_{2h}(C)\,= 4\, (\, K^{2h} - 1 ) = 0\ .
\lb{directP2h}
\ee
\eod

Since $D\,$ maps isomorphically the $\bU\,$ GR ${\mathfrak S}_{2h}\,$ to a $2h$-dimensional subring of the centre,
${\mathfrak S}_{2h} \ \stackrel{D}{\longrightarrow} \ D({\mathfrak S}_{2h}) \subset {\cal Z}_q\,,$
the algebra of the corresponding central elements $D(V_p^\epsilon )\,$ provides, in turn, a convenient description of the Grothendieck fusion.
As a representation of $\bU\,,$ $\pi_f\,$ (with Drinfeld image $C$ (\ref{Tr1})) coincides with the IR $V_2^+\,$
(see (\ref{specK-Vp})). It is not difficult to derive the expressions for the Drinfeld images of all the IR of $\bU\,.$
This is done in Appendix B.3 (see Proposition B.1), following \cite{FGST1, FHT7}. In principle, it is possible to find
the $\bU\,$ GR ring structure from the explicit expressions (\ref{DrVp1}). We shall follow however another path.

Albeit the GR of $\bU\,$ is finite, the Fock space representation makes it natural to express its multiplication rules
in terms of the {\em infinite} number of representations ${{\cal F}_p}\,$ which are of $su(2)$ type:
\be
D({\cal F}_p) \, .\, D({\cal F}_{p'}) = \sum_{{p^{\prime\prime} = |p - p'| + 1}\atop{p^{\prime\prime}-p-p'= 1\, mod\, 2}}^{p+p'-1}\,
D({\cal F}_{p^{\prime\prime}})\ ,\qquad\quad p=1, 2,\dots \ .
\lb{n5}
\ee
The justification of (\ref{n5}) takes into account the well known fact that an analogous decomposition holds for tensor products
of the (irreducible) representations ${\cal F}_p\,$ for generic $q\,;$ {\em in the GR context} it should remain true after specializing
$q\,$ to a root of unity as well. Note that the GR content of ${\cal F}_{Nh+p}\,$ for $N\in {\Z}_+\,,\ 1\le p\le h\,$ which replaces the
precise indecomposable structure (\ref{shexseqN}),
\be
{\cal F}_{Nh+p} = (N+1)\, V^{\epsilon(N)}_p + N\, V^{-\epsilon(N)}_{h-p}
\lb{GRpb}
\ee
obeys the following "parity rule": one always has an odd number of irreducible $\bU\,$ modules of type $V^+\,$
and an even number of modules of type $V^-\,.$

Assuming that (\ref{n5}) holds, we shall make use of the following corollary of Proposition B.1.

\medskip

\noindent
{\bf Corollary 5.1~} {\em The Drinfeld images of the $\bU\,$ IR
\be
d^\epsilon_p := D (V^\epsilon_p) = ({\rm Tr}_{\pi_{V^\epsilon_p}} K^{-1}\otimes id ) \, {\cal M} \in {\cal Z}\ ,\quad
1\le p\le h\ ,\quad \epsilon = \pm
\lb{Dr-Vp}
\ee
satisfy}
\be
d^+_1 = 1 \ ,\qquad d^+_2 = C\,,\qquad d^{- \epsilon}_p = -\, K^h d_p^{\,\epsilon} = T_h (\frac{C}{2})\, d^{\,\epsilon}_p\ .
\lb{Drinfeld12}
\ee

\medskip

From (\ref{n5}) for $p' = 2\,$ and (\ref{Drinfeld12}) one concludes that $D({\cal F}_p)\,$ are functions of $C\,$ satisfying both
the recurrence relations and the initial conditions for the Chebyshev polynomials of the second kind $U_p (x)\,,$ defined by
\be
U_{m+1} (x) = x\, U_m (x) - U_{m-1} (x) \,,\quad m\ge 1\,,\quad U_0 (x) = 0\,,\quad U_1 (x) = 1
\lb{recurseUm}
\ee
and hence,
\be
D ({\cal F}_p ) = U_p (C)\ ,\qquad p\in {\Z}_+\ .
\lb{Dr-VP}
\ee
It follows from (\ref{recurseUm}) that $U_m(x)\,$ are monic polynomials of $\deg\, U_m = m-1$ and
\be
U_m (2\,\cos t) = \frac{\sin m t}{\sin t}\ ,\qquad U_2 (x) = x\ ,\qquad U_m (2) = m\ .
\lb{Um}
\ee
Using (\ref{GRpb}) for $N=0\,$ and $N=1\,,$ one sees that the Drinfeld images (\ref{Dr-Vp}) of the $\bU\,$ IR are given by
\be
d^+_p = U_p (C)\ ,\qquad
d^-_p = \frac{1}{2}\, (U_{h+p} (C) - U_{h-p} (C) )\ ,\qquad 1\le p\le h\ .
\lb{DR-gen}
\ee
By (\ref{Cheby1}) and (\ref{Um}), the trigonometric relation  $2 \sin t \cos m \, t = \sin (m+1)\, t - \sin (m-1)\, t\,$ is equivalent to
\be
2\,T_m (\frac{x}{2}) = U_{m+1} (x) - U_{m-1} (x) \ ,
\lb{TU}
\ee
so that the condition (\ref{P2hT2h}), (\ref{directP2h}) is converted in terms of $U_m(x)$ to the equality
\be
T_{2h} (\frac{C}{2}) = 1\quad\Leftrightarrow\quad U_{2h+1}(C) - U_{2h-1}(C) - 2 = 0\ .
\lb{T2h=1}
\ee
Eq.(\ref{T2h=1}) ensures the consistency between (\ref{DR-gen}) and the IR content of ${\cal F}_{2h+1}$ (\ref{GRpb}):
\be
U_{2h+1}(C) = D ({\cal F}_{2h+1}) = 3\, D (V^+_p) + 2\, D (V^-_{h-p}) = U_{2h-1}(C) + 2\, U_1(C)\ .
\lb{F2h+1}
\ee
One can check that the fusion of (\ref{F2h+1}) with $U_2(C)\,$ justifies, step by step, the consistency of the
representation (\ref{DR-gen}) for any ${\cal F}_{Nh+p}\,,\ N\ge 2\,,$ i.e. no additional conditions appear.
As $U_m(x)\,,\ m\in {\mathbb Z}_+\,$ span the polynomial ring ${\mathbb C} [x]\,,$ the $\bU\,$ GR is equivalent
to the quotient ring of ${\mathbb C} [C]\,$ modulo the ideal generated by the polynomial (\ref{T2h=1}) \cite{FGST1}.

It is elementary to derive from (\ref{n5}) and (\ref{GRpb}) the multiplication rules for the GR images
(in terms of the $\bU\,$ IR) which, as it has been shown in \cite{FGST1}, read
\ba
&&D(V_p^\epsilon)\, .\, D(V_{p'}^{{\epsilon}'}) =
\sum_{{s = |p - p'| + 1}\atop{s-p-p' = 1\, mod\, 2}}^{p+p'-1}\,
D({\widehat{V}}^{\epsilon{\epsilon}'}_{s})\,,\qquad 1\le p\,,\,p' \le h\,,\ \ \epsilon\,,\,\epsilon' = \pm\,\,,\qquad\nn\\
&&{\widehat{V}}_s^\epsilon = \left\{
\begin{array}{ll}
V^\epsilon_s&{\rm for}\ 1\le s\le h\\
V^\epsilon_{2h-s} + 2\,V^{-\epsilon}_{s-h}&{\rm for}\ h+1\le s\le 2h-1
\end{array}
\right.\quad .
\lb{GRres}
\ea
Indeed, Eq.(\ref{n5}) imply directly (\ref{GRres}) for $\epsilon = \epsilon' = +\,,$ and the cases when $\epsilon\,,\ \epsilon'$ or both
are of opposite sign follow from these by multiplying them with $T_h (\frac{C}{2})\,,$ see (\ref{Drinfeld12}), taking into account that
$(T_h (\frac{C}{2}))^2 = 1\,,$ cf. (\ref{P2hT2h}) and (\ref{directP2h}).
For a proof that (\ref{GRres}) imply in turn (\ref{n5}), see \cite{FHT7}.

\smallskip

Eq.(\ref{P2h=0}) can be regarded as the characteristic equation of
the Casimir $C\,$ as an operator on the subalgebra of the centre $D({\mathfrak S}_{2h}) \subset {\cal Z}$ generated by
the Drinfeld images of the $\bU\,$ IR. As the eigenvalues $\b_p = \b_{2h-p}\,$ are doubly degenerate for $1\le p \le h-1\,,$
\be
P_{2h} (C) = (C-2)(C+2)\prod_{p=1}^{h-1} (C - \b_p)^2 = 0\ ,\qquad \b_p = q^p + q^{-p}\ ,
\lb{P2h-2}
\ee
the spectral decomposition of $C\,$ is of Jordan type:
\be
C = 2\, e_0 - 2\, e_h + \sum_{p=1}^{h-1} (\b_p\, e_p + w_p)\ ,\quad (C-\b_p)\, e_p = w_p\ ,\quad (C-\b_p)\, w_p = 0\ .\qquad\quad
\lb{spC}
\ee
The primitive idempotents $e_s$ and nilpotents $w_p$ obey
\ba
&&e_r e_s = \d_{rs} e_r\ ,\quad e_r w_p = \d_{rp} w_p\ ,\quad w_p w_{p'}= 0\ ,\quad 0\le r,s \le h\ ,\ 1\le p,p' \le h-1
\nn\\
&&\Rightarrow\qquad f(C) = f(2)\, e_0 + f(-2)\, e_h + \sum_{p=1}^{h-1} (f(\b_p)\, e_p + f'(\b_p)\, w_p)\ .
\lb{Cew}
\ea
In particular, the coefficients of the idempotents $e_p\,,\ 1\le p\le h-1$ in the expansion of $U_s (C)\,$ are equal to
\be
U_s (\b_p) = U_s (2 \cos \frac{p \,\pi}{h} ) = \frac{\sin\frac{s p \,\pi}{h}}{\sin\frac{p \,\pi}{h}} = \frac{[s\, p]}{[p]} \ .
\lb{UpC}
\ee

\smallskip

The {\em unitary} WZNW model only includes {\em integrable} affine algebra representations \cite{DFMS}.
In the $\widehat{su}(2)_k\,$ case, the corresponding shifted weights are in the interval $1\le p \le h-1 \ (\equiv k+1)\,.$ It has been known from the early studies \cite{PS, FK}
that the fusion of the corresponding  "physical representations" of $U_q(s\ell(2))\,$
(for $q=e^{\pm i \frac{\pi}{h}}$) can be recovered from the ordinary $su(2)$ fusion by appropriately factoring out
representations of zero quantum dimension. As representations of $\bU\,,$ the latter form the ideal of {\em Verma modules}
\cite{FGST1, FGST2}. The latter are $h$-dimensional and include the two IR
${\cal V}_h^\epsilon := V_h^\epsilon\,,\ \epsilon = \pm\,$
as well as other $2h-2\,$ indecomposable representations with subquotient structure
\be
0\ \rightarrow\ V^\epsilon_p\ \rightarrow\ {\cal V}_p^\epsilon\ \rightarrow V^{-\epsilon}_{h-p}\ \rightarrow\ 0\ ,
\quad p=1,\dots, h-1\ .
\lb{Verma}
\ee
In the GR $\,{\cal V}_p^\epsilon\,$ and ${\cal V}_{h-p}^{-\epsilon}\,$ cannot be distinguished so
it is appropriate to use the notation
\be
{\cal V}_s := V^+_s + V^-_{h-s}\ ,\quad 0\le s \le h\qquad
(\,V^\pm_0 = \{ 0 \}\ ;\ \ {\cal V}_0 = V^-_h\ ,\ \ {\cal V}_h = V^+_h\,)\ .
\lb{VermaGR}
\ee

That ${\cal V}_s\,$ form an ideal in ${\mathfrak S}_{2h}\,$ is quite easy to prove using (\ref{GRres}), and ${\rm qdim} \, {\cal V}_s = 0\,$
follows from (\ref{specK-Vp}) since $[s] - [h-s] = 0\,.$ On the other hand, the Drinfeld images of the $h+1\,$ representations
(\ref{Verma}) are spanned by $e_0\,,\, e_h$ and $\{ w_p \}_{p=1}^{h-1}\,$ only, i.e. the corresponding coefficients of $\{ e_p \}_{p=1}^{h-1}\,$
in (\ref{Cew}) vanish. Indeed, by (\ref{Verma}) and (\ref{DR-gen}),
\ba
&&D({\cal V}_0) = D(V_h^-) = \frac{1}{2}\, U_{2h} (C)\ ,\qquad D({\cal V}_h) = D(V_h^+) = U_h(C)\ ,\nn\\
&&D({\cal V}_s) = D(V^+_s) + D(V^-_{h-s}) = \frac{1}{2}\,(U_s(C) + U_{2h-s}(C))\ ,\quad 1\le s \le h-1\qquad\qquad
\lb{DVerma}
\ea
and (\ref{UpC}) gives
\ba
&&U_{2h} (\b_p) = 0 =  U_h (\b_p)\ ,\quad 1\le p\le h-1\ ,
\nn\\
&&U_s (\b_p) + U_{2h-s} (\b_p) = \frac{[s\, p] + [(2h-s)p]}{[p]} = 0\ ,\quad 1\le p\,, s\le h-1\ .\qquad
\lb{Ups}
\ea
The canonical images of $D(V^+_p)\,$ in the $(h-1)$-dimensional quotient with respect to the Verma modules' ideal
are therefore of the form
\be
d_p = \sum_{s=1}^{h-1} U_p(\b_s)\, e_s = \sum_{s=1}^{h-1} \frac{[ p\, s ]}{[s]}\, e_s \ ,\quad 1\le p\le h-1
\lb{dpUp}
\ee
(note that the coefficient $\frac{[ p\, s ]}{[s]} \equiv [p]_{q^s}\,$ to $e_s\,$ in the expansion (\ref{dpUp}) of $d_p\,$
is just the quantum dimension of $V^+_p\,$ evaluated at $q^s$).
The algebra of $d_p\,$ follows from (\ref{Cew}) and the easily verifiable relation
\be
[ps]\, [p' s]  = [s ]\, \sum_{{r = |p-p'|+1}\atop{step\, 2}}^{p+p'-1} \, [r s]\ ,
\quad 1\le p,\, p' \le h-1
\lb{su2rel}
\ee
by taking into account that, for $p+p' > h\,$ (and $1\le s \le h-1$), the terms
with $r\ge h\,$ either vanish or cancel with the mirror ones w.r. to $h\,,$ due to
\be
[h s] = 0\ ,\qquad [(h + m) s] + [(h - m) s] = 0\ ,\quad m=1,2,\dots\ .
\lb{cancel}
\ee
Thus, the upper limit of the summation in (\ref{su2rel}) doesn't actually exceed $h-1\,$ and one reproduces the fusion rules
of the primary fields of weights $0\le \l,\,\mu \le k\,$ in the unitary $\widehat{su}(2)_k\,$ WZNW model
\be
d_\l\,  d_\mu = \sum_{{\nu = |\l-\mu|}\atop{step\,2}}^{k-|k-\l-\mu|} \, d_\nu
\lb{fusion-su2-I}
\ee
for $ p=\l +1\,,\ p' = \mu +1\,,\ h=k+2\,$ \cite{DFMS}.

\smallskip

The centre of $\bU\,$ is $(3h-1)$-dimensional, being spanned by the idempotents $e_s\,,\ 1\le s \le h\,$
and nilpotents $w^\pm_p\,,\ 1\le p\le h-1$ such that $w^+_p + w^-_p = w_p\,$ \cite{FGST1, FHT7}. The elements
$w^\pm_p\,$ do not belong to the algebra of the Casimir operator; to obtain them one needs to introduce, in
addition to the (Drinfeld images of) $q$-traces over the IR (\ref{canCh}), certain {\em pseudotraces} \cite{GT}.

\subsection{Extended chiral $\widehat{su}(2)_k\,$}

The structure of the zero modes' Fock space (\ref{FFp-dim}) suggests that for $n=2\,$ the chiral state space (\ref{space}) takes the form
\be
{\cal H} = \oplus_{p=1}^\infty {\cal H}_p \otimes {\cal F}_p\ ,
\lb{HpFp2}
\ee
where $p\,$ is the shifted weight labelling the corresponding representation of the ${\widehat{su}}(2)\,$
affine algebra and $\bU\,,$ respectively. Involving the full list of dominant weights, the space (\ref{HpFp2})
(on which the quantum group covariant field $g(z)\,$ acts) is much bigger than the one of the unitary model
\cite{GW} which only has a finite number of sectors corresponding to integrable affine weights, $1\le p \le h-1\,.$

In accord with (\ref{HpFp2}), we have to assume that primary fields $\phi_p(z)\,$ (\ref{Ward}) with conformal dimensions
$\Delta_p = \frac{p^2-1}{4h}\,$ (\ref{conf-dim-L}) exist for all integer $p \ge 1\,.$ Their exchange (generalizing (\ref{braidR}))
inside an $N$-point conformal block satisfying the KZ equation (\ref{KZW-N}) gives rise to a "monodromy representation" of the braid group
of $N\,$ strands ${\cal B}_N\,$ determined by choosing appropriately the principal branches and analytically continuing along homotopy classes of paths.
The braid group ${\cal B}_N\,$ admits a presentation with generators $B_i\,,\ i=1, \dots , N-1$ subject to Artin's relations
\be
B_i B_{i+1} B_i = B_{i+1} B_i B_{i+1}\ ,\qquad B_i B_j = B_j B_i \quad{\rm for}\quad |i-j| > 1\ .
\lb{Bgroup}
\ee
We shall recall below, without derivation, the results obtained in \cite{STH, MST, HP} for the corresponding
representations of ${\cal B}_4\,$ on the conformal blocks of four operators $\phi_p(z_a)\,,\ p\ge 1$
(as in this case $B_3 = B_1\,,$ the braid group actually reduces to ${\cal B}_3 \subset {\cal B}_4$).
It turns out that they are similar (dual) to those of an infinite dimensional extension $\tU\,$ of the restricted quantum group
which we proceed to review first.

\subsubsection{Lusztig's extension $\tU\,$ of the restricted quantum group $\bU$}

Introduce, following Lusztig \cite{L1, L}, the "divided powers"
\be
E^{(n)} = \frac{1}{[n]!}\, E^n\ ,\quad F^{(n)} = \frac{1}{[n]!}\, F^n\qquad{\rm for}\qquad n\ge 1\ .
\lb{divpowEF}
\ee
Their action on the basis (\ref{base2}) follows from (\ref{Uqprop2}):
\be
E^{(r)} |p,m{\cal i} = \left[{p-m-1}\atop{r}\right]\, |p,m+r{\cal i}\ ,\quad
F^{(s)} |p,m{\cal i} = \left[{m}\atop{s}\right]\,\, |p,m-s{\cal i}\ .
\lb{UqpropL}
\ee
Here the (Gaussian) $q$-binomial coefficients $\left[{a}\atop{b}\right]\,$ defined, for $a\in {\Z}\,,\, b\in {\Z}_+\,,$ as
\ba
&&\left[ {a\atop b}\right]:= \prod_{t=1}^{b} \frac{q^{a+1-t} - q^{t-a-1}}{q^t-q^{-t}}\ , \qquad \left[ {a\atop 0}\right] := 1\qquad
\lb{Gbinom}\\
&&\left(\, \left[ {a\atop b}\right] = \frac{[a]!}{[b]![a-b]!}\quad {\rm for}\quad a \ge b \ge 0\ ,\qquad
\left[ {a\atop b}\right] = 0\quad {\rm for}\quad b > a \ge 0 \,\right)\nn
\ea
are {\em polynomials in $q\,$ and $q^{-1}$} with integer coefficients\footnote{Hence, for $q$ a root of unity they are just polynomials in $q\,.$}.
The following general formula is valid for $M\in {\Z}\,,\ N\in {\Z}_+\,,\ 0\le a, b\le h-1\,$ (see Lemma 34.1.2 in \cite{L}),
\be
\left[ {Mh+a\atop Nh+b}\right] = (-1)^{(M+1)Nh + aN - bM}\, \left[ {a\atop b}\right]\, \left( {M\atop N}\right)\ ,
\lb{q-bin1}
\ee
where $\left( {M\atop N}\right) \in {\Z} \,$ is an {\em ordinary} binomial coefficient.

It is sufficient to add just $E^{(h)}\,$ and $F^{(h)}\,$ to $E, F\,$ and $K^{\pm 1}\,$ in order to generate Lusztig's $\tU\,$ algebra.
Their powers and products give rise to an infinite sequence of new elements -- in particular,
\be
\left( E^{(h)} \right)^n = \frac{[nh]!}{([h]!)^n}\, E^{(nh)} = \left( \prod_{\ell =1}^n
\left[ {\ell\, h}\atop{h} \right] \right)\, E^{(nh)} = (-1)^{\left(n\atop 2\right) h} n!\, E^{(nh)}\ .
\lb{Edpn}
\ee
The representations of the extended QUEA $\tU\,$ in the Fock space ${\cal F}\,$ (\ref{FFp-dim}) are easily described by the following

\medskip

\noindent
{\bf Proposition 5.2~}
{\em

\noindent
(a) The irreducible $\bU\,$ modules ${\cal F}_p\,,\ 1\le p\le h\,$ extend to irreducible
$\tU$ modules, with $E^{(h)}\,$ and $F^{(h)}\,$ acting trivially.

\noindent
(b) The fully reducible $\bU$ modules ${\cal F}_{Nh}\,,\ N\ge 2\,$ give rise to irreducible $\tU\,$ modules.

\noindent
(c) For $1\le p\le h-1\,,\ N=1,2,\dots\,$ the spaces ${\cal F}_{Nh+p}\,$ are indecomposable $\tU\,$ modules.
Their structure is given by a short exact sequence similar to (\ref{shexseqN}),
\be
0\ \ \rightarrow\ F_{N+1 ,\, p} \ \ \rightarrow\ \ {\cal F}_{Nh+p}\ \
\rightarrow\ \  {\tilde F}_{N ,\,h-p}\ \ \rightarrow\ \ 0\ ,
\lb{shex-eqN}
\ee
where this time the submodule
\be
F_{N+1 ,\, p} = \oplus_{n=0}^N\ Span \, \{\, |Nh+p , nh+m {\cal i}\,\}_{m=0}^{p-1}
\lb{FN+1p}
\ee
and the corresponding subquotient
\be
{\tilde F}_{N ,\,h-p} = {\cal F}_{Nh+p} \, / F_{N+1 ,\, p}
\lb{FNh-p}
\ee
are both irreducible with respect to $\tU\,.$
}
\medskip

\noindent
{\bf Proof~} Using (\ref{Uqprop2}) and the relation $\left[{n}\atop h\right]=0\,$
for $n<h\,,$ we find
\be
E^{(h)} | p , m {\cal i} \, =\, 0\, =\, F^{(h)} | p , m {\cal i}\qquad{\rm for}
\quad p\le h\ ,
\lb{EhFhzero}
\ee
proving (a). On the other hand, $E^{(h)}\,$ and $F^{(h)}\,,$ shifting the label $m\,$ by $\pm h\,$
combine otherwise disconnected equivalent (in particular, of the same parity) irreducible
$\bU\,$ submodules or subquotients into a single irreducible representation of $\tU\,.$ Together with (\ref{Uqprop2}), the relation
\ba
&&E^{(h)} | Nh+p , nh+m \rangle = \left[{(N-n)h+p-m-1}\atop{h} \right]\,
| Nh+p , (n+1)h+m \rangle =\nn\\
&&= (-1)^{(N-n+1)h+p-m-1} \, (N-n)\, | Nh+p , (n+1)h+m \rangle
\lb{Eh1}
\ea
where $0\le n \le N\,,\ 0\le m\le p-1\le h-1\,$ and the similar relation for $F^{(h)}\,$
\ba
&&F^{(h)} | Nh+p , nh+m \rangle = \left[{nh+m}\atop{h} \right]\,
| Nh+p , (n-1)h+m \rangle =\nn\\
&&= (-1)^{(n+1)h+m} \, n\, | Nh+p , (n-1)h+m \rangle
\lb{Fh1}
\ea
imply (b), for $p=h\,,$ and the first (submodule) part of (c), for $p<h\,.$
The second part of (c) is obtained by using again (\ref{Uqprop2}) as well as (\ref{Eh1}), (\ref{Fh1}) but
this time for $0\le n \le N-1\,,\ 1\le p \le m\le h-1\,.$
\eod

\smallskip

According to the "parity rule" (\ref{GRpb}), each IR of $\tU\,$ combines an odd number of irreducible $\bU\,$ modules of type $V^+\,$
and an even number of modules $V^-\,.$

\subsubsection{KZ equation and braid group representations}

In addition to the KZ equation, an ${\widehat{su}}(2)_k$ conformal block is subject to M\"obius and $SU(2)\,$ invariance conditions.
The components of a primary field $\phi_p(z)\,$ form a $p$-dimensional irreducible $SU(2)\,$ multiplet $V_p\,$ so that
their $4$-point conformal block $w^{(p)}\,$ belongs to the space ${\rm Inv}\, V_p^{\otimes 4}\,$ (which itself is $p$-dimensional).
Realizing each $V_p\,$ as a space of polynomials of degree $p-1\,$ in a variable $\zeta_a\,,\, a=1,2,3,4\,,$
the $4$-point $SU(2)$-invariants appear as homogeneous polynomials of degree $2(p-1)\,$ in the differences $\zeta_a-\zeta_b\,.$
One can express, accordingly, $w^{(p)}\,$ in terms of an amplitude $f^{(p)}\,$ that depends on two invariant cross ratios $\xi\,$ and $\eta\,,$ writing
\ba
&&\langle\,\phi_p (z_1)\, \phi_p (z_2)\, \phi_p (z_3)\, \phi_p (z_4)\,\rangle =:
w^{(p)} (\zeta_1, z_1 ; \dots ; \zeta_4 , z_4 )= D_p\, (\underline{\zeta} ,\uz ) \, f^{(p)}(\xi, \eta)\ ,\nn\\
&&\zeta_{ab} = \zeta_a - \zeta_b\ ,\quad z_{ab} = z_a-z_b\ ,\quad
\xi = \frac{\zeta_{12}\zeta_{34}}{\zeta_{13}\zeta_{24}}\ ,\quad
\eta = \frac{z_{12}z_{34}}{z_{13}z_{24}}\ ,\nn\\
&&D_p \, ( \underline{\zeta} ,\uz  ) = \left(\frac{z_{13}z_{24}}{z_{12}z_{34}z_{14}z_{23}} \right)^{2\Delta_p} (\zeta_{13}\zeta_{24})^{p-1}
\lb{xi}
\ea
where $f^{(p)}(\xi, \eta)\,$ is a polynomial in $\xi\,$ of degree not exceeding $p-1\,.$
The polarized Casimirs are represented by second order differential operators in the
isospin variables and the KZ system (\ref{KZW-N}) is equivalent to the following partial
differential equation for $f^{(p)}(\xi, \eta)$:
\ba
&&\left(\, h\,\eta\, (1-\eta)\,\frac{\partial}{\partial \eta} - (1-\eta)\, C^{(p)}(\xi) + \eta\, C^{(p)} (1-\xi) \, \right) \, f^{(p)}(\xi, \eta) = 0\ ,\lb{KZf}\\
&&C^{(p)} (\xi) :=(p-1)( p-(p-1)\,\xi) - (\xi + 2\,(p-1)(1-\xi) )\,\xi\frac{\partial}{\partial \xi} + \xi^2(1-\xi)\frac{\partial^2}{\partial \xi^2}\ .
\nn
\ea
A {\em regular basis} of the $p\,$ linearly independent solutions
\be
\{\, f^{(p)}_\mu = f^{(p)}_\mu (\xi , \eta)\,,\quad \mu=0,1,\dots , p-1 \}
\lb{fxii}
\ee
of Eq.(\ref{KZf}) has been constructed in \cite{STH} in terms of appropriate multiple
contour integrals. We shall describe below the explicit braid group action on the conformal blocks $w^{(p)}_\mu = D_p\, f^{(p)}_\mu\,$ (\ref{xi}).
The braid generators $b_i\,,\ i=1,2,3\,$ act by an anti-clockwise rotation at angle $\pi\,$ of the pair of world sheet variables
$(z_i\,,\, z_{i+1} )\,$ and a simultaneous exchange $\zeta_i \leftrightarrow \zeta_{i+1}\,.$
Then $w^{(p)}_\mu (\underline{\zeta} ,\uz )\ \rightarrow\ w^{(p)}_\l (\underline{\zeta} ,\uz )\,(B^{(p)}_i)^\l_{~\mu}\,$
while the invariant amplitudes $f^{(p)}_\mu (\xi, \eta)\,$ transform as
\ba
&&b_1 \,( = b_3)\, :\
f^{(p)}_\mu (\xi, \eta)\ \rightarrow\ (1-\xi)^{p-1}(1-\eta)^{4\Delta_p}
f^{(p)}_\mu(\frac{\xi}{\xi-1},\frac{e^{-i\pi}\eta}{1-\eta}) =\nn\\
&&=f^{(p)}_\l(\xi,\eta)\,(B^{(p)}_1)^\l_{~\mu}\ ,\nn\\
&&b_2\, :\
f^{(p)}_\mu (\xi, \eta)\ \rightarrow\ \xi^{p-1}\eta^{4\Delta_p}
f^{(p)}_\mu(\frac{1}{\xi},\frac{1}{\eta}) = f^{(p)}_\l(\xi,\eta)\,(B^{(p)}_2)^\l_{~\mu}\ ,
\lb{B1B3B2}
\ea
respectively. The $p\times p\,$ braid matrices $B^{(p)}_i\,,\, i=1,2\,$ are (lower, resp. upper) triangular:
\ba
&&(B^{(p)}_1)^\l_{~\mu} =  (-1)^{p-\l-1} q^{\l (\mu +1)-\frac{p^2-1}{2}} \left[{\l\atop\mu}\right] = (B^{(p)}_3)^\l_{~\mu}\ ,
\quad\l\,,\,\mu = 0,1,\dots , p-1\ ,\nn\\
&&(B^{(p)}_2)^\l_{~\mu} = (B^{(p)}_1)^{p-\l-1}_{~p-\mu-1} = (-1)^\l q^{(p-\l-1) (p-\mu )-\frac{p^2-1}{2}} \left[{p-\l-1\atop p-\mu-1}\right]\lb{B1B2}\\
&&(\, B^{(p)}_2 = F^{(p)} \, B^{(p)}_1 \,F^{(p)}\ ,\quad (F^{(p)})^\l_{~\mu} = \d^{\l}_{p-1-\mu} \ ,\quad  (F^{(p)})^2 = \id_p\, )\ .
\nn
\ea
By contrast, the commonly used "$\!s$-basis" braid matrices (where $B^{(p)}_1 = B^{(p)}_3\,$ is assumed to be diagonal)
do not exist in this case, yielding singularities for $p\ge h\,.$

It is instructive to arrange the emerging $p$-dimensional representation spaces ${\cal S}_p\,$ of ${\cal B}_4\,$ spanned by
$w^{(p)}_\mu (\underline{\zeta} ,\uz )\,,\ \mu=0,1,\dots , p-1\,$ in arrays similar to
${\cal F}_p\,$ in the zero modes' Fock space depicted on Figure 1 above.

\medskip

\noindent
{\bf Proposition 5.3~}
{\em The $p$-dimensional ${\cal B}_4\,$ modules ${\cal S}_p\,$ have a structure
dual to that of the $\tU\,$ modules ${\cal F}_p\,$ described in Proposition 5.2, in the following sense.

\noindent
The representation spaces ${\cal S}_p\,$ are irreducible

\noindent
(a) for $1\le p\le h\,,$ as well as

\noindent
(b) for $p=Nh\,,\ N\ge 2\,.$

\noindent
(c) For $1\le p \le h-1\,,\ N=1,2,\dots\,$ the module ${\cal S}_{Nh+p}\,$ is indecomposable, with structure given by
the exact sequence
\be
0\ \ \rightarrow\ S_{N,h-p} \ \ \rightarrow\ \ {\cal S}_{Nh+p}\ \ \rightarrow\ \  {\widetilde S}_{N+1, p}\ \ \rightarrow\ \ 0\ .
\lb{shex-eqS}
\ee
Here the $N (h-p)$-dimensional invariant subspace
\be
S_{N,h-p} = \oplus_{n=0}^{N-1}\ Span \, \{\,f_\mu^{(Nh+p)}\, \}_{\mu=nh+p}^{(n+1)h-1}
\lb{Sh-p}
\ee
and the corresponding $(N+1)p$-dimensional quotient ${\tilde S}_{N+1, p}\,$ are both irreducible under the action of the braid group.
}

\medskip

\noindent
{\bf Proof~} Only the case (c) needs some work.
The fact that the subspace $S_{N, h-p}\subset {\cal S}_{Nh+p}\,$ (\ref{Sh-p}) is ${\cal B}_4$ invariant
follows from the observation that the entries of the $(Nh+p)$-dimensional matrices (\ref{B1B2}) satisfy
\ba
&&(B_1)^{mh+\a}_{~nh+\b}\ \sim \ \left[{{mh+\a}\atop{nh+\b}}\right] \ \sim\
\left[{\a\atop \b}\right] \left({m\atop{n}}\right) \ ,\nn\\
&&(B_2)^{mh+\a}_{~nh+\b}\ \sim \ \left[{{(N-m)h+p-\a-1}\atop{(N-n-1)h+h+p-\b-1}}\right] \ \sim\ \nn\\
&&\ \sim\ \left[{p-\a-1\atop{h+p-\b-1}}\right] \left({N-m\atop{N-n-1}}\right)\ ,
\lb{B=0}
\ea
cf. (\ref{q-bin1}), and hence vanish for $0\le \a \le p-1\,,\ p\le \b \le h-1\,$ and $\,0\le m \le N\,,\ 0\le n \le N-1$
(since $\b>\a\ge 0\,$ and $h+p-\b-1 > p-\a-1 \ge 0\,,$ see (\ref{Gbinom})).
An inspection of the same expressions (\ref{B=0}) for $0\le \b \le p-1\,$ allows to conclude that
the subspace $S_{N, h-p}\,$ has no ${\cal B}_4$ invariant complement in ${\cal S}_{Nh+p}\,$ which is thus indeed indecomposable.
It is also straightforward to verify that the quotient space
\be
{\widetilde S}_{N+1, p} \ = \ {\cal S}_{Nh+p}\, / S_{N, h-p}
\lb{factS}
\ee
carries another IR of ${\cal B}_4\,.$ The "duality" of the indecomposable representations ${\cal V}_{Nh+p}\,$ (of $\tU\,$)
and ${\cal S}_{Nh+p}\,$ (of ${\cal B}_4\,$) is summed up by the observation that each of them
contains, in the GR sense, two irreducible components of the same dimensions,
but the arrows of the exact sequences (\ref{shex-eqN}) and (\ref{shex-eqS}) are reversed. \eod

\bigskip

The ${\cal B}_4\,$ invariance and irreducibility of the subspaces $Span \, \{\,f_{(n+1)h-1}^{((N+1)h-1)}\, \}_{n=0}^{N-1}\,$
(or $S_{N,1} \subset {\cal S}_{(N+1)h-1}\,,$ in our notation (\ref{Sh-p})) has been noted by A. Nichols in
\cite{N1}\footnote{The scope of the paper \cite{N1} is actually broader, including also fractional levels.}.
Their dimension is equal to $N\,;$ this fact is nicely visualized by reversing the arrows on Figure 1 where these sets correspond
to the upper tips of the yellow and blue (or white and black, in BW print) squares.
They possess an internal $su(2)\,$ structure where the action of the $su(2)\,$ generators $e\,$ and $f\,$
is given by that of $E^{(h)}\,$ (\ref{Eh1}) and $F^{(h)}\,$ (\ref{Fh1}), respectively, under the identification
\ba
&&f_{(n+1) h-1}^{((N+1)h-1)} \equiv v^N_n := | (N+1) h -1 , (n+1) h -1 \rangle \ ,\quad n=0,\dots , N-1 \ ,\nn\\
&&e \, v^N_n = (-1)^{(N-n+1)h-1} (N-n-1)\, v^N_{n+1}\ ,\qquad f \, v^N_n = (-1)^{nh-1} n\, v^N_{n-1}\ ,\nn\\
&&h := [e, f]\ ,\qquad \left( h- (-1)^{Nh} (2n-N+1)\right)\, v^N_n = 0\ .
\lb{su2}
\ea
The corresponding $N\times N\,$ {\em reduced} braid matrices $\left((B^{red}_i)^n_{~m} := (B_i)^{(n+1)h-1}_{~(m+1)h-1}\right)\,$ have
remarkable properties \cite{N1}. As one can easily deduce from (\ref{B=0}) and (\ref{q-bin1}), they are proportional to matrices with integer entries;
moreover, the corresponding {\em monodromy matrices} $B_i^2\,,\ i=1,2\,$ are equal (up to a sign, for $N\,$ even {\em and} $h\,$ odd) to the unit one:
\ba
&&(B^{red}_1)^n_{~m} = q^{\frac{1}{2}(N+1)^2 h^2} (-1)^{N+1+(n+m)h+n} \left( {n\atop m}\right)\ ,\nn\\
&&B^{red}_2 = F^{red} B_1^{red} F^{red}\ ,\quad (F^{red})^n_{~m} = \d^n_{N-1-m}\ ,\quad n,m=0,\dots , N-1\ ,\nn\\
&&(B_i^{red})^2 = (-1)^{(N+1)h}\, \id_N\ ,\qquad i=1,2\ .
\lb{Bred}
\ea
Explicitly, the first few rows of $B^{red}_1\,$ are given by
\be
(-1)^{N+1} q^{- \frac{1}{2}(N+1)^2 h^2}\, B^{red}_1 = \begin{pmatrix} 1&0&0&0&\dots\cr (-1)^{h+1}&-1&0&0&\dots\cr
1&(-1)^h 2&1&0&\dots\cr (-1)^{h+1} &-3&(-1)^{h+1} 3 & -1&\dots \cr \dots&\dots&\dots&\dots&\dots\end{pmatrix}
\lb{B1red}
\ee
(for $N\le 3\,,$ just take the relevant upper left corner submatrix).

Thus, for all natural $N\,$ there exist $N$-plets of non-unitary, local chiral primary fields $\phi^{(n)}_{(N+1)h-1} (z)\,$
of $su(2)\,$ "spin" $j=\frac{N-1}{2}\,,$ isospin $I = \frac{N+1}{2}\, h - 1\,$ and conformal dimension
$\Delta_{(N+1)h-1} = \frac{((N+1)h-1)^2-1}{4h} = \frac{I(I+1)}{h} = \frac{(N+1)^2}{4}\, h - \frac{N+1}{2}\,$
(all these numbers are integers for $N\,$ odd).
The presence of additional $su(2)\,$ quantum numbers in non-unitary extended WZNW  (and minimal) models
has been confirmed by other methods, see e.g. \cite{N2}\footnote{Cf. also \cite{S1, ST13}
for another approach to the problem.}. Such models are examples of {\em logarithmic}
conformal field theory (LCFT) characterized by Jordan block (indecomposable, and hence, non-hermitean)
structure of the dilation operator $L_0\,$ \cite{Gu}. The latter fact explains
the possible appearance of logarithms in conformal blocks noticed first in \cite{RozS}.
(A representative collection of recent papers and reviews on LCFT can be found in \cite{GRR}.)

The singlet field $\phi^{(0)}_{2h-1} (z)\,$ (the conformal block of which spans the $1$-dimen-sional subspace
$S_{1,1} \subset {\cal S}_{2h-1}\,$) has isospin $I\,$ and conformal dimension both equal to $h-1 = k+1\,,$
\ba
&&2I+1 = 2h-1 \qquad\Rightarrow\qquad I = h-1\ ,\nn\\
&&\Delta_{2h-1} = \frac{(2h-1)^2 -1}{4h}\  \left( \equiv \frac{I(I+1)}{h} \right) \ = h-1
\lb{DI}
\ea
and hence provides a natural candidate for a local extension of the chiral (current) algebra.
As the conformal dimensions $\Delta_{2Nh-p}\,$ and $\Delta_p\,$ are integer spaced,
\be
\Delta_{2Nh-p} = \frac{(2Nh-p)^2-1}{4h} = N(Nh-p) + \Delta_p\ ,\quad 1\le p \le h-1\ ,
\ee
it is the "mirror" counterpart of the unit operator ($p=1$) under the duality $p\ \leftrightarrow\ 2h-p\,.$

The locality of $\phi^{(0)}_{2h-1} (z)\,$ implies that the corresponding conformal block
$w^{(2h-1)}_{h-1} = w^{(2h-1)}_{h-1}\,( \underline{\zeta} ,\uz  )$ (\ref{xi})
is a rational function of $z_{ij}\,.$ This means, in turn, that $f^{(2h-1)}_{h-1} (\xi , \eta )\,$ is a polynomial
in $\eta\,$ of order not exceeding $4 \Delta_{2h-1}\,$ \cite{MST} such that
\be
f^{(2h-1)}_{h-1} (1-\x , 1-\eta ) = f^{(2h-1)}_{h-1} (\x , \eta ) =
\x^{2(h-1)} \eta^{4(h-1)} f^{(2h-1)}_{h-1} (\frac{1}{\x} , \frac{1}{\eta} )\ .
\lb{rat}
\ee
The corresponding solution of Eq.(\ref{KZf}) has been found in \cite{HP}:
\ba
&&f^{(2h-1)}_{h-1}(\x ,\eta ) = (\eta (1-\eta ))^{h-1}\, p_{h-1} (\x ,\eta )\ ,\quad
p_{h-1} (\x ,\eta ) = \sum_{m=0}^{2(h-1)}\sum_{n=0}^{h-1} C^{h-1}_{mn} \x^m \eta^n\ ,\nn\\
&&C^I_{mn}=(-1)^{I+m+n} {{I}\choose{m+n-I}} {{m+n}\choose{n}} {{3I-m-n}\choose{I-n}}\ .
\lb{polyKZ}
\ea
A characteristic property of $f^{(2h-1)}_{h-1}\,$ is that it {\em belongs} to the regular basis of ${\cal S}_{2h-1}\,.$
Writing the braid invariance requirement in the form
\be
(b_i - 1)\, f_{inv}^{(2h-1)} = 0\ ,\ \ i=1,2\ ,\quad f_{inv}^{(2h-1)} = s^\mu f^{(2h-1)}_\mu \ ,\ \ \l,\mu = 1,\dots , 2h-1\ ,
\lb{w-br-inv}
\ee
we verify that the common eigenvector problem has the predicted solution,
$f_{inv}^{(2h-1)} = f_{h-1}^{(2h-1)}$:
\be
(B_i^{(2h-1)})^\l_{~\mu} s^\mu = s^\l \ , \ \ i=1,2\qquad {\rm for} \qquad s^\mu = \d^\mu_{h-1}\ .
\ee
Note that, as the matrices $B_1^{(p)}\,$ and $B_2^{(p)}\,$ (\ref{B1B2}) do not commute,
they possess common invariant eigenvectors only in special cases.

\medskip

\noindent
{\bf Remark 5.2~}
All polynomial solutions of the KZ equation (\ref{KZf}) for {\em integrable} weights $0\le p\le h-1\,$
giving rise to local extensions of chiral current algebra ${\widehat{su}}(2)_{h-2}\,$ have been found in \cite{MST}.
The list corresponds to the $D_{2\ell+2}\,$ series in the ADE classification
of modular invariant partition functions \cite{CIZ1},
\ba
&&D_{2\ell+2}\,:\quad h= 4\ell+2\ ,\quad p=4\ell+1 = h-1 \quad (\,\Delta_{4\ell +1} = \ell\,)\ ,\qquad\nn\\
&&f_{inv}^{(4\ell+1)} = f_{inv}^{(4\ell+1)} (\x , \eta ) = (\x - \eta )^{4\ell}\ ,\quad \ell\in{\mathbb N}
\lb{Deven-inv}
\ea
and a few exceptional cases occurring for
\ba
&&E_6\,:\quad h=12\ ,\quad  p = 7 \quad (\, \Delta_7 = 1\, )\ ,\nn\\
&&f_{inv}^{(7)} = f_{inv}^{(7)} (\x , \eta ) = (\x - \eta )^2 \left( (\x^2-\eta)^2-4\,\x \eta (1-\x)^2 \right)
\lb{E6}
\ea
and
\be
E_8\,:\quad h=30\ ,\quad  p = 11\,,\, 19\,,\, 29 \quad (\, \Delta_{11} = 1\,,\, \Delta_{19} = 3\,,\, \Delta_{29} = 7\,)\ .\nn
\lb{E8}
\ee
It can be easily verified \cite{HP} that the regular basis components of (\ref{Deven-inv}) are
\be
D_{2\ell+2}\,:\quad f_{inv}^{(4\ell +1)} = s^\mu f^{(4\ell +1)}_\mu\ ,\quad s^\mu = \frac{(-1)^\mu}{[\mu +1]}\ ,\quad \mu = 0,\dots , 4\ell \ ;
\lb{Deven1}
\ee
to prove that $(B_i^{(4\ell+1)})^\l_{~\mu} s^\mu = s^\l\,,\ i=1,2\ $ (for $h=4\ell+2$), one makes use of a well known
$q$-binomial identity\footnote{We have in mind the one obtained by setting $z=-1\,$ in  the equality
$$\prod_{m=0}^{\l} (1+q^{2m} z) = \sum_{\mu = 0}^{\l+1} q^{\l \mu}\left[ {\l+1}\atop\mu \right] z^\mu\qquad {\rm for}\qquad \l \ge 0\,$$
which is elementary to derive by induction in $\l\,$ (see 1.3.1(c) and 1.3.4 in \cite{L}).}
written in the form
\be
\sum_{\mu = 0}^{4\ell} (-1)^\mu q^{\l (\mu +1)}\left[ {\l +1}\atop{\mu +1}\right] = 1
\qquad {\rm for}\qquad 0\le \l\le 4\ell\ ,\ \  q=e^{- i\frac{\pi}{4\ell+2}}\ .
\lb{Deven2}
\ee
Solving the common eigenvector problem in the $E_6\,$ case ($h=12\,, p = 7\,,$ cf. (\ref{E6})),
one gets $f_{inv}^{(7)} = s^\mu f^{(7)}_\mu\,$ with
\be
E_6\,:\quad s^0 = s^6 = 1\ ,\ \ s^1 = s^5 = -\frac{1}{[2]}\ ,\ \ s^2 = s^4 = \frac{1}{[3]}\ ,\ \ s^3 = - \frac{3}{[3][4]}\ .
\ee

\section{From chiral to \boldmath$2D\,$ WZNW model}

\setcounter{equation}{0}
\renewcommand\theequation{\thesection.\arabic{equation}}

\subsection{The right chiral sector}

It is usually assumed that, instead of solving anew the quantization problem, the exchange relations for the right sector quantities
can be recovered in a straightforward way from those for the left sector. This is true in general, yet the change of chirality needs some care.
Writing the quantum analog of (\ref{LR}) in the form $g (x\,, \bar x) = g (x)\,\bar g (\bar x)\,$ for $x = x^+\,,\, \bar x = x^-\,$
and following the reasoning for the classical case considered in Section 3.5.4, one concludes that the exchange relations
for $\bar g (\bar x)\,$ are obtained from the left sector ones by just inverting the order of terms in matrix products.\footnote{The heuristic
derivation uses the fact that the constant $R$-matrix (\ref{R}) evaluated at the inverse deformation parameter (\ref{height-h}),
$q\ \to \ q^{-1}\,$ equals the inverse matrix, $\ R^{-1}_{12}\,$ (equivalently, ${\hat R}_{12}\, \to \, {\hat R}_{21}^{-1}$).
The exchange relations for $\bar g (\bar x)\,$ contain however the original $R$-matrix (at the original value of $q$).}
One can then verify directly that their quasi-classical expansions match the corresponding PB brackets.
We shall display in what follows all the relevant right sector exchange relations in terms of the bar fields.
Our guiding principle in the choice of quantization conventions is the implementation of locality and monodromy invariance
of the $2D\,$ field and of the quantum group covariance of its chiral components.

\subsubsection{Constant $R$-matrix exchange relations for the right sector}

Starting with the left sector equalities (\ref{ggR}), (\ref{Rx}) and following the procedure described above, we obtain the exchange relations
\ba
&&g_1(x_1)\, g_2 (x_2) = g_2(x_2) \, g_1(x_1)\, (R_{12}\, \theta (x_{12}) + R_{21}^{-1}\, \theta (x_{21}))\qquad\Rightarrow\nn\\
&&\bar g_2(\bar x_2)\, \bar g_1(\bar x_1) = (R_{12}\, \theta (\bar x_{12}) + R_{21}^{-1}\, \theta (\bar x_{21}))\, \bar g_1(\bar x_1) \, \bar g_2(\bar x_2) \qquad\Leftrightarrow\nn\\
&&\bar g_1(\bar x_1) \, \bar g_2(\bar x_2) = (R_{12}^{-1}\, \theta (\bar x_{12}) + R_{21}\, \theta (\bar x_{21}))\, \bar g_2(\bar x_2)\, \bar g_1(\bar x_1) \ ,
\lb{ggbarLR}
\ea
where
\be
x_i = x_i^+\ ,\quad \bar x_i = x_i^-\ ,\quad i=1,2\ , \qquad -2\pi < x_{12} , \bar x_{12} <2\pi\ .
\lb{xxbar}
\ee

The next step is to derive the exchange relations including the general monodromy matrix $\bar M\,$ defined by
\be
\bar g (\bar x + 2\pi) = \bar M \, \bar g (\bar x)\qquad (\,\bar M = M_R^{-1}\,) \ .
\lb{defbarM}
\ee
The consistency of the last exchange relation in (\ref{ggbarLR}) for $0 < {\bar x}_{12} < 2\pi\,$ requires
\ba
&&\bar g_1(\bar x_1)\,\bar g_2(\bar x_2 + 2\pi) = R_{21}\,\bar g_2(\bar x_2+2\pi)\,\bar g_1(\bar x_1) \ ,\qquad {\rm i.e.}\nn\\
&&\bar g_1(\bar x_1)\,\bar M_2\,\bar g_2(\bar x_2) = R_{21}\,\bar M_2\,\bar g_2(\bar x_2)\,\bar g_1(\bar x_1) =
R_{21}\,\bar M_2\,R_{12}\,\bar g_1(\bar x_1)\,\bar g_2(\bar x_2)\qquad\qquad\nn\\
&&\Rightarrow\qquad R_{12}^+\,\bar g_1(\bar x)\, \bar M_2 = \bar M_2\,R_{12}^- \, \bar g_1 (\bar x) \qquad\ (\, R^-_{12} = R_{12}\,,\ R^+_{12} = R_{21}^{-1}\, )\ .\qquad
\lb{Mbargq}
\ea
The latter exchange relation can be derived alternatively from the one for the left sector, (\ref{Mgq})
by using again the procedure described in the beginning of this subsection.
From (\ref{Mexch}) one obtains in a similar way the reflection equation for the bar sector,
\be
M_1\, R_{12}\, M_2\, R_{21} = R_{12}\, M_2\, R_{21}\, M_1\quad\Rightarrow\quad
\bar M_1\, R_{21}\, \bar M_2\, R_{12} = R_{21}\, \bar M_2\, R_{12}\, \bar M_1\ .
\lb{Mbarexch}
\ee
The same rule suggests that the factorization of $\bar M\,$ into Gauss components (the right sector counterpart of (\ref{M+-q})) reads
\be
\bar M = q^{\frac{1}{n}-n}\,\bar M^{-1}_- \bar M_+\ ,\quad {\rm diag}\, \bar M_+  = {\rm diag}\, \bar M_-^{-1}\qquad (\, \bar M_\pm = M_{R\pm}^{-1}\,)\ .
\lb{M+-qbar}
\ee
Before discussing the "quantum coefficient" in the definition of $\bar M\,,$ we shall first note that the (homogeneous -- and hence, normalization independent)
exchange relations for $M_\pm\,$ (\ref{Mpmq}) imply {\em the same} relations for $\bar M_\pm\,,$
\ba
&&R_{12} M_{\pm 2} M_{\pm 1} = M_{\pm 1} M_{\pm 2} R_{12}\ ,\quad R_{12} M_{+2} M_{-1} = M_{-1} M_{+2} R_{12}\qquad\Rightarrow\nn\\
&&R_{12} \bar M_{\pm 2} \bar M_{\pm 1} = \bar M_{\pm 1} \bar M_{\pm 2} R_{12}\ ,\quad R_{12} \bar M_{+2} \bar M_{-1} = \bar M_{-1} \bar M_{+2} R_{12}
\lb{Mpmq-bar}
\ea
and thus provide, by the FRT construction, another copy of the QUEA, {\em identical} to that for the left sector.
Further, from (\ref{Mg}) one obtains
\be
g_1(x)\, R_{12}^\mp\, M_{\pm 2} = M_{\pm 2}\, g_1(x) \quad\Rightarrow\quad \bar M_{\pm 2}\, R_{12}^\mp\, \bar g_1(\bar x) = \bar g_1(\bar x)\, \bar M_{\pm 2}\ .
\lb{Mgbar}
\ee
By taking (\ref{M+-qbar}) into account, (\ref{Mbarexch}) follows from (\ref{Mpmq-bar}) and (\ref{Mbargq}), from (\ref{Mgbar}).

\smallskip

We shall now argue that the overall coefficient $q^{\frac{1}{n}-n}\,$ in (\ref{M+-qbar})
(the {\em inverse} to the factor $e^{-2\pi i {\bar\Delta}}\,$ in (\ref{gzM}))
is consistent with the QUEA invariance of the {\em "bra"} vacuum vector (\ref{Uqlvac}) implying\footnote{Recall that, by
(\ref{Mpmq-bar}), the diagonal elements of ($M_\pm\,$ and) $\bar M_\pm\,$ are expressed in terms of Cartan generators while
the off-diagonal ones contain step operators of the same type, either raising or lowering; see Section 4.3
for the FRT construction of the QUEA.}
\be
\lvac (\bar M_\pm)^\a_{~\b} = \varepsilon ((\bar M_\pm)^\a_{~\b})\, \lvac = \d^\a_\b \, \lvac\ .
\lb{lvac-inv}
\ee
To this end we multiply the bar sector equality in Eq.(\ref{gzM}) by ${\bar z}^{2\bar\Delta}\,$ and take into account the definition
of "bra" (or "out") states
\be
(\, \langle \bar \Delta\! \mid \ = \,)\ \lim_{\bar z \to \infty} {\bar z}^{2\bar\Delta}\, \lvac \bar g (\bar z)
\equiv e^{- 4 \pi i {\bar\Delta}} \lim_{\bar z \to \infty} {\bar z}^{2\bar\Delta}\, \lvac \bar g (e^{-2\pi i} \bar z)\ ,
\lb{def-bra-out}
\ee
see e.g. Eq.(4.70c) in \cite{FSoT} (or Eqs. (6.4), (6.5) in \cite{DFMS})).

Following a line of reasoning similar to the one in the beginning of Section 4.5, we shall further assume that the quantized chiral field
$\bar g(\bar z)$ splits as in (\ref{guaq}) and that the right chiral state space is again a direct sum of subspaces created from the vacuum
by identical homogeneous polynomials in the corresponding zero modes $\bar a_j = (\bar a^\a_j)\,$ and
generalized CVO $\,\bar u^j = (\bar u^j_B(\bar z))\,,$ respectively:
\be
\bar g^\a_B(\bar z) =  \bar a^\a_j\otimes \bar u^j_B(\bar z)\ ,\qquad \bar{\cal H}\,
=\, \bigoplus_{\bar p}\,\bar{\cal F}_{\bar p}\otimes \bar{\cal H}_{\bar p} \ .
\lb{guaqbar}
\ee
The monodromy matrix of the field $\bar u(\bar z) = (\bar u^j_B(\bar z))$ is, by definition, diagonal,
\be
e^{- 2\pi i {\bar L}_0} \bar u^j_B(\bar z)\, e^{2\pi i {\bar L}_0}
= e^{-2\pi i \bar\Delta} \, \bar u^j_B (e^{-2\pi i}\,\bar z) = \bar u^i_B (\bar z)(\bar M_{\bar p})^j_i\ .
\lb{uuMpqbar}
\ee
On the space $\bar{\cal H}\,$ (\ref{guaq}), $\bar M_{\bar p}\,$ is "inherited" by the zero modes, in the sense that
\be
\bar a^\a_j \otimes \bar u^i_B (\bar z) (\bar M_{\bar p})^j_i =
\bar a^\a_j (\bar M_{\bar p})^j_i\otimes \bar u^i_B (\bar z) = \bar M^\a_{~\b}\, \bar a^\b_i \otimes \bar u^i_B (\bar z)\ .
\lb{inhMpbar}
\ee
This happens since $\bar u^j_B (\bar z)\,$ and $\bar a^\a_j\,$ satisfy identical exchange relations with the commuting
operators ${\bar p_i}\,,\ i=1,\dots,n\,$ (where $\sum_{i=1}^n \Bp_i = 0\ \Rightarrow\ \prod_{i=1}^n q^{\Bp_i} = 1$),
\ba
&&{\bar p_i}\, \bar u^j_B (\bar z) = \bar u^j_B (\bar z)\, (\bar p_i + \d^j_i - \frac{1}{n})\ ,\qquad
{\bar p_i}\, \bar a_j^\a  = \bar a^\a_j\, (\bar p_i + \d_{ij} - \frac{1}{n})\qquad\Rightarrow\nn\\
&&q^{\bar p_{i\ell}}\,\bar u^j_B (\bar z) = \bar u^j_B (\bar z)\, q^{\bar p_{i\ell} + \d_i^j - \d_\ell^j}\ ,\qquad
q^{\bar p_{i\ell}}\, \bar a^\a_j = \bar a^\a_j\, q^{\bar p_{i\ell} + \d_{ij} - \d_{\ell j}}
\lb{barMp}
\ea
and hence, both $\bar{\cal F}_{\bar p}\,$ and $\bar{\cal H}_{\bar p}\,$ are eigenspaces of
${\bar p_i}\,$ corresponding to the same common eigenvalues. We set, accordingly
\be
\bar M \,  {\bar a} = \bar a\,  \bar M_{\bar p}\ ,\qquad (\bar M_{\bar p})^j_i = q^{2 \bar p_i + 1 - \frac{1}{n}}\,\d^j_i\ ,\qquad q^{\bar p_i}\vac = q^{\frac{n+1}{2}-i}\vac
\lb{barMMp}
\ee
and assume that the field $\bar u^j_B (\bar z)\,$ and the zero modes $\bar a^\a_j\,$ act on the (bra or ket) vacuum as their left sector counterparts do, i.e.
\ba
\bar u^i_B(\bar z)  \vac = 0 = \bar a^\a_i \vac \quad &{\rm for }&\quad n\ge i\ge 2\ ,\quad {\rm resp.}\nn\\
\lvac \bar u^i_B(\bar z) = 0 = \lvac \bar a^\a_i\quad &{\rm for}&\quad 1\le i\le n-1\ .
\lb{bara-onvac}
\ea
Applying (\ref{uuMpqbar}) to the vacuum we see that its consistency is guaranteed by (\ref{barMMp}) (and in particular, by the "quantum normalization factor" of $\bar M_{\bar p}$) since
\be
e^{-2\pi i \bar \Delta} \vac \ \equiv q^{n-\frac{1}{n}} \vac \ = q^{2 \bar p_1 + 1 - \frac{1}{n}} \vac \ .
\lb{Mpbar-cons}
\ee
It is easy to verify that if $i_\mu \ne i_\nu$ for $\mu\ne\nu\,,$ then $\prod_{\mu=1}^n\, q^{-2p_{i_\mu}} = \id\,$ and hence,
\be
\bar a_{i_1}^{\a_1}\,q^{2{\bar p}_{i_1} +1-{1\over n}}\,
\bar a_{i_2}^{\a_2}\,q^{2{\bar p}_{i_2} +1-{1\over n}}\ldots a^{i_n}_{\a_n} \,q^{2{\bar p}_{i_n} +1-{1\over n}}\,  =\,
\bar a_{i_1}^{\a_1} \bar a_{i_2}^{\a_2}\ldots \bar a_{i_n}^{\a_n}
\lb{qsumbar}
\ee
so that
\be
(\bar M \bar a)^{\a_1}_{i_1}\dots (\bar M \bar a)^{\a_n}_{i_n} \equiv (\bar a\, \bar M_{\bar p})^{\a_1}_{i_1}\dots (\bar a\, \bar M_{\bar p} )^{\a_n}_{i_n} = \bar a^{\a_1}_{i_1}\dots \bar a^{\a_n}_{i_n}\ .
\lb{qbarsum}
\ee
The exchange relations of $\bar a\,$ with ${\bar M}_\pm\,$ are identical to these of ${\bar g}\,$ (\ref{Mgbar}):
\be
\bar M_{\pm 2}\, R_{12}^\mp\, \bar a_1 = \bar a_1\, \bar M_{\pm 2}\ .
\lb{Mabar}
\ee
For $n=2\,$ the zero mode parts of Eqs. (\ref{bara-onvac}) and (\ref{barMp}) read
\be
{\bar a}^\a_2 \vac = 0\ ,\quad \lvac {\bar a}^\a_1 = 0\ ;\qquad q^{\hat{\bar p}}\, {\bar a}^\a_1 =
{\bar a}^\a_1\, q^{\hat{\bar p}+1}\ ,\quad q^{\hat{\bar p}}\, {\bar a}^\a_2 =
{\bar a}^\a_2\, q^{\hat{\bar p}-1}\ ,
\lb{bara-vac}
\ee
respectively, and the transposition is defined as
\ba
&&(q^{\bar p})' = q^{\bar p}\ ,\qquad ({\bar a}^\a_i )' = {\tilde{\bar a}}^i_\a
:= {\bar a}^\b_j \epsilon^{ji} \e_{\b\a}\ ,\quad {\rm i.e.}\lb{transp-bar}\\
&&({\bar a}^1_1)' = q^{-\frac{1}{2}}\,{\bar a}^2_2\ ,\quad
({\bar a}^2_1)' = - q^{\frac{1}{2}}\,{\bar a}^1_2\ ,\quad ({\bar a}^1_2)'
= - q^{-\frac{1}{2}}\,{\bar a}^2_1\ ,\quad ({\bar a}^2_2)' = q^{\frac{1}{2}}\,{\bar a}^1_1\ .
\nn
\ea
Comparing (\ref{transp-bar}) with (\ref{transp2}), we deduce that the inner products of vectors
$\mid\!\bar p , \bar m \rangle := ({\bar a}^1_1)^{\bar m} ({\bar a}^2_1)^{\bar p - 1 -\bar m} \vac\,$
are obtained from (\ref{bilin2}) by complex conjugation:
\be
\lb{bilin2bar}
\langle {\bar p}', {\bar m}' | {\bar p} , {\bar m} \rangle = \d_{{\bar p} {\bar p}'}\, \d_{{\bar m} {\bar m}'}\,
q^{-{\bar m}({\bar m}+1-{\bar p})} [{\bar m}]! [{\bar p}-{\bar m}-1]!\ .
\ee
The quantum group transformation properties of the bar zero modes (cf. (\ref{AdXa1}) for their left sector
counterparts) follow from the exchange relations (\ref{Mabar}) which are equivalent to
$S(\bar M_{\pm 2})\,\bar a_1\,\bar M_{\pm 2} = R_{12}^\mp\, \bar a_1\,,$ or
\ba
&&{\bar k}\, {\bar a}^1_i\, {\bar k}^{-1} = q^{-\frac{1}{2}} {\bar a}^1_i\ ,\quad
{\bar k}\, {\bar a}^2_i\, {\bar k}^{-1} =q^{\frac{1}{2}} {\bar a}^2_i\ ,\nn\\
%% &&{\bar a}^1_i  {\bar E} = q {\bar E}\, {\bar a}^1_i + {\bar a}^2_i\ ,\quad {\bar a}^2_i {\bar E} = q^{-1} {\bar E}\, {\bar a}^2_i\ ,\nn\\
&&q\,{\bar E}\, {\bar a}^1_i = {\bar a}^1_i  {\bar E} - {\bar a}^2_i\ ,\quad {\bar E}\, {\bar a}^2_i = q\, {\bar a}^2_i {\bar E} \ ,\nn\\
&&[ {\bar F} , {\bar a}^1_i ] = 0\ ,\quad [ {\bar F} , {\bar a}^2_i ] = - {\bar K}^{-1} {\bar a}^1_i\qquad\Leftrightarrow\nn\\
&&Ad_{\bar X}^{-1} ({\bar a}_i^\a ) \equiv \sum_{({\bar X})} S({\bar X_1})\,{\bar a}_i^\a \, {\bar X}_2 =
({\bar X}^f)^\a_{~\s}\, {\bar a}_i^\s\ ,\quad \bar X\in \bbU\ .
\lb{AdXa-bar}
\ea
The $2\times 2\,$ matrices ${\bar X}^f \, (\, = {\bar E}^f , {\bar F}^f , {\bar k}^f\, )\,$ in (\ref{AdXa-bar})
coincide with those given in (\ref{bUf}), and the relevant coproducts are displayed in (\ref{coalg2}), (\ref{dk2}).
The action of ${\bar X}$ on ${\bar a}^\a_i\,$ is the same as that of $\s (X)$ on $a^i_\a\,$ where
$\s\,$ is the $\bbU$-algebraic {\em homomorphism}
\be
\s (X) = S(X')\ ,\quad{\rm i.e.}\quad \s (E) = - q^{-1} F\ ,\quad \s (F) = - q E\ ,\quad \s (k) = k^{-1}\ ,
\lb{sig}
\ee
cf. (\ref{EFH}) (supplemented by  $k'=k$) and (\ref{coalg2}), (\ref{dk2}).

\subsubsection{Dynamical $R$-matrix exchange relations for the right sector}

The comparison between the left and right diagonal monodromy matrices, (\ref{Mpq}) and (\ref{barMMp}) (for $\bar a = a_R^{-1}\,$ and $\bar p = p_R$)
indicates that while $q_R = q_L^{-1}\,,$ we should assume, when passing from the left to the right sector,
that $q^{p_L} \to q^{p_R}\equiv q^{\bar p}\,.$ The origin of this rule can be traced back to the $p$-dependent
symplectic forms for the Bloch waves and the zero modes,
(\ref{OB}) and (\ref{Oq}) with $M_p\,$ as defined in (\ref{uuMp}), which change sign when we only
change the sign of $k\,$ but {\em not} that of $\frac{p}{k}\,.$\footnote{This observation is confirmed after
performing a careful examination of both the extended and unextended forms, including $\o_q^{\rm ex}(p)\,$ (\ref{oexqp}) and
$\o_q(p)\,$ (\ref{unextoq}), with $f_{j\ell}(p)\,$ given by (\ref{f01}).}

Another important feature of the left-right correspondence (the classical counterpart of which has been mentioned in Remark 3.7) is that the left and right dynamical $R$-matrices
{\em need not} coincide, as functions of the respective variables $p\,$ and $\bar p\,,$ in the presence of the chiral zero modes.
One can take advantage of this fact to make the bar sector zero modes' exchange relations {\em identical} to the left sector
ones (\ref{ExRaa1}) by setting the "bar" dynamical $R$-matrix $\hat{\bar R}_{12}(\bar p)\,$ equal to the {\em transposed} matrix (\ref{RRp2}):
\be
R_{12}\, \bar a_1\, \bar a_2\, =\, \bar a_2\, \bar a_1\, \bar R_{12}(\bar p)\quad\Leftrightarrow\quad
\hat R_{12}\, \bar a_1\, \bar a_2\, =\, \bar a_1\, \bar a_2\, \hat{\bar R}_{12}(\bar p)\ , \quad \hat{\bar R}_{12}(\bar p) = {^t\!{\hat R}}_{12} (\bar p)\ .
\lb{ExRaabar}
\ee
To show that (\ref{ExRaa1}) and (\ref{ExRaabar}) actually coincide (\,for $p \leftrightarrow \bar p\,$), one uses the symmetry of the constant braid operator $\hat R = P R\,$ corresponding to (\ref{R}),
as well the property (\ref{Rpvv}) of the dynamical one (which is in general {\em not} symmetric) together with the exchange relations (\ref{barMp}) between $\bar a^\a_j\,$ and $q^{\bar p_i}\,.$

We shall now describe how the exchange relations (\ref{ExRaabar}) can be obtained. Let ${\hat R}^\a_{12} (p)\,$ be an arbitrary solution of the dynamical YBE (\ref{QDYBE}) from the set (\ref{RRp2})
(for a certain choice of $\a_{ij}(p_{ij})\,$ satisfying (\ref{canRp})). One first shows that, following the rules above describing the left-right correspondence of $p$-dependent quantities, one derives
\be
({\hat R}^\a)^{-1}_{21} (p_R)\, a_{R1}\, a_{R2}\, = \, a_{R1}\, a_{R2}\, \hat R_{21}^{-1}\quad\Leftrightarrow\quad
\hat R_{12}\, \bar a_1\, \bar a_2\, =\, \bar a_1 \,\bar a_2 \,{\hat R}^\a_{12} (\bar p)\ .
\lb{exRpR}
\ee
Then it remains to just note that transposing the matrix (\ref{RRp2}) (having in mind our preferred one for which $\a_{ij}(p_{ij}) = 1$) is equivalent to choosing
\be
\a_{ij}(p_{ij}) = \a (p_{ij}) = \frac{[p_{ij} + 1]}{[p_{ij}-1]}\ .
\lb{a-a}
\ee
The quasi-classical expansion
\be
\a (p_{ij})^{\pm 1} = \frac{[p_{ij} \pm 1]}{[p_{ij}\mp1]} =
\frac{1\pm \tan\frac{\pi}{k}\, \cot (\frac{\pi}{k}\,p_{ij})}{1\mp \tan\frac{\pi}{k}\, \cot (\frac{\pi}{k}\,p_{ij})}
= 1 \pm 2\,\frac{\pi}{k} \cot (\frac{\pi}{k}\,p_{ij}) + O(\frac{1}{k^2})
\lb{a-b}
\ee
shows that this choice of $\a_{ij}(p_{ij})\,$ changes the sign of the diagonal terms in the classical dynamical
$r$-matrix (\ref{dyn-r-matr}), (\ref{f01})
(for $\b(p) = 0$ and $\b(p) = 2\,\cot p\,$ Eq.(\ref{f-alpha}) gives
$f_{j\ell}(p) = \pm i \, \frac{\pi}{k}\, {\rm cot} \left( \frac{\pi}{k}\, p_{j\ell} \right)$, respectively).

\medskip

\noindent
{\bf Remark 6.1~} The unique {\em symmetric} matrix in the family (\ref{RRp2}) is not rational, the corresponding $\a_{ij}(p_{ij})\,$ being given by the square root of
(\ref{a-a}).\footnote{As already mentioned (in the comments after (\ref{aa2})), we prefer to consider our algebra over
the field of rational functions of $q^{p_j}\,.$}
This choice has been used, for $n=2\,,$ in \cite{BF} (see Eq.(2.22) therein) in connection with
the $U_q(s\ell(2))\ 6j$-symbol interpretation of the entries of $\hat R(p)\,$ \cite{F1, AF, BBB}. As
\be
\sqrt{\frac{[p_{ij} + 1]}{[p_{ij} - 1]}} = 1 + \frac{\pi}{k}\, \b (\frac{\pi}{k}\,p_{ij}) +
O(\frac{1}{k^2})\qquad{\rm for}\qquad \b(p) = \cot p\ ,
\lb{sqrt-a}
\ee
it follows from (\ref{f01}) that the respective $r_{12}(p)\,$ (\ref{dyn-r-matr}) has no diagonal terms, i.e. $f_{j\ell}(p) = 0\,.$

\smallskip

We shall assume in what follows that (\ref{ExRaabar}) holds which implies that ${\bar a}^\a_i\,$ satisfy exchange relations
{\em identical} to those for $a^i_\a\,,$ (\ref{aa2}).

\smallskip

The exchange relations of the "bar" chiral fields $\bar u (\bar x)\,$ corresponding to (\ref{ExRaabar}) (and reproducing together with them (\ref{ggbarLR})) are
\be
\bar u_1(\bar x_1) \, \bar u_2(\bar x_2) = ({\bar R}^{-1}_{12} (\bar p)\, \theta (\bar x_{12}) + {\bar R}_{21} (\bar p)\, \theta (\bar x_{21}))\, \bar u_2(\bar x_2)\, \bar u_1(\bar x_1) \ .
\lb{uubarLR}
\ee
If $\bar u(\bar x)\,$ is the "Bloch wave (or CVO) part" of the respective chiral field with general monodromy matrix $\bar g(\bar x)\,$ (i.e., if it is accompanied by the bar zero modes' matrix),
the dynamical $R$-matrix ${\bar R}_{12} (\bar p)\,$ in (\ref{uubarLR}) should be the same as in (\ref{ExRaabar}).

If however we only consider (left and right sector) fields with {\em diagonal} monodromy, then ${\bar R}_{12} (\bar p)\,$
should match the one for the left sector, (\ref{uuRp}) in order the field $u^A_j(x)\otimes {\bar u}^j_B (\bar x)\,$ to be local
(in this case $p = \bar p$).\footnote{As discussed in Section 4.5.3, this could be only sensible
if there was a way to truncate the common spectrum of (shifted) weights to integrable dominant ones ($p_{i \, i+1}\ge 1\,,\ p_{1n}\le h-1$).}

\subsection{Back to the $2D$ field}

\subsubsection{Locality and quantum group invariance}

As the left and right (or, bar) variables commute, the local commutativity
of the $2D\,$ quantum field $g(x,\bar x) = g(x)\,\bar g(\bar x)\,,$
\be
g_1(x_1,\bar x_1)\, g_2(x_2,\bar x_2) = g_2(x_2,\bar x_2)\, g_1(x_1,\bar x_1)\qquad {\rm for}\qquad x_{12}\, \bar x_{12} > 0
\lb{qloc-q}
\ee
follows from Eq.(\ref{ggbarLR}) (the quantum counterpart of (\ref{gloc})).

Further, Eqs. (\ref{Mgbar}) imply that the entries of the $2D\,$ field commute with those of $\bar M_{\pm} M_{\pm}\,,$
\ba
&&\bar M_{\pm 2}\,M_{\pm 2}\, g_1(x,\bar x) = \bar M_{\pm 2}\,( g_1(x)\, R_{12}^\mp\, M_{\pm 2} )\,\bar g_1(\bar x) = \nn\\
&&= g_1(x)\, (\bar M_{\pm 2}\,R_{12}^\mp\,\bar g_1(\bar x))\,M_{\pm 2} = g_1(x,\bar x)\,\bar M_{\pm 2}\,M_{\pm 2}
\lb{2Dinv}
\ea
(we have used the mutual commutativity of operators in different
sectors\footnote{As $[ (M_\pm)^\a_{~\b} , (\bar M_\pm)^\g_{~\d} ] = 0\,,$
only the {\em matrix} multiplication is important here, not the order of the left and
the right matrix elements: $(\bar M_\pm M_\pm)^\a_{~\b}  \equiv
(\bar M_\pm)^\a_{~\s} (M_\pm)^\s_{~\b} = (M_\pm)^\s_{~\b} (\bar M_\pm)^\a_{~\s}$.});
see (\ref{Mpm2dg}) for a classical analog of this relation.
Having in mind a realization of the $2D\,$ operator theory in the tensor product
of the chiral state spaces ${\cal H}\otimes {\bar{\cal H}}\,,$ we can rewrite (\ref{2Dinv}) as
\be
\left((M_\pm)^\s_{~\b}\otimes (\bar M_\pm)^\a_{~\s}\right) \, g^A_{~\rho}(x)\otimes \bar g^\rho_{~B} (\bar x) =
g^A_{~\rho}(x)\otimes \bar g^\rho_{~B} (\bar x)\,\left((M_\pm)^\s_{~\b}\otimes (\bar M_\pm)^\a_{~\s}\right)
\lb{2Dinv-ind}
\ee
and, as $M_\pm\,$ and $\bar M_\pm\,$ satisfy identical exchange relations, interpret their (matrix) product as the {\em opposite} coproduct in the natural coalgebra structure (\ref{Hopf-FRT}).
The above property reflects the "quantum group invariance" of the $g(x,\bar x)\,.$

\smallskip

In order to discuss the periodicity of the $2D\,$ field (or, which amounts to the same,
its monodromy invariance), we have to be able
to impose the constraint of equal left and right monodromy (\ref{constrC}) at the quantum level. In gauge theories this procedure
corresponds to finding an appropriate "physical" subspace of the extended space of states which, in the case of {\em general}
monodromies, is of the form
\be
{\cal H}\otimes \bar{\cal H} =
\oplus_{p, \bar p}\, {\cal H}_p \otimes {\cal F}_p \otimes {\bar{\cal F}}_{\bar p} \otimes {\bar{\cal H}}_{\bar p}
\lb{HHbar}
\ee
(see (\ref{guaq}), (\ref{space}), (\ref{guaqbar})). We shall study this problem in what follows by exploring in detail the
"$2D\,$ zero modes' kernel" $Q^i_j = a^i_\a\otimes \bar a^\a_j\,$ (acting on the spaces ${\cal F}_p \otimes {\bar{\cal F}}_{\bar p}$)
which is responsible for the "gauge" quantum group symmetry. We shall only notice here that, since the exchange relations of the
chiral zero modes with $M_\pm\,$ and ${\bar M}_\pm\,$ (\ref{aMpm}), (\ref{Mabar}) are the same as those of the chiral fields
(\ref{Mg}), (\ref{Mgbar}), a relation similar to (\ref{2Dinv-ind}) holds for $Q^i_j\,$ as well:
\be
[\, (M_\pm)^\s_{~\b}\otimes (\bar M_\pm)^\a_{~\s} \,,\, a^i_\rho\otimes \bar a^\rho_j \, ] = 0\ , \qquad {\rm or}\qquad
[\, \Delta' (M_\pm) \,,\, Q \,] = 0\ .
\lb{Qinv}
\ee
It is also easy to verify that for $p = \bar p\,$ the left and right monodromies cancel so that
$u^A_B(z, \bar z) := u^A_j(z)\otimes{\bar u}^j_B(\bar z)\,$ is single valued:
\ba
&&e^{2\pi i (\Delta - \bar \Delta)} u^A_B(e^{2\pi i} z , e^{-2\pi i} \bar z) \, v =
(M_p)^\ell_j\, u^A_\ell(z) \otimes \bar u^m_B(\bar z) (\bar M_{\bar p})^j_m \, v = \nn\\
&&= u^A_j(z)\, q^{-2p_j -1+\frac{1}{n}}\otimes \bar u^j_B(\bar z)\, q^{2p_j +1-\frac{1}{n}}\,v
=  u^A_B(z , \bar z) \, v\ ,\quad \forall\, v \in {\cal H}_p\otimes{\bar{\cal H}}_p\ .\qquad\qquad
\lb{mon-inv-u}
\ea
(We have used (\ref{uuMpq}), (\ref{Mpq}), (\ref{ExRup}), (\ref{uuMpqbar}) and (\ref{barMMp}).)

\smallskip

Hence, deducing the diagonality ($p = \bar p$) and the truncation of $p\,$ to integrable weights from
the properties of $Q^i_j\,,$ we would have a bridge from the canonically quantized to the unitary WZNW model.
We shall first show how this idea can be realized in the $n=2\,$ case, and then try to extend
the results to general $n\,.$

\subsubsection{Physical factor space of the unitary $2D\,$ model for $n=2$}

We shall construct in the present section, for $n=2\,,$ a truncated (finite dimensional) Fock representation of the
$\bbU$-invariant bilinear combinations of left and right zero modes and obtain, as a result,
a description of the unitary $2D\,$ WZNW model as a rational CFT in a gauge-field-theory-like setting.

Before discussing the action of the WZNW field $g(z,\bar z)\,$ on the extended state space (\ref{HHbar}) we shall tackle the
intermediate problem concerning the $2D\,$ zero modes acting on the tensor product of chiral Fock spaces
${\cal F}\otimes {\bar{\cal F}} = \oplus_{p, \bar p}\,{\cal F}_p \otimes {\bar{\cal F}}_{\bar p}\,$ \cite{FHT2, FHT3, Goslar, DT, FH1}.
To this end, as mentioned above, we have to introduce the matrix of operators
\be
Q = ( Q^i_j ) =
\begin{pmatrix}Q^1_1&Q^1_2\cr Q^2_1&Q^2_2\end{pmatrix} \equiv
\begin{pmatrix} A&B\cr C&D\end{pmatrix}\ ,\quad
Q^i_j = a^i_\a\otimes \bar a^\a_j\ .
\lb{Qmatr}
\ee
It is convenient to write down the left and right sector zero modes' exchange relations in the form (\ref{detc-n2}), (\ref{altEx}) which only
involves the constant (but not the dynamical) $R$-matrix and also reflects the determinant conditions
$\det (a) = [\hat p]\ ,\ \det (\bar a ) = [\hat{\bar p}]\,,$
\ba
&&q^{\frac{1}{2}}\,a^i_\rho a^j_\s {\hat R}^{\rho\s}_{~\a\b} = a^j_\a a^i_\b - q^{1-\hp_{ij}}
\e_{\a\b}\ ,\qquad a^j_\a a^i_\b \, \varepsilon^{\a\b} = [\hp_{ij}+1]\ ,
\nn\\
&&q^{\frac{1}{2}} {\hat R}^{\rho\s}_{~\a\b}\,{\bar a}_i^\a {\bar a}_j^\b = {\bar a}_j^\rho {\bar a}_i^\s - q^{1-\hat{\bar p}_{ij}} \e^{\rho\s}\ ,
\qquad \varepsilon_{\a\b}\,{\bar a}_j^\a {\bar a}_i^\b  = [{\hat{\bar p}}_{ij}+1]\quad (\,i\ne j\,)\ ,\qquad\lb{altExbar}\\
&&q^{\frac{1}{2}}\,a^i_\rho a^i_\s {\hat R}^{\rho\s}_{~\a\b} = a^i_\a a^i_\b \ ,\quad
q^{\frac{1}{2}} {\hat R}^{\rho\s}_{~\a\b}\,{\bar a}_i^\a {\bar a}_i^\b = {\bar a}_i^\rho {\bar a}_i^\s \quad\Leftrightarrow\quad
a^i_\rho a^i_\s \e^{\rho\s} = 0 =\e_{\a\b} \,{\bar a}_i^\a {\bar a}_i^\b
\nn
\ea
(here, as usual, $\hp = {\hat p}_{12}\,,\ {\hat{\bar p}} = {\hat{\bar p}}_{12}$).
With the help of (\ref{altExbar}) we are able to show that
\ba
&&B A = (a^1_\rho\otimes \bar a^\rho_2)\, (a^1_\s\otimes \bar a^\s_1 ) =
a^1_\rho a^1_\s\otimes \bar a^\rho_2 \,\bar a^\s_1 =
a^1_\rho a^1_\s \otimes ( q^{\frac{1}{2}}\,{\hat R}^{\rho\s}_{~\a\b} \bar a^\a_1 \bar a^\b_2 + q^{1-\hat{\bar p}} \e^{\rho\s}) =\nn\\
&&=a^1_\a a^1_\b \otimes {\bar a}_1^\a {\bar a}_2^\b  =
(a^1_\a\otimes {\bar a}_1^\a) \, (a^1_\b \otimes {\bar a}_2^\b ) = A B
\lb{BAp}
\ea
and similarly, $C A = A C\,,\ B D = D B\,,\ C D = D C\,,$ i.e. the off-diagonal elements of the matrix $Q\,$
commute with the diagonal ones.

On the other hand, we obtain
\ba
&&B C = (a^1_\a\otimes \bar a^\a_2)\, (a^2_\b\otimes \bar a^\b_1 ) = a^1_\a a^2_\b \otimes \bar a^\a_2 \bar a^\b_1 =
( q^{\frac{1}{2}}\,a^2_\rho a^1_\s {\hat R}^{\rho\s}_{~\a\b} + \e_{\a\b} q^{\hp +1} ) \otimes \bar a^\a_2 \bar a^\b_1 = \quad\nn\\
&&= a^2_\rho a^1_\s \otimes ({\bar a}_1^\rho {\bar a}_2^\s - q^{\hat{\bar p} +1} \e^{\rho\s}) + q^{\hp +1} \otimes [{\hat{\bar p}}+1] = \lb{BCN}\\
&&= a^2_\rho a^1_\s \otimes {\bar a}_1^\rho {\bar a}_2^\s - [\hat  p +1] \otimes q^{\hat{\bar p} +1} + q^{\hp +1} \otimes [{\hat{\bar p}}+1] = \nn\\
&&= C B + \frac{N-N^{-1}}{q-q^{-1}}\ ,\quad N^{\pm 1} :=  - \, q^{\pm\hat p} \otimes q^{\mp\hat{\bar p}}\ .
\nn
\ea
Eq.(\ref{ExRapn2}) and its right sector counterpart (\ref{bara-vac}) imply
\be
N B = q^2 B N\ ,\qquad N C = q^{-2} C N\ .
\lb{NBC}
\ee
Similarly, for the diagonal elements of $Q\,$ (\ref{Qmatr}) we find
\ba
&&A D = (a^1_\a\otimes \bar a^\a_1)\, (a^2_\b\otimes \bar a^\b_2 ) = a^1_\a a^2_\b \otimes \bar a^\a_1 \bar a^\b_2 =
( q^{\frac{1}{2}}\,a^2_\rho a^1_\s {\hat R}^{\rho\s}_{~\a\b} + \e_{\a\b} q^{\hp +1} ) \otimes \bar a^\a_1 \bar a^\b_2 = \quad\nn\\
&&= a^2_\rho a^1_\s \otimes ({\bar a}_2^\rho {\bar a}_1^\s - q^{1-\hat{\bar p}} \e^{\rho\s}) - q^{\hp +1} \otimes [{\hat{\bar p}}-1] = \lb{ADL}\\
&&= a^2_\rho a^1_\s \otimes {\bar a}_2^\rho {\bar a}_1^\s - [\hat p +1] \otimes q^{1-\hat{\bar p}} - q^{\hp +1} \otimes [{\hat{\bar p}}-1] =\nn\\
&&= D A + \frac{L-L^{-1}}{q-q^{-1}}\ , \quad L^{\pm 1} := - \, q^{\pm{\hat p}} \otimes q^{\pm{\hat{\bar p}}}
\nn
\ea
as well as
\be
L A = q^2 A L\ ,\qquad L D = q^{-2} D L\ .
\lb{ADp}
\ee

To summarize, the entries of the operator matrix $Q\,$ (\ref{Qmatr}) generate two commuting $U_q(s\ell(2))\,$ algebras.
The first one contains the off-diagonal elements $B\,$ and $C\,$ as well as the operators $N^{\pm 1}\,,$ and the other the diagonal
ones, $A\,$ and $D\,,$ together with $L^{\pm 1}\,.$

\smallskip

As a unitary rational CFT, the WZWN model on a compact group only involves integrable representations of the corresponding affine
algebra. In the $\widehat{su}(2)_k\,$ case these correspond to shifted affine weights with $1\le p \le k+1 = h-1\,.$ We shall
sketch in what follows how such a physical space can be defined within the extended state space ({\ref{HHbar}). As a first step we
consider the tensor product of quotient zero modes algebra ${\cal M}^{(h)}_q$ (\ref{ah}), (\ref{qhpij}) and its right sector
counterpart ${\bar {\cal M}}^{(h)}_q\,,$ determined by the additional relations
\be
(a^i_\a)^h = 0 = ({\bar a}^\b_j )^h \qquad (\, i,j,\a,\b = 1,2\,)\ ,\qquad q^{2h \hat p} = \id
= q^{2h \hat {\bar p}}\ .
\lb{ABCDh}
\ee
The corresponding "restricted" Fock representation
\be
{\cal F}^{(h)} \otimes {\bar{\cal F}}^{(h)} = {\cal M}^{(h)}_q \otimes {\bar{\cal M}}^{(h)}_q\,\vac
\lb{Fock-h2}
\ee
forms a $h^4$-dimensional subspace of ${\cal F}\otimes {\bar{\cal F}}\,.$ (${\cal F}^{(h)}\,$ contains the IR
$\, {\cal F}_p \simeq V^+_p\,$ for $1\le p \le h\,$ as well as the irreducible quotients of ${\cal F}_{h+p}\,$ isomorphic to
$V^+_{h-p}\,$ for $1\le p\le h-1\,,$ cf. (\ref{shexseq}) so its dimension is $2\, (1+\dots + h-1) + h = h^2\,.$)

\smallskip

As we shall show below, Eqs. (\ref{ABCDh}) imply that the the four entries of the operator matrix $Q\,$ (\ref{Qmatr}) generate two
commuting restricted $\bU\,$ algebras (\ref{Uq-res}). The vacuum representation of the one formed by the diagonal elements $A\,$
and $D\,$ (\ref{ADL}), (\ref{ADp}) defines the zero modes' projection of the unitary $2D\,$ WZNW $SU(2)_k\,$ model physical
space in ${\cal F}^{(h)} \otimes {\bar{\cal F}}^{(h)}\,.$ Indeed, introducing
\be
A_1 = a^1_1 \otimes {\bar a}^1_1\ ,\quad A_2 = a^1_2 \otimes {\bar a}^2_1\qquad \Rightarrow \quad A_2 A_1 = q^2A_1 A_2
\lb{A1A2}
\ee
(the implication follows from the last two equalities (\ref{altExbar}) which are equivalent to
$a^i_2 a^i_1 = q\, a^i_1 a^i_2\,$ and ${\bar a}^2_i {\bar a}^1_i = q \,{\bar a}^1_i {\bar a}^2_i\,,$
respectively) and similarly for $B, C\,$ and $D\,,$ one derives the relations
\be
A^h = 0 = D^h\ ,\quad L^{2h} = \id\ ;\qquad B^h = 0 = C^h\ ,\quad N^{2h} = \id\ .
\lb{ADLh}
\ee
The calculation is based on the $q$-binomial identity
\be
A_2 A_1 = q^2 A_1 A_2\quad\Rightarrow\quad (A_1+A_2)^m = \sum_{r=0}^m \left({m\atop r}\right)_+ A_1^r A_2^{m-r}
\lb{qbin}
\ee
where
\ba
&&\left({m\atop r}\right)_+ = \frac{(m)_+!}{(r)_+! (m-r)_+!}\ ,\qquad (r+1)_+! = (r+1)_+ (r)_+!\ ,\qquad (0)_+! = 1\ ,\nn\\
&&(r)_+ := \frac{q^{2r}-1}{q^2-1} = q^{r-1} [r]\qquad\Rightarrow\qquad \left({m\atop r}\right)_+ = q^{r(m-r)} \left[{m\atop r}\right]
\ea
implying, in particular,
\be
A^h = (A_1 + A_2)^h = A_1^h + \sum_{r=1}^{h-1} \left({h\atop r}\right)_+ A_1^r A_2^{h-r} + A_2^h = 0\ .
\lb{Ah}
\ee
From Eqs. (\ref{a-vac}), (\ref{bara-vac}) and (\ref{a2.n}), (\ref{barMMp}) we obtain further
\ba
&&D \vac = 0\ ,\quad \lvac A = 0\ ,\quad L \vac = - q^2 \,\vac\ ,\nn\\
&&B \vac = 0 = C \vac \ ,\quad \lvac B = 0 = \lvac C\ ,\quad N \vac = - \vac \ .\qquad
\lb{Dvacetc}
\ea
Hence, the vacuum representation of the $\bU\,$ triple formed by the operators $B, C\,$ and $N\,$ (commuting with  $A, D\,$ and $L\,,$ see
(\ref{BAp})) is equivalent to $V_1^-\,.$ Applying powers of $A\,$ on the vacuum, we generate a $h$-dimensional representation of
$\bU\,$ equivalent to the Verma module ${\cal V}_1^-\,$ (\ref{Verma}) (for $E \to A\,,\ F \to D\,,\ K \to L$). Indeed, defining
\be
\mid m \rangle := \frac{A^{m}}{[m]!} \vac \ ,\quad \ m=0,\dots , h-1\ , \lb{m-vect} \ee we derive \be A \mid m \rangle
= [m+1] \mid m + 1 \rangle\ ,\quad D \mid m \rangle = [m+1] \mid m -1\rangle\ ,\quad (L + q^{2(m+1)}) \mid m \rangle = 0\
\lb{ADm}
\ee
(assuming that $D\! \vac = 0\,,$ see the first Eq.(\ref{Dvacetc})). It follows from (\ref{specK-Vp}) that the
$1$-dimensional submodule spanned by the vector $| h-1 \rangle\,$ is isomorphic to the IR $V^-_1\,$
(note that $ A\! \mid h-1 \rangle = 0 = D\! \mid h-1 \rangle $), and the $(h-1)$-dimensional
irreducible subquotient spanned by the vectors $\mid m \rangle\,$ for $m=0,\dots, h-2\,,$ to $V_{h-1}^+\,.$

Assuming that $(X\otimes Y)' = X'\otimes Y'\,,$ Eqs. (\ref{transp2}) and (\ref{transp-bar}) imply
\ba
&&L' = L\ ,\quad N' = N \qquad {\rm as\ well\ as} \qquad ( Q^i_j)' = \epsilon_{i\ell}\, \epsilon^{jm} Q^\ell_m\ ,\qquad {\rm i.e.}\nn\\
&&A' = (Q^1_1)' = Q^2_2 = D\ ,\qquad  B' = (Q^1_2)' = - Q^2_1 = - C \ .\qquad\quad
\lb{transpQ}
\ea
(Note that the transposition (\ref{transpQ}) differs from (\ref{EFH}).) Applying (\ref{ErFr}),
we obtain (for $P\,$ playing the auxiliary role of the Casimir operator $C$)
\ba
&&0 = \l^2 A D \vac = (P - q^{-1} L - q L^{-1}) \vac = (P + q + q^{-1}) \vac \quad\Rightarrow\qquad\lb{DmAm}\\
&&D^m A^m \vac = \l^{-2m} \prod_{s=1}^{m} (q^{2s+1} + q^{-2s-1} - q - q^{-1} ) \vac = [m+1]\,([m]!)^2 \vac
\nn
\ea
and finally,
\be
\langle m' \mid m \rangle = [m+1]\,\d_{m m'}\ , \quad  \langle m' \mid := \lvac \frac{D^{m'}}{[m']!}\ ,\quad \ m=0,\dots , h-1\ .
\lb{m'm}
\ee
We see, in particular, that the vector $| h-1 \rangle\,$ spanning the $1$-dimensional submodule $V^-_1\,$ is
orthogonal to all vectors in the Verma module.

The fact that the Gram matrix $diag\, (1 , [2 ] , \dots , [h-1], 0)$ of the vectors $\{ {|m\rangle} \}_{m=0}^{h-1}\,$ is real
(in contrast with (\ref{bilin2}), (\ref{bilin2bar})) allows to introduce a Hermitean structure on their complex span
\cite{DT}.\footnote{In \cite{DT} the nilpotency ($A^h=0$\,) of the operator $A\,$ is used to define a BRST-type operator by {\em generalized}
(as $h>2$) homology methods.} To this end we define a sesquilinear (antilinear in the first argument and linear in the second) inner
product $( \,.\, |\, .\, )\,$ which coincides with the bilinear one (\ref{DmAm}) on the real span of (\ref{m-vect}). The
corresponding {\em antilinear} antiinvolution (hermitean conjugation of operators $X \to X^\dagger$) defined by
$( u | X^\dagger v) = (X u | v )\,$ is given by
\be
D^\dagger = A\ ,\qquad L^\dagger = L^{-1}\qquad (\, q^\dagger = q^{-1}\,)\ .
\lb{HermF}
\ee
It thus differs from the transposition (\ref{transpQ}) when applied to $L\,,$ still leaving the relations (\ref{ADL}), (\ref{ADp}) invariant.

\smallskip

We shall denote by ${\cal F}'\,$ the $h$-dimensional (complex) vector space spanned by $\{ {|m\rangle} \}_{m=0}^{h-1}\,$ and
endowed with the (semi)positive inner product described above, and by ${\cal F}''\,$ its $1$-dimensional null subspace
${\mathbb C}\,|h-1\rangle\,.$ By construction, ${\cal F}'\,$ is the subspace of the tensor product of left and right
Fock spaces ${\cal F}\otimes {\bar{\cal F}}\,$ generated from the vacuum by the diagonal elements of the
matrix $Q\,$ (\ref{Qmatr}). We shall show below that the action of $Q\,$ on it is {\em monodromy invariant}, in the sense that
\be
Q_M\, v = Q \, v \equiv \begin{pmatrix}
A&0\cr 0&D\end{pmatrix}\, v\qquad \forall\ v \in {\cal F}'\ ,\qquad (Q_M)^i_j :=
(a\, M)^i_\a \otimes ({\bar M}^{-1} \bar a )^\a_j\ .
\lb{QMQm}
\ee
Indeed, using (\ref{aMMpa}), (\ref{Mpq}) and (\ref{barMMp}), we obtain
\ba
&&(Q_M)^i_j = ( M_p \, a )^i_\a \otimes ({\bar a}\, {\bar M}_{\bar p})^\a_j = Q^i_j\, (q^{-2 p_i} \otimes q^{2 {\bar p}_j})\ ,\nn\\
&&Q = \begin{pmatrix} A&B\cr C&D\end{pmatrix}\ \rightarrow\ Q_M = - \begin{pmatrix} A&0\cr 0&D\end{pmatrix}
\begin{pmatrix} N^{-1}&0\cr 0&N\end{pmatrix} - \begin{pmatrix} L^{-1}&0\cr 0&L\end{pmatrix}
\begin{pmatrix} 0&B\cr C&0\end{pmatrix}\ . \qquad\qquad
\ea
Eq.(\ref{QMQm}) now follows from
\be B \,v = C\, v = 0\ ,\qquad N^{\pm 1}\, v = - v \qquad \forall\ v \in {\cal F}'\ .
\lb{BCNm}
\ee

The relation (\ref{Qinv}) (valid for general $n$) implies that every vector $v\in{\cal F}'\,$ is $\bbU$-invariant,
$X \, v = \varepsilon (X)\, v\,,$ where $X\in\bbU\,$ is given by the Fock representation of the opposite coproduct:
\be
\left((M_\pm)^\s_{~\b}\otimes (\bar M_\pm)^\a_{~\s}\right) =
\pi_{\cal F} \otimes \pi_{\bar{\cal F}}\, \Delta' ((M_\pm)^\a_{~\b})\ .
\lb{MbMDp}
\ee
Indeed, (\ref{MbMDp}) shows that (\ref{Qinv}) is equivalent to
\be
[\, \pi_{\cal F} \otimes \pi_{\bar{\cal F}}\, \Delta'(X) \,,\, Q^i_j \, ] = 0 \qquad \forall\, X \in \bbU
\lb{QDp}
\ee
which can be alternatively substantiated for $n=2$: by using the relations (\ref{AdXa1}),
(\ref{AdXa-bar}) and the coproduct formulae (\ref{coalg2}), (\ref{dk2}), one can easily verify that the operators
$Q^i_j = a^i_1\otimes {\bar a}^1_j + a^i_2\otimes {\bar a}^2_j\,$ commute with
\be
k\otimes {\bar k}\ ,\quad K\otimes {\bar E} + E \otimes \id\ ,\quad \id \otimes {\bar F} + F \otimes {\bar K}^{-1}\ .
\lb{kKEF}
\ee
Thus the $\bbU$-invariance of all vectors in ${\cal F}'\,$ follows from the invariance of the vacuum vector.

We thus have a finite dimensional toy model realizing typical ingredients of the axiomatic approach to gauge theories (see e.g.
\cite{BLOT, Str}) -- an extended state space ${\cal F}^{(h)} \otimes {\bar{\cal F}}^{(h)}\,,$ a pre-physical subspace
${\cal F}'\,$ on which the scalar product is positive semidefinite, a subspace of zero-norm vectors ${\cal F}''\,,$ and a physical subquotient
\be
{\cal F}^{phys} = {\cal F}' / {\cal F}'' \simeq \oplus_{p=1}^{h-1} {\cal F}^{phys}_p\ ,\qquad {\cal F}^{phys}_p :=
{\mathbb C}\,| p-1\rangle =  {\mathbb C}\, A^{p-1} \vac\ \ .
\lb{Fph}
\ee
In this picture the entries $Q^i_j\,$ of the operator matrix (\ref{Qmatr}) play the role of observables and $\bbU\,,$ of
the (generalized) gauge symmetry leaving them invariant, see (\ref{QDp}).

\smallskip

It follows from the above that it is consistent to present the $2D\,$ field corresponding to the unitary rational CFT
${\widehat{su}}(2)_k\,$ WZNW model in the following {\em diagonal} form:
\be
g^A_B(z,\bar z) = \sum_{j=1}^2 u^A_j (z)\,\otimes\, Q^j_j\,\otimes\, {\bar u}^j_B ({\bar z})\ , \quad{\rm acting \ on}
\quad {\cal H}^{phys} = \oplus_{p=1}^{h-1} {\cal H}_p \otimes\, {\cal F}_p^{phys} \otimes {\bar{\cal H}}_p\ \ .
\lb{2Dg}
\ee
(The fact that $p = \bar p\,$ follows from the triviality of the action of the off-diagonal entries of $Q\,$ on ${\cal F}'\,$
(\ref{BCNm}).) Note that the monodromy invariance of $Q\,$ (\ref{QMQm}) ensures the periodicity (\ref{gzzbar-per}) of
$g(z,\bar z)\,$ on ${\cal H}^{phys}$:
\be
(Q_M - Q)\,{\cal F}_p^{phys} = 0\quad\Rightarrow\quad
\left( g(e^{2\pi i} \, z, e^{-2\pi i} \, \bar z) - g(z,\bar z) \right)\, {\cal H}^{phys} \, = 0\ .
\lb{2dper}
\ee
Recalling that $M=M_L\,,\ {\bar M}^{-1} = M_R\,$ (cf. also (\ref{gzM})), one can assert that
Eq.(\ref{2dper}) is the quantum implementation of the constraint (\ref{O2alt}) of equal left and right monodromy matrices.

\smallskip

The physical representation space ${\cal F}^{phys}\,$ reproduces the structure of the ${\widehat{su}}(2)_k\,$ fusion ring
(\ref{fusion-su2-I}) generated by the integrable representations of the affine algebra \cite{V, P, DFMS} in the following way.
The (binary) {\em fusion} matrices $F_h^{(\l)}\,$ corresponding to a primary field of weight $\l$
that can be extracted from the action of the operator $(A+D)^\l\,$ for $\l = 0, 1, \dots , k\,$
in the basis $|m\rangle\,$ (\ref{m-vect}) have Perron-Frobenius eigenvalue
$[\l +1]\,$ and provide a representation of the ring (\ref{fusion-su2-I}).

The simplest non-trivial example is given by the step operator (for $\l=1$) when the characteristic polynomial $C_h(x)\,$ of the
$(h-1)\times (h-1)\,$ fusion matrix
\be
F_h^{(1)} = \begin{pmatrix} 0&1&0&\dots&0&0\cr 1&0&1&\dots&0&0\cr
0&1&0&\dots&0&0\cr \dots&\dots&\dots&\dots&\dots&\dots\cr
0&0&0&\dots&0&1\cr 0&0&0&\dots&1&0 \end{pmatrix}
\lb{F1}
\ee
satisfies, as a function of its index, the recurrence relation and initial conditions
\be
C_{h+1}(x) = - x C_h(x) - C_{h-1}(x)\ ,\quad C_2(x) = - x\ ,\quad C_3(x) = x^2-1\ .
\lb{char-eq-F1}
\ee
It follows from (\ref{recurseUm}) that, for $h\ge 2\,,$ $C_h(x) = U_h(-x)\,$
where the polynomials $U_m(x)\,$ are defined in (\ref{Um}). Hence, the eigenvalues of the real symmetric matrix
(\ref{F1}) coincide with the roots $x_j = 2 \cos\frac{\pi j}{h}\,,\ j = 0,\dots, h-1\,$ of $U_h(x)\,.$
In particular, the maximal (Perron-Frobenius) eigenvalue of $F_h^{(1)}\,$ is $2 \cos\frac{\pi}{h} = [2]\,.$

\smallskip

The above results shed light on the mechanism by which the quantum group, albeit remaining "hidden" in the $2D\,$ model,
leaves its imprints on the fusion rules.

\subsubsection{The $Q$-algebra for general $n\,$}

For $n>2\,$ a reduction of the $Q$-algebra to the fusion ring of the unitary $2D\,$ model
(that would generalize the one for $n=2\,$ displayed above) is not known. The first difficulty stems from the
$n$-linear determinant condition which is not easily coupled with the exchange relations for $n>2\,.$
We shall list in what follows those properties of the $Q$-algebra which we are able to derive presently from the
exchange relations of $a^i_\a\,$ and $\Ba_i^\a\,.$

\smallskip

We shall assume that the exchange relations of ${\bar a}^\a_i\,$ originate from (\ref{ExRaabar}) and hence, are
identical to those for $a^i_\a\,$ (\ref{aa2}):
\ba
&&\Ba_j^\b \Ba_i^\a\,[{\hat \Bp}_{ij}-1]
= \Ba_i^\a \Ba_j^\b\,[{\hat \Bp}_{ij}] - \,\Ba_i^\b\, \Ba_j^\a \,q^{{\epsilon}_{\a\b}{\hat \Bp}_{ij}} \quad
(\,{\rm for}\quad i\ne j \quad {\rm and}\quad\alpha\ne\beta\, )\ ,\nn\\
&&[\Ba_j^\alpha , \Ba_i^\alpha ] = 0\ ,\qquad \Ba_i^\alpha \Ba_i^\beta =
q^{{\epsilon}_{\alpha\beta}}\, \Ba_i^\beta \Ba_i^\alpha\ ,\qquad \a,\b, i,j=1,\dots, n\ .
\lb{aa2barn}
\ea
The commutation relations of $p_j\,$ with $a^i_\a\,$ and their action on the vacuum are given in (\ref{pacomm}) and (\ref{a2.n}), respectively;
the analogous formulae for the bar quantities are contained in (\ref{barMp}), (\ref{barMMp}) and (\ref{bara-onvac}).

\smallskip

Define the $2D\,$ zero mode $n\times n\,$ matrix of quantum group invariant operators as in (\ref{Qmatr}):
\be
Q = (Q^i_j)\ ,\quad Q^i_j = a^i_\a\otimes \bar a^\a_j \ (\, \equiv \sum_{\a = 1}^n a^i_\a\otimes \bar a^\a_j\,)\ .
\lb{defQn}
\ee
As it follows from the chiral exchange relations (\ref{aa2}), (\ref{aa2barn}),
the quadratic exchange relations for the entries of $Q\,$ involve two {\em dynamical} $R$-matrices:
\ba
&&\hat R_{12}(p) \, a_1 \, a_2 = a_1 \, a_2 \, \hat R_{12}\ , \qquad
\hat R_{12}\, \bar a_1\, \bar a_2\, = \, \bar a_1\, \bar a_2\, {\hat {\bar R}}_{12}(\bar p)\qquad \Rightarrow \nn\\
&&\qquad\qquad\qquad{\hat R}_{12} (p)\,Q_1\,Q_2 = Q_1\,Q_2\,{\hat {\bar R}}_{12}(\bar p)\ .
\lb{RQQ0}
\ea
Using the definition (\ref{defQn}) and the relation (\ref{RQQ0}) one can derive the following properties of
the $Q$-operators which we shall formulate without proof:

\medskip

\noindent
{\bf Lemma 6.1~} {\em The entries of $Q\,$
belonging to {\underline{the same row or column}} commute:}
\be
[Q^j_i , Q^\ell_i ] = 0 = [Q^i_j , Q^i_\ell ]\ .
\lb{QQcomm}
\ee

\smallskip

\noindent
{\bf Lemma 6.2~} {\em The entries of $Q\,$ that belong to {\underline{different rows and columns}} satisfy}
\ba
&&( [p_{ij}-1]\otimes [\Bp_{\ell m} ]- [p_{ij} ] \otimes [\Bp_{\ell m}-1])\, Q^i_\ell \, Q^j_m =\nn\\
&&= [p_{ij}-1]\otimes [\Bp_{\ell m} ]\,Q^j_\ell\, Q^i_m - [p_{ij} ]\otimes [\Bp_{\ell m} -1 ]\,  Q^i_m\, Q^j_\ell\ ,
\quad i\ne j\,,\ \ell\ne m\ .\qquad\qquad
\lb{QQijlm}
\ea

\smallskip

\noindent
{\bf Lemma 6.3~} {\em The relations $\ ( a^i_\a )^h = 0 = ( \Ba_j^\a )^h\quad \forall\ 1\le i, j, \a \le n\,$ imply}
\be
(Q^i_j)^h = 0\ .
\lb{Qh0}
\ee

\medskip

We have to return to the complete description of the $Q$-algebra and its
Fock representation for $n\ge 3\,$ in a future work.

\section{Discussion and outlook}

\setcounter{equation}{0}
\renewcommand\theequation{\thesection.\arabic{equation}}

The WZNW model has been and continues to be a source of inspiration and a laboratory for (conformal) quantum field theory
since its inception thirty years ago. A prime example of a
rational conformal field theory \cite{MS2, DFMS} whose partition functions
are expressed in terms of modular forms yielding a beautiful
(ADE) classification \cite{CIZ1}, it also appears as an integrable model giving
rise to an early physical realization of quantum groups and their generalizations (in particular by Ocneanu \cite{PZ});
the rich and well developed subject of boundary CFT \cite{C, GTT} naturally arises from this model;
its non-compact counterpart serves as a testing ground for the theory of a black hole \cite{Gaw1};
it teaches us via dualities about $4$-dimensional CFT \cite{AGT, DLH}.

Rather than aiming at a broad coverage of the field the present review concentrates on a particular subject: the canonical
Lagrangian approach, starting from the classical theory with its Poisson-Lie symmetry and going to its quantization involving
the tricky properties of the quantum universal enveloping algebra $U_q(s\ell(n))\,$ at $q\,$ a root of unity. We encounter
on the way and treat in detail topics deserving a greater popularity than they currently enjoy: the first order Hamiltonian
formalism of the Polish school (see \cite{G} and the references therein) and of B. Julia \cite{JS},
Gaw\c{e}dzki's derivation of the constant $r$-matrix relations for the chiral field \cite{G, FG1},
our treatment of the $su(n)\,$ zero modes (classical and quantum \cite{HIOPT, FHIOPT, FHT6}), to mention a few.

The canonical quantization of the quantum group covariant chiral WZNW
field naturally gives rise to an extended state space \cite{STH, HST}.
As exemplified in the ${\widehat{su}}(2)_k\,$ case \cite{HP, HF}, such
a theory has the features of a logarithmic CFT \cite{RozS, Gu}
characterized by the non-diagonalizability of the conformal energy operator $L_0\,.$
The "Kazhdan-Lusztig correspondence" advocated in \cite{FGST1, FGST2}
states the equivalence of the (non-semisimple) representation categories of certain
LCFT fusion algebras and (restricted) quantum groups. Following this idea, we have explored \cite{FHT7} the structure
of the $n=2\,$ zero modes' Fock space as a $\bU\,$ module.
The fusion algebra of the unitary model has been recovered within this wider framework.

In order not to postpone indefinitely the completion of this review, we did not discuss here more direct approaches
to the unitary model through a weak $C^*\,$ Hopf algebra \cite{BNS} or Ocneanu's double triangle algebra \cite{PZ}.
Their relation to the extended quantum group symmetric picture considered here deserves further exploration.
It would be also interesting to relate the Q-algebra of Section 6.2 with the "affine local plactic algebra" of
\cite{KS} (see also \cite{Ko2, Wa2012}).

\section*{Acknowledgments}
\addcontentsline{toc}{section}{Acknowledgements}

We have profited from insights and ideas during our collaboration with Anton Alekseev,
Michel Dubois-Violette, Krzysztof Gaw\c{e}dzki, Alexei Isaev, Oleg Ogievetsky, Roman Paunov,
Todor Popov, Pavel Pyatov and Yassen Stanev.
We are also indebted to J\'anos Balog, L\'aszl\'o Feh\'er, Ctirad Klim\v{c}\'{\i}k, L\'aszl\'o Palla
and Valya Petkova for useful discussions.
PF acknowledges the support of the Italian Ministry of University and Research (MIUR).
LH thanks INFN, Sezione di Trieste, and both he and IT thank ICTP and SISSA/ISAS, Trieste
for hospitality during multiple visits in the course of this work.

\newpage

%%%%%%%%%%%%%%%%%%%%%%%%%%%%%%%%%%%%%%%%%%%%%%%%%%%%%%%%%%%%%%%%%%%%%%%%%%%%%%%%%%%%%%%

\section*{Appendix A. Semisimple Lie algebras}
\addcontentsline{toc}{section}{Appendix A. Semisimple Lie algebras}

\setcounter{equation}{0}
\renewcommand\theequation{A.\arabic{equation}}

Here we shall introduce some relevant notions and fix our conventions about semisimple Lie algebras
(see e.g. \cite{FulH, FS, Hum, Serre}).

\smallskip

Let ${\cal G_{\mathbb C}}\,$ be the complexification of the Lie algebra ${\cal G}$ of a compact
semisimple Lie group $G\,.$ We shall use throughout this paper the notation ${\rm tr}$ for the Killing form.
It is proportional to the {\em matrix trace} ${\rm Tr} = {\rm Tr}_\pi$ in any (non-trivial) finite dimensional
irreducible representation $\pi\,$ of ${\cal G}\,,$
\be
{\rm tr}\,(XY) \equiv (X , Y ) := \frac{1}{2\, g^\vee}\, {\rm Tr} \,( ad (X)\, ad (Y) ) =
\frac{1}{N(\pi )}\,{\rm Tr}\,(\pi (X)\, \pi (Y))
\lb{Kill}
\ee
for all $X, Y \in {\cal G}\,.$ Here $ad = ad_{\cal G}$ is the adjoint representation of ${\cal G}\ $
($ad(X)\, Y = [X,Y]\,,\ \dim\, (ad_{\cal G}) = \dim {\cal G}$), $\ g^\vee$ is the {\em dual Coxeter number}
defined in (\ref{gCox}) below,
\be
N(\pi ) = C_2 (\pi )\,\frac{{\rm dim}\, \pi}{{\rm dim}\, {\cal G}}
\lb{secD}
\ee
is the {\em second order Dynkin index} of the representation $\pi\,$ and
$C_2(\pi)$ is the corresponding second order Casimir invariant.
Eqs. (\ref{Kill}) and (\ref{secD}) are consistent since
\be
N(ad) = C_2(ad) = 2\, g^\vee\ ,
\lb{NC2g}
\ee
see (\ref{Cpiad}).

For a pair $\{ T_a \}\,,\ \{ t^b \}\,$ of dual bases of ${\cal G}_{\mathbb C}$ (such that
${\rm tr}\, (T_a \, t^b ) = \d^a_b\,$) we define the Killing metric tensor $\eta_{ab}\,$ (\ref{etaab})
and its inverse, $\eta^{ab}\,$ as
\be
\eta_{ab} = {\rm tr}\, (T_a T_b )\,,\quad
\eta^{ab} = {\rm tr}\, (t^a t^b )\quad\Leftrightarrow\quad t^a = \eta^{ab} T_b\ .
\lb{Killeta}
\ee
Conversely, for a given semisimple ${\cal G_{\mathbb C}}\,,$ its (unique) compact real form ${\cal G}\,$
can be characterized by the fact that $(\eta_{ab})\,$ is negative definite on it.
A {\em Cartan-Weyl basis} of ${\cal G}_{\mathbb C}\,$ is given by $\{ T_a \} = \{ h_i , e_{\a} \}\,$ where
$h_i\,,\ i = 1,2,\dots , r\equiv{\rm{rank}}\, {\cal G}_{\mathbb C}\,$ span a Cartan subalgebra
${\mathfrak h}\subset {\cal G}_{\mathbb C}\,$ and $e_{\a}$ are the step operators labeled by the roots $\a\,$ of ${\cal G}_{\mathbb C}\,.$
If we define a Hermitean conjugation on ${\mathcal G}_{\mathbb C}$ acting on the Cartan-Weyl generators as
$h_i^* = h_i \,,\ e_\a^* = e_{-\a}\,,$ then its compact form consists of the {\em antihermitean} elements;
hence, ${\mathcal G}$ is the real span of
\be
i h_i\ ,\quad i (e_\a + e_{-\a} )\ ,\quad e_\a - e_{-\a}\ ,\quad i=1,\dots ,r\,,\ \a>0 \ .
\lb{compf}
\ee
Denote by $\{ \a_j \}_{j=1}^r\,$ the simple roots and by
$\a^\vee := \frac{2}{(\a | \a )}\,\a\,$ the coroot corresponding to $\a\,.$ Let $(~ \mid ~ )\,$ be
the Euclidean metric induced by the Killing form on the ($r$-dimensional) {\em real} linear span of all roots;
then $(\a | \b^\vee) \in {\mathbb Z}\,$ for all pairs of roots $\a$ and $\b\,$ (see e.g. \cite{FS}).
A root is either positive or negative, depending on the (common) sign of the non-zero integer coefficients in its expansion into simple roots.
The {\em Gauss decomposition} of ${\cal G}_{\mathbb C}\,$ as a vector space reads
\be
{\cal G}_{\mathbb C} = {\cal G}_+ \oplus {\mathfrak h} \oplus {\cal G}_-\ ,\quad {\cal G}_\pm = span\, \{ e_\a\,,\, \pm\, \a > 0 \}\ ,
\lb{Gauss}
\ee
where all the three direct summands are in fact Lie subalgebras (${\cal G}_\pm$ are nilpotent and the {\em Borel
subalgebras} ${\mathfrak b}_\pm := {\mathfrak h} \oplus {\cal G}_\pm$ are solvable).
In the {\em Chevalley normalization} of the step operators characterized by
\be
[e_\a , e_{-\a} ] =: h_\a \ ,\qquad {\rm tr}\,(h_\a h_\b ) = (\a^\vee | \, \b^\vee )
\lb{hee}
\ee
which we shall adopt here, the components $\eta_{ij} = {\rm tr}\, (h_i h_j)\,,\ \eta_{i\a} = {\rm tr}\, (h_i e_\a)$ and
$\eta_{\a\b} = {\rm tr}\, (e_\a e_\b )\,$ of the Killing metric tensor read
\be
\eta_{ij}= (\a^\vee_i |\, \a^\vee_j )\ ,\quad \eta_{i\a} = 0\ ,\quad\eta_{\a\b} = \frac{2}{( \a |\, \a )}\, \d_{\a\,, -\b}\quad (\, \Rightarrow\
\eta^{\a\b} = \frac{( \a |\, \a )}{2}\, \d_{\a\,, -\b}\,)
\lb{CCWC}
\ee
while the Lie commutation relations assume the form
\ba
&&[ h_i , h_j ] = 0\ ,\qquad [h_i , e_{\a} ] = (\a |\, \a_i^\vee )\, e_\a\qquad\Rightarrow\quad
[h_i , e_{\pm j} ] = \pm c_{ji} e_{\pm j}\nn\\
&&{\rm for}\qquad  c_{i\!j} := (\a_i |\, \a_j^\vee )\equiv 2\, \frac{(\a_i |\, \a_j )}{(\a_j |\, \a_j )}\ ,\qquad e_{\pm j} := e_{\pm \a_j}\ ,\nn\qquad\\
&&{\rm and}\quad [e_i , e_{-j} ] = \d_{ij}\, h_j\ ,
\lb{CWbasis}
\ea
where $(c_{i\!j})\,$ is the {\em Cartan matrix}.
The Lie algebra ${\cal G}_{\mathbb C}\,$ admits a presentation in terms of generators and relations: it is generated by the $3r\,$ generators
$\{ h_i , e_i , e_{-i} \}_{i=1}^r$ (forming the {\em Chevalley basis}), subject to the Lie bracket relations
in (\ref{CWbasis}) and the {\em Serre relations}
\be
(ad(e_{\pm i}))^{1-c_{ji}} e_{\pm j} = 0 = \sum_{\ell = 0}^{1-c_{ji}} (-1)^\ell \left({{1-c_{ji}}\atop{\ell}} \right)
e_{\pm i}^\ell e_{\pm j}\, e_{\pm i}^{1-c_{ji}-\ell}= 0\,,\quad i\ne j\ .
\lb{Serre2}
\ee
(the second relation using the associative product of step operators takes place in the {\em universal enveloping algebra} $U({\cal G}_{\mathbb C})$).

The {\em fundamental weights} $\Lambda^j\,$ defined by
\be
( \Lambda^j \mid \a_\ell^\vee )\, =\, \d^j_\ell\ ,\quad j,\ell = 1,\dots , r
\lb{fundw}
\ee
form another basis $\{ \L^j \}_{j=1}^r\,$ referred to as the {\em Dynkin basis},
and the coefficients of a weight $\L\,$ with respect to it, as {\em Dynkin labels}.
The canonical duality $\,h \in {\cal G}_{\mathbb C} \leftrightarrow {{\cal G}_{\mathbb C}}^*\,$
established by the Killing form assumes, in particular,
\be
h_\a \leftrightarrow \a^\vee :\quad \a^\vee(h) = {\rm tr}\, (h_\a h)\quad\forall\ h \in {\mathfrak h}
\quad \Rightarrow\quad h_i \leftrightarrow \a_i^\vee\ ,\quad h^j \leftrightarrow \L^j \ .
\lb{cdual}
\ee
The orthogonality of the Dynkin and coroot basis vectors (\ref{fundw}) implies that
$\sum_{j=1}^r ( x |\, \L^j )\, \a_j^\vee = x = \sum_{j=1}^r ( x |\, \a_j^\vee )\,  \L^j\,$ for any $x\in{\cal G}_{\mathbb C}\,.$
Putting, in particular, $x=\L^i\,,\ x=\a_i$ and $x= \a^\vee$ in this relation, we obtain
\be
\L^i = \sum_{j=1}^r (\L^i | \L^j )\, \a_j^\vee\ ,\quad \a_i = \sum_{j=1}^r c_{i\!j}\, \L^j
\quad{\rm and}\quad \a^\vee = \sum_{j=1}^r (\a^\vee |\, \L^j )\, \a_j^\vee \ ,
\lb{usef}
\ee
respectively. From the first formula in (\ref{usef}) one derives the Cartan components of the inverse Killing metric tensor
\be
\eta^{ij} = (\L^i |\, \L^j )\ ,
\lb{etaup}
\ee
and the last one implies that the Cartan element $h_\a$ (\ref{hee})
dual to an arbitrary (i.e. not necessarily simple) coroot is expressed as
\be
h_\a = \sum_{j=1}^r (\a^\vee |\, \L^j ) \, h_j\quad \Rightarrow\quad [ h_\a , e_{\pm\a} ] = \pm 2\, e_{\pm \a}\ .
\lb{h-a}
\ee
Linear combinations of simple roots (coroots, weights) with integral coefficients form the {\em root (coroot, weight) lattice}.
The coefficients $\{ a_i \}_{i=1}^r$ in the expansion of the {\em highest root} $\theta = \sum_{i=1}^r a_i\, \a_i\,$
are called the {\em Kac labels}, and the positive integer $g := 1+ \sum_{i=1}^r a_i\,,$
the {\em Coxeter number} of ${\cal G}_{\mathbb C}\,$).
The elements of the weight lattice, called {\em integral weights},
are the possible (in general, degenerate) eigenvalues of $\pi (h_i)\,$ for any finite dimensional representation $\pi\,$ of ${\cal G}\,.$
The {\em dominant} (integral) weights $\L\,$ are the weights whose Dynkin labels are non-negative integers,
\be
\L = \sum_{i=1}^r \l_i \,\L^i\ ,\quad \l_i = (\L \mid \a^\vee_i ) \in {\Z}_+\ ,\quad i=1,\dots ,r\ .
\lb{dintw}
\ee
They are in one-to-one correspondence with the (non-degenerate) {\em highest weights} of the
{\em irreducible} representations $\pi_\L\,$ of ${\cal G}\,,$
\be
(\pi_\L (h_i) - \l_i ) \mid \L \rangle = 0 = \pi_\L (e_\a) \mid \L \rangle\ ,\qquad i=1,\dots ,r\,,\quad \a>0\ .
\lb{HWpi}
\ee
The highest root $\theta\,$ is the highest weight vector of the adjoint representation $ad$ of ${\cal G}\,.$
The expansion of $\theta^\vee\,$ in terms of the simple coroots $\{ \a^\vee_{i}\}_{i=1}^r\,,$
\be
\theta^\vee \equiv \frac{2}{(\theta |\, \theta )}\,\,\theta = \sum_{i=1}^r\, a^\vee_{i} \a_i^\vee\ ,
\lb{dCL}
\ee
defines the {\em dual Kac labels} $\{ a^\vee_i \}_{i=1}^r$ and the dual Coxeter number
\be
g^{\vee} := 1 + \sum_{i=1}^r a^\vee_{i}\ .
\lb{gCox}
\ee
From now on we shall fix $(\theta |\, \theta ) = 2\,$ so that $\theta^\vee \equiv \theta\,.$
For $s\ell (n) = A_{n-1}$ all $a_i^\vee\,,\ i=1,\dots ,n-1$ are equal to $1$ so that $g^\vee_{s\ell(n)} = n\,.$

\smallskip

The quadratic Casimir operator $C_2 = \eta^{ab}\, T_a\, T_b$ belonging to $U({\cal G}_{\mathbb C})$ commutes
with all the elements of ${\cal G}_{\mathbb C}\,$ and so is proportional to the unit operator $\id_\pi$
in any irreducible representation $\,\pi\,,$ i.e. $\pi(T_a)\, \pi(t^a) = C_2 (\pi )\, \id_\pi\,.$
On the other hand, using the definition of the dual bases and (\ref{Kill}), we obtain
\be
N(\pi )\, {\rm tr}\,( T_a\, t^a ) = {\rm Tr}\, (\pi(T_a)\, \pi(t^a)) =
N(\pi )\, \d^a_a = N(\pi )\, {\rm dim}\, {\cal G}\ .
\lb{NC2}
\ee
Taking into account that ${\rm Tr}\,\id_\pi = {\rm dim} \,\pi\,,$ we find that the second order
Dynkin index $N(\pi)$ is related to the Casimir eigenvalue $C_2(\pi)$ by (\ref{secD}).

By (\ref{etaup}) and (\ref{CCWC}), $C_2$ assumes the form
\ba
&&C_2 = \eta^{ab}\, T_a\, T_b =
\sum_{i,j = 1}^r\, (\L^i |\, \L^j )\,h_i\, h_j +
\sum_{\a > 0}\,\frac{(\a |\, \a )}{2}\, (e_\a \, e_{-\a} + e_{-\a} e_\a ) =\nn\\
&&=\, \sum_{i=1}^r h^i h_i + \sum_\a e^\a e_\a\ ,\quad h^i :=
\sum_{j=1}^r (\L^i |\, \L^j )\, h_j\ ,\quad e^\a := \frac{(\a |\, \a )}{2}\, e_{-\a}\ .\qquad\qquad
\lb{CasCW}
\ea
Computing $\pi_\L(C_2)$ on the highest weight vector $\mid \L \rangle\,$ of a given IR for $\L$
given by (\ref{dintw}), we obtain
\ba
&&C_2 (\pi_\L ) = \sum_{i,j = 1}^r\, (\L^i |\, \L^j )\,\l_i \l_j +
\sum_{\a > 0}\,\frac{(\a |\, \a )}{2}\,\sum_{j=1}^r (\a^\vee |\, \L^j) \l_j =\nn\\
&&= (\L |\, \L ) + \sum_{\a>0} (\L |\,\a ) = (\L |\, \L + 2 \rho )\ ,
\lb{c2piL}
\ea
where
\be
\rho := \frac{1}{2}\,\sum_{\a >0} \a\, = \sum_{i =1}^r \L^i
\lb{Wv}
\ee
is the {\em Weyl vector}. In particular, for the eigenvalue of the Casimir in the adjoint representation
(with highest weight $\L= \theta$) one reproduces (\ref{NC2g}):
\be
C_2\, (ad) = (\theta |\, \theta + 2\rho ) = (\theta |\, \theta )\, (1 + \sum_{i=1}^r (\theta^\vee |\, \L^i )) =
(\theta |\, \theta ) \, g^{\vee}  = 2\, g^\vee
\lb{Cpiad}
\ee
(see (\ref{dCL}) and (\ref{gCox})). On the other hand,
the matrices $f_{a \, .}^{~~ . }$ given by the structure constants
are nothing but the generators of the adjoint representation. This allows to relate them to the dual Coxeter number.
Indeed, using (\ref{Kill}), (\ref{secD}), (\ref{Killeta}) and (\ref{Cpiad}), we find
\be
{\rm Tr}\, ( ad(T_a)\, ad(T_b) ) = i^2 \,f_{as}^{~~t} f_{bt}^{~~s} = 2\, g^\vee\,\eta_{ab}\ .
\lb{adff}
\ee

The dimension of an IR $\pi_\L$ is given by the {\em Weyl dimension formula}
\be
{\rm dim}\,{\pi_{\L}} = \prod_{\a >0} \frac{(\L + \rho \,|\, \a )}{(\rho\, |\, \a )}\ .
\lb{Weyldim}
\ee

The {\em Weyl group} of a root system is the finite group generated by the simple reflections $s_i := s_{\a_i}\,,\ i=1,\dots, r\,$
where $\, s_\alpha (\beta) = \beta - 2\,\frac{(\beta \mid \alpha )}{(\alpha \mid \alpha )}\, \alpha\,.$
It is a {\em Coxeter group} with generators $s_i$ subject to the relations $(s_i s_j )^{m_{ij}} = 1\,,$ where
\be
m_{ij} = \left\{
\begin{array}{ll}
\, 1&, \quad i=j\\
\, 2&, \quad \#(i,j) = 0\\
\, 3&, \quad \#(i,j) = 1\\
\, 4&, \quad \#(i,j) = 2\\
\, 6&, \quad \#(i,j) = 3\\
\end{array}
\right.
\lb{Wrels}
\ee
and $\#(i,j)$ is the number of bonds joining the $i^{th}$ and $j^{th}$ vertex of the Dynkin diagram.

The {\em fundamental Weyl chamber} consists of the vectors $\L = \sum_{i=1}^r p_{\a_i} \L^i$ in the weight space
forming the cone $(\L |\, \a^\vee_i ) \equiv p_{\a_i} \ge 0\,,\ i = 1,\dots , r\,,$ and the
(level $k$) {\em positive Weyl alcove}, a subset of it,
is the simplex whose points are restricted by the additional requirement $( \L |\, \theta) \le k\,.$
They serve as fundamental domains of the corresponding Weyl group and {\em affine} Weyl group, respectively.

\smallskip

It is easy to see that for $s\ell (r+1) = A_r\,$ the nontrivial Eqs.(\ref{Wrels}) (i.e., those for $i\ne j$) reduce to the braid relations (\ref{braidR})
for $s_i\,,\ i=1,\dots ,r\,,$ in accord with the fact that the corresponding Weyl group is the symmetric group ${\mathcal S}_{r+1}\,.$
In this case it is convenient to use the standard {\em barycentric} parametrization of the roots and weights by imbedding them in an $n$-dimensional
Euclidean space with a distinguished orthonormal basis $\{\varepsilon_s\,,\, s=1,\dots , r+1 \equiv n \}$
such that the simple roots and the fundamental weights assume the form
\ba
&&\a_\ell = \e_\ell - \e_{\ell +1}\ ,\quad 1\le \ell \le n-1\ ,\quad (\e_r |\, \e_s ) = \d_{rs}\ ,
\quad 1\le r,s \le n\ ,\nn\\
&&\L^i = (1-\frac{i}{n}) \sum_{j=1}^i \e_j - \frac{i}{n} \sum_{j=i+1}^n \e_j\ ,\quad (\L^i |\, \a_\ell ) = \d^i_\ell\ ,
\quad 1\le i,\ell \le n-1\ .\qquad\qquad
\lb{al}
\ea
The set of positive roots then admits a double index labeling\,,
\be
\a_{ij} = \sum_{\ell = i}^{j-1} \a_{\ell} = \e_i - \e_j\,,\quad 1\le i < j\le n\qquad
(\, \a_\ell \equiv \a_{\ell\, \ell +1}\, )
\lb{slnroots}
\ee
and the highest root is $\,\theta = \a_{1 n} = \e_1 - \e_n = \Lambda^1 + \Lambda^{n-1}\,.$
As the weight and root systems lie in the hyperplane orthogonal to the vector
$\e := \sum_{s=1}^n \e_s\,$ (one can easily verify that $ ( \a_{ij} |\, \e ) = 0 = (\L^m | \, \e )$ for all
$1\le i<j \le n\,,\ 1\le m\le n-1$), any weight $\L = \sum_{i =1}^r \l_i\L^i$ can be expressed in terms of
the barycentric coordinates $\ell_j \,,\ j = 1,..., r+1\,$ such that
\be
\L = \sum_{i =1}^r \l_i\, \L^i  = \sum_{j=1}^{r+1} \ell_j\, \e_j\ ,\qquad
( \L \,| \e ) = 0\quad\Rightarrow\quad \sum_{j=1}^{r+1} \ell_j = 0\ .
\lb{baryA}
\ee
The Dynkin labels $\{ \l_i \}_{i= 1}^r$ and $\{ \ell_j \}_{j=1}^{r+1}\,$ can be found from each other by
\be
\l_i = \ell_i - \ell_{i+1} \ ,\qquad \ell_j = \sum_{m=j}^r \l_m - \frac{1}{r+1}\sum_{m=1}^r m\, \l_m\ .
\lb{lambda-ell}
\ee
It would be useful to present explicit formulas for the barycentric coordinates of some important dominant weights
$\L$. One has, in particular,
\ba
&&\ell_j(\rho) = \frac{n+1}{2} - j\ ,\qquad\ \ \ell_j (\pi_f) = \d_{j1} - \frac{1}{n}\ ,\nn\\
&&\ell_j (\pi_s) = 2 \left( \d_{j1} - \frac{1}{n}\right)\ ,\quad
\ell_j (\pi_a) = \d_{j1} + \d_{j2} - \frac{2}{n}\ ,\quad\nn\\
&&\ell_j (\pi_{\bar s}) = 2 \left( \frac{1}{n} - \d_{jn} \right)\ ,\quad
\ell_j (\pi_{\bar a}) = \frac{2}{n} - \d_{j, n-1} - \d_{j n}\qquad\lb{llab}\ea
for the labels of the Weyl vector $\rho = \sum_{i=1}^r \L^i$ (\ref{Wv}) and of the highest weights
of the defining representation, $\L^1$, of its symmetric and antisymmetric powers, $2\L^1$ and $\L^2$,
and of their conjugate representations, $2 \L^{n-1}$ and $\L^{n-2}$, respectively.
The eigenvalue of the quadratic Casimir operator (\ref{c2piL}) in the IR with highest weight $\L$ (\ref{baryA})
can be then expressed as
\be
C_2(\pi_\L) = (\L \mid \L + 2\rho) = \sum_{j=1}^n \ell_j (\ell_j + 2\ell_j(\rho))
= \sum_{j=1}^n \ell_j (\ell_j - 2j)\ .
\lb{C2L}
\ee
We get, in particular, $C_2(\pi_f) = \frac{n^2-1}{n}$ so that,  from (\ref{secD}),
\be
N({\pi_f}) = C_2({\pi_f})\,\frac{{\rm dim}\, \pi_f}{{\rm dim}\, {sl(n)}} = \frac{n^2-1}{n}\,.\,\frac{n}{n^2-1} = 1 \ .
\lb{Npif}
\ee
It follows that in the fundamental representation of ${\cal G} = su(n)$ the Killing trace $\,{\rm tr}\,$
(\ref{Kill}) coincides with the usual matrix trace ${\rm Tr}\,.$

On the other hand, for $s\ell(n)$ all $a^\vee_i = 1\,,$ hence $g^{\vee} = n\,, $ so for the adjoint representation
$C_2\, (ad) = 2 n = N (ad)\,,$ cf. (\ref{dCL}), (\ref{gCox}), (\ref{Cpiad}) and (\ref{NC2g}).
The corresponding level $k$ positive Weyl alcove contains dominant weights (\ref{dintw}) satisfying in addition
\be
( \L |\, \theta) \equiv \sum_{j, \ell = 1}^{n-1} \l_j \,a^\vee_\ell \, (\L^j |\, \a^\vee_\ell) = \sum_{j=1}^{n-1} \l_j = \ell_1 -\ell_n \le k\ .
\lb{Wslnlambda}
\ee

As all the roots of $s\ell(n) = A_{n-1}$ have equal length square,
the corresponding $(n-1)\times (n-1)$ Cartan matrix $c^{(n)} = (c_{ij})$ (\ref{CWbasis}) is symmetric:
\be
c_{ij} = (\a_i | \a_j)\ ,\quad c_{ii} = 2 \ , \quad c_{i\,i\pm 1} = -1 \ , \quad c_{ij} = 0 \quad \mbox{for} \quad \vert i-j \vert > 1\ .
\lb{Cq}
\ee
It is easy to see that  $\det\, c^{(n)} = n\,$ as it obeys
\be
\det\, c^{(n)} = 2\,\det\, c^{(n-1)} -\, \det\, c^{(n-2)}\ ,\quad \det\, c^{(2)} = 2\ ,\quad \det\, c^{(3)} = 3\ .
\lb{detcn}
\ee
We have, furthermore
\be
\eta_{ij} = c_{ij} \ ,\qquad \eta^{ij} = (\L^i | \L^j ) = {\rm min}\, (i,j) - \frac{ij}{n}
\lb{etas} \ee
so that
\ba
&&h_i = \sum_{j=1}^{n-1} c_{ij}\, h^j = 2 h^i - h^{i-1} - h^{i+1}\qquad\Leftrightarrow\nn\\
&&h^i = \sum_{j=1}^i j\, ( 1-\frac{i}{n} )\, h_j + \sum_{j=i+1}^{n-1} i\,( 1 - \frac{j}{n})\, h_j\quad .
\lb{h_ih^i}
\ea

\newpage

\section*{Appendix B. Hopf algebras}
\addcontentsline{toc}{section}{Appendix B. Hopf algebras}

\setcounter{equation}{0}
\renewcommand\theequation{B.\arabic{equation}}

\subsection*{B.1. The Hopf algebra $U_q(s\ell(n))$}
\addcontentsline{toc}{subsection}{B.1. The Hopf algebra $U_q(s\ell(n))$}

We shall spell out the definition of the QUEA $U_q ({\mathcal G})$ as a Hopf algebra for ${\mathcal G} = A_r = s\ell_{r+1}\,.$
It is customary in mathematical textbooks to take first $q$ as just a central indeterminate and consider at a later stage various
{\em specializations} of $q$ as a (complex) deformation parameter. The definition below follows \cite{CP},
a comprehensive text on the subject (see in particular Definition-Proposition 9.1.1 therein),
where the "rational form" $U_q({\cal G})$ is introduced
as an associative algebra over ${\mathbb Q}(q)\,,$ the field of rational functions of $q\,.$
The $n$-fold "cover" $U_q^{(n)}(s\ell(n))$ defined
by adjoining to $U_q(s\ell(n))$ the invertible elements $k_i\,,\ i=1,\dots , n-1$ (\ref{dk})
then corresponds to the {\em simply-connected} rational form \cite{CP}.

\smallskip

The {\em Chevalley basis} of $U_q (A_r)$ contains $r$ group-like generators $K_i$ and their inverses $K_i^{-1}$
(such that $K_i K_i^{-1} = K_i^{-1} K_i = \id$) which correspond to the
classical Cartan generators, and $2r$ Lie algebra-like ones, the raising and lowering operators
$E_i$ and $F_i$, corresponding to the simple roots. They obey the following CR,
\ba
&&K_i \, E_j \, K_i^{-1} = q^{c_{ij}} \, E_j \ , \quad
K_i \, F_j \, K_i^{-1} = q^{-c_{ij}} \, F_j \ ,\nn\\
&&[E_i , F_j] = \delta_{ij} \, \frac{K_i - K_i^{-1}}{q-q^{-1}} \ , \qquad i,j = 1,\ldots ,r
\lb{CRq}
\ea
(here $(c_{ij})$ is the $A_r$ Cartan matrix (\ref{Cq})) and {\it $q$-Serre relations} (that are only non-trivial for $r>1$):
\ba
&&E_i^2 \, E_j + E_j \, E_i^2 = [2]\, E_i \, E_j \, E_i \ , \qquad
F_i^2\, F_j + F_j \, F_i^2 = [2]\, F_i \, F_j\, F_i  \nn\\
&&{\rm for} \quad \vert i-j\vert = 1\ ,\qquad
[E_i , E_j] = 0 = [F_i , F_j] \quad \mbox{for} \quad \vert i-j \vert > 1 \ .\qquad
\lb{Sq}
\ea

The definition of an arbitrary Hopf algebra ${\mathfrak A}$ involves the coproduct (an algebra homomorphism
$\Delta:\, {\mathfrak A} \to {\mathfrak A} \otimes {\mathfrak A}$), the counit
(a homomorphism $\varepsilon:\, {\mathfrak A} \to \C$) and the antipode
(an antihomomorphism $S:\, {\mathfrak A} \to {\mathfrak A}$).
The compatibility conditions on the coalgebra structures read
\ba
&&({id} \otimes \Delta ) \,\Delta =  (\Delta \otimes {id}) \,\Delta \ ,\nn\\
&&({id} \otimes \varepsilon )\,\Delta (X) = (\varepsilon \otimes {id})\,\Delta (X) = X\ ,\nn\\
&&\lb{compHopf}
m\,({id} \otimes S) \, \Delta (X) = m(S \otimes {id}) \, \Delta (X) = \varepsilon (X) \, \id\ .
\ea
The first property is called {\em coassociativity}. In the third relation,
$m$ is just the multiplication in the algebra considered as a map
$m:\ {\fU}\otimes {\fU} \to {\fU}\,,\ m (X\otimes Y) = XY \quad \forall X,Y\in {\fU}\,.$

In the case of $U_q (A_r)$ we define these structures on the generators $\{ K_i , E_i , F_i \}\,,$ $i=1,\ldots , r$ as follows:
\be
\Delta (K_i) = K_i \otimes K_i \ , \ \Delta (E_i) = E_i \otimes K_i + \id\, \otimes E_i\ ,\
\Delta (F_i) = F_i \otimes \id\, + K_i^{-1} \otimes F_i \ ,
\lb{copr}
\ee
\be
\varepsilon (K_i) = 1 \ , \quad \varepsilon (E_i) = \varepsilon (F_i) = 0 \ ,
\lb{coun}
\ee
\be S(K_i) = K_i^{-1} \ , \quad  S(E_i) = -E_i \, K_i^{-1} \ , \quad S(F_i) = -K_i \, F_i\ .
\lb{antip}
\ee

A Hopf algebra ${\mathfrak A}\,$ is said to be {\em {cocommutative}} if the coproduct
$\Delta (X) = \sum_{(X)} X_1\otimes X_2\,$ is equal to its opposite ${\Delta}' (X)= \sum_{(X)} X_2\otimes X_1\,,$
see (\ref{DDp})\footnote{The
universal enveloping algebra $U({\cal G})$ of any classical Lie algebra is non-commutative but cocommutative.
The deformed QUEA $U_q({\cal G})$ is however neither commutative nor cocommutative.}.
It is said to be {\em {almost cocommutative}} if there exists
an invertible element ${\cal R}\in{\mathfrak A}\otimes {\mathfrak A}\,$ called {\em universal $R$-matrix}
which intertwines $\Delta (X)$ and its opposite, ${\Delta}'(X) = {\cal R}\, \Delta (X)\, {\cal R}^{-1}\,,$ see (\ref{intR}).
In this case the element
\be
{\cal M}:= {\cal R}_{21} {\cal R}\ \in \fU\otimes\fU
\lb{univM}
\ee
is called the (universal) {\em monodromy matrix}. Exchanging the order of the terms in the tensor products we obtain
that ${\cal M}\,$ commutes with the coproduct:
\be
{\Delta}(X) = {\cal R}_{21}\, \Delta' (X)\, {\cal R}_{21}^{-1} \equiv {\cal R}_{21}\, {\cal R}\, \Delta (X)\, {\cal R}^{-1}\, {\cal R}_{21}^{-1}\quad
\Rightarrow\quad [\,{\cal M}\,,\, \Delta(X)\,] = 0\ .
\lb{UM}
\ee
An almost cocommutative ${\mathfrak A} = ({\mathfrak A}\,, {\cal R})$
is {\em {quasitriangular}} if ${\cal R}\,$ satisfies, in addition,
\be
(\Delta \otimes id) {\cal R} = {\cal R}_{13} {\cal R}_{23}\ ,\quad
(id \otimes \Delta ) {\cal R} = {\cal R}_{13} {\cal R}_{12}\ .
\lb{qtr}
\ee
Any of these two relations implies that ${\cal R}$ solves the Yang-Baxter equation
\be
{\cal R}_{12} {\cal R}_{13} {\cal R}_{23} = {\cal R}_{23} {\cal R}_{13} {\cal R}_{12}
\lb{YBE-R}
\ee
(and also fixes the normalization of ${\cal R}\,$); for example, the definition of ${\cal R}\,$ and the first equation (\ref{qtr})
(equivalent to $(\Delta' \otimes id) {\cal R} = {\cal R}_{23} {\cal R}_{13}$) imply
\be
{\cal R}_{12} {\cal R}_{13} {\cal R}_{23} = {\cal R}_{12} (\Delta \otimes id) {\cal R} =
((\Delta' \otimes id) {\cal R} )\, {\cal R}_{12} = {\cal R}_{23} {\cal R}_{13} {\cal R}_{12}\ .
\lb{derYB}
\ee
The following relations also hold:
\ba
&&(\e\otimes id) {\cal R} = \id = (id\otimes \e) {\cal R}\ ,\nn\\
&&(S\otimes id) {\cal R} = {\cal R}^{-1} = (id\otimes S^{-1}) {\cal R}\quad\Rightarrow\quad
(S\otimes S) {\cal R}^{\pm 1} = {\cal R}^{\pm 1}\ .\quad
\lb{R-rel}
\ea
If $({\mathfrak A}\,, {\cal R})$ is quasitriangular, so is $({\mathfrak A}\,, {\cal R}_{21}^{-1})\,.$

\smallskip

Universal $R$-matrices ${\cal R}$ for quantum deformations of $U({\cal G})$ for any simple ${\cal G}$
can be found by considering in the place of $U_q({\cal G})$ a "topological" version of it and
appropriately completing the tensor square which requires, however, a non-algebraic setting.
One can consider, as a replacement of $U_q({\cal G})$ for $q=e^t\,,$
the {\em topologically free ${\C} [[t]]$ algebra} (i.e. the algebra over the formal power series in $t\,$)
$U_t = U_t({\cal G})$ generated, in the case ${\cal G} = A_r\,,$ by $\{ E_i, F_i, H_i \}_{i=1}^r$
subject to relations (\ref{CRq}) -- (\ref{antip}) (with $K_i$ replaced by $e^{hH_i}$),
and use an appropriate completion of the tensor product $U_t\otimes U_t\,.$
The universal $R$-matrix ${\cal R}$ (obtained by Drinfeld \cite{D} for $U_t (A_1)\,,$ by Rosso \cite{R}
for $U_t (A_r)\,,$ and by Kirillov Jr. and Reshetikhin \cite{KR90} and, independently, by Levendorskii and Soibelman
\cite{LS} for $U_t ({\cal G})\,$ where ${\cal G}\,$ is a general simple complex Lie algebra) is a product
of similar terms for any $s\ell_2$ triple, appropriately ordered by using a quantum analog of the Weyl group.

For $U_t({s\ell(2)})$ the corresponding universal $R$-matrix has the form
\be
{\cal R} =  \sum_{\nu = 0}^{\infty} \frac{ q^{-\frac{\nu (\nu -1)}{2}}(-\l)^\nu}{[\nu ]!}\,
F^\nu\otimes E^\nu\, q^{- \frac{1}{2} H\otimes H} \ .
\lb{RUq2}
\ee
Clearly, the infinite series in $\nu$ reduces to a finite sum in any finite dimensional representation of $U_t$
of "classical type" (i.e. such that $E$ and $F$ are nilpotent). It is easy to verify, in
particular, in the $n=2$ case that (\ref{RUq2}) reproduces (\ref{R2}) for $E^f$ and $F^f$ given by (\ref{bUf}) and
\be
\left( q^{H}\right)^f = q^{H^f} = \begin{pmatrix}q&0\cr0&q^{-1}\end{pmatrix}\ ,\qquad
[H^f] = H^f = \, \begin{pmatrix}1&0\cr 0&-1\end{pmatrix}\ .
\lb{Hf}
\ee
For general $n\,,$ the matrix $R_{12}$ (\ref{R}) can be obtained in a similar way
from the universal $R$-matrix ${\cal R}$ for $U_t (s\ell (n))\,.$

\smallskip

For $q$ a root of unity (as it is in our case, (\ref{height-h})), {\em finite dimensional}
quasitriangular quotients of $U_q({\cal G})\,$ exist so that the construction of their ${\cal R}$-matrix becomes purely algebraic.

\subsection*{B.2. The Drinfeld double}
\addcontentsline{toc}{subsection}{B.2. The Drinfeld double}

We are going to briefly recall here, following \cite{D, RS, Ka, Ma}, the construction of the Drinfeld double
$D(\fU)\,$ of a (finite dimensional) Hopf algebra $\fU\,.$ Any double is quasitriangular and factorizable;
moreover, there is a canonical expression for its universal $R$-matrix ${\cal R}_D\,.$ We shall apply further
the general theory to the finite dimensional quotients of the Borel subalgebras in $U^{(2)}_q(s\ell(2))\,.$

\smallskip

Formally, the Drinfeld double $D(\fU)\,$ is the {\em bicrossed product} of the dual ${\fU}^*\,$
taken with the {\em opposite} coproduct, and ${\fU}\,$ itself (see Chapter IX of \cite{Ka}): $D(\fU) := ({\fU}^*)^{cop} \bowtie \fU\,.$
The Hopf structure on $({\fU}^*)^{cop}\,$ is defined, for $X, Y \in \fU\,,\ F, G \in \fU^*\,,\ \Delta(X) = \sum_{(X)} X_{(1)}\otimes X_{(2)}\,$ etc., by
\ba
&&(F\, G) (X) = (F \otimes G )\, (\Delta (X)) \left(\equiv \sum_{(X)} F (X_{(1)})\, G (X_{(2)})\right)\ ,
\nn\\
&&\Delta (F) (X \otimes Y) \left(\equiv \sum_{(F)} F_{(1)} (X)\, F_{(2)} (Y) \right) = F (Y X)\ ,
\lb{U*op}\\
&&\id\, (X) = \e (X)\ ,\qquad \e (F) = F (\id)\ ,\qquad S(F) (X) = F (S^{-1}(X))\ .
\nn
\ea
From practical point of view, the following properties of the double $D(\fU)\,$ are sufficient
to reproduce its general structure as a quasitriangular Hopf algebra.

\begin{itemize}
\item
As a vector space, the double $D(\fU)\,$ is just the tensor product ${\fU}^*\otimes{\fU}\,.$

\item
As a coalgebra, the double $D(\fU) = ({\fU}^*)^{cop} \otimes \fU\,.$
The tensor product of coalgebras ${\mathfrak B}\,$ and $\fU \,$ with coproducts
$\Delta_{\mathfrak B} (F) = \sum_{(F)} F_{(1)}\otimes F_{(2)}\,$ and $\Delta_{\fU} (X) = \sum_{(X)} X_{(1)}\otimes X_{(2)}\,,$ respectively,
is a coalgebra with counit $\e_{\mathfrak B\otimes\fU} (F \otimes X) := \e_{\mathfrak B} (F)\, \e_{\fU} (X)\,$ and coproduct\footnote{Note
the flip between $F_{(2)}\,$ and $X_{(1)}\,$ which makes (\ref{tens-pr-coalg}) differ from
$\Delta_{\mathfrak B} (F) \,\otimes\, \Delta_{\fU} (X)\ .$}
\be
\Delta_{\mathfrak B\otimes\fU} (F \otimes X) := \sum_{(F), (X)} F_{(1)}\otimes X_{(1)} \otimes F_{(2)} \otimes X_{(2)}\ .
\lb{tens-pr-coalg}
\ee

\item
The multiplication in $D(\fU)\,$ is defined as
\be
(F\otimes X) \,.\, (G \otimes Y)  = \sum_{(X)} F\, G(S^{-1} (X_{(3)}) \,?\, X_{(1)} )\, \otimes \, X_{(2)}\, Y\ ,
\lb{mult-gen}
\ee
where
$$
\sum_{(X)} X_{(1)}\otimes X_{(2)}\otimes X_{(3)} = (id\otimes \Delta)\, \Delta (X) = ( \Delta\otimes id)\, \Delta (X)\,
$$
and the $?\,$ sign in the right-hand side stands for the missing argument of the functional.
Identifying $\fU\,$ and its dual with Hopf subalgebras of $D(\fU)\,,$ e.g. $\fU\simeq \id\otimes\fU \subset D(\fU)\,,$
we derive from (\ref{mult-gen}) the following constraint on the mixed multiplication in $D(\fU)$:
\be
X\,.\, F  = \sum_{(X)} F(S^{-1} (X_{(3)}) \,?\, X_{(1)} )\, X_{(2)}\ ,\quad \forall\, X\in\fU\,,\, F\in \fU^*\ .
\lb{mult-pm}
\ee

\item
If $e_i  \in \fU\,$ and $e^j \in \fU^*\,$ are dual linear bases of $\fU\,$ and $\fU^*$, respectively,
the $R$-matrix ${\cal R}_D\,$ of the double $D(\fU)\,$ is given by the (basis independent) expression
\be
{\cal R}_D = \sum_i e_i\otimes e^i  \,\in\, D(\fU) \otimes D(\fU)\qquad (\, e^j (e_i) = \d^j_i \,)\ .
\lb{RDA}
\ee
\end{itemize}

\smallskip

We shall now apply all this to the Hopf algebras $U_q(\bo_\pm)\,$ where
\ba
U_q(\bo_+)\, :\quad &&F k_+ = q\, k_+ F\ ,\quad\ F^h = 0\ ,\quad k_+^{4h} = \id\ ,\nn\\
&&\D(F) = F\otimes \id + k_+^{-2} \otimes F\ ,\quad \D(k_+) = k_+\otimes k_+\ ,\lb{B+ex}\\
&&\e(F) = 0\ ,\quad \e(k_+) = 1\ ,\quad S(F) = - k_+^2 F\ ,\quad S(k_+) = k_+^{-1}\nn
\ea
and
\ba
U_q(\bo_-)\, :\quad &&k_- E = q\, E \, k_- \ ,\quad E^h = 0\ ,\quad k_-^{4h} = \id\ ,\nn\\
&&\D(E) = E\otimes k_-^2 + \id\otimes E\ ,\quad \D(k_-) = k_-\otimes k_-\ ,\lb{B-ex}\\
&&\e(E) = 0\ ,\quad \e(k_-) = 1\ ,\quad S(E) = - E k_-^{-2} \ ,\quad S(k_-) = k_-^{-1}\nn
\ea
are the Borel subalgebras of the QUEA $\bbU\,$ defined in Section 5.2.2.

\smallskip

It is not difficult to prove that $(U_q(\bo_\pm)^*)^{cop} \simeq U_q(\bo_\mp)\,.$\footnote{The duality of the quantized Borel subalgebras
is a well known fact \cite{D}.}
To this end, we identify e.g. the elements $k_-\,$ and $E\,$ with the following functionals (defined by their values on certain
PBW basis of $U_q(\bo_+)$):
\ba
&&k_- (f_{\nu n}) := \d_{\nu\, 0} \,q^{-\frac{n}{2}}\ ,\quad E (f_{\nu n}) := -\, \d_{\nu 1}\,\frac{1}{\l} \qquad\
(\, \id (f_{\nu n}) = \e ( f_{\nu n}) = \d_{\nu\, 0}\,)\nn\\
&&{\rm for}\qquad f_{\nu n} := F^\nu k_+^n \in U_q(\bo_+)\ ,\quad 0\le n\le 4h-1\ ,\quad 0\le\nu\le h-1\ .\qquad\qquad
\lb{PBW+}
\ea
Applying the first relation (\ref{U*op}), one derives by induction the general relation
\be
(E^\mu k_-^m )(f_{\nu n}) = \d_{\mu\nu} \frac{[\mu ]!}{(-\l)^\mu}\,q^{\frac{\mu(\mu-1)-mn}{2}}
\lb{d+}
\ee
which can be used to prove, with the help of the other definitions in (\ref{U*op}), that Eqs. (\ref{B-ex}) hold.

In accord with (\ref{RDA}), the $R$-matrix for the $16 h^4$-dimensional double $\ D(U_q(\bo_+))\,$ is given by
\be
{\cal R}_D= \sum_{\nu=0}^{h-1} \sum_{n = 0}^{4h-1} f_{\nu n} \otimes e^{\nu n}
\lb{Rdouble}
\ee
with $f_{\nu n}\,$ as defined in (\ref{PBW+}) and
\be
e^{\mu m} = \frac{(-\l)^\mu q^{-\frac{\mu (\mu -1)}{2}}}{4h\,[\mu ]!}
\sum_{r=0}^{4h-1} q^{\frac{mr}{2}}\, E^\mu k_-^r \quad\quad
(\, e^{\mu m} (f_{\nu n}) = \d^\mu_\nu \d^m_n \,)
\lb{dual+}
\ee
forming the dual PBW basis of  $U_q (\bo_-)\,.$ Finally, the mixed relations
\be
[ k_+ , k_- ] = 0\ ,\quad k_+ E = q\, E\, k_+\ ,\quad F\, k_- = q\, k_- F\ ,\quad [ E , F ] = \frac{k^{2}_- - k_+^{-2} }{q-q^{-1}}
\lb{B-mix}
\ee
which are derived from (\ref{mult-pm}), show that
\be
D(U_q(\bo_+)) = \bbU \otimes U_q ({\mathfrak h}) \ ,\qquad U_q ({\mathfrak h}) = \{ \kappa^m \}_{m=0}^{4h-1}\ , \quad \kappa := k_+ k_-^{-1}
\lb{DBU}
\ee
where $U_q ({\mathfrak h})\,$ belongs to the centre of the double. Hence, the quotient with respect to the relation $\kappa = \id\,$ (i.e.,
$k_+ = k_- =: k\,$) is isomorphic to $\bbU\,.$ Accordingly, the same substitution in (\ref{Rdouble}) reproduces the $R$-matrix (\ref{RbD}).

\smallskip

Interchanging the roles of the two Borel subalgebras (\ref{B+ex}) and (\ref{B-ex}) we obtain the same result (\ref{DBU}) for $D(U_q(\bo_-))\,.$
Of course, the corresponding $R$-matrix of the double differs from (\ref{Rdouble}); the universal $R$-matrix of $\bbU\,$ we obtain from it
coincides with (\ref{RbD21}).

{\subsection*{B.3. Factorizable Hopf algebras and the Drinfeld map}
\addcontentsline{toc}{subsection}{B.3. Factorizable Hopf algebras and the Drinfeld map}

A (finite dimensional) Hopf algebra $\fU$ is called factorizable, if there exists a universal monodromy matrix
\be
{\cal M} = {\cal R}_{21} {\cal R} = \sum_{i} m_i\otimes m^i\, \in {\mathfrak A}\otimes {\mathfrak A}
\lb{Mm}
\ee
such that both $\{ m_i \}\,$ and $\{ m^i \}\,$ form bases of $\fU$. Alternatively,
a factorizable Hopf algebra $\fU\,$ is such for which the Drinfeld map $\hat D\,$ (\ref{Dr-map})
$$
\hat D\, :\ {\mathfrak A}^*\ \rightarrow\ {\mathfrak A}\ ,\qquad
\phi\ \mapsto\ \hat D (\phi) := (\phi\otimes id ) ({\cal M}) = \sum_{i} \phi (m_i)\otimes m^i
$$
is a linear isomorphism, i.e. $\hat D({\mathfrak A}^*) = {\mathfrak A}\,$ and $\hat D\,$ is invertible
(the equivalence of the two definitions is a simple exercise in linear algebra).
The opposite extreme is the case of {\em triangular} Hopf algebra for which ${\cal R}_{21} = {\cal R}^{-1}\,$ and hence, ${\cal M} = \id\otimes\id\,.$
(Cf. Remark 3.2 for the infinitesimal notions of factorizability and triangularity, respectively, of a Lie bialgebra
defined by means of a classical $r$-matrix \cite{RS}.)

\smallskip

The space of ${\mathfrak A}$-characters (\ref{Ch-Ad*inv}) (functionals obeying $\phi (x y) = \phi (S^2(y) x)$),
is an algebra under the multiplication
\be
(\phi_1 . \phi_2 )(x) := (\phi_1 \otimes \phi_2) \,(\Delta (x))
\qquad \forall\, \phi_1\,,\, \phi_2 \in {\mathfrak C}{\mathfrak h}
\lb{V-Ch-homo}
\ee
which, for ${\mathfrak A}\,$ quasitriangular, is commutative \cite{D3}:
\ba
&&(\phi_2 . \phi_1 )(x) = (\phi_1 \otimes \phi_2) \,(\Delta' (x)) = (\phi_1 \otimes \phi_2) \,({\cal R}\, \Delta (x)\, {\cal R}^{-1}) = \nn\\
&&= (\phi_1 \otimes \phi_2) \,(((S^2 \otimes S^2) \,{\cal R}^{-1})\, {\cal R}\, \Delta (x)\, ) = (\phi_1 \otimes \phi_2) \,(\Delta (x))
= (\phi_1 . \phi_2 )(x)\ .\qquad\quad
\lb{char-commut}
\ea
(We use consecutively the definition of ${\cal R}\,$ (\ref{intR}), the one of ${\mathfrak A}$-characters
and apply the last equation (\ref{R-rel}).) Denote by ${\cal Z}\,$ the centre of ${\mathfrak A}\,,$ and by
${\mathfrak A}^\Delta\,$ the subalgebra of ${\mathfrak A}\otimes {\mathfrak A}\,$ consisting of
elements $B\,$ such that $[ B\, ,\, \Delta (x) ] = 0\ \ \forall x\in {\mathfrak A}\,.$
Drinfeld has shown in Proposition 1.2 of \cite{D3} that
\be
\phi \in {\mathfrak C}{\mathfrak h}\,,\quad B\in {\mathfrak A}^\Delta\,\quad \Rightarrow\quad
(\phi\otimes id ) (B) \in {\cal Z}\ .
\lb{Ch-AD-Z}
\ee
As ${\cal M} \in {\mathfrak A}^\Delta\,$ (cf. (\ref{UM})), the restriction of the
Drinfeld map $\hat D\,$ to ${\mathfrak A}$-characters sends them into central elements.
Moreover, it provides a (commutative)  algebra homomorphism
${\mathfrak C}{\mathfrak h}\ \rightarrow\ {\cal Z}\,$ (Proposition 3.3 of \cite{D3}),
\be
\hat D (\phi_1 . \, \phi_2 ) =  \hat D (\phi_1 ) \, \hat D ( \phi_2 )\qquad \forall\, \phi_1\,,\, \phi_2
\in {\mathfrak C}{\mathfrak h}
\lb{D-homom}
\ee
which, for ${\mathfrak A}\,$ factorizable, is an isomorphism (Theorem 2.3 of \cite{Sch01}).
So in this case we have an alternative description of the algebra of the characters ${\mathfrak C}{\mathfrak h}$
in terms of more tractable objects, the elements of the centre ${\cal Z}\,.$

\smallskip

It follows from (\ref{canch}) that all $q$-traces (\ref{canCh}) are ${\mathfrak A}$-characters.
The map from the GR ${\mathfrak S}\,$ of $\,{\mathfrak A}\,$ to the subalgebra of ${\mathfrak C}{\mathfrak h}$ generated by the $q$-traces
\be
{\hat S} :\ {\mathfrak S}\ \rightarrow \ {\mathfrak C}{\mathfrak h}\ ,\qquad
V \ \stackrel{\hat S}{\mapsto} \ Ch_V^g \in {\mathfrak C}{\mathfrak h}
\lb{Shat}
\ee
is a ring homomorphism since
\be
Ch_{V_1 + V_2}^g = Ch_{V_1}^g  + \, Ch_{V_2}^g\ ,\quad Ch_{V_1\otimes V_2}^g = Ch_{V_1}^g . \, Ch_{V_2}^g
\lb{V1V2}
\ee
where the multiplication of characters is defined in (\ref{V-Ch-homo}).
The proof uses the identity (\ref{tens-ring}), the group-like property of the balancing element
$g\,$ (\ref{balance}) implying $\Delta (g^{-1} x) = (g^{-1}\otimes g^{-1} ) \Delta (x)\,$
and the equality ${\rm Tr}\, (A\otimes B) = {\rm Tr} A \,\, {\rm Tr} B\,.$

Applying further the Drinfeld map (\ref{Dr-map}) to the $q$-traces we obtain a commutative
ring homomorphism from the GR ${\mathfrak S}$ to the centre ${\cal Z}$ of $\fU\,,$
\be
{\hat D} \circ {\hat S} = D :\ {\mathfrak S}\ \rightarrow \ {\cal Z}\ ,\qquad
D (V) := {\hat D} (Ch_V^g) \ \in\ {\cal Z}\ .
\lb{DPhi}
\ee
Indeed, denoting by $V_1 . V_2$ the tensor product $V_1\otimes V_2$ in the GR sense,
Eqs. (\ref{DPhi}), (\ref{V1V2}) and (\ref{D-homom}) imply
\be
D (V_1 . V_2) = {\hat D} (Ch_{V_1\otimes V_2}^g) = {\hat D} (Ch_{V_1}^g . \, Ch_{V_2}^g ) = D (V_1) D(V_2)\ .
\lb{D-homom1}
\ee
Thus, the GR representation theory of $\fU$ is equivalent to the ring structure
of the {\em Drinfeld images} $D(V)$ of its IR in the centre ${\cal Z}\,.$

\medskip

\noindent
{\bf Proposition B.1~} (\cite{FGST1, FHT7}) {\em The Drinfeld images of the $\bU\,$ IR
\be
d^\epsilon_p := D (V^\epsilon_p) = \sum_{i} ({\rm Tr}_{\pi_{V^\epsilon_p}} ( K^{-1} m_i ))\otimes m^i \,\in {\cal Z}\ ,
\quad 1\le p\le h\ ,\quad \epsilon = \pm
\lb{Dr-VpA}
\ee
(for ${\cal M} = \sum_{i} m_i\otimes m^i\,$ (\ref{Mm}) taken from (\ref{Mmatr}))) are given by}
\ba
&&d^+_p = \sum_{s=0}^{p-1} \sum_{\mu = 0}^s \l^{2\mu} q^{(\mu +p-2s-1)(\mu+1)}\,
\left[{{\mu+p-s-1}\atop{\mu}}\right] \left[{{s}\atop{\mu}}\right] \, F^\mu E^\mu K^{\mu +p-2s-1}\,,\nn\\
&&d^-_p = -\, K^h\, d^+_p = T_h (\frac{C}{2}) \, d^+_p \ .
\lb{DrVp2}
\ea

\medskip

\noindent
{\bf Proof~} To evaluate the traces in (\ref{Dr-VpA}), one first derives the relation
\be
{\rm Tr}_{\pi_{V^\epsilon_p}} E^\mu F^\nu K^j = \d^{\mu \nu}\,\epsilon^{j+\mu} ([\mu]!)^2 \sum_{s=0}^{p-1} q^{j(2s-p+1)}
\left[{{\mu +p-s-1}\atop{\mu }}\right] \left[{{s}\atop{\mu }}\right]
\lb{TrVa}
\ee
which follows from
\ba
&&E^\mu F^\mu K^j\, | p, m{\cal i}^\epsilon = \frac{1}{\l^{2\mu}}\ q^{jH}\,\prod_{s=0}^{\mu -1}
(C - q^{-2s-1} K - q^{2s+1} K^{-1}  )\,  | p, m{\cal i}^\epsilon = \nn\\
&&= \epsilon^{j+\mu}\, q^{j(2m-p+1)}\,
\prod_{s=0}^{\mu-1} \frac{ q^p+q^{-p}-q^{2(m-s)-p}-q^{p-2(m-s)} }{\l^{2}}\,| p, m{\cal i}^\epsilon =\nn\\
&&=\epsilon^{j+\mu}\, q^{j(2m-p+1)}\,\prod_{s=0}^{\mu -1}[p-m+s] [m-s]\,| p, m{\cal i}^\epsilon
\lb{EFHa}
\ea
(one uses (\ref{ErFr}), (\ref{specK-Vp}) and (\ref{EFK-eps})).
In view of (\ref{Mmatr}) and (\ref{TrVa}), the computation of the Drinfeld images $d^\epsilon_p= D (V^\epsilon_p)\,$
(\ref{Dr-VpA}) reduces to
\ba
\lb{DrVp1}
&&d^\epsilon_p = \frac{1}{2h} \sum_{\mu=0}^{h-1} \sum_{m,n=0}^{2h-1} \frac{\l^{2\mu}\, q^{\mu }}{([\mu ]!)^2}\,
q^{mn + \mu (n-m)}\left( {\rm Tr}_{V^\epsilon_p} (E^\mu F^\mu K^{m-1})\right)\, F^\mu E^\mu K^n =\\
&&=\frac{1}{2h} \sum_{\mu=0}^{h-1} \sum_{m,n=0}^{2h-1} \epsilon^{\mu+m-1} q^{m(n-\mu)+\mu(n+1)} \l^{2\mu} \times\nn\\
&&\times\, \sum_{s=0}^{p-1} q^{(m-1)(2s-p+1)}\left[{{\mu +p-s-1}\atop{\mu}}\right] \left[{{s}\atop{\mu}}\right] \, F^\mu E^\mu K^n\ .
\nn
\ea
For $\epsilon = +1\,,$ taking the sum over $m\,$ makes the summation in $n\,$ automatic. Taking $\epsilon = -1\ ( = q^h )\,$
is equivalent to multiplying the result for $\epsilon = +1\,$ by $-\, K^h\,,$ arriving eventually at (\ref{DrVp2}). \eod

\bigskip

\noindent{\bf Remark B.1~}
Recall that the left sector monodromy matrix $M\,$ (\ref{calcM2}) is related to the universal
one ${\cal M}\,$ (\ref{Mmatr}) by (\ref{piidM}). A similar, but not identical,
relation exists for the right sector monodromy matrix $\bar M\,$ as well.

There is one more algebra of ${\mathfrak A}$-characters \cite{D3} given by the functionals
\be
\overline{{\mathfrak C}{\mathfrak h}}  := \{\, \bar\phi \in {\mathfrak A}^*\, |\ \bar\phi (x y) = \bar\phi (y S^2(x))
\ \ \forall\, x,y\in {\mathfrak A} \}\ ,
\lb{Ch-Ad*inv-bar}
\ee
the corresponding Drinfeld map being defined as
\be
{\mathfrak A}^*\ \rightarrow\ {\mathfrak A}\,,\qquad \bar\phi\ \mapsto\ (id  \otimes \bar\phi) ({\cal M})
\lb{Dr-map-bar}
\ee
(cf. (\ref{Ch-Ad*inv}) and (\ref{Dr-map}), respectively). The $q$-traces, now given by\footnote{Note that the
balancing element $g\,$ itself enters (\ref{canCh-bar}) and not its inverse as in (\ref{canCh}).}
\be
{\overline {Ch}}_V^g\, (x) := {\rm Tr}_{\pi_V} (g \, x)\qquad \forall\, x\in {\mathfrak A}\ ,
\lb{canCh-bar}
\ee
belong to ${\overline{{\mathfrak C}{\mathfrak h}}}$ (\ref{Ch-Ad*inv-bar}) since
\be
{\overline {Ch}}_V^g\, (y\,S^2(x)) = {\rm Tr}_{\pi_V} (g \, y\, S^2(x)) = {\rm Tr}_{\pi_V} (g \,y\, g\, x\, g^{-1}\!) = {\overline {Ch}}_V^g\, (xy)\ .
\lb{canch-bar}
\ee
We shall show below that the bar monodromy $\bar M\,$ is related
to the universal monodromy matrix for the right sector copy of $\bU\,$
by a map of the type (\ref{Dr-map-bar}).

As the exchange relations (\ref{Mpmq-bar}) for the Gauss components of the left and right monodromy matrices coincide, we
can parametrize them in the same way as we did for the left sector, using the FRT construction described in Section 4.3.
The expression (\ref{M+-qbar}) for right sector monodromy matrix is, however, different from (\ref{M+-q})
so it is not a surprise that $\bar M\,$ does not coincide with (\ref{calcM2}):
\ba
&&q^{\frac{3}{2}} \bar M = {\bar M}_-^{-1} {\bar M}_+  =
\begin{pmatrix}{\bar k}^{-1}&0\cr - \l\, {\bar E} {\bar k}^{-1} & {\bar k} \end{pmatrix}
\begin{pmatrix} {\bar k}^{-1}& -\l\, {\bar F} {\bar k} \cr 0&{\bar k} \end{pmatrix}
=\nn\\
&&= \begin{pmatrix}{\bar K}^{-1} & -q \l\, {\bar F} \cr - \l\,
{\bar E} {\bar K}^{-1} & q\l^2 {\bar E} {\bar F}+ {\bar K}
\end{pmatrix}\ .
\lb{calcMbar2}
\ea
By a calculation similar to (\ref{calcM}) one shows that ${\bar M}\,$ (\ref{calcMbar2}) is proportional to
\ba
&&(id \otimes \pi_f)\,{\cal M} = \lb{calcMbar}\\
&&= \frac{1}{2h}\,\sum_{m,\,n=0}^{2h-1}\,
\begin{pmatrix} q^{(m+1)n} {\bar K}^m& -\l\, q^{m(n-1)+1} {\bar F} {\bar K}^m \cr
-\l\, q^{(m+1)n} {\bar E} {\bar K}^m &(q^{(m-1)n} + \l^2 q^{m(n-1)+1} {\bar E} {\bar F})\, {\bar K}^m \end{pmatrix} =
q^{\frac{3}{2}} \bar M\nn
\ea
which implies indeed that the right sector monodromy realizes the alternative version (\ref{Dr-map-bar})
of the Drinfeld map. In accord with this, applying (\ref{canCh-bar}) for the defining representation $\pi_f\,$
reproduces the Casimir (\ref{C}),
\ba
&&{\rm Tr}\, ({\bar K}^f (id\otimes \pi_f)\,{\cal M}) =
{\rm Tr} \left\{\begin{pmatrix}q&0\cr0&q^{-1} \end{pmatrix}\,
\begin{pmatrix}{\bar K}^{-1} & -q \l\, {\bar F} \cr - \l\, {\bar E} {\bar K}^{-1} & q\l^2 {\bar E} {\bar F} + {\bar K} \end{pmatrix}  \right\} =\nn\\
&&\nn\\
&&= \l^2 {\bar E} {\bar F} + q^{-1} {\bar K} + q{\bar K}^{-1}  = \bar C \in {\bar{\cal Z}}
\lb{Tr2}
\ea
viewed now as an element of the centre ${\bar{\cal Z}}\,$ of the {\em right} copy of $\bU\,$ (cf. (\ref{Tr1}) for
the similar computation in the case of the left sector).

%%%%%%%%%%%%%

\newpage

\section*{Appendix C. The quantum determinant \boldmath$\det (M)$}
\addcontentsline{toc}{section}{Appendix C. The quantum determinant \boldmath$\det (M)$}

\setcounter{equation}{0}
\renewcommand\theequation{C.\arabic{equation}}

The exposition below follows \cite{FH2, HF2}. To understand the meaning of the second relation (\ref{detaM})
$\,\det (a) = \det (a M)\,,$ we  shall first point out that
\be
a_1 M_1\, a_2 M_2\, \dots a_n M_n = a_1 a_2 \dots a_n \, ({\hat R}_{12} {\hat R}_{23} \dots {\hat R}_{n-1\, n} M_n )^n
\lb{aMn}
\ee
(the proof of (\ref{aMn}) as well as that of (\ref{MRn}) below can be found in \cite{FH2}). Defining
\be
\det (a\, M) := \frac{1}{[n]!}\, \epsilon_{i_1 \ldots i_n}\, (a\, M)_{\b_1}^{i_1} \ldots (a\, M)_{\b_n}^{i_n}\, \e^{\b_1 \dots \b_n}\ ,
\lb{detaM1}
\ee
using (\ref{aMn}) and the first relation (\ref{det-intertw}), we obtain
\be
\det (a M) = \det (a) \det (M)
\lb{det-mult}
\ee
with the following expression for the determinant of the monodromy matrix:
\be
\det (M) := \frac{1}{[n]!}\,\e_{\a_1 \dots \a_n}\,
\left[ ({\hat R}_{12} {\hat R}_{23} \dots {\hat R}_{n-1\, n} M_n )^n\right]^{\a_1 \dots \a_n}_{~\b_1 \dots \b_n}\,\e^{\b_1\dots \b_n}\ .
\lb{detM}
\ee
One can further rearrange (\ref{detM}) in terms of the Gauss components of the monodromy matrix, using
\be
({\hat R}_{12} {\hat R}_{23} \dots {\hat R}_{n-1\, n} M_n )^n
= q^{1-n^2} ({\hat R}_{12} \dots {\hat R}_{n-1\, n})^n  M_{+ n}\dots M_{+ 1} M_{- 1}^{-1} \dots M_{- n}^{-1}\ .
\lb{MRn}
\ee
The first relation (\ref{Mpmq}) (rewritten as ${\hat R}_{12}M_{\pm 2}M_{\pm 1}=M_{\pm 2}M_{\pm 1}{\hat R}_{12}$) implies
\be
\lb{AMMA}
A_{1n}\, M_{\pm n}\dots M_{\pm 1} = M_{\pm n}\dots M_{\pm 1}\, A_{1n}
\ee
where $A_{1n}$ is the constant quantum antisymmetrizer (\ref{A1n}), and Eq.(\ref{AMMA}) leads, in turn, to
\ba
&&\e_{\a_1 \dots \a_n}\,(M_\pm )^{\a_n}_{~\b_n}\dots (M_\pm )^{\a_1}_{~\b_1}  = \det (M_\pm) \,\e_{\b_1\dots \b_n}\ ,\nn\\
&&(M_\pm )^{\a_n}_{~\b_n}\dots (M_\pm )^{\a_1}_{~\b_1}\,\e^{\b_1\dots \b_n} = \det (M_\pm) \,\e^{\a_1 \dots \a_n}
\lb{detMpmvar}
\ea
where we define originally
\be
\det (M_\pm) := \frac{1}{[n]!}\,\e_{\a_1 \dots \a_n}\,(M_\pm )^{\a_n}_{~\b_n}\dots (M_\pm )^{\a_1}_{~\b_1} \,\e^{\b_1\dots \b_n}\ .
\lb{detMpmvar1}
\ee
(The line of reasoning is similar to the one used in the proof of Proposition 4.1.)
Due to the triangularity of $M_\pm\,,$ the only nontrivial terms in the sum (\ref{detMpmvar1}) are
the $n!$ products of its (commuting) diagonal elements, hence
\be
\det (M_\pm) = \prod_{\a =1}^n (M_\pm )^{\a}_{~\a} = 1
\lb{detMpmvar2}
\ee
(cf. (\ref{MpmD1})).
Since
\be
\det (M_\pm^{-1}) = \det (S(M_\pm)) = \det (M_\pm)^{-1}\, = 1
\lb{detM-1}
\ee
(where $S$ is the antipode (\ref{Hopf-FRT})) and, due to (\ref{eqs-eps}),
\be
\varepsilon_{\alpha_1\dots \s_i \s_{i+1}\dots \alpha_n} {\hat R}_{~\alpha_i \alpha_{i+1}}^{\s_i \s_{i+1}}=
- q^{1+\frac{1}{n}}\,\e_{\a_1 \dots \a_n}\ ,\quad i=1,\dots, n-1
\ee
so that the $q^{1-n^2}$ prefactor in (\ref{MRn}) is exactly compensated by
\be
\e_{\a_1 \dots \a_n}\!\left[({\hat R}_{12} {\hat R}_{23} \dots {\hat R}_{n-1\, n} )^n\right]^{\a_1 \dots \a_n}_{~\b_1 \dots \b_n}
= (- q^{1+\frac{1}{n}})^{(n-1)n}\,\e_{\b_1 \dots \b_n}
= q^{n^2-1}\,\e_{\b_1 \dots \b_n}\ ,
\lb{epsRij}
\ee
we obtain from (\ref{detM}), (\ref{MRn}) and (\ref{detMpmvar}), (\ref{detM-1}) that
\be
\det (M) = \det (M_+) \,\det (M_-)^{-1}\, = 1\ .
\lb{MMMpm}
\ee
Eqs. (\ref{det-mult}) and (\ref{MMMpm}) ensure the validity of the second relation (\ref{detaM}).

\smallskip

Here we shall content with an illustration, calculating $\det (M)$ for $n=2$ by using (\ref{detM}).
From (\ref{detc-n2-1}), (\ref{R2}), (\ref{calcM}) and (\ref{piidM}) we obtain
\be
{\det} (M) = \frac{1}{[2]}\,\e_{\a\b} \left( {\hat R}_{12} M_2 {\hat R}_{12} M_2 \right)^{\a\b}_{~\rho\s} \e^{\rho\s} =
\frac{1}{[2]}\,(2\, q^{-1}  - \l^2\, [E,F] K + \l\, K^2 ) = 1\ ,
\lb{detqMn=2}
\ee
as prescribed by (\ref{MMMpm}) \cite{HF2}.

\newpage

%%%%%%%%%%%%%%% BIBLIOGRAPHY - (~CMP style) %%%%%%%%%%%%%%%%%%%

\end{document}